\documentclass{svmult}
\usepackage{graphicx}
\usepackage[square,numbers]{natbib}
\usepackage{revsymb}

\def\lco{La$_2$CuO$_4$}
\def\lbcoate{La$_{1.875}$Ba$_{0.125}$CuO$_4$}
\def\lcod{La$_2$CuO$_{4+\delta}$}

\def\lno{La$_2$NiO$_4$}

\def\lsnod{La$_{2-x}$Sr$_x$NiO$_{4+\delta}$}
\def\lsco{La$_{2-x}$Sr$_x$CuO$_4$}
\def\lbco{La$_{2-x}$Ba$_x$CuO$_4$}
\def\lnsco{La$_{1.6-x}$Nd$_{0.4}$Sr$_x$CuO$_4$}
\def\ybco{YBa$_2$Cu$_3$O$_{6+x}$}
\newcommand{\YBCO}[1]{YBa$_2$Cu$_3$O$_{#1}$}
\def\bscco{Bi$_2$Sr$_2$CaCu$_2$O$_{8+\delta}$}
\def\tbco{Tl$_2$Ba$_2$CuO$_{6+\delta}$}
\def\scoc{Sr$_2$CuO$_2$Cl$_2$}
\def\ncco{Nd$_{2-x}$Ce$_x$CuO$_4$}
\def\pcco{Pr$_{2-x}$Ce$_x$CuO$_4$}

\def\qaf{${\mathbf Q}_{\rm AF}$}
\def\tco{$T_{\rm co}$}
\def\tc{$T_{\rm c}$}
\def\tn{$T_{\rm N}$}

\begin{document}
\title{Neutron Scattering Studies of Antiferromagnetic Correlations in
Cuprates} 
\titlerunning{Neutron Scattering}
\author{John M.~Tranquada}
\institute{Physics Department, Brookhaven National Laboratory, Upton,
NY 11973, USA}
\maketitle
\begin{abstract}
Neutron scattering studies have provided important information about the
momentum and energy dependence of magnetic excitations in cuprate
superconductors.  Of particular interest are the recent indications of
a universal magnetic excitation spectrum in hole-doped cuprates.  That
starting point provides motivation for reviewing the antiferromagnetic
state of the parent insulators, and the destruction of the ordered state
by hole doping.  The nature of spin correlations in stripe-ordered phases
is discussed, followed by a description of the doping and temperature
dependence of magnetic correlations in superconducting cuprates.  After
describing the impact on the magnetic correlations of perturbations such
as an applied magnetic field or impurity substitution, a brief summary of
work on electron-doped cuprates is given.  The chapter concludes with a
summary of experimental trends and a discussion of theoretical
perspectives.
\end{abstract}
%


\section{Introduction}
\label{sc:intro}

Neutron scattering has played a major role in characterizing the nature
and strength of antiferromagnetic interactions and correlations in the
cuprates.  Following Anderson's observation \cite{ande87} that \lco, the
parent compound of the first high-temperature superconductor, should be a
correlated insulator, with moments of neighboring Cu$^{2+}$ ions
anti-aligned due to the superexchange interaction, antiferromagnetic
order was discovered in a neutron diffraction study of a polycrystalline
sample \cite{vakn87}.  When single-crystal samples became available,
inelastic studies of the spin waves determined the strength of the
superexchange, $J$, as well as weaker interactions, such as the coupling
between CuO$_2$ layers.  The existence of strong antiferromagnetic spin
correlations above the N\'eel temperature, \tn, has been
demonstrated and explained.  Over time, the quality of such
characterizations has improved considerably with gradual evolution in the
size and quality of samples and in experimental techniques.

Of course, what we are really interested in understanding are the
optimally-doped cuprate superconductors.  It took much longer to get a
clear picture of the magnetic excitations in these compounds, which
should not be surprising given that there is no static magnetic order,
the magnetic moments are small, and the bandwidth characterizing the
magnetic excitations is quite large.  Nevertheless, we are finally at a
point where a picture of universal behavior, for at least two families of
cuprates, is beginning to emerge.  Thus, it seems reasonable to start our
story with recent results on the excitation spectrum in superconducting
\ybco\ and \lsco, and the nature of the spin gap that appears below the
superconducting transition temperature, \tc.  (Note that these are
hole-doped superconductors, which is where most of the emphasis will be
placed in this chapter.)  An important result is that this spectrum looks
quite similar to that measured for \lbcoate, a compound in which \tc\ is
depressed towards zero and ordered charge and spin stripes are observed. 
The nature of stripe order and its relevance will be discussed later.

Following the initial discussion of results for the superconductors in
\S\ref{sc:super}, one can have a better appreciation for the
antiferromagnetism of the parent insulators, presented in
\S\ref{sc:AF}.  The destruction of antiferromagnetic order by hole
doping is discussed in \S\ref{sc:dest}. In \S\ref{sc:stripe}, evidence
for stripe order, and for other possible ordered states competing with
superconductivity, is considered.  \S\ref{sc:doping} discusses how
the magnetic correlations in superconducting cuprates evolve with
hole-doping and with temperature.  While doping tends to destroy
antiferromagnetic order, perturbations of the doped state can induce
static order, or modify the dynamics, and these effects are discussed in
\S\ref{sc:perturb}.  A brief description of work on electron-doped
cuprates is given in \S\ref{sc:e-doped}.  The chapter concludes with a
discussion, in \S\ref{sc:disc}, of experimental trends and theoretical
perspectives on the magnetic correlations in the cuprates. It should be
noted that there is not space here for a complete review of neutron
studies of cuprates; some earlier reviews and different perspectives
appear in Refs.~\cite{kast98,bour98,regn98,hayd98,maso01,brom03}.

\begin{figure}[t]
\centerline{\includegraphics[width=4in]{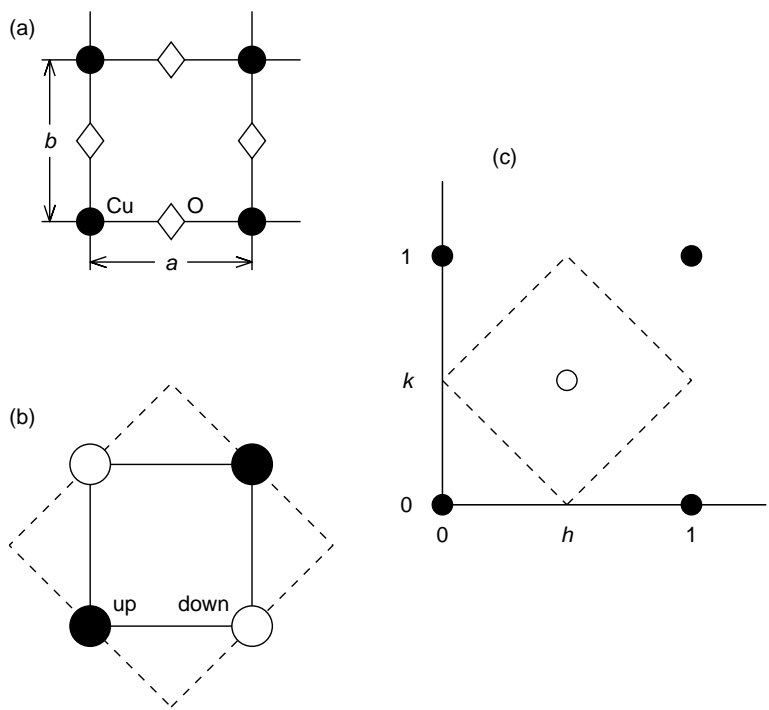}}
\medskip
\caption{(a) CuO$_2$ plane, indicating positions of the Cu and O atoms
and identifying the lattice parameters, $a$ and $b$.  (b) Sketch of
antiferromagnetic order of Cu moments, with filled (empty) circles
representing up (down) spins.  Solid line indicates the chemical unit
cell; dashed line denotes the magnetic unit cell.  (c)  Filled circles:
Bragg peak positions in reciprocal space corresponding to the chemical
lattice.  Empty circle: magnetic Bragg peak due to antiferromagnetic
order.  Dashed line indicates the antiferromagnetic Brillouin zone.}
\label{fg:AF_cell}
\end{figure}

Before getting started, it is useful to first establish some notation.  A
general wave vector ${\bf Q}=(h,k,l)$ is specified in units of the
reciprocal lattice, $(a^*,b^*,c^*) = (2\pi/a,2\pi/b,2\pi/c)$.  The CuO$_2$
planes are approximately square, with a Cu-Cu distance of
$a\approx b\sim3.8$~\AA.  Antiferromagnetic order of Cu moments
($S=\frac12$) in a single plane causes a doubling of the unit cell and
is characterized by the wave vector ${\bf Q}_{\rm AF} =
(\frac12,\frac12,0)$, as indicate in Fig.~\ref{fg:AF_cell}; however, the
relative ordering of the spins along the $c$ axis can cause the
intensities of
$(\frac12,\frac12,L)$ superlattice peaks to have a strongly modulated
structure factor as a function of $L$.  For the magnetic excitations, we
will generally be interested in their dependence on ${\bf Q}_{\rm 2D} =
{\bf Q} - (\frac12,\frac12,L)$ associated with an individual CuO$_2$
plane and ignoring the $L$ dependence.

The magnetic scattering function measured with neutrons can be written as
\cite{shir02,squi96}
\begin{equation}
  {\cal S}({\mathbf Q},\omega) = \sum_{\alpha,\beta}
   \left(\delta_{\alpha,\beta} - Q_\alpha Q_\beta/Q^2\right)
   {\cal S}^{\alpha\beta}({\mathbf Q},\omega),
  \label{eq:sdef}
\end{equation}
where
\begin{equation}
  {\cal S}^{\alpha\beta}({\mathbf Q},\omega) = 
    {1\over2\pi}\int_{-\infty}^{\infty} dt\, 
    e^{-i\omega t} \sum_{\mathbf r} e^{i{\mathbf Q}\cdot{\mathbf r}}
    \langle S_{\mathbf 0}^\alpha(0) S_{\mathbf r}^\beta(t)\rangle.
\end{equation}
Here $S_{\mathbf r}^\beta(t)$ is the $\beta$ ($= x$, $y$, $z$) component
of the atomic spin at lattice site $\mathbf r$ and time $t$, and the
angle brackets, $\langle\ldots\rangle$ denote an average over
configurations.   For inelastic scattering, it is possible to relate
${\cal S}({\mathbf Q},\omega)$ to the imaginary part of the dynamical
spin susceptibility, $\chi''({\mathbf Q},\omega)$ via the
fluctuation-dissipation theorem,
\begin{equation}
  {\cal S}({\mathbf Q},\omega) = {\chi''({\mathbf Q},\omega)\over
    1-e^{-\hbar\omega/k_{\mathrm B}T}}.
\end{equation}
Another useful quantity is the ``local'' susceptibility
$\tilde{\chi}''(\omega)$, defined as
\begin{equation}
  \tilde{\chi}''(\omega) = \int d{\bf Q}_{\rm 2D} \,\,
    \chi''({\bf Q},\omega).
\end{equation}
Experimentally, the integral is generally not performed over the entire
first Brillouin zone, but rather over the measured region about \qaf.

\section{Magnetic excitations in hole-doped superconductors}
\label{sc:super}

\subsection{Dispersion}

Most of the neutron scattering studies of cuprate superconductors have
focused on two families: \lsco\ and \ybco.  The simple reason for this is
that these are the only compounds for which large crystals have been
available.  For quite some time it appeared that the magnetic spectra of
these two families were distinct.  In \lsco, the distinctive feature was
incommensurate scattering, studied at low energies
($<20$~meV) \cite{cheo91,aepp97,yama98a}, whereas for \ybco\ the
attention was focused on the commensurate scattering (``41-meV'' or
``resonance'' peak \cite{ross92,mook93,fong96,dai01,sidi04}) that grows in
intensity (and shifts in energy
\cite{bour99}) as the temperature is cooled below \tc.  A resonance peak
was also detected in \bscco\ \cite{fong99,meso00,he01} and in \tbco
\cite{he02}.  Considerable theoretical attention has been directed
towards the resonance peak and its significance ({\it e.g.}, see
\cite{esch00,aban01,kee02}).  The fact that no strongly
temperature-dependent excitation at \qaf\ was ever observed in \lsco\
raised questions about the role of magnetic excitations the cuprate
superconductivity.

\begin{figure}[t]
\centerline{\includegraphics[width=4in]{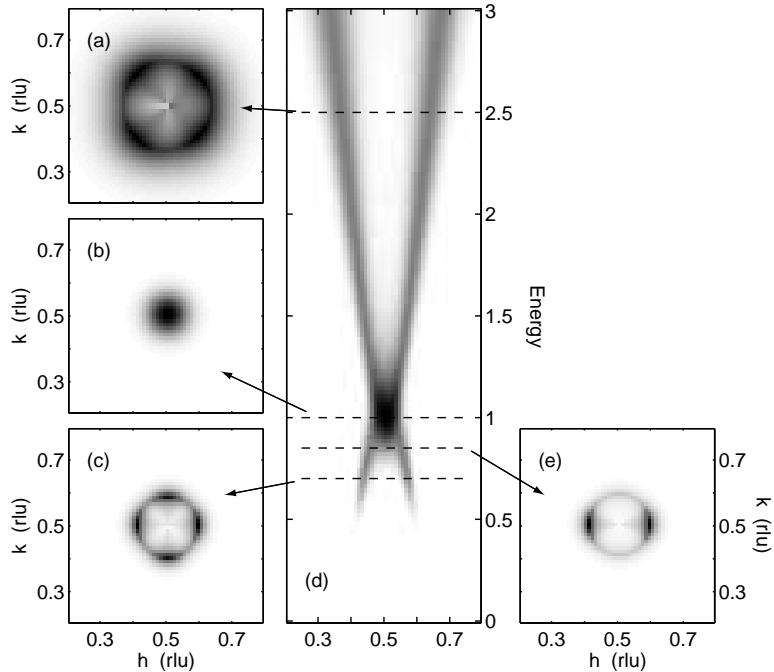}}
\medskip
\caption{Schematic plots intended to represent neutron scattering
measurements of $\chi''({\bf Q},\omega)$ in superconducting
YBa$_2$Cu$_3$O$_{6+x}$ at $T\ll T_{\rm c}$.  Panels (a), (b), and (c)
represent the distribution of scattering in reciprocal space about \qaf\
at relative energies indicated by the dashed lines in (d), for
a twinned sample.  (d) $\chi''$ along ${\bf Q}=(h,\frac12,L)$ as a
function of energy (normalized to the saddle-point energy, which is
doping dependent).  (a-d) modeled after \protect\cite{hayd04,stoc05}.  (e)
Anisotropic distribution of scattering inferred for a detwinned,
single-domain sample of \YBCO{6.85}, after \protect\cite{hink04}.}
\label{fg:sc_disp}
\end{figure}

While considerable emphasis has been placed on the resonance peak, it has
been clear for quite some time that normal-state magnetic excitations in
under-doped \ybco\ extend over a large energy range
\cite{bour97,dai99}, comparable to that in the
antiferromagnetic parent compound
\cite{hayd96b,rezn96}.  The first clear signature that the excitations
below the resonance are incommensurate, similar to the low-energy
excitations in \lsco\ \cite{cheo91,yama98a}, was obtained by Mook~{\textit
et al.} \cite{mook98} for \YBCO{6.6}.  That these incommensurate
excitations disperse inwards towards the resonance energy was
demonstrated in \YBCO{6.7} by Arai {\it et al.} \cite{arai99} and in
\YBCO{6.85} by Bourges {\it et al.} \cite{bour00}.  More recent
measurements have established a common picture of the dispersion
\cite{hayd04,pail04,rezn04,stoc05,ito02}.

A schematic of the measured dispersion is shown in Fig.~\ref{fg:sc_disp},
with the energy normalized to that of the commensurate excitations,
$E_r$.  (Note that the distribution of intensity is not intended to
accurately reflect experiment, especially in the superconducting state.) 
The figure also indicates the {\bf Q} dependence of magnetic scattering
at fixed excitation energies.  For $E<E_r$, measurements on
crystallographically-twinned crystals indicate a four-fold intensity
pattern, with maxima at incommensurate wave vectors displaced from \qaf\
along [100] and [010] directions.  For $E>E_r$, Hayden {\it et al.}
\cite{hayd04} infer for their \YBCO{6.6}\ sample a four-fold structure
that is rotated by 45$^\circ$ compared to low energies, whereas Stock
{\it et al.} \cite{stoc05} find an isotropic ring of scattering for
\YBCO{6.5}.  (These differences are minor compared to the overall level
of agreement.)  The spectrum with a finite spin gap is applicable to
measurements below \tc; the gap fills in above \tc, where it also becomes
difficult to resolve any incommensurate features.

\begin{figure}[t]
\centerline{\includegraphics[width=3.5in]{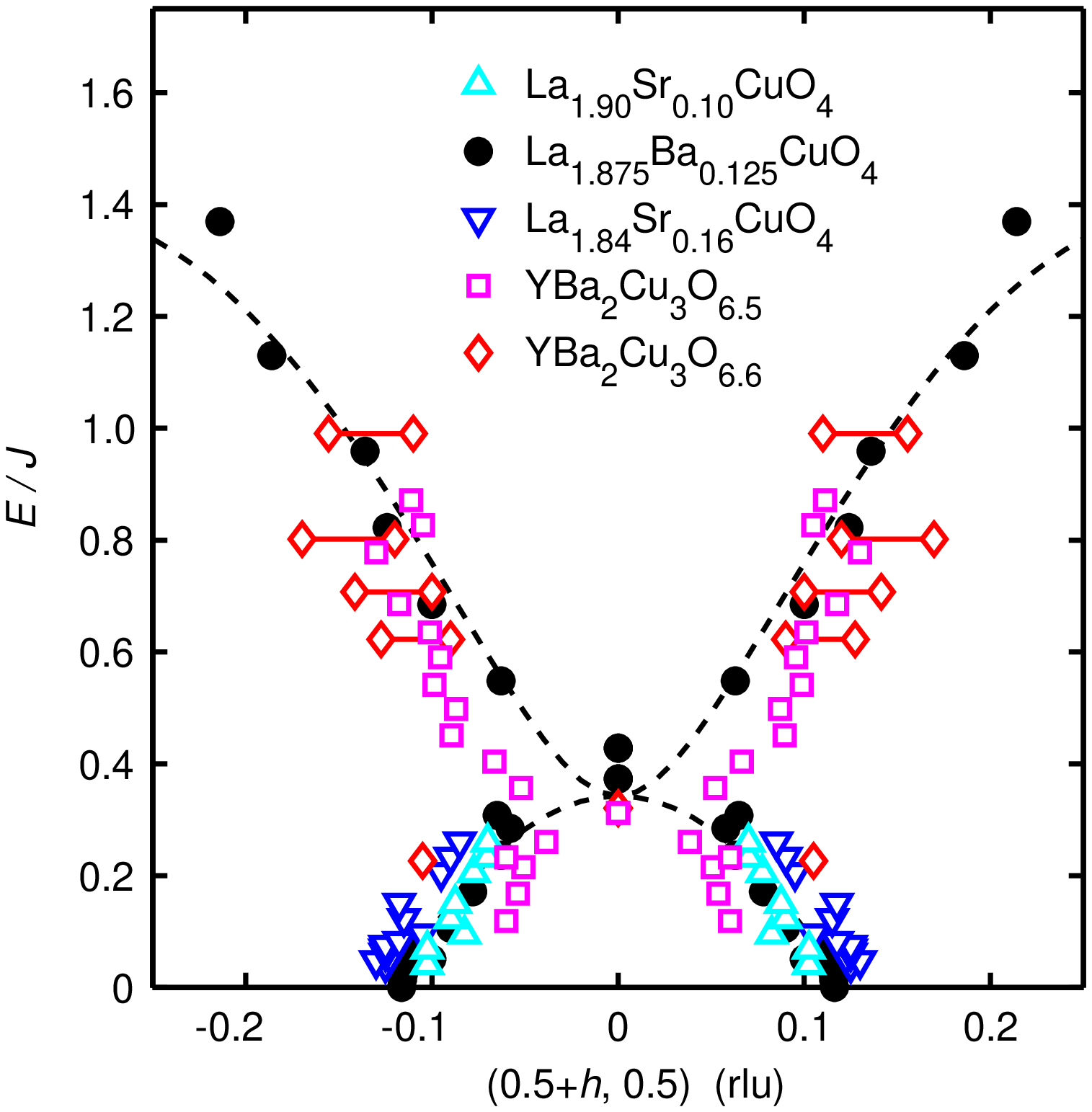}}
\medskip
\caption{Comparison of measured dispersions along ${\bf Q}_{\rm 2D} =
(0.5+h,0.5)$ in \lsco\ with $x=0.10$ (up triangles) and 0.16 (down
triangles) from Christensen {\it et al.} \cite{chri04}, in \lbcoate\
(filled circles) from
\cite{tran04}, and in
\ybco\ with $x=0.5$ (squares) from Stock {\it et al.} \cite{stoc05} and
0.6 (diamonds) from Hayden {\it et al.} \cite{hayd04}.  The energy 
has been scaled by the superexchange energy $J$ for the appropriate
parent insulator as given in Table~\ref{tab:AF}.  For \YBCO{6.6}, the
data at higher energies were fit along the [1,1] direction; the doubled
symbols with bars indicate two different ways of interpolating the
results for the [1,0] direction.  The upwardly-dispersing dashed curve
corresponds to the result of Barnes and Riera \cite{barn94} for a 2-leg
spin ladder, with an effective superexchange of $\sim\frac23 J$; the
downward curve is a guide to the eye. }
\label{fg:lbco_disp}
\end{figure}

Fig.~\ref{fg:lbco_disp} shows a direct comparison of measurements on
\lsco\ \cite{chri04} and under-doped \ybco\ \cite{hayd04,stoc05}, with
energy scaled by the superexchange energy, $J$ (see Table~\ref{tab:AF} in
\S\ref{sc:AF}).  Also included in the figure are results for \lbco\ with
$x=\frac18$ \cite{fuji04,tran04}, a compound of interest because of the
occurrence of charge and spin stripe order\cite{tran95a} (to be
discussed later) and a strongly suppressed \tc.  At the lowest energies,
the spin excitations rise out of incommensurate magnetic
(two-dimensional) Bragg peaks.  Besides the presence of Bragg peaks, the
magnetic scattering differs from that of
\ybco\ in the absence of a spin gap.  The results for optimally-doped
\lsco, with $x=0.16$, interpolate between these cases, exhibiting the
same inward dispersion of the excitations towards \qaf\ (measured up to 40
meV) and a spin gap of intermediate magnitude in the superconducting
state \cite{chri04}.  The degree of similarity among the results shown in
Fig.~\ref{fg:lbco_disp} is striking, and suggests that the magnetic
dispersion spectrum may be universal in the cuprates
\cite{chri04,tran05a}.

For optimally doped \ybco, the measured dispersive excitations are
restricted to a narrower energy window, as shown in
Fig.~\ref{fg:ybco_disp}.  Nevertheless, excitations are observed to
disperse both downward and upward from $E_r$, and the qualitative
similarity with dispersions at lower doping is obvious.

\begin{figure}[t]
\centerline{\includegraphics[width=3.5in]{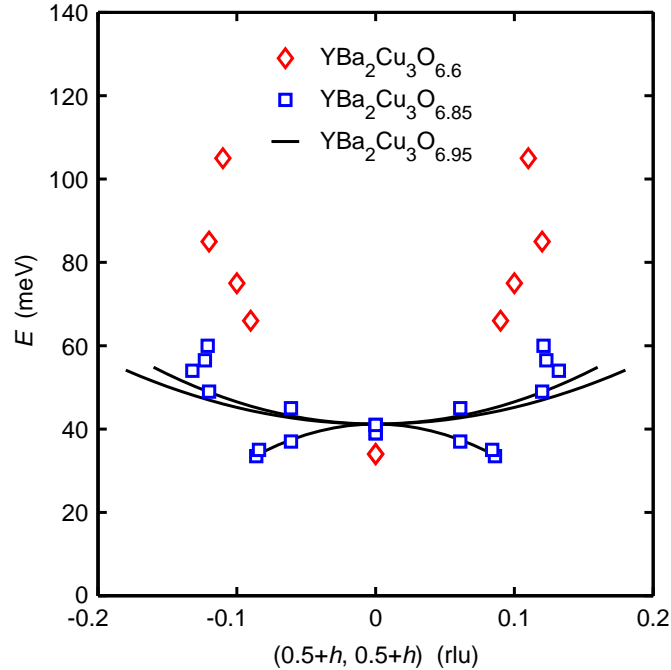}}
\medskip
\caption{Comparison of measured dispersions along ${\bf Q}_{\rm 2D} =
(0.5+h,0.5+h)$ in the superconducting state of \ybco\ with $x=0.6$
(diamonds) from Hayden {\it et al.}
\cite{hayd04} and 0.85 (squares) from Pailh\`es {\it et al.}
\cite{pail04}.  The solid lines represent the model dispersion (and
variation in dispersion) compatible with measurements on \YBCO{6.95}\
from Reznik {\it et al.} \cite{rezn04}. }
\label{fg:ybco_disp}
\end{figure}

Anisotropy of the magnetic scattering as a function of ${\bf Q}_{\rm 2D}$
can be measured in specially detwinned samples of \ybco, as the crystal
structure has an anisotropy associated with the orientation of the CuO
chains.  (Note that it is a major experimental challenge to detwin samples
of sufficient volume to allow a successful inelastic neutron scattering 
study.)  An initial study of a partially detwinned sample of \YBCO{6.6}
by Mook {\it et al.} \cite{mook00} demonstrated that, for the
incommensurate  scattering at an energy corresponding to 70\%\ of the
saddle point, the intensity is quite anistropic, with maxima along the
${\bf a}^\ast$ direction (perpendicular to the orientation of the CuO
chains).  A recent study of an array of highly detwinned crystals of
\YBCO{6.85} by Hinkov {\it et al.} \cite{hink04} found substantial
anisotropy in the peak scattered intensity for an energy of 85\%\ of the
saddle point, but also demonstrated that scattered intensity at that
energy forms a circle about \qaf\ [see Fig.~\ref{fg:sc_disp}(e)]. 
Measurements on a partially-detwinned sample of \YBCO{6.5} by Stock {\it
et al.} \cite{stoc04,stoc05} suggest a strong anisotropy in the
scattered intensity at $0.36E_r$, but essentially perfect isotropy for
$E>E_r$.

\subsection{Spin gap and ``resonance'' peak}
\label{sec:spin_gap}

\begin{figure}[t]
\centerline{\includegraphics[width=3in]{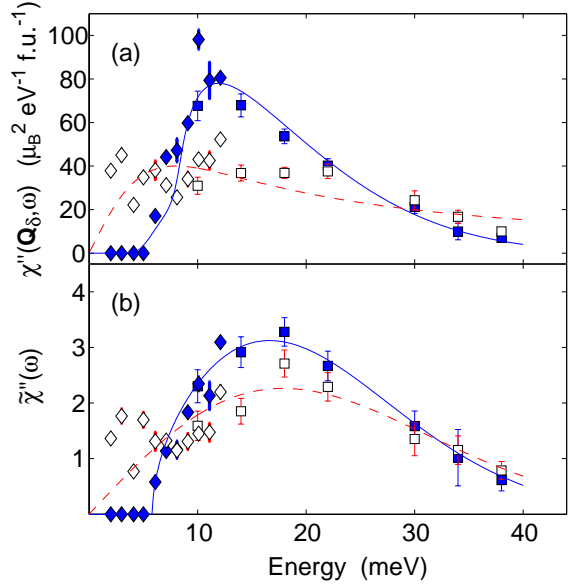}}
\medskip
\caption{(a) The fitted peak intensity of $\chi''({\bf Q}_\delta,\omega)$
(where ${\bf Q}_\delta$ is the peak position) for \lsco\ with $x=0.16$. 
(b) The local susceptibility, $\tilde{\chi}''(\omega)$.  Filled symbols:
$T<T_{\rm c}$; open symbols: $T>T_{\rm c}$.  Results are a combination of
data from time-of-flight measurements (squares) and triple-axis
measurements (diamonds).  Lines are guides to the eye. From Christensen
{\it et al.}
\cite{chri04}.}
\label{fg:lsco_gap}
\end{figure}

For optimally-doped cuprates, the most dramatic change in the magnetic
scattering with temperature is the opening of a spin gap, with
redistribution of spectral weight from below to above the gap.  A clear
example of this has been presented recently by Christensen {\it et
al.} \cite{chri04} for \lsco\ with $x=0.16$; their results are shown in
Fig.~\ref{fg:lsco_gap}.  For the energy range shown, the scattering is
incommensurate in {\bf Q}, with the dispersion indicated in
Fig.~\ref{fg:lbco_disp}.  In the normal state, the amplitude of $\chi''$
heads to zero only at $\omega = 0$; in the superconducting state,
weight is removed from below a spin gap energy of $\Delta_s\approx8$~meV,
and shifted to energies just above $\Delta_s$.  This is apparent both for
the plot of the peak amplitude of $\chi''$ in Fig.~\ref{fg:lsco_gap}(a),
and for the {\bf Q}-integrated $\chi''$ in (b); within the experimental
uncertainty, the spectral weight below 40 meV is conserved on cooling
through \tc\ \cite{chri04}.  Another important feature of the spin gap is
that its magnitude is independent of {\bf Q} \cite{lake99}.  This is of
particular interest because it is inconsistent with a weak-coupling
prediction of $\chi''$ for a $d$-wave superconductor, assuming that the
spin response comes from quasiparticles \cite{lu92}.

The behavior is similar near optimal doping in \ybco\
\cite{bour00,pail04,rezn04,dai01}, with the difference being that the
spin gap energy of $\sim33$~meV is much closer to $E_r=41$~meV.  The
strongest intensity enhancement below \tc\ occurs at $E_r$, where 
$\chi''$ is peaked at \qaf; however, there is also enhanced intensity
at energies a bit below and above $E_r$, where $\chi''$ is incommensurate
\cite{bour00,pail04}.  The spin gap $\Delta_s$ decreases and broadens with
underdoping, so that the region over which $\chi''\approx 0$ is no more
than a few meV for \YBCO{6.5}\ \cite{stoc04,bour97,ito02}.

Besides the temperature dependence, there is also a similar behavior of
the enhanced intensity for these two cuprate families in response to an
applied magnetic field.  As the cuprates are type-II superconductors
with a very small lower critical field, an applied magnetic field can
enter a sample as an array of vortices.  Dai {\it et al.} \cite{dai00}
showed that application of a magnetic field of 6.8 T along the $c$
axis of \YBCO{6.6} at $T\ll T_{\rm c}$ caused a reduction of the intensity
at $E_r$ by $\sim30$\%.  A study of \lsco\ with $x=0.18$ found that
application of a 10-T field along the $c$-axis caused a reduction of the
intensity maximum at 9 meV of about 25\%\ (with an increase in $Q$ width)
and a shift of some weight into the spin gap \cite{tran04b}.  The
field-induced increase of weight within the spin gap of \lsco\
($x=0.163$) was first observed by Lake {\it et al.} \cite{lake01}.

\begin{figure}[t]
\centerline{\includegraphics[width=3in]{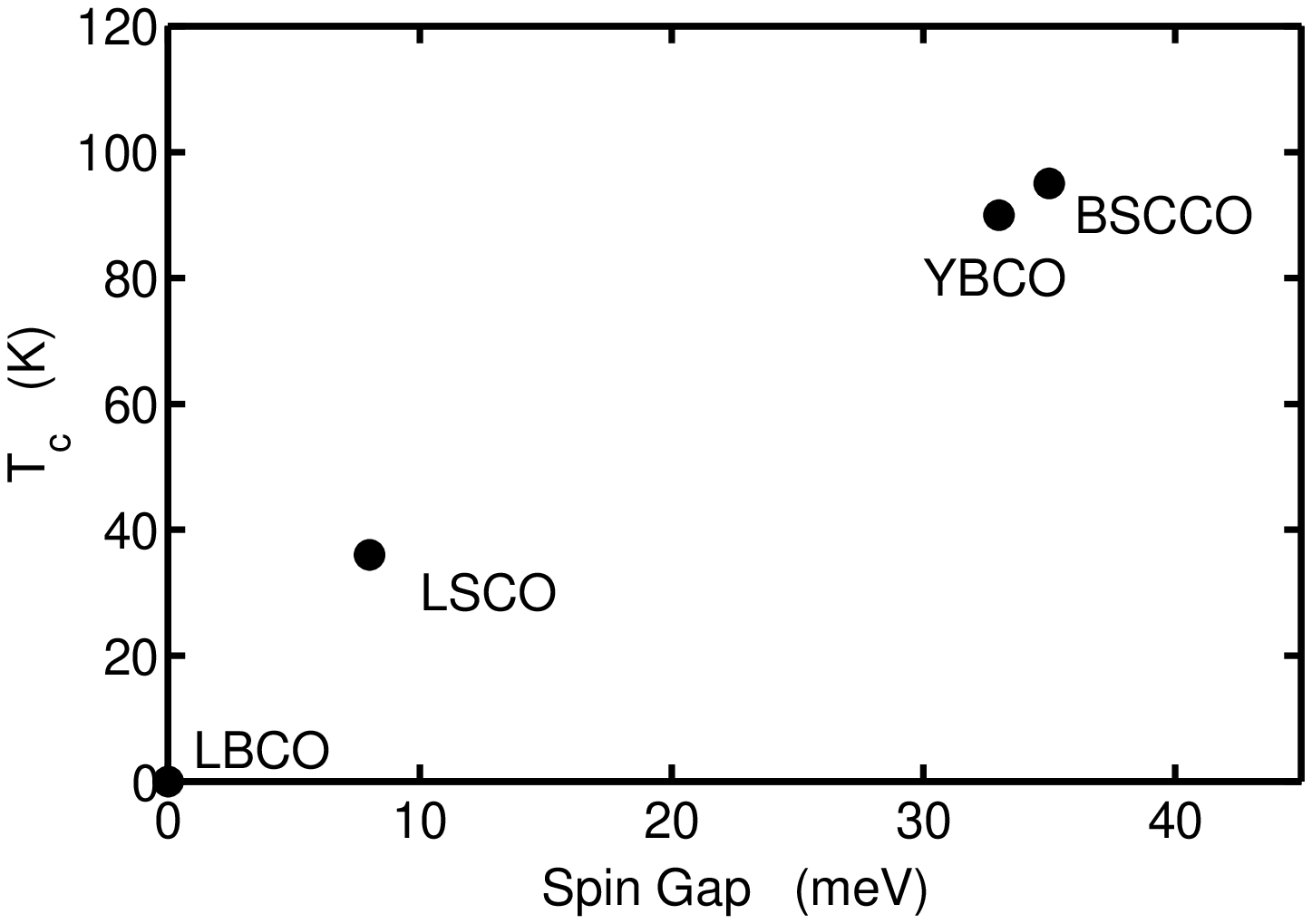}}
\medskip
\caption{Plot of \tc\ vs.\ spin-gap energy, $\Delta_s$, in cuprates near
optimal doping. LSCO: \lsco\ ($x=0.16$) from \cite{chri04}; YBCO:
\YBCO{6.85} from \cite{bour00}; BSCCO: \bscco\ with $\Delta_s$ estimated
by scaling $E_r$ with respect to \ybco, from \cite{fong99}; LBCO: \lbco\
($x=\frac18$) from \cite{fuji04}. }
\label{fg:tc_gap}
\end{figure}

By focusing on $\Delta_s$ rather than $E_r$, it is possible to identify a
correlation between magnetic excitations and \tc\ that applies to a
variety of cuprates.  Figure~\ref{fg:tc_gap} shows a plot of \tc\ as a
function of the spin-gap energy for several different cuprates near
optimal doping.  This trend makes clear that the magnetic excitations are
quite sensitive to the superconductivity, but, by itself, it does not
resolve the issues of whether or how magnetic correlations may be involved
in the pairing mechanism.

\subsection{Discussion}

From the results presented above, it now appears that there may be a
universal magnetic excitation spectrum for the cuprates.  On entering the
superconducting state, a gap in the magnetic spectrum develops, with
spectral weight redistributed from below to above the gap.  The magnitude
of the spin gap is correlated with \tc.

A long-standing question concerns the role of magnetic excitations in the
mechanism of high-temperature superconductivity, and some varying
perspectives are presented in later chapters of this book.  An underlying
issue concerns the nature of the magnetic excitations themselves.  Given
that \lsco\ and \ybco\ exhibit antiferromagnetically ordered phases when
the hole-doping of the CuO$_2$ planes goes to zero, one approach is to 
look for a connection between the magnetic correlations in
the superconducting and in the correlated-insulator phases.  On the other
hand, the magnetic response of common metallic systems (such as chromium)
is tied to the low-energy excitations of electrons from filled to empty
states, across the Fermi surface.  This motivates attempts to interpret
the magnetic excitations in terms of electron-hole excitations.  It is not
clear that these contrasting approaches can be reconciled with one another
\cite{ande97}, but, in any case, there are presently no consensual
criteria for selecting one approach over another.

An experimentalist's approach is to consider the correlations in the
superconducting cuprates in the context of related systems.  Thus, in the
following sections we consider experimental results for antiferromagnetic
cuprates, other doped transition-metal-oxide systems, perturbations to
the superconducting phase, and the doping dependence of the magnetic
correlations in the superconductors.  A comparison of theoretical
approaches is better discussed within the full context of experimental
results.

\section{Antiferromagnetism in the parent insulators}
\label{sc:AF}

\subsection{Antiferromagnetic order}

In the parent insulators, the magnetic moments of the copper atoms order
in a 3D N\'eel structure.   Powder neutron diffraction studies first
demonstrated this for \lco\ \cite{vakn87}, and later for \ybco\
\cite{tran88}.  The magnetic moments tend to lie nearly parallel to the
CuO$_2$ planes.  The details of the magnetic structures are tied to the
crystal structures, so we will have to consider these briefly.

\begin{figure}[t]
\centerline{\includegraphics[width=2in]{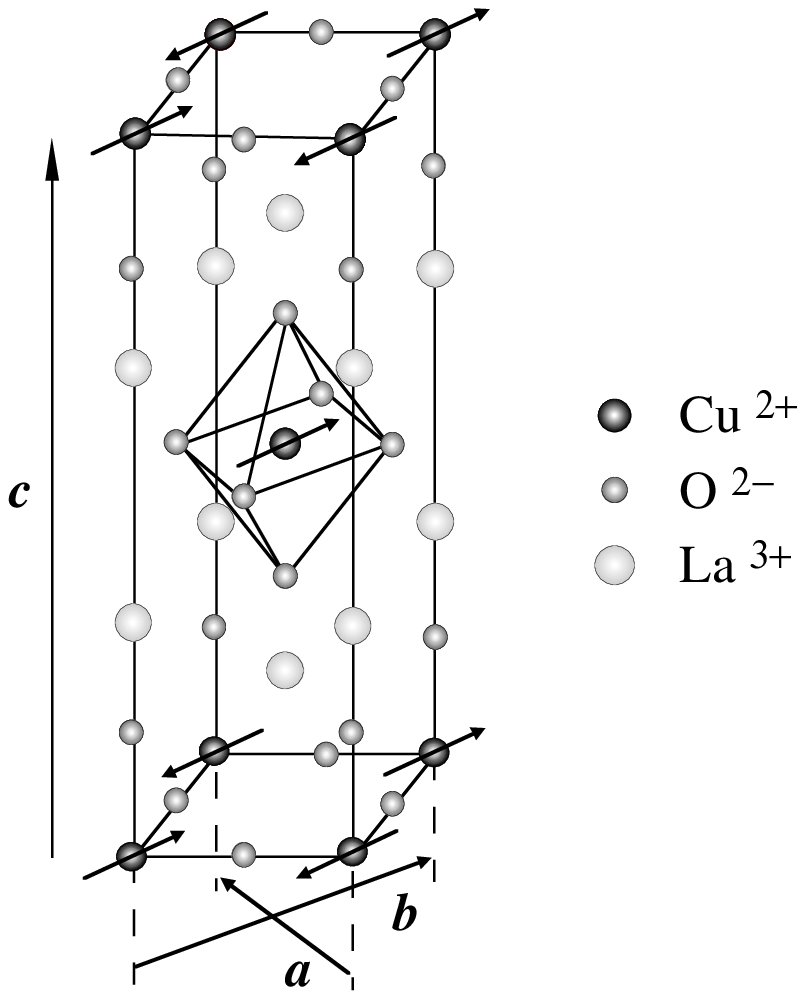}
\hspace{2cm}\includegraphics[width=1in]{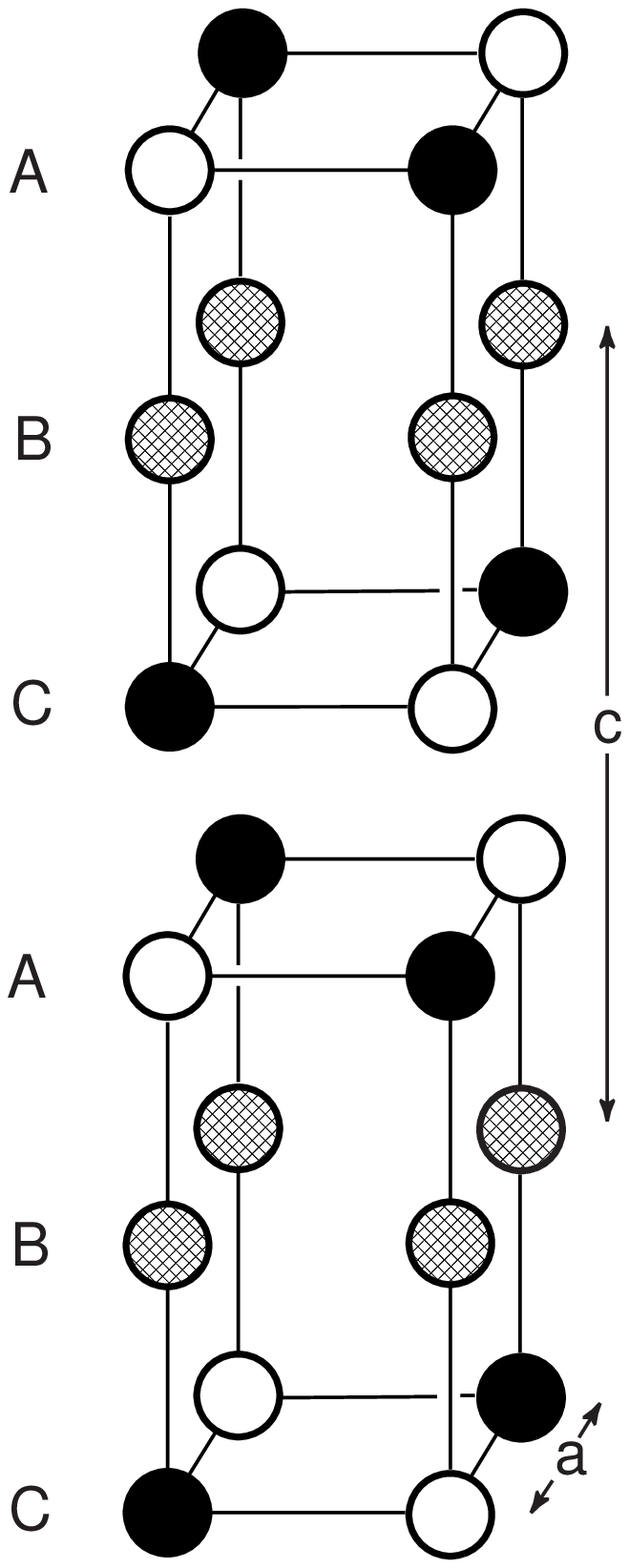}}
\medskip
\caption{Left: crystal structure of \lco.  Arrows indicate orientation of
magnetic moments on Cu sites in the antiferromagnetic state. After
Lee {\it et al.} \protect\cite{lee99}.  Right: Magnetic structure of
\YBCO{6}.  Circles: Cu atoms; lines: paths bridged by oxygen. Filled and
empty circles represent Cu$^{2+}$ sites with opposite spin orientations;
hatched circles denote non-magnetic Cu$^{1+}$ sites.  After Tranquada
{\it et al.} \protect\cite{tran88a}.}
\label{fg:LCO_struc}
\end{figure}

The crystal structure of \lco\ is presented in Fig.~\ref{fg:LCO_struc}. 
The CuO$_2$ planes are stacked in a body-centered fashion, so that the
unit cell contains two layers.  Below 550~K each CuO$_6$ octahedron
rotates about a [110] axis of the high-temperature tetragonal cell. 
Neighboring octahedra within a plane rotate in opposite directions,
causing a doubling of the unit cell volume and a change to orthorhombic
symmetry, with the ${\mathbf a}_{\mathrm O}$ and ${\mathbf b}_{\mathrm O}$
axes rotated by $45^\circ$ with respect to the Cu-O bond directions.  In
the orthorhombic coordinates, the octahedral tilts are along the
${\mathbf b}_{\mathrm O}$ direction (but $b_{\mathrm O}>a_{\mathrm O}$,
contrary to naive expectations).

In the antiferromagnetic phase of \lco, the spins point along the $b_{\rm
O}$ axis, and they have the stacking
sequence shown in Fig.~\ref{fg:LCO_struc}
\cite{vakn87,frel87}.  As the octahedral tilts break the tetragonal
symmetry of the planes, they allow spin-orbit effects, in the form of
Dzyalozhinsky-Moriya (DM) antisymmetric exchange, to cause a slight
canting of the spins along the
$c$ axis.  This canting is in opposite directions for neighboring planes,
resulting in no bulk moment, but a modest magnetic field can flip the
spins in half of the planes, yielding a weakly ferromagnetic state
\cite{kast88}.  The tendency to cant in the paramagnetic state above
\tn\ leads to a ferromagnetic-like susceptibility at high temperatures
and a cusp at \tn\ \cite{thio94}.  Studies of quasi-1D cuprates have made
it clear that the DM (and additional) interactions are quite common
\cite{zhel98}; however, the tetragonal CuO$_2$ planes of other layered
cuprate antiferromagnets cause the effects of the DM interaction to
cancel out, so that there is no canting \cite{coff91,stei96}. 

In the early diffraction studies, the \lco\ powder samples contained some
excess oxygen and the first crystals had contamination from flux or the
crucible, thus resulting in a reduced ordering temperature.  (It is now
known that excess oxygen, in sufficient quantity, can segregate to form
superconducting phases \cite{well97}.)  It was eventually found that by
properly annealing a crystal one can obtain a sample with $T_{\rm N} =
325$~K \cite{keim92a}.  The ordered magnetic moment is also sensitive to
impurity effects.  In a study of single crystals with different annealing
treatments, Yamada {\it et al.} \cite{yama87} found that the ordered
moment is correlated with \tn, with a maximum Cu moment of 0.60(5) 
$\mu_{\rm B}$ [apparently determined from the intensity of the (100)
magnetic reflection alone].

The magnetic coupling between layers in \lco\ is quite weak because each
Cu sees two up spins and two down spins at nearly the same distance in a
neighboring layer.  The small orthorhombic distortion of the lattice
removes any true frustration, resulting in a small but finite coupling. 
There are, however, several other cuprate antiferromagnets with a similar
centered stacking of layers, but with tetragonal symmetry (see
Table~\ref{tab:AF}).  Yildirim {\it et al.} \cite{yild94a} showed that the
long-range order (including spin directions) can be understood when one
takes into account zero-point spin fluctuations, together with the proper
exchange anisotropies \cite{yild94b}.

\begin{table}[t]
\centering
\caption{Compilation of some neutron scattering results for a number of
layered cuprate antiferromagnets.  $m_{\rm Cu}$ is the average ordered
moment per Cu atom at $T\ll T_{\rm N}$.  The superexchange energy $J$
corresponds to the value obtained from the spin wave velocity after
correction for the quantum-renormalization factor $Z_c=1.18$. For crystal
symmetry, O = orthorhombic, T = tetragonal.}
\label{tab:AF}       
\begin{tabular}{lcccccl}
\hline\noalign{\smallskip}
Compound & \tn\ & $m_{\rm Cu}$ & $J$ & Crystal & Layers & Refs.  \\
         & (K)  & ($\mu_{\rm B}$) &  (meV) & Symmetry & per cell & \\
\noalign{\smallskip}\hline\noalign{\smallskip}
\lco\ & 325(2) & 0.60(5) & 146(4) & O & 1 &\cite{yama87,keim92a,cold01}\\
Sr$_2$CuO$_2$Cl$_2$ & 256(2) & 0.34(4) & 125(6) & T & 1 &
\cite{vakn90,grev95,toku90} \\
Ca$_2$CuO$_2$Cl$_2$ & 247(5) & 0.25(10) &  & T & 1 & \cite{vakn97} \\
Nd$_2$CuO$_4$ & 276(1) & 0.46(5)  & 155(3) & T & 1 &
\cite{mats90,skan93,lynn01,bour97c} \\ 
Pr$_2$CuO$_4$ & 284(1) & 0.40(2) & 130(13) & T & 1 &
\cite{suma95,mats90} \\
\YBCO{6.1} & 410(1) & 0.55(3) & 106(7) & T & 2 &
\cite{casa94,hayd96b} \\
TlBa$_2$YCu$_2$O$_7$ & $>350$ & 0.52(8) & & T & 2 & \cite{mizu88} \\
Ca$_{0.85}$Sr$_{0.15}$CuO$_2$ & 537(5) & 0.51(5) & & T & $\infty$ &
\cite{vakn89} \\
\noalign{\smallskip}\hline
\end{tabular}
\end{table}

The parent compounds of the electron-doped superconductors, Nd$_2$CuO$_4$
and Pr$_2$CuO$_4$, have somewhat more complicated magnetic structures. Nd
moments and induced moments on Pr couple to the order in the CuO$_2$
planes, resulting in noncollinear magnetic structures and spin
reorientation transitions as a function of temperature; these are
described in the review by Lynn and Skanthakumar \cite{lynn01}.  The
magnetic structures and transitions have been explained by Sachidanandam
{\it et al.} \cite{sach97} by taking account of the single-ion anisotropy
and crystal-field effects for the rare-earth ions.  Further discussion is
given by Petitgrand {\it et al.} \cite{peti99}.

The crystal structure of \ybco\ contains pairs of CuO$_2$ layers
(bilayers).  There is also a third layer of Cu atoms, but in \YBCO{6}\
these are non-magnetic Cu$^{1+}$ ions.  (Added oxygen goes into this
layer, forming the CuO chains of \YBCO{7}.)  The magnetic structure of
\YBCO{6} is indicated in Fig.~\ref{fg:LCO_struc}.  Because of the relative
antiferromagnetic ordering of the bilayers, together with a spacing that
is not determined by symmetry, there is a structure factor for the
magnetic Bragg peaks that depends on $Q_z$.  This structure factor also
affects the spin-wave intensities, as will be discussed.

It is not possible to determine the spin direction from
zero-field diffraction measurements due to the tetragonal symmetry of the
lattice and inevitable twinning of the magnetic domains.  Nevertheless,
Burlet {\it et al.} \cite{burl98} were able to determine the spin
direction by studying the impact of a magnetic field applied along a
$[1,-1,0]$ direction of a \YBCO{6.05}\ single crystal.  They found that in
zero field, the spins must lie along [100] or [010] directions (parallel
to the Cu-O bonds), and that the applied field rotates them towards
[110]. This result has been confirmed by electron-spin resonance studies
of \ybco\ with a small amount of Gd sustituted for Y \cite{jano99}.

A complication in studies of magnetic order involving some of the first
crystals of \ybco\ arose from inadvertent partial substitution of Al ions
onto the Cu(1) (``chain'') site.  The Al presumably came from the use of
crucibles made of Al$_2$O$_3$.  Kadowaki {\it et al.} \cite{kado88},
performing one of the first single-crystal diffraction studies on a
\ybco\ sample with \tn\ of 405~K, found that below 40~K a new set of
superlattice peaks appeared, indicating a doubling of the magnetic unit
cell along the $c$ axis.  It was later demonstrated convincingly, by
comparing pure and Al-doped crystals, that the low-temperature doubling
of the magnetic period only occurs in crystals with Al
\cite{casa94,brec95}.  A model explaining how the presence of Al on Cu
chain sites can change the magnetic order was developed by Andersen and
Uimin \cite{ande97b}.

To evaluate the ordered magnetic moment, it is necessary to have
knowledge of the magnetic form factor.  In all of the early studies of
antiferromagnetic order in cuprates it was assumed that the spin-density
on a Cu ion is spherical; however, this assumption is far from being
correct.  The magnetic moment results from the half-filled $3d_{x^2-y^2}$
orbital, which deviates substantially from sphericity. The proper,
anisotropic form factor was identified by Shamoto {\it et al.}
\cite{sham93} and shown to give an improved description of magnetic Bragg
intensities for \YBCO{6.15}.  An even better measurement of magnetic Bragg
peaks was done on small crystal of \YBCO{6.10}\ by Casalta {\it et al.}
\cite{casa94}.  They obtained a Cu moment of 0.55(3) $\mu_{\rm B}$.  Use
of the proper form factor is important for properly evaluating the
magnetic moment, as there is always a gap between $Q=0$ (where the
magnitude of the form factor is defined to be 1) and the $Q$ value of the
first magnetic Bragg peak.  It does not appear that anyone has gone back
to reevaluate the magnetic diffraction data on other cuprates, such as
\lco\ or Sr$_2$CuO$_2$Cl$_2$ using the anisotropic form factor.

The maximum observed Cu moments are consistent with a large reduction due
to zero-point spin fluctuations as predicted by spin-wave theory.  The
moment $m$ is equal to $g\langle S\rangle \mu_{\rm B}$, where a typical
value of the gyromagnetic ratio $g$ is 2.1.  Without zero-point
fluctuations, one would expect $m\approx 1.1$~$\mu_{\rm B}$. 
Linear spin-wave theory predicts $\langle S\rangle=0.303$ \cite{line70},
giving $m\approx 0.64$~$\mu_{\rm B}$, a bit more than the largest observed
moments.  Further reductions can occur due to hybridization effects
\cite{lore05,capr05}.  

\begin{figure}[t]
\centerline{\includegraphics[width=3in]{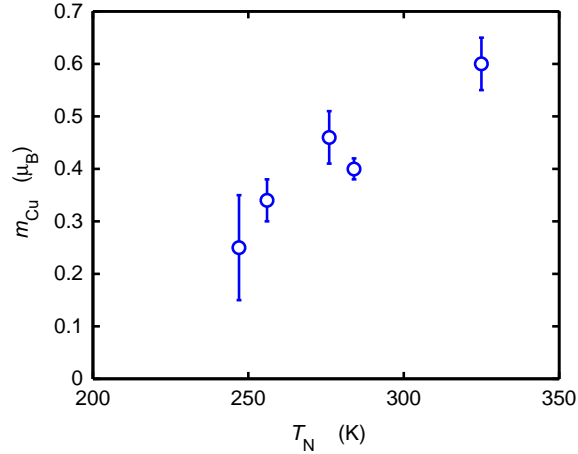}}
\medskip
\caption{Ordered magnetic moment per Cu atom vs.\ \tn\ for the first five
compounds in Table~\ref{tab:AF}, all of which have a similar
body-centered stacking of CuO$_2$ layers.}
\label{fg:mcu}
\end{figure}

The ordered moments of the oxy-chlorides listed in Table~\ref{tab:AF}
seem surprisingly small.  While this might be due to hybridization
effects, it is interesting to note that there is a correlation between
$m_{\rm Cu}$ and \tn\ for the first five antiferromagnets in the table,
which all share a similar body-centered stacking of the CuO$_2$ planes. 
The correlation is illustrated in Fig.~\ref{fg:mcu}.  The ratio 
$T_{\rm N}/J$ is expected to be sensitive to the interlayer exchange $J'$
\cite{yasu05}, and $J'$ varies substantially among these compounds;
however, I am not aware of any predicted dependence of
$m_{\rm Cu}$ on $J'$.  A correlation between $m_{\rm Cu}$ and $T_{\rm
N}/J$ has been reported for quasi-1D antiferromagnets, but such a
correlation is expected in that case \cite{koji97}.

\subsection{Spin waves}

\begin{figure}[t]
\centerline{\includegraphics[width=2.5in]{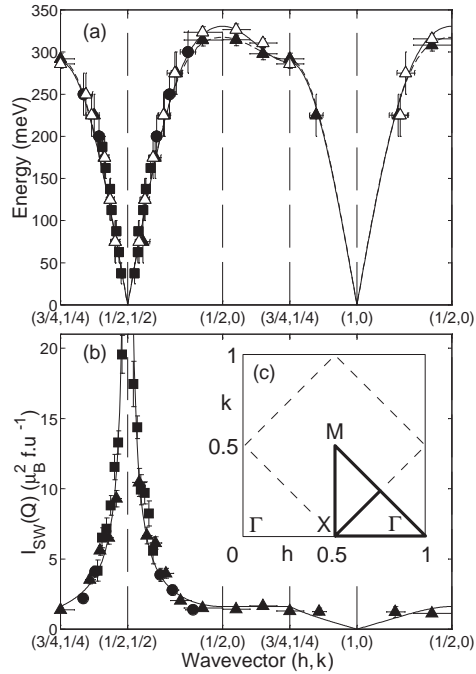}}
\medskip
\caption{(a) Spin-wave dispersion in \lco\ along high-symmetry directions
in the 2D Brillouin zone, as indicated in (c); $T=10$~K (295~K): open
(filled) symbols.  Solid (dashed) line is a fit to the 10-K (295-K) data.
(b) Spin-wave intensity vs.\ wave vector.  Line is prediction of linear
spin-wave theory.  From Coldea {\it et al.} \cite{cold01}.}
\label{fg:lco_disp}
\end{figure}

The starting point for considering magnetic interactions in the cuprates
is the Heisenberg hamiltonian:
\begin{equation}
 H = J \sum_{\langle i,j\rangle} {\bf S}_i\cdot {\bf S_j},
\end{equation}
where $\langle i,j\rangle$ denotes all nearest-neighbor pairs,
each included once.  This hamiltonian can be derived in second-order
perturbation from a Hubbard model for a single, half-filled band of
electrons.  Such a model includes a nearest-neighbor hopping energy $t$
and the Coulomb repulsion energy $U$ for two electrons on the same site;
in terms of these parameters, $J=4t^2/U$ \cite{ande59}.  Spin-wave theory
can be applied to the Heisenberg hamiltonian to calculate the dispersion
of spin fluctuations about \qaf\ \cite{oguc60}.  At low energies the spin
waves disperse linearly with ${\bf q} = {\bf Q} - {\bf Q}_{\rm AF}$ (see
Fig.~\ref{fg:lco_disp}), having a velocity $c = \sqrt{8}SZ_c Ja/\hbar$,
where $Z_c\approx 1.18$ \cite{sing89} is a quantum-renormalization
factor.  Thus, by measuring the spin-wave velocity, one can determine
$J$.  

Spin-wave measurements have been performed for a number of cuprates, and
some results for $J$ are listed in Table~\ref{tab:AF}.  (Complementary
measurements of $J$ can be obtained by two-magnon Raman scattering
\cite{lyon88}.)  To calculate the values of $J$ from spectroscopically
determined parameters, one must consider at least a 3-band Hubbard model
\cite{emer88}.  Recent {\it ab initio} cluster calculations
\cite{oost96,muno00} have been able to achieve reasonable agreement with
experiment.  While the magnitude of $J$ in layered cuprates is rather
large, it is not extreme; a value of $J=226(12)$~meV has been measured
for Cu-O chains in SrCuO$_2$ \cite{zali04}.

To describe the experimental dispersion curves in greater detail, one
must add more terms to the spin hamiltonian. For example, in a {\it tour
de force} experiment, Coldea {\it et al.} \cite{cold01} have measured the
dispersion of spin waves in \lco\ along high-symmetry directions of the 2D
Brillouin zone, as shown in Fig.~\ref{fg:lco_disp}.  The observed
dispersion along the zone boundary, between $(\frac12,0)$ and
$(\frac34,\frac14)$, is not predicted by the simple Heisenberg model.  To
describe this, they consider the additional terms that appear when the
perturbation expansion for the single-band Hubbard model is extended to
fourth order.  The most important new term involves 4-spin cyclic
exchange about a plaquette of four Cu sites \cite{taka77,roge89,macd88}. 
Coldea and coworkers were able to fit the data quite well with the added
parameters [see lines through data points in Fig.~\ref{fg:lco_disp}(a)],
obtaining, at 10 K, $J=146(4)$ meV and a cyclic exchange energy
$J_c=61(8)$~meV \cite{cold01}.  (Superexchange terms coupling sites
separated by two hops are finite but negligible.)

If, instead of expanding to higher order, one extends the Hubbard model
to include hopping between next-nearest-neighbor Cu sites, one can
calculate a superexchange term $J'$ between next-nearest neighbors that
is on the order of 10\%\ of $J$ \cite{morr98b,anne89}.  It turns out,
however, that fitting the measured dispersion with only $J$ and $J'$
requires that $J'$ correspond to a ferromagnetic interaction
\cite{cold01}, which is inconsistent with the model predictions.

\begin{figure}[t]
\centerline{\includegraphics[width=3in]{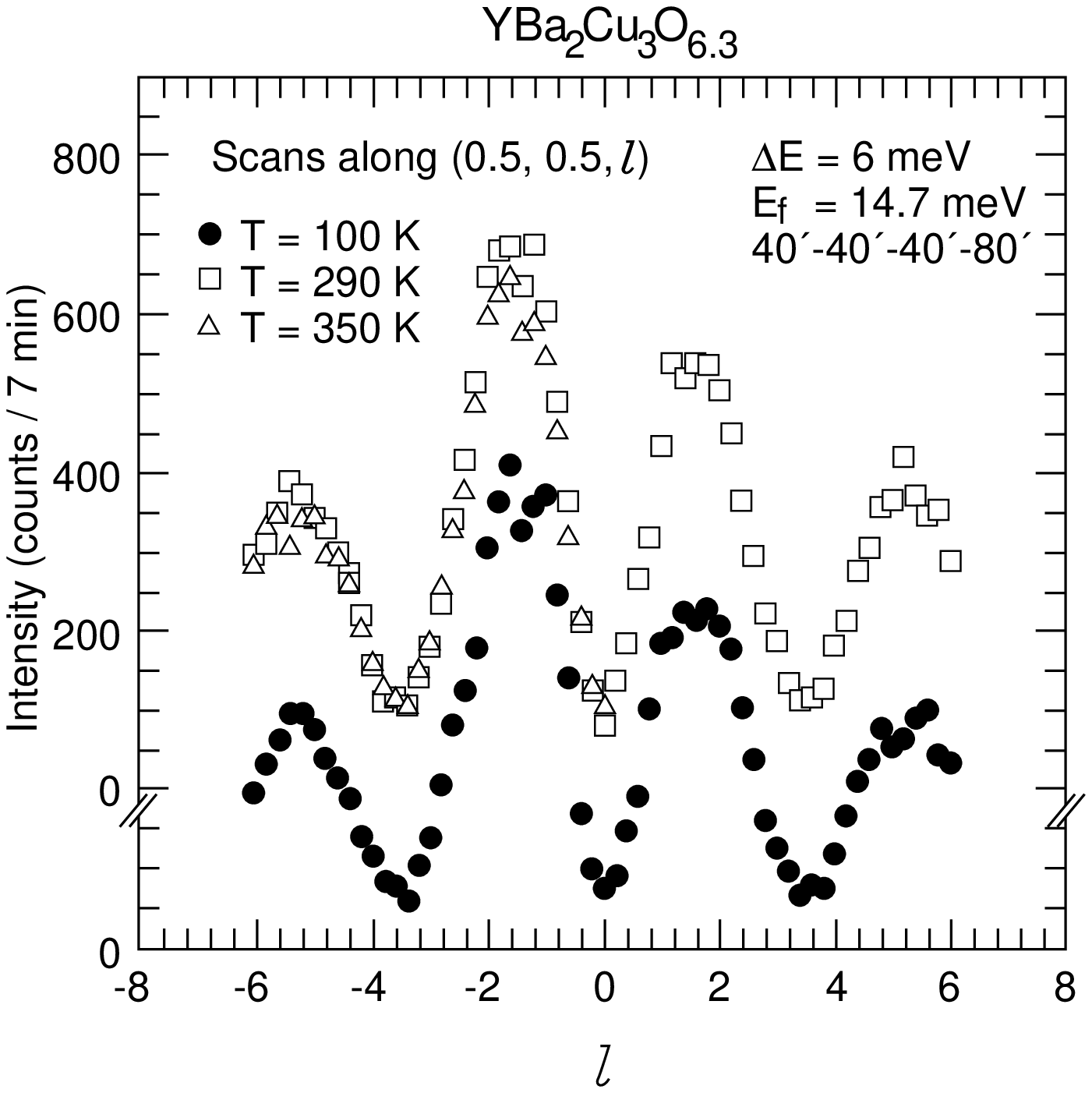}}
\medskip
\caption{Scans along the quasi-2D antiferromagnetic scattering rod ${\bf
Q}=(\frac12,\frac12,l)$ at a fixed energy transfer of 6 meV for a
crystal of \YBCO{6.3}\ with $T_N = 260(5)$~K.  The sinusoidal modulation
is due to the inelastic structure factor. The asymmetry in the scattering
as a function of $l$ is an effect due to the experimental resolution
function \cite{shir02}; the decrease in intensity at large $|l|$ is due to
the magnetic form factor. After Tranquada {\it et al.} \cite{tran89}.}
\label{fg:ybco_mod}
\end{figure}

In \ybco, the effective exchange coupling between Cu moments in
nearest-neighbor layers is substantial. Its effect is to split the
spin waves into acoustic and optic branches, having odd and even
symmetry, respectively, with respect to the bilayers.  The structure
factors for these excitations are \cite{tran89}
\begin{eqnarray}
  g_{\rm ac} & = & \sin(\pi zl), \label{eq:gac} \\
  g_{\rm op} & = & \cos(\pi zl),
\end{eqnarray}
where $z = d_{\rm Cu-Cu}/c$ is the relative spacing between Cu moments
along the $c$ axis within a bilayer ($d_{\rm Cu-Cu} \approx 3.285$~\AA
\cite{jorg90}); the intensity of the spin-wave scattering is proportional
to $g^2$.  An example of the intensity modulation due to the acoustic-mode
structure factor in the antiferromagnetic state is indicated by the
filled circles in Fig.~\ref{fg:ybco_mod}.

The energy gap for the optical magnons has been measured to be
approximately 70~meV \cite{rezn96,hayd96b}.  Experimental results for the
spin wave dispersion and the spectral weight are shown in
Fig.~\ref{fg:ybco_disp}.  The magnitude of the gap indicates that the
intra-bilayer exchange is 11(2) meV \cite{rezn96,hayd96b}.

\begin{figure}[t]
\centerline{\includegraphics[width=2.7in]{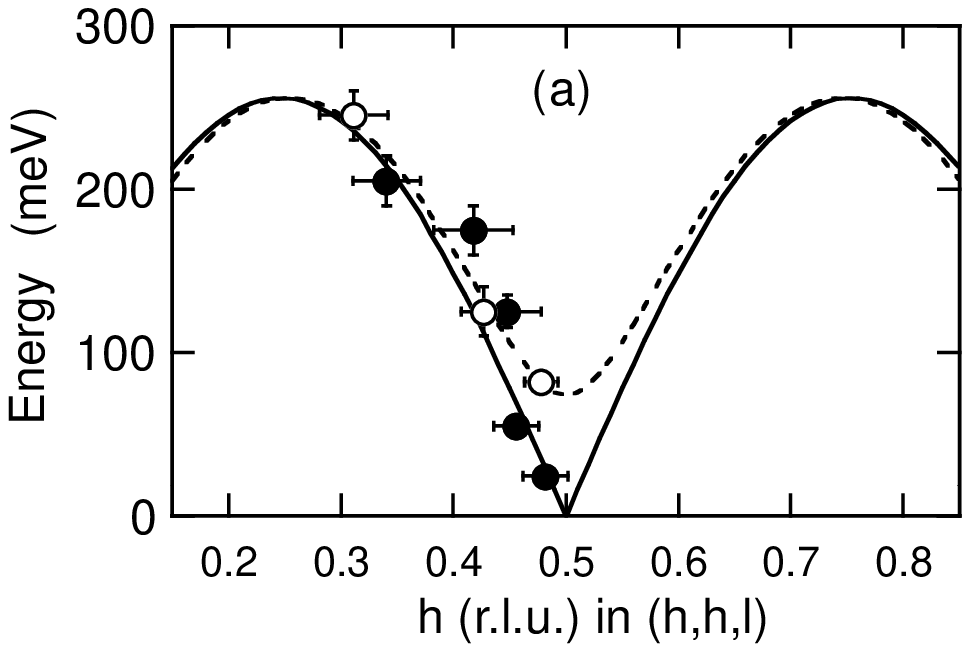}
\includegraphics[width=2.2in]{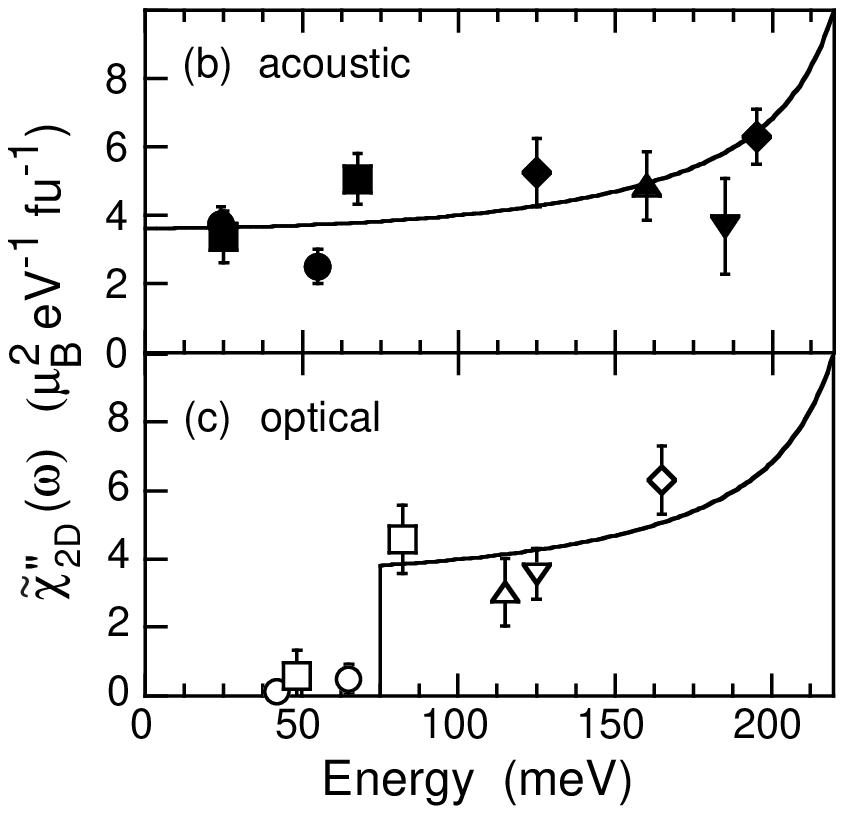}
}
\medskip
\caption{Experimental results for \YBCO{6.15}.  (a) Dispersion of
acoustic (filled symbols, solid line)  and optic (open symbols, dashed
line) spin-wave modes.  (b) {\bf Q}-integrated dynamic susceptibility,
$\tilde{\chi}''(\omega)$ for the acoustic, and (c) optic modes. After
Hayden {\it et al.}
\cite{hayd96b}.}
\label{fg:ybco_disp}
\end{figure}

At low energies, there are other terms that need to be considered.  There
need to be anisotropies, with associated spin-wave gaps, in order to fix
the spin direction; however, an atom with $S=\frac12$ cannot have
single-ion anisotropy.  Instead, the anisotropy is associated with the
nearest-neighbor superexchange interaction.  Consider a pair of
nearest-neighbor spins, ${\bf S}_i$ and ${\bf S}_j$, within a CuO$_2$
plane, with each site having tetragonal symmetry.  The Heisenberg
Hamiltonian for this pair can be written
\begin{equation}
  H_{\rm pair} = J_\| S_i^\| S_j^\| + J_\bot S_i^\bot S_j^\bot +
     J_z S_i^z S_j^z,
\end{equation}
where $\|$ and $\bot$ denote directions parallel and perpendicular to the
bond within the plane, and $z$ is the out-of-plane direction.  Yildirim
{\it et al.} \cite{yild94b} showed that the anisotropy can be explained
by taking into account both the spin-orbit and Coulomb exchange
interactions.  To discuss the anisotropies, it is convenient to define
the quantities $\Delta J \equiv J_{\rm av} - J_z$, where $J_{\rm av}
\equiv (J_\| + J_\bot)/2$, and $\delta J_{\rm in} \equiv J_\| - J_\bot$
\cite{yild94a}.  For the cuprates, $J_{\rm av} \gg \Delta J > \delta
J_{\rm in} > 0$.  The out-of-plane anisotropy, $\alpha_{\rm XY} = \Delta
J/J_{\rm av}$, causes the spins to lie, on average, in the $x$-$y$ plane,
and results in a spin-wave gap for out-of-plane fluctuations.  The
in-plane anisotropy
$\delta J_{\rm in}/J_{\rm av}$, contributing through the quantum
zero-point energy \cite{yild94a,shen82}, tends to favor alignment of the
spins parallel to a bond direction, and causes the in-plane spin-wave
mode to have a gap.  The effective coupling between planes (which can
involve contributions from several interactions \cite{yild94a}) leads to
(weak) dispersion along $Q_z$.

For stoichiometric \lco, the out-of-plane spin gap is 5.5(5) meV,
corresponding to $\alpha_{\rm XY} = 1.5\times 10^{-4}$ \cite{keim93}. The
in-plane gap of 2.8(5) meV has a contribution from anisotropic exchange
of the Dzyaloshinsky-Moriya type \cite{pete88,shek92}, as well as from
$\delta J_{\rm in}$.  No dispersion along $Q_z$ has been reported.

For antiferromagnetic \ybco, an out-of-plane gap of about 5 meV has been
observed \cite{tran89,vett89,ross90}, indicating an easy-plane anisotropy
similar to that in \lco. No in-plane gap has been resolved; however,
the in-plane mode shows a dispersion of about 3 meV along $Q_z$ 
\cite{tran89,vett89,ross90}.  The latter dispersion is controlled by the
effective exchange between Cu moments in neighboring bilayers through the
nonmagnetic Cu(1) sites, which is on the order of $10^{-4}J$.

\subsection{Spin dynamics at $T>T_{\rm N}$}

That strong 2D spin correlations survive in the CuO$_2$ planes for
$T>T_{\rm N}$ initially came as a surprise \cite{shir87}.  Such behavior
was certainly uncommon at that point.  Detailed studies have since been
performed measuring the instantaneous spin correlation length $\xi$ as a
function of temperature in \lco \cite{birg99} and in \scoc
\cite{grev94,grev95}. The correlation length is obtained using an
experimental trick to integrate the inelastic scattering over excitation
energy, and using the formula
\begin{equation}
  {\cal S}({\mathbf q}_{\mathrm 2D}) = \int d\omega\, 
   {\cal S}({\mathbf q}_{\mathrm 2D},\omega) 
   = {{\cal S}(0) \over 1 + q_{\mathrm 2D}^2\xi^2}.
\end{equation}
Here, ${\mathbf q}_{\mathrm 2D}$ is the momentum-transfer component
parallel to the planes, and the scattering is assumed to be independent
of momentum transfer perpendicular to the planes.  (The experimental
energy integration is imperfect, but, by proper choice of incident
neutron energy, does capture most of the critical scattering.)

To theoretically analyze the behavior of the correlation length,
Chakravarty, Halperin, and Nelson \cite{chak89} evaluated the 2D quantum
nonlinear $\sigma$ model using renormalization group techniques; their
results were later extended to a higher-order approximation by Hasenfratz
and Niedermayer \cite{hase91}.  The essential result is that 
\begin{equation}
  \xi/a \sim e^{2\pi\rho_s/T},
\end{equation}
where the spin stiffness $\rho_s$ is proportional to $J$ (see
\cite{birg99} for a thorough discussion).  The experimental results are
in excellent agreement with theory, with essentially no adjustable
parameters.  The unusual feature of $\xi(T)$ is the exponential, rather
than algebraic, dependence on temperature; nevertheless, note that it is
consistent with achieving long-range order at $T=0$.  The robustness of
the experimentally-observed spin correlations is due to the large value
of $J$, comparable to 1500~K, and the weak interlayer exchange, $J'$.  The
3D ordering temperature can be estimated as \cite{chak88}
\begin{equation}
  kT_{\mathrm N} \approx J'\left({m\over m_0}\right)^2
  \left({\xi\over a}\right)^2,
\end{equation}
where $m/m_0=0.6$ is the reduction of the ordered moment due to quantum
fluctuations.  Because of the small $J'$, the correlation length can
reach the order of $100a$ before ordering occurs \cite{birg99}.

Although \scoc\ has essentially the same structure as \lco, its
tetragonal symmetry leads one to expect, classically, that the net
interlayer exchange should be zero; however, an analysis by Yildirim {\it
et al.} \cite{yild94a} has shown that a finite interaction of appropriate
size results when quantum zero-point energy is taken into account. 
Because of its relatively low \tn\ of 257 K, it has been possible to
detect in \scoc\ a crossover to XY-like behavior about 30 K above \tn, as
reported in a $^{35}$Cl NMR study \cite{suh95}.  This behavior results
from the small easy-plane exchange anisotropy common to the layered
cuprates \cite{cucc03}.  Using neutrons to study the same material, it was
possible to shown that the characteristic fluctuation rate in the
paramagnetic state follows the behavior $\Gamma=\omega_0\sim\xi^{-z}$
with $z=1.0(1)$ \cite{kim01}, consistent with dynamic scaling theory for
the 2D Heisenberg antiferromagnet.\cite{hohe77}

There has been less work done on the paramagnetic phase of \ybco, as the
inelastic structure factor, Eq.~\ref{eq:gac}, complicates the
experimental trick for energy integration.  There are also complications
to studying \ybco\ samples at elevated temperatures, as oxygen can easily
diffuse into and out of crystals as one heats much above room temperature.
In any case, Fig.~\ref{fg:ybco_mod} shows that the bilayers
remain correlated at $T>T_{\rm N}$ \cite{tran89}.

\section{Destruction of antiferromagnetic order by hole doping}
\label{sc:dest}

The long-range antiferromagnetic order of \lco\ is completely destroyed
when 2\%\ of Sr (measured relative to Cu concentration) is doped into the
system \cite{nied98}.  Adding holes effectively reduces the number of Cu
spins, so one might consider whether the reduction of order is due to
dilution of the magnetic lattice.  For comparison, an extensive study of
magnetic dilution has been performed by Vajk {\it et al.} \cite{vajk02}
on La$_2$Cu$_{1-z}$(Zn,Mg)$_z$O$_4$, where Cu is substituted by
nonmagnetic Zn and/or Mg.  They found that long-range antiferromagnetic
order was lost at the classical 2D percolation limit of $z\approx41$\%. 
Thus, holes destroy magnetic order an order of magnitude more rapidly
than does simple magnetic dilution.

\begin{figure}[t]
\centerline{\includegraphics[width=2.3in]{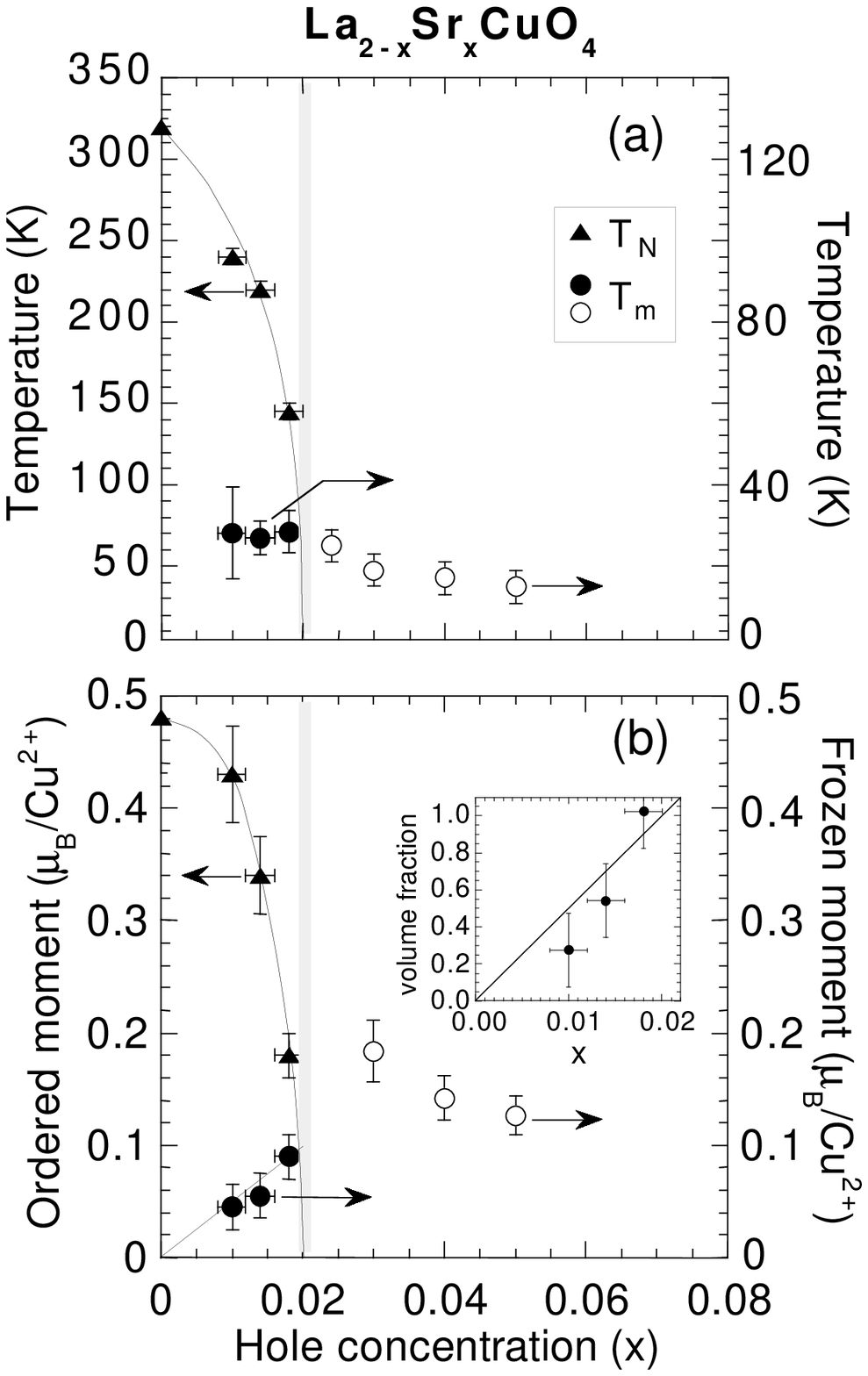}
\includegraphics[width=2.4in]{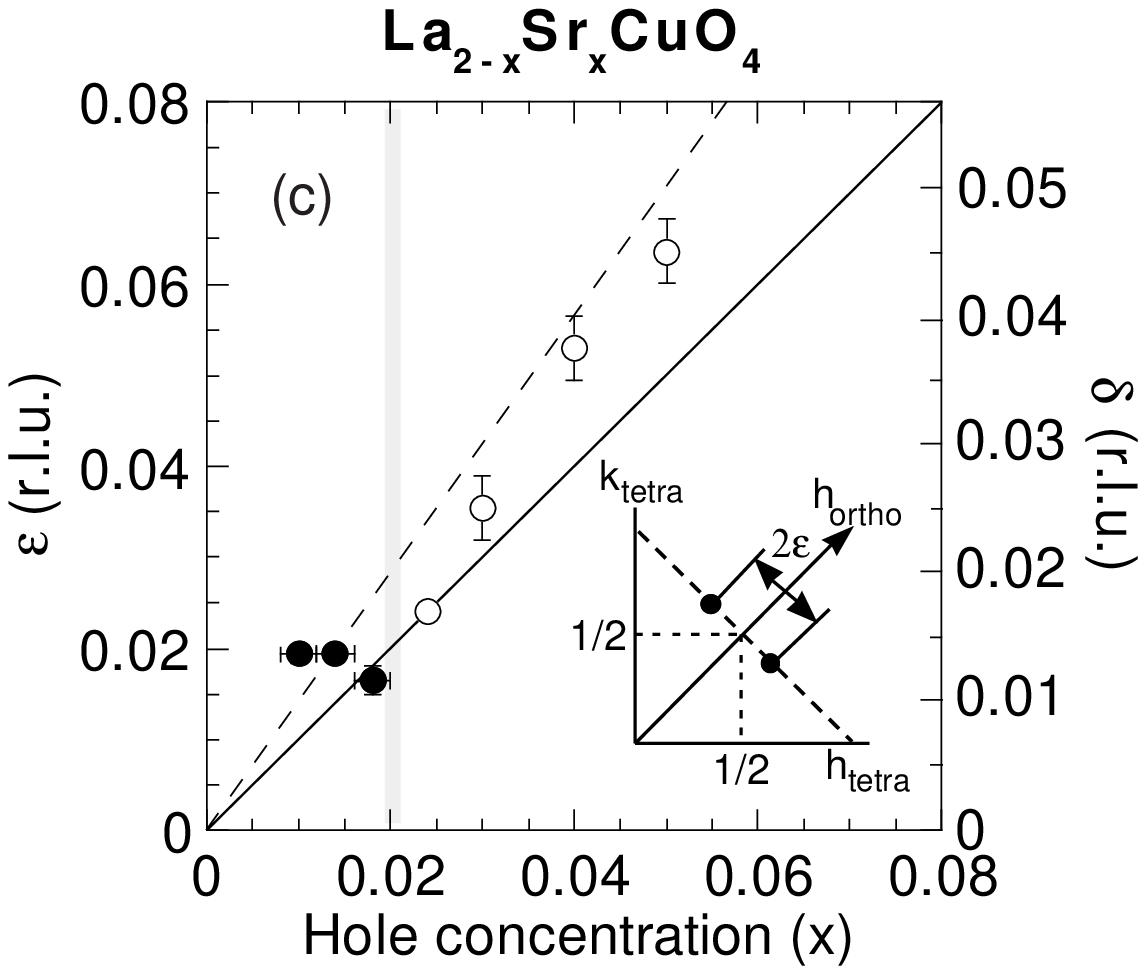}
}
\medskip
\caption{Results for lightly-doped \lsco.  (a) Magnetic transition
temperatures for commensurate (triangles) and incommensurate (circles)
order vs.\ hole concentration. (b) Commensurate ordered moment at
$T=30$~K and incommensurate frozen moment at $T=4$~K vs.\ hole
concentration.  Inset shows estimated volume fraction of incommensurate
phase.  (c) Variation of the incommensurability $\epsilon$ vs.\ hole
concentration; $\delta=\epsilon/\sqrt{2}$.  Solid and broken lines
correspond to $\epsilon=x$ and $\delta=x$, respectively. Inset shows the
positions of the incommensurate superlattice peaks in reciprocal space.
After Matsuda {\it et al.} \cite{mats02}, including results from
\cite{waki00,mats00b,waki01b,fuji02c}.}
\label{fg:lsco_ps}
\end{figure}

The reduction of the N\'eel temperature at small
but finite doping is accompanied by a strong depression of the
antiferromagnetic Bragg intensities, together with an anomalous loss of
intensity at $T<30$~K \cite{mats02}.  Matsuda {\it et al.} \cite{mats02}
showed recently that the latter behavior is correlated with the onset of
incommensurate magnetic diffuse scattering below 30~K.  In tetragonal
coordinates, this scattering is peaked at
$(\frac12,\frac12,0)\pm\frac1{\sqrt{2}}(\epsilon,\epsilon,0)$.  To be
more accurate, it is necessary to note that the crystal structure is
actually orthorhombic, with the unit-cell axes rotated by $45^\circ$; the
magnetic modulation is uniquely oriented along the $b_{\rm o}^\ast$
direction [see inset of Fig.~\ref{fg:lsco_ps}(c)].  

The doping dependence of the transition temperature, ordered moments, and
incommensurability are shown in Fig.~\ref{fg:lsco_ps}.  The facts that
(a) the volume fraction of the incommensurate phase grows with $x$ for
$x\le0.02$ [inset of Fig.~\ref{fg:lsco_ps}(b)] and (b) the
incommensurability does not change for $x\le0.02$ strongly suggest that
an electronic phase separation occurs \cite{mats02}.  Thus, it appears
that commensurate antiferromagnetic order is not compatible with hole
doping.  The disordered potential due to the Sr dopants may be
responsible for the finite range of doping over which the N\'eel state
appears to survive.  The diagonally-modulated, incommensurate
spin-density-wave phase induced by doping survives up to the onset of
superconductivity at $x\approx0.06$ \cite{fuji02c}, and it corresponds to
what was originally characterized as the ``spin-glass'' phase, based on
bulk susceptibility studies \cite{chou95,waki00b}.

Further evidence for electronic phase separation comes from studies of
oxygen-doped \lcod\ (for a review, see \cite{tran98b}).  The oxygen
interstitials are mobile, in constrast to the quenched disorder of Sr
substitution, and so they can move to screen discrete electronic phases.
For $\delta<0.06$, a temperature-dependent phase separation is observed
between an oxygen-poor antiferromagnetic phase and an oxygen-rich
superconducting phase \cite{jorg88,hamm90}; further miscibility gaps are
observed between superconducting phases at higher oxygen content
\cite{well97}.  A sample with $\delta\approx0.05$ and quenched disorder
(due to electrochemical oxygenation in molten salt) exhibited reduced
N\'eel and superconducting transition temperatures \cite{gnez04}.  The
interesting feature in this case was the observation of a decrease in the
antiferromagnetic order with the onset of the superconductivity,
suggesting a competition between the two phases.

In \ybco\ the situation is somewhat more complicated, as the doping of
the planes is coupled to the tetragonal-orthorhombic (T-O) transition
\cite{rena87,male89,cava90} that occurs in the vicinity of $x=0.3$--0.4,
depending on the degree of annealing.  In the tetragonal phase, an
isolated oxygen atom entering the ``chain'' layer simply converts
neighboring Cu(1) sites from Cu$^{1+}$ to Cu$^{2+}$; holes are created
when chain segments form \cite{tran88,uimi92}.  The transfer of holes from
the chains to the planes must be screened by displacements in the Ba-O
layer that sits between, and a large jump in this screening occurs at
the T-O transition \cite{rena87,male89,cava90}.  Thus, one tends to find
a discontinuous jump from a very small planar hole density in the
antiferromagnetic phase just below the T-O transition to a significant
density ($\sim0.05$ holes/Cu) just above.

Antiferromagnetic order has been observed throughout the tetragonal phase
of \ybco, with \tn\ decreasing rapidly as the T-O transition (at
$x\approx0.4$) is approached \cite{tran88,ross88}.  A study of a set of
carefully annealed powder samples, for which the T-O transition occurred
at $x\approx0.2$, indicated antiferromagnetic order in the orthorhombic
phase at $x=0.22$ and 0.24 with $T_{\rm N}=50(15)$~K and 20(10)~K,
respectively.  For tetragonal crystals with $x\sim0.3$, a drop in the
antiferromagnetic Bragg intensity has been observed below $\approx30$~K
\cite{tran89,ross88}; as the Bragg intensity decreased, an increase in
diffuse intensity along the 2D antiferromagnetic rod (with an acoustic
bilayer structure factor) was found.  This latter behavior might be
related to the apparent phase separation in \lsco\ with $x<0.02$
\cite{mats02} discussed above.

The best study of a single-crystal sample just on the orthorhombic side
of the T-O boundary is on \YBCO{6.35}, a sample with $T_c = 18$~K
\cite{stoc05b}.  Quasielastic diffuse scattering is observed at the
antiferromagnetic superlattice positions.  The peak intensity of this
central mode grows on cooling below $\sim30$~K, but the energy width
decreases below $T_c$.  These results indicate their is no coexistence of
long-range antiferromagnetic order in the superconducting phase. The
spin-spin correlation length is short ($\sim8$ unit cells), suggesting
segregation of hole-poor and hole-rich regions \cite{stoc05b}.

A possibly-related response to doping is observed in the bilayer system
La$_{2-x}$(Sr,Ca)$_x$CaCu$_2$O$_{6+\delta}$.  Studies of crystals with
$x=0.1$--0.2 reveal commensurate short-range antiferromagnetic order that
survives to temperatures $>100$~K \cite{ulri02,huck05}, despite evidence
from optical conductivity measurements for a substantial hole density in
the CuO$_2$ planes \cite{wang03}.  Thus, there seems to be a local phase
separation between hole-rich regions and antiferromagnetic clusters
having an in-plane diameter on the order of 10 lattice spacings.

\section{Stripe order and other competing states}
\label{sc:stripe}

\subsection{Charge and spin stripe order in nickelates}

To understand cuprates, it seems sensible to consider the behavior of
closely related model systems.  One such system is \lsnod, a material that
is isostructural with \lsco.  It is obtained by replacing $S=\frac12$
Cu$^{2+}$ ions ($Z=29$) with $S=1$ Ni$^{2+}$ ions ($Z=28$).  One might
consider the nickelates to be uninteresting as they are neither
superconducting nor metallic (for $x<0.9$) \cite{cava91,shin02}; however,
the insulating behavior is inconsistent with the predictions of band
theory, and it is important to understand why.

\begin{figure}[t]
\centerline{
\raisebox{1in}{\includegraphics[width=1.5in]{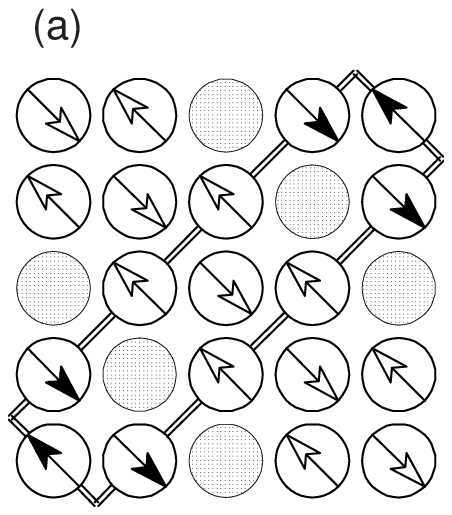}}
\hspace{0.7in}\includegraphics[width=2.4in]{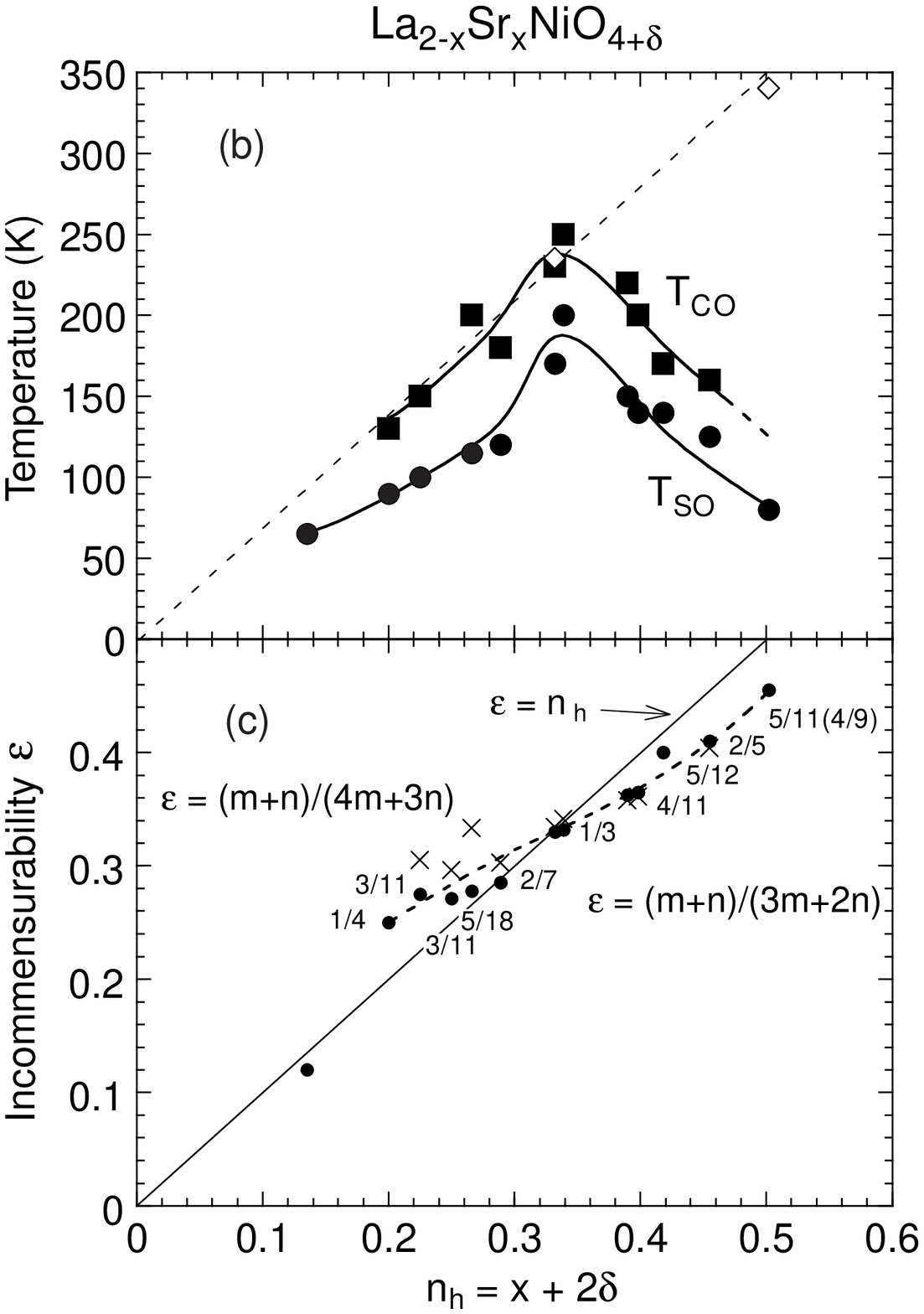}
}
\medskip
\caption{(a) Cartoon of diagonal stripe order in an NiO$_2$ plane (only
Ni sites indicated) for $n_{\rm h}\approx1/4$.  Magnetic unit cell is
indicated by double lines, shaded circles indicate charge stripes with a
hole density of one per Ni site.  (b) Transition temperatures for charge
order, $T_{\rm co}$ (squares), and spin order, $T_{\rm so}$ (circles),
measured by neutron diffraction.  Open diamonds: transition temperatures
from transport measurements \cite{cheo94}.  (c) Incommensurability vs.\
$n_{\rm h}$.  Circles (crosses) results at low temperature (high
temperature, near $T_{\rm so}$). Fraction labels are approximate
long-period commensurabilities. (b) and (c) after Yoshizawa {\it et al.}
\cite{yosh00}, including results from
\cite{tran94a,sach95,tran96a,woch98}.
}
\label{fg:lsno_pd}
\end{figure}

Pure \lno\ is an antiferromagnetic insulator \cite{yama92} that is easily
doped with excess oxygen, as well as by Sr substitution for La
\cite{tran98b}.  Doping the NiO$_2$ planes with holes reduces \tn\ more
gradually than in the cuprates \cite{tran94b}.  It is necessary to dope
to a hole concentration of $n_{\rm h} = x+2\delta\sim0.2$ before the
commensurate antiferromagnetic order is replaced by stripe order
\cite{tran94b,tran94a,sach95}.  Figure~\ref{fg:lsno_pd}(a) shows a
schematic of diagonal stripe order appropriate for $n_h\approx1/4$. 
The charge stripes, with a hole filling of one per Ni site, act as
antiphase domain walls for the magnetic order, so that the magnetic
period is twice that of the charge order.  The nature of the stripe order
has been deduced from the positions and intensities of the superlattice
peaks \cite{tran98b,tran95b}.  The characteristic wave vector for
spin order is 
${\bf q}_{\rm so} = {\bf Q}_{\rm AF}\pm\frac1{\sqrt{2}}
(\epsilon,\epsilon,0)$
and that for charge order is 
${\bf q}_{\rm co} = \frac1{\sqrt{2}}(2\epsilon,2\epsilon,0)+(0,0,1)$. 
When the symmetry of the average lattice does not pick a unique
orientation, modulations rotated by 90$^\circ$ will also be present in
separate domains. The fact that diagonal stripe order has a unique
modulation direction within each domain has been confirmed by electron
diffraction
\cite{li03}. Evidence for significant charge modulation has also been
provided by nuclear magnetic resonance studies
\cite{yosh99,abu99}.  The charge-ordering transition is always observed
to occur at a higher temperature than the spin ordering, as shown in
Fig.~\ref{fg:lsno_pd}(b), with the highest ordering temperatures
occurring for $x=1/3$ \cite{cheo94,lee97}.

The magnetic incommensurability $\epsilon$, is inversely
proportional to the period of the magnetic modulation.  It increases
steadily with doping, as shown in Fig.~\ref{fg:lsno_pd}(c), staying close
to the line $\epsilon=n_{\rm h}$, indicating that the hole-density within
the charge stripes remains roughly constant but the stripe spacing
decreases with doping.  For a given sample, the incommensurability
changes with temperature, tending towards $\epsilon=1/3$ as $T\rightarrow
T_{\rm co}$ \cite{ishi04}.  In a sample with ordered oxygen interstitials,
$\epsilon$ has been observed to pass through lock-in plateaus on warming,
indicating a significant coupling to the lattice \cite{woch98}.

The impact of hole-doping on the magnetic interactions has been
determined from measurements of the spin-wave dispersions for crystals
with $x\approx1/3$ \cite{bour03,boot03,woo05}.  Analysis shows that the
superexchange $J$ within an antiferromagnetic region is 27.5(4) meV
\cite{woo05}, which is only a modest reduction compared to $J=31(1)$~meV
in undoped \lno\
\cite{yama91}.  The effective coupling across a charge stripe is found to
be $\approx0.5J$, a surprisingly large value.  In the spin-wave
modelling, it was assumed that there is no magnetic contribution from the
charge stripes; however, it is not obvious that this is a correct
assumption.  Combining an $S=\frac12$ hole with an $S=1$ Ni ion should
leave at least an $S=\frac12$ per Ni site in a domain wall.  Recently,
Boothroyd {\it et al.} \cite{boot03b} have discovered quasi-1D magnetic
scattering that disperses up to about 10 meV and becomes very weak above
100 K.  This appears to correspond to the spin excitations of the charge
stripes.

Inelastic neutron scattering measurements at $T>T_{\rm co}$ indicate that
incommensurate spin fluctuations survive in the disordered state
\cite{lee02,bour03}, implying the existence of fluctuating stripes.  This
result is consistent with optical conductivity studies
\cite{kats96,home03} which show that while the dc conductivity approaches
that of a metal above room temperature, the dynamic conductivity in the
disordered state never shows the response of a conventional metal.

The overall message here is that a system very close to the cuprates
shows a strong tendency for charge and spin to order in a manner that
preserves the strong superexchange interaction of the undoped parent
compound.  It is certainly true that Ni$^{2+}$ has $S=1$ while Cu$^{2+}$
has $S=\frac12$, and this can have a significant impact on the strength
of the charge localization in the stripe-ordered nickelates
\cite{tran98c}; however, the size of the spin cannot, on its own, explain
why conventional band theory breaks down for the nickelates.  The
electronic inhomogeneity observed in the nickelates suggests that
similarly unusual behavior might be expected in the cuprates.

\subsection{Stripes in cuprates}
\label{sc:cu_stripes}

\begin{figure}[t]
\centerline{
\includegraphics[width=3in]{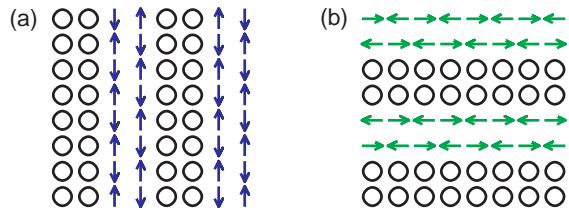}
}
\medskip
\caption{Cartoons of equivalent domains of (a) vertical and (b)
horizontal bond-centered stripe order within a CuO$_2$ plane (only Cu
sites shown).  Note that the magnetic period is twice that of the charge
period.  The charge density along a stripe is one hole for every two
sites in length. }
\label{fg:vert}
\end{figure}

Static charge and spin stripe orders have only been observed in a couple
of cuprates, \lbcoate\ \cite{fuji04} and \lnsco\ \cite{tran95a,ichi00} to
be specific.  The characteristic 2D wave vector for spin order is 
${\bf q}_{\rm so} = {\bf Q}_{\rm AF}\pm(\epsilon,0)$ and that for charge
order is ${\bf q}_{\rm co} = (2\epsilon,0)$.  A cartoon of stripe order
consistent with these wave vectors is shown in Fig.~\ref{fg:vert}(a); the
inferred charge density within the charge stripes is approximately one
hole for every two sites along the length of a stripe.  The magnetic unit
cell is twice as long as that for charge order.  It should be noted that
the phase of the stripe order with respect to the lattice has not been
determined experimentally, so that it could be either bond-centered, as
shown, or site-centered.

In a square lattice, the domains of vertical and horizontal stripes shown
in Fig.~\ref{fg:vert} are equivalent; however, each breaks the rotational
symmetry of the square lattice.  In fact, static stripe order has only
been observed in compounds in which the average crystal structure for
each CuO$_2$ plane exhibits a compatible reduction to rectangular
symmetry.  This is the case for the low-temperature-tetragonal (LTT)
symmetry (space group $P4_2/ncm$) of \lbcoate\ and \lnsco\ \cite{axe94},
where orthogonal Cu-O bonds are inequivalent within each plane, but
the special direction rotates by 90$^\circ$ from one layer to the next. 
Because planes of each orientation are equally represented in the LTT
phase, both stripe domains are equally represented.  The correlation
between lattice symmetry and stripe ordering is especially clear in
studies of the system La$_{1.875}$Ba$_{0.125-x}$Sr$_x$CuO$_4$ by Fujita
and coworkers \cite{fuji02,fuji02b,kimu03}. 

When diffraction peaks from orthogonal stripe domains are present
simultaneously, one might ask whether the diffraction pattern is also
consistent with a checkerboard structure (a superposition of orthogonal
stripe domains in the same layer) \cite{fine04}.  There are a number of
arguments against a checkerboard interpretation.   (1) The observed
sensitivity of charge and spin ordering to lattice symmetry would have no
obvious explanation for a checkerboard structure, with its 4-fold
symmetry.  (2) For a pattern of crossed stripes, the positions of the
magnetic peaks should rotate by 45$^\circ$ with respect to the
charge-order peaks.  One would also expect to see additional charge-order
peaks in the [110] direction.  Tests for both of these possibilities
have come out negative \cite{tran99b}.  It is possible to imagine other
two-dimensional patterns that are consistent with the observed
diffraction peaks \cite{fine04}; however, the physical justification for
the relationship between the spin and charge modulation becomes unclear
in such models.  (3) The intensity of the charge-order scattering is
strongly modulated along $Q_z$, with maxima at $l=n\pm\frac12$, where $n$
is an integer.  This behavior is straightforwardly explained in terms of
unidirectional stripe order tied to local lattice symmetry, with Coulomb
repulsion between stripes in equivalent (next-nearest-neighbor) layers
\cite{vonz98}.  For a checkerboard pattern, one would expect correlations
between nearest-neighbor layers, which would give a $Q_z$ dependence
inconsistent with experiment.

There has also been a report of stripe-like charge order and
incommensurate spin fluctuations in a \ybco\ sample with a nominal
$x=0.35$ \cite{mook02}.   The superconducting transition for this sample,
having a midpoint at 39 K and a width of 10 K, is a bit high to be
consistent with the nominal oxygen content \cite{stoc05b}; this may
indicate some inhomogeneity of oxygen content in the very large
melt-grown crystal that was studied.  Weak superlattice peaks
attributed to charge order corresponding to vertical stripes with
$2\epsilon =0.127$ retain finite intensity at room temperature.  The
difference in magnetic scattering at 10 K relative to 100 K shows a
spectrum very similar to that in Fig.~\ref{fg:lbco_disp}, with
$E_r\approx23$ meV and $\epsilon\approx0.06$.  
While these experimental results are quite intriguing, it would be
desirable to confirm them on another sample.  In any case, it is
interesting to note that a recent muon spin rotation ($\mu$SR) study by
Sanna {\it et al.}
\cite{sann04} identified local magnetic order at low temperatures in
\ybco\ with $x\le0.39$, and coexistence with superconductivity for
$x\ge0.37$.

Elastic incommensurate scattering consistent with stripe order has been
observed in stage-4 \lcod\ with $T_c=42$~K, although charge order has not
been detected \cite{lee99}.  An interesting question is whether static
stripe order coexists homogeneously with high-temperature
superconductivity in this sample.  The fact that the 4-layer staging of
the oxygen interstitials creates two inequivalent types of CuO$_2$ layers
suggests the possibility that the order parameters for stripe order and
superconductivity might have their maxima in different sets of layers. 
A $\mu$SR study of the same material found that only about 40\%\ of the
muons detected a local, static magnetic field \cite{savi02}.  While it
was argued that the best fit to the time dependence of the zero-field
muon spectra was obtained with an inhomogeneous island model, the data
may also be compatible with a model of inhomogeneity perpendicular to
the planes. 

Beyond static order, we can consider the excitations of a stripe-ordered
system.  It has already been noted in \S8.2.1 that the magnetic
excitations of \lbcoate\ at low temperature exhibit a similar dispersion
to good superconductors without stripe order.   The overall spectrum is
only partially consistent with initial predictions of linear spin-wave
theory for a stripe-ordered system \cite{bati01,krug03,carl04}; however,
it is reasonably reproduced by calculations that consider weakly-coupled
two-leg spin ladders (of the type suggested by Fig.~\ref{fg:vert})
\cite{vojt04,uhri04} or that treat both spin and electron-hole
excitations of a stripe-ordered ground state \cite{seib05}. 

\begin{figure}[t]
\centerline{
\includegraphics[width=2in]{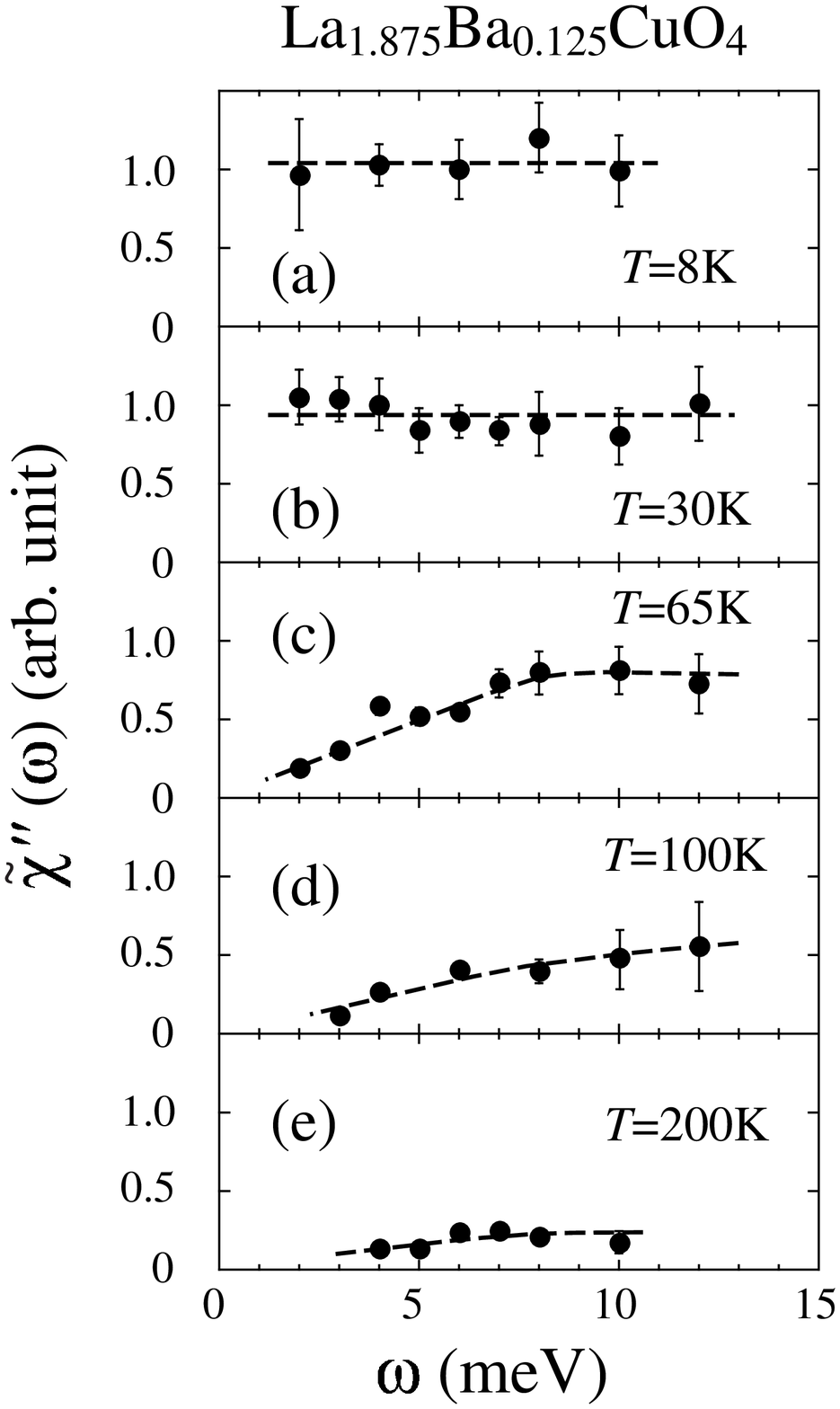}
\hspace{0.5in}
\raisebox{0.3in}{\includegraphics[width=2in]{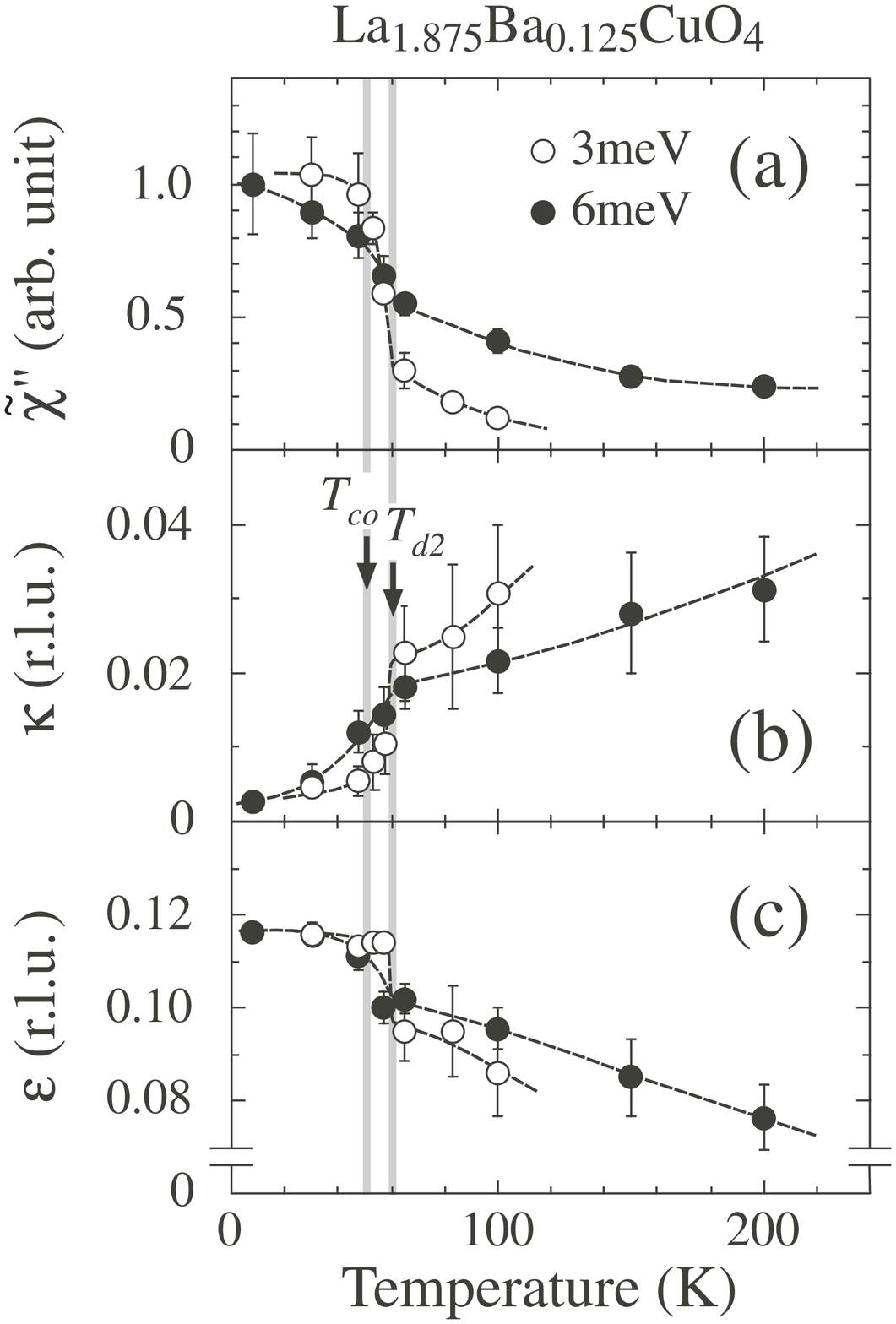}}
}
\medskip
\caption{Results for low-energy inelastic magnetic scattering in
\lbcoate.  Left: local susceptibility,
$\tilde{\chi}''(\omega)$, as a function of $\hbar\omega$ for temperatures
below [(a) and (b)] and above [(c)-(e)] the charge-ordering temperature,
$T_{\rm co}$.  Right: Temperature dependence, for $\hbar\omega=3$ and 6
meV, of (a) local susceptibility, (b) $\kappa$, half-width in {\bf Q} of
the incommensurate peaks, (c) incommensurability $\epsilon$.  Vertical
lines denote $T_{\rm co}$ and $T_{\rm d2}$, the transition to the LTT
phase.  The dashed lines are guides to the eye.  From Fujita {\it et al.}
\cite{fuji04}. }
\label{fg:lbco_chi}
\end{figure}

The temperature dependence of the magnetic scattering at low energies
($\le10$~meV) has been reported by Fujita {\it et al.} \cite{fuji04};  
Fig.~\ref{fg:lbco_chi} shows some of the results.  On the left, one can
see that, in the stripe-ordered state ($T=8$ and 30 K), the {\bf
Q}-integrated dynamic susceptibility is independent of frequency and
temperature.  Such behavior is consistent with expectations for spin
waves.  In the disordered phase (65 K and above),
$\tilde{\chi}''(\omega)$ heads linearly to zero at zero frequency;
however, at 10 meV the decrease with temperature is relatively slow.  The
temperature dependence of
$\tilde{\chi}''(\omega)$ at 3 meV and 6 meV is shown in more detail on
the right side, in panel (a).  There is a rapid drop above $T_{\rm co}$
at 3 meV, but the change at 6 meV is more gradual.  There is a substantial
increase in {\bf Q} width of the incommensurate peaks at the transition,
as shown in (b).  Interestingly, there is also a significant drop in
incommensurability at the transition, shown in (c), with a continuing
decrease at higher temperatures.  Similar results for \lnsco\ with
$x=0.12$ were obtained by Ito {\it et al.} \cite{ito03}.  The jump in
$\epsilon$ on cooling through \tco\ may be related to commensurability
effects in the stripe-ordered state.

The results in the disordered state ($T>60$~K) of \lbcoate\ look similar
to those found in the normal state of \lsco\ \cite{aepp97}.  The
continuous variation of the magnetic scattering through the transition
suggests that the nature of the underlying electronic correlations is the
same on both sides of the transition.  The simplest conclusion seems to
be that dynamic stripes are present in the disordered state of \lbcoate\
and in the normal state of \lsco.  The similarity of the magnetic spectrum
to that in
\ybco\ (see Fig.~\ref{fg:lbco_disp}) then suggests that dynamic stripes
may be common to under- and optimally-doped cuprates.

\subsection{Spin-density-wave order in chromium}

Chromium and its alloys represent another system that has been proposed as
a model for understanding the magnetic excitations in superconducting
cuprates \cite{fawc94}.  Pure Cr has a body-centered-cubic structure and
exhibits antiferromagnetic order that is slightly incommensurate
\cite{fawc88}.  Overhauser and Arrott \cite{over60} first proposed that
the order was due to a spin-density-wave instability of the conduction
electrons.  Lomer \cite{lome62} later showed that the amplitude of the
SDW order could be understood in terms of approximate nesting of separate
electron and hole Fermi surfaces.  The ordering can be modified by
adjusting the Fermi energy through small substitions of neighboring 3d
elements.  For example, adding electrons through substitution of less
than a percent of Mn is enough to drive the ordering wave vector to
commensurability, whereas reducing the electron density with V causes the
incommensurate ordering temperature to head to zero at a concentration of
about 3.5\%\
\cite{fawc94}.

The magnetic excitations in pure Cr have a very large spin-wave velocity
\cite{sinh77,fuku96b}, similar to the situation for cuprates.  The
results seem to be qualitatively consistent with calculations based on
Fermi-surface nesting \cite{kane01,fish96}.  A study of paramagnetic
Cr$_{0.95}$V$_{0.05}$ at low temperature \cite{hayd00} has revealed
incommensurate excitations at low energy that broaden with increasing
energy.  $\chi''$ has a peak at about 100 meV, but remains substantial up
to at least 400 meV.  A comparison of the magnitude of the experimental
$\chi''$ with {\it ab initio} calculations \cite{stau99} indicates a
substantial exchange enhancement over the bare Lindhard susceptibility
\cite{hayd00}.

Given that Cr is cubic, there are three equivalent and orthogonal nesting
wave vectors.  Within an ordered domain, the ordering wave vector
consists of just one of these three possibilities.  Along with the SDW
order, there is also a weak CDW order that appears.  A neutron diffraction
study showed that the intensity of the CDW peaks scales as the square of
the intensity of the SDW peaks, indicating that the CDW is a secondary
order parameter and that the ordering transition is driven by the
magnetic ordering \cite{pynn76}.  It is natural to compare this behavior
with that found in stripe ordered cuprates.  The behavior in the latter
is different, with the charge order peaks appearing at a higher
temperature than those for spin order in \lnsco\ with $x=0.12$
\cite{tran96b}.  That result indicates that either charge ordering alone,
or a combination of charge and spin energies, drive the initial ordering
\cite{zach98}, so that stripe order is distinct from SDW order.

There are certainly some similarities between the magnetic excitations of
Cr alloys and those of optimally-doped cuprates.  The fact that the
magnetism in Cr and its alloys is caused by Fermi-surface nesting has led
many people to assume that a similar mechanism might explain the
excitations of superconducting cuprates, as discussed elsewhere in this
book.  Some arguments against such an approach have been presented in
Sec.~VI of \cite{kive03}.

\subsection{Other proposed types of competing order}

New types of order beyond spin-density waves or stripes have been proposed
for cuprates.  One is $d$-density-wave (DDW) order, which has been
introduced by Chakravarty {\it et al.} \cite{chak01} to explain the
$d$-wave-like pseudogap seen by photoemission experiments on underdoped
cuprates. (A related model of a staggered-flux phase was proposed by Wen
and Lee \cite{wen96} with a similar motivation; however,
their model does not have static order.) The model of DDW order involves
local currents that rotate in opposite directions about neighboring
plaquettes within the CuO$_2$ planes.  The orbital currents should induce
weak, staggered magnetic moments oriented along the $c$ axis.  Because of
the large size of the current path in real space, the magnetic form factor
should fall off very rapidly with ${\bf Q}_{\rm 2D}$ in reciprocal
space.  Mook {\it et al.}
\cite{mook02b} have done extensive measurements in search of the proposed
signal in \ybco\ with several values of $x$.  The measurements are
complicated by the fact that large crystals are required to achieve the
necessary sensitivity, while the largest crystals available are
contaminated by a significant amount of Y$_2$BaCuO$_5$.  Stock {\it et
al.} \cite{stoc02} studied a crystal of
\YBCO{6.5}\ with unpolarized neutrons, and concluded that no ordered
moment could be seen to a sensitivity of $\sim0.003$~$\mu_{\rm B}$. 
Using polarized neutrons, Mook {\it et al.} \cite{mook04} have seen, in
the spin-flip channel, a weak peak at \qaf\ on top of a large
background.  Without giving an error bar, they suggest that the
associated moment might be 0.0025~$\mu_{\rm B}$.  They concluded that
``the present results provide indications that orbital current phases are
not ruled out'' \cite{mook04}.

Varma \cite{varm97} has proposed a different model of ordered orbital
currents, in which the currents flow between a single Cu ion and its four
coplanar O neighbors.  This state breaks time-reversal and rotational
symmetry but not translational symmetry.  Thus, magnetic scattering from
the $c$-axis-oriented orbital moments should be superimposed on nuclear
Bragg scattering from the crystal lattice.  Information on the nature of
the orbital currents is contained in a strongly {\bf Q}-dependent
structure factor.  The only practical way to detect such a small magnetic
signal on top of the strong nuclear peaks would be
with polarized neutrons.  Lee {\it et al.} \cite{lee99b} performed
extensive polarized-beam studies on \lsco\ and \ybco\ single crystals. 
They found no positive evidence for the proposed magnetic moments, with a
sensitivity of 0.01~$\mu_{\rm B}$ in the case of 3D order, and
0.1~$\mu_{\rm B}$ in the case of quasi-2D order.  Simon and Varma
\cite{simo02} have since proposed a second pattern of orbital currents
that would have a different magnetic structure factor from the original
version.  Positive results in \ybco\ corresponding to this second pattern
have recently been reported by Fauqu{\'e} {\it et al.} \cite{fauq05}.

\section{Variation of magnetic correlations with doping and temperature in
cuprates}
\label{sc:doping}

\subsection{Magnetic incommensurability vs.\ hole doping}

\begin{figure}[t]
\centerline{
\includegraphics[width=2.15in]{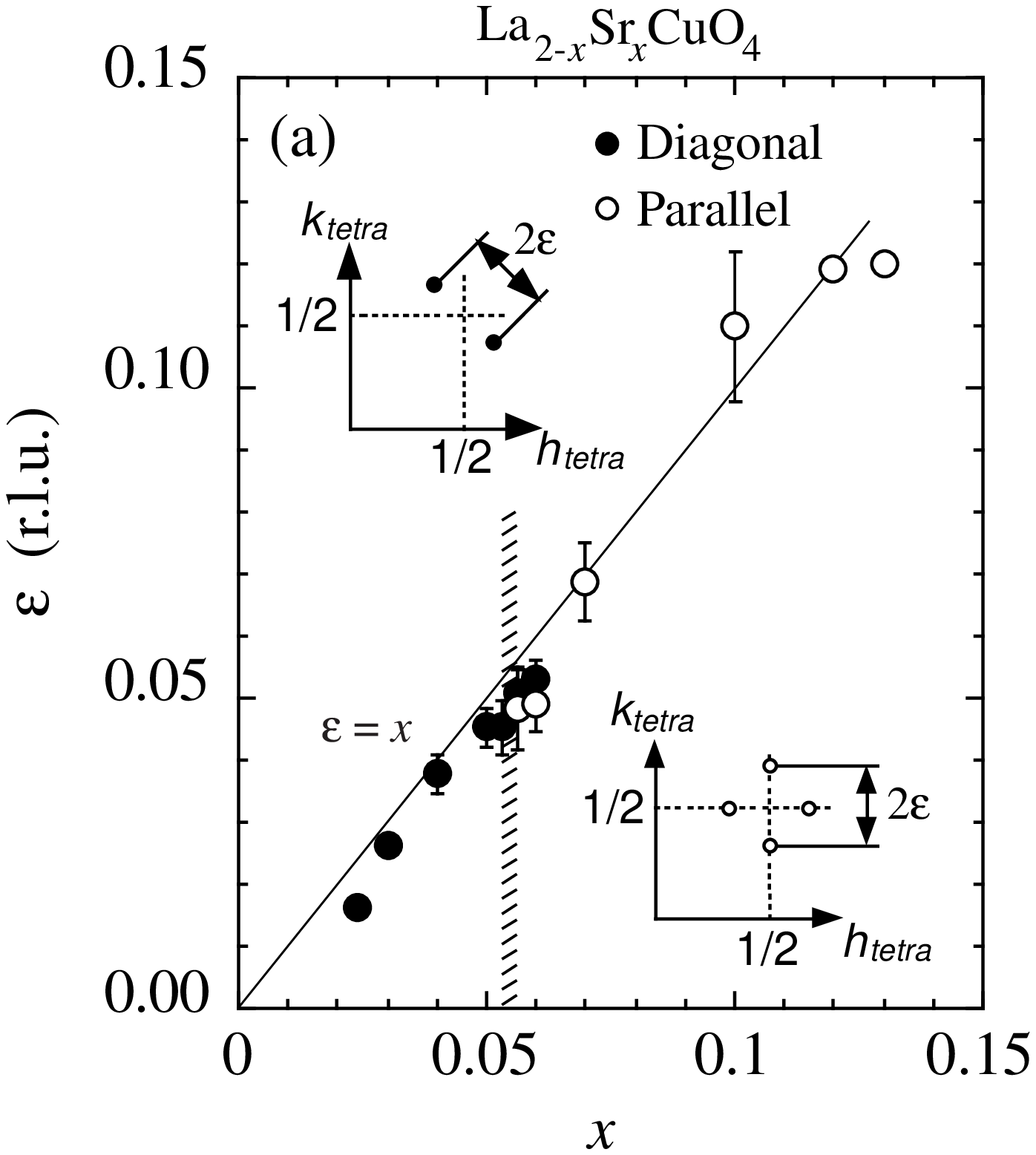}
\hspace{0.2in}
\raisebox{7pt}{\includegraphics[width=2.15in]{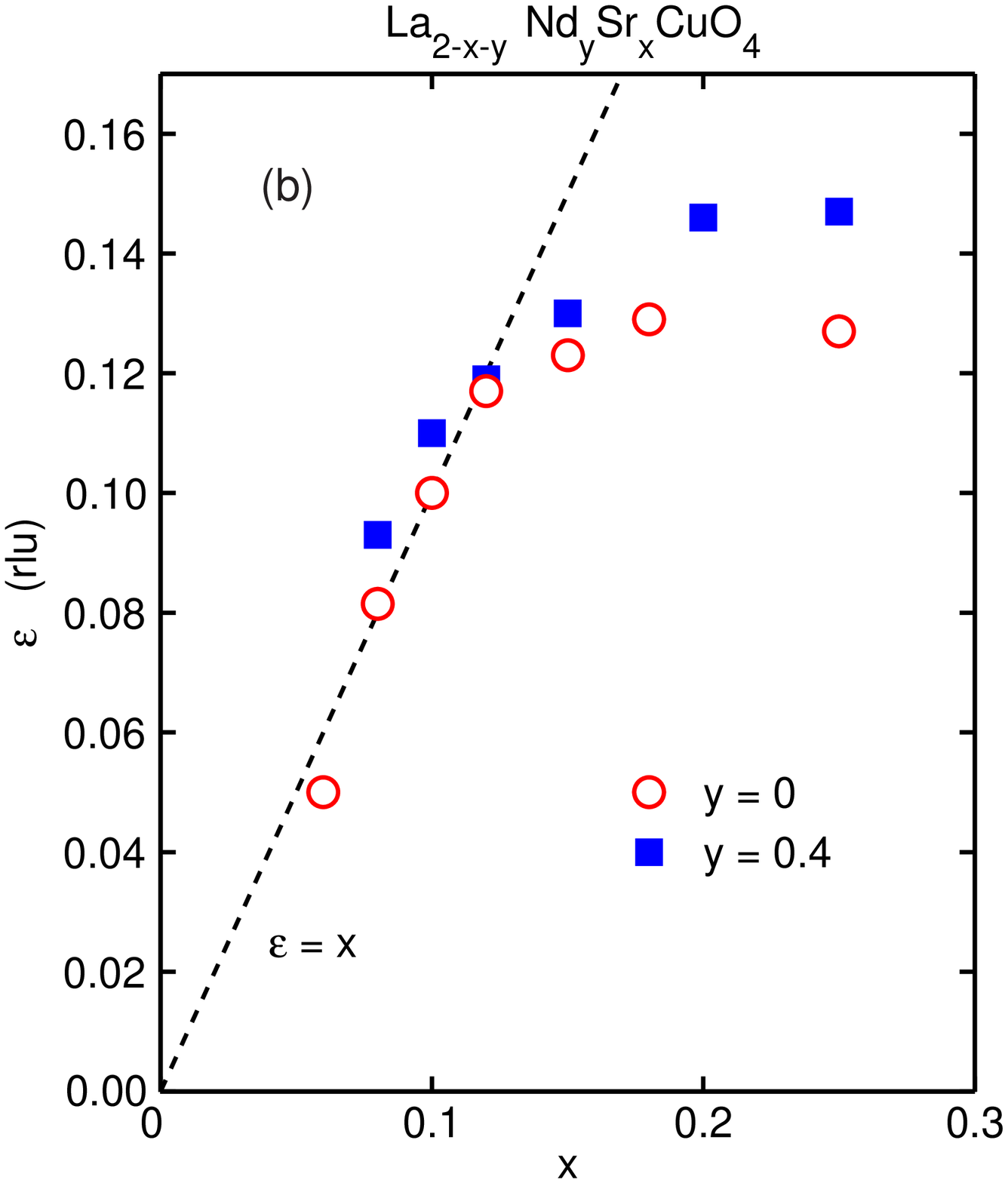}}
}
\medskip
\caption{Variation of the magnetic incommensurability $\epsilon$ [as
defined in the insets of (a)] for (a) lightly-doped \lsco, and (b)
\lsco\ with and without Nd-codoping.  In (a) the filled (open) symbols
correspond to diagonal (bond-parallel, or vertical) spin stripes. 
Adapted from Fujita {\it et al.} \cite{fuji02c}.  In (b), open
circles are from measurements of excitations at $E\sim3$~meV and
$T\approx T_{\rm c}$ in
\lsco\ from Yamada {\it et al.} \cite{yama98a}; filled squares are from
elastic scattering on \lnsco\ from Ichikawa {\it et al.} \cite{ichi00}.}
\label{fg:eps_vs_x}
\end{figure}

The doping dependence of the low-energy magnetic excitations in
superconducting \lsco\ have been studied in considerable detail
\cite{yama98a,fuji02c}.  In particular, the {\bf Q} dependence has been
characterized.  We already saw in \S\ref{sc:dest}\ that the destruction
of antiferromagnetic order by hole doping leads to diagonal spin
stripes.  As shown in Fig.~\ref{fg:eps_vs_x}(a), the magnetic
incommensurability $\epsilon$ grows roughly linearly with $x$ across the
``spin-glass'' regime.  (The results in this region are from elastic
scattering.)  At
$x\approx 0.055$, there is an insulator to superconductor transition, and
along with that is a rotation of the incommensurate peaks (as shown in the
insets), indicating a shift from diagonal to vertical (or bond-parallel)
stripes \cite{fuji02c}.  The rotation of the stripes is not as sharply
defined as is the onset of the superconductivity---there is a more gradual
evolution of the distribution of stripe orientations as indicated by the
measured peak widths, especially around the circle of radius $\epsilon$
centered on \qaf.  Interestingly, the magnitude of $\epsilon$ continues
its linear $x$ dependence through the onset of superconductivity.

In the superconducting phase, $\epsilon$ continues to grow with doping up
to $x\sim\frac18$, beyond which it seems to saturate, as indicated by the
circles \cite{yama98a} in Fig.~\ref{fg:eps_vs_x}(b).  Interestingly, the
same trend in incommensurability is found for static stripe order in
Nd-doped \lsco\ \cite{ichi00}, as indicated by the filled squares in the
same figure.  The differences in wave vector for a given $x$ may reflect
a change in the hole density of the charge stripes when they become
statically ordered in the anisotropic lattice potential of the LTT phase.

While low-energy incommensurate scattering is also observed in overdoped
\lsco, Wakimoto {\it et al.} \cite{waki04} have found that the magnitude
of $\chi''$, measured at $E\sim6$ meV, drops rapidly for $x>0.25$,
becoming neglible by $x=0.30$.  The decrease in the magnitude of $\chi''$
is correlated with the fall off in \tc.  Interestingly, these results
suggest that the superconductivity weakens as magnetic correlations
disappear.

\begin{figure}[t]
\centerline{
\includegraphics[width=2.3in]{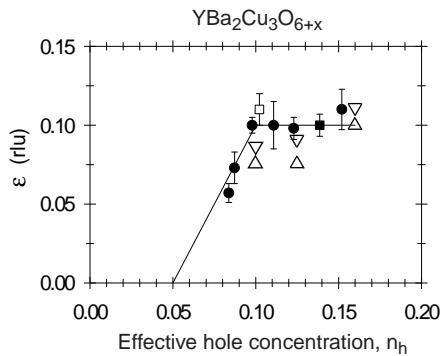}
}
\medskip
\caption{Magnetic incommensurability in \ybco\ (circles and squares)
measured just above the spin-gap energy at $T\ll T_{\rm c}$, with 
$n_{\rm h}$ estimated from $T_{\rm c}$, from Dai {\it et al.}
\cite{dai01}.  The triangles indicate the incommensurability at 20 meV
(upper) and 30 meV (lower) for \lsco\ with $x=0.10$ and 0.16
\cite{chri04} and \lbcoate\ \cite{tran04}.}
\label{fg:ybco_eps}
\end{figure}

In \ybco, the incommensurability of the magnetic excitations at 
$E<E_{\rm r}$ is resolvable only for $T\ll T_{\rm c}$.  The presence of a
substantial spin gap in the superconducting state, together with the
dispersion of the magnetic excitations, makes it difficult to compare
directly the results for \ybco\ with the behavior of \lsco\ shown in
Fig.~\ref{fg:eps_vs_x}(b).  Dai {\it et al.} \cite{dai01} have determined
$\epsilon$ at energies just above the spin gap; the results for \ybco\
are represented by circles and squares in Fig.~\ref{fg:ybco_eps}.  For
comparison, the triangles indicate the effective incommensurabilities
found at energies of 20 and 30 meV in
\lsco\ with $x=0.10$ and 0.16 \cite{chri04} and in \lbcoate\
\cite{tran04}.  The trends in the two different cuprate families seem
to be similar when one accounts for the dispersion.  (Comparable
behavior in \ybco\ and \lsco\ was also noted by Balatsky and Bourges
\cite{bala99}.)

\subsection{Doping dependence of energy scales}

The doping dependence of $E_{\rm r}$ in \ybco\ and \bscco\ has received
considerable attention.  In optimally-doped and slightly under-doped
\ybco, the scattering at $E_{\rm r}$ (for $T<T_{\rm c}$) is relatively
strong and narrow in {\bf Q} and $\omega$.  Of course, when the
intensity is integrated over {\bf Q} and $\omega$ one finds that it
corresponds to a small fraction of the total expected sum-rule weight
\cite{kee02}; it is also a small fraction of the 
the total spectral weight that is actually measured (which is much
reduced from that predicted by the sum rule \cite{lore05}).

\begin{figure}[t]
\centerline{
\includegraphics[width=2.15in]{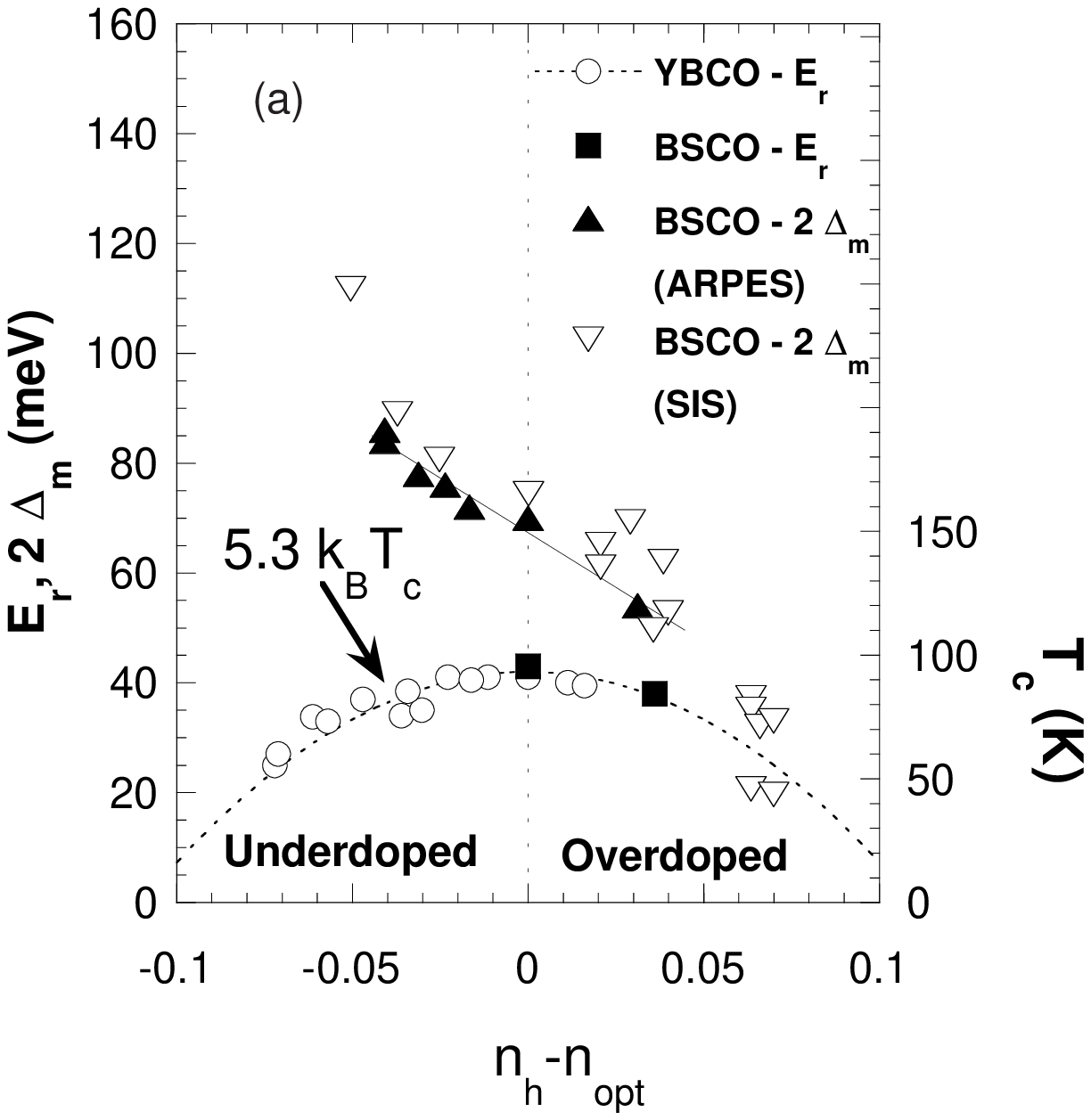}
\hspace{0.2in}
\raisebox{14pt}{\includegraphics[width=2.15in]{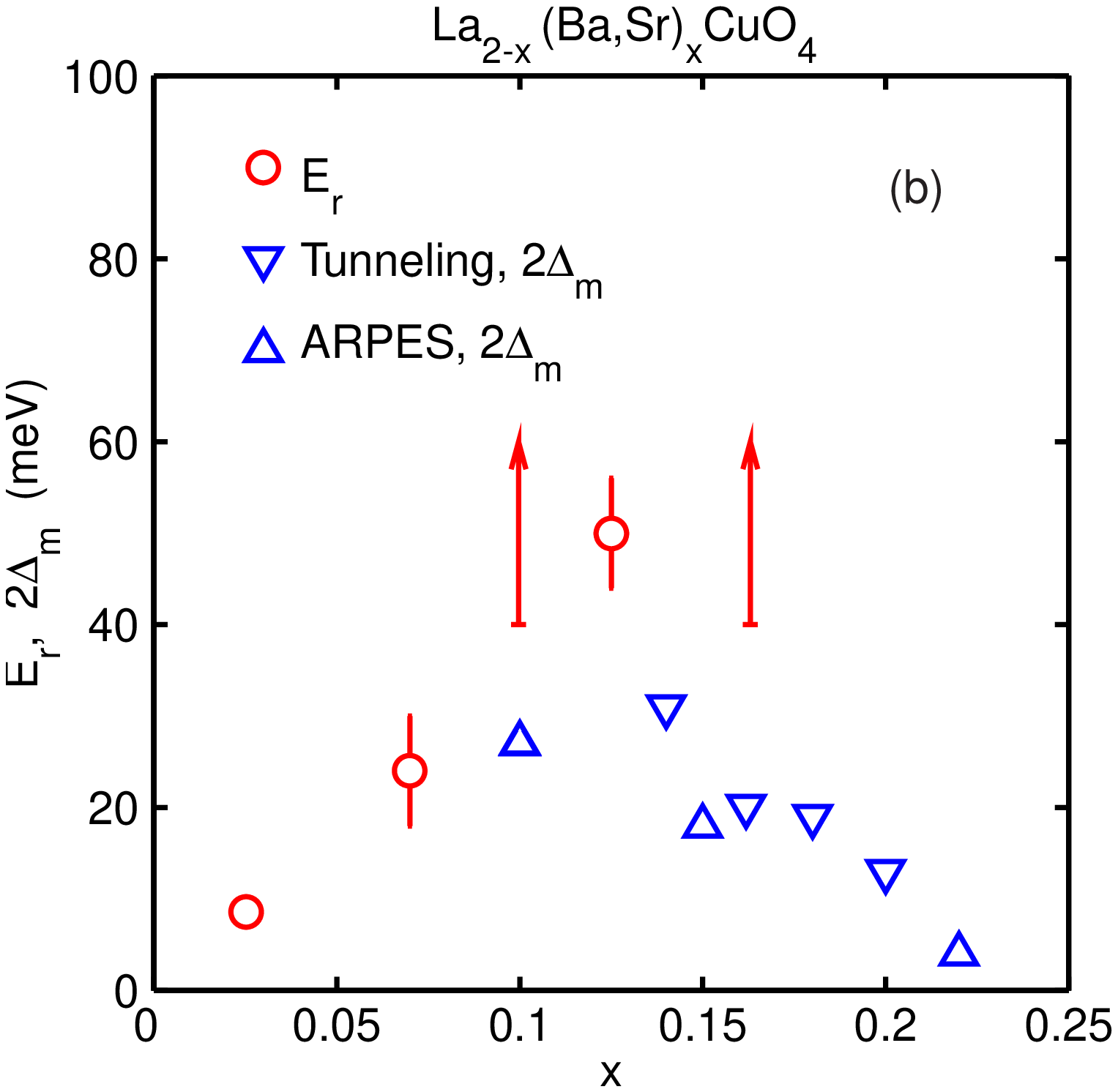}}
}
\medskip
\caption{(a) Summary of results for the resonance energy $E_{\rm r}$ from
neutron scattering measurements on \ybco\ (open circles) and \bscco\
(filled squares), and twice the maximum of the superconducting gap,
$2\Delta_{\rm m}$, from angle-resolved photoemission (ARPES, filled
triangles) and tunneling (open triangles) measurements on \bscco, taken
from Sidis {\it et al.} \cite{sidi04}.  (b)  $E_r$ (circles) for
\lbcoate\
\cite{tran04} and estimated for \lsco\ from measurements on $x=0.024$
\cite{mats00b},
$x=0.07$
\cite{hira01}, $x=0.10$ and 0.16 \cite{chri04};
$2\Delta_{\rm m}$ for \lsco\ from tunneling (downward triangles)
\cite{naka98b,kato05} and ARPES (upward triangles) \cite{ino02}. }
\label{fg:Er}
\end{figure}

Figure~\ref{fg:Er}(a) presents a summary, from Sidis {\it et al.}
\cite{sidi04}, of experimental results for
$E_{\rm r}$ from neutron scattering and for twice the superconducting gap
maximum,
$2\Delta_{\rm m}$, from other techniques.  For these materials, the
resonance energy is found to scale with \tc\ and fall below 
$2\Delta_{\rm m}$.  Unfortunately, a major deviation from these trends
occurs in \lsco\ [see Fig.~\ref{fg:Er}(b)], where $E_{\rm r}$ tends
to be larger than $2\Delta_{\rm m}$, and any constant of proportionality
between $E_{\rm r}$ and $kT_{\rm c}$ is considerably larger than the
value of $\sim5$ found for \ybco.

\begin{figure}[t]
\centerline{
\includegraphics[width=2.6in]{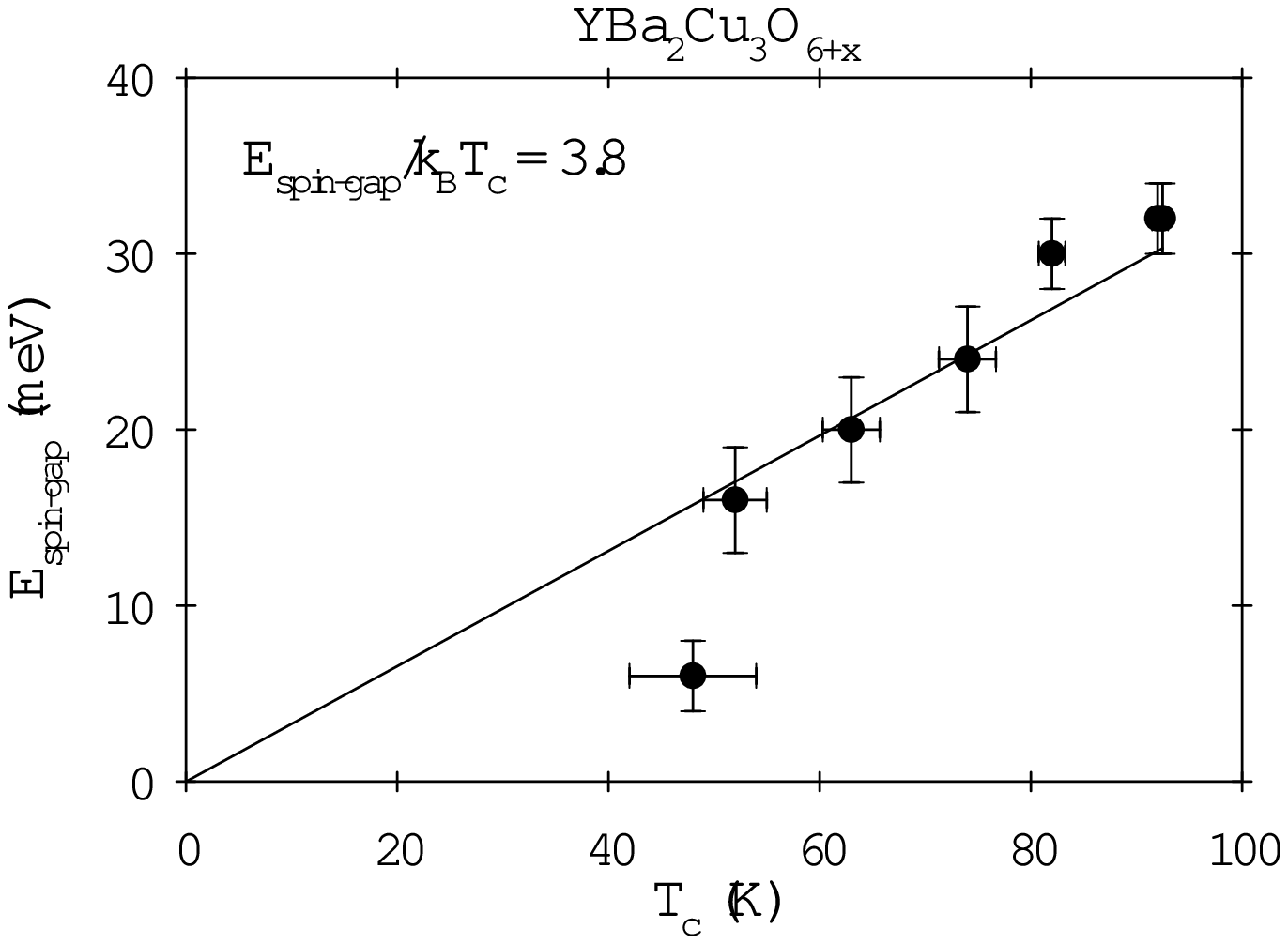}
}
\medskip
\caption{Spin-gap energy vs.\ $T_{\rm c}$ for \ybco\ from Dai {\it et
al.} \cite{dai01}.}
\label{fg:ybco_spin_gap}
\end{figure}

As discussed in \S\ref{sec:spin_gap}, there may be a more general
correlation between the spin-gap energy and \tc. 
Figure~\ref{fg:ybco_spin_gap} shows the variation of the spin-gap energy
with \tc\ for a range of dopings in \ybco\ as obtained by Dai {\it et
al.} \cite{dai01}.  The correlation seen there looks very much like that
shown in Fig.~\ref{fg:tc_gap} for different cuprate families at optimum
doping.  For \lsco, a true spin gap is not observed for $x<0.14$
\cite{lee00}, and this might have a connection with the rapid
disappearance of the spin gap in \ybco\ for $x<0.5$ \cite{dai01}. 

\subsection{Temperature-dependent effects}

A detailed study of the thermal evolution of the magnetic excitations
($E\le15$ meV) in La$_{1.86}$Sr$_{0.14}$CuO$_4$ was reported by Aeppli
{\it et al.} \cite{aepp97}.  Fitting the {\bf Q} dependence of the
incommensurate scattering with a lorentzian-squared peak shape, they
found that $\kappa$, the $Q$-width as a function of both frequency and
temperature, can be described fairly well by the formula
\begin{equation}
  \kappa^2 = \kappa_0^2 + {1\over a^2}\left[\left({kT\over E_0}\right)^2
   + \left({\hbar\omega\over E_0}\right)^2\right],
\end{equation}
where $\kappa_0=0.034$~\AA$^{-1}$, $a$ is the lattice parameter, and
$E_0=47$~meV.  For $T\ge T_{\rm c}$, the low-frequency limit of
$\chi''({\bf Q}_\delta,\omega)/\omega$ (where ${\bf Q}_\delta$ is a peak
position) varies with temperature essentially as $1/T^2$.  They argued
that these behaviors are consistent with proximity to a quantum critical
point, and that the type of ordered state that is being approached at low
temperature is the stripe-ordered state.

In a study of \lsco\ crystals at somewhat higher doping ($x=0.15$, 0.18,
and 0.20), Lee {\it et al.} \cite{lee03} found evidence for a spin
pseudogap at $T\ge T_{\rm c}$.  The pseudogap (with a hump above it) was
similar in energy to the spin gap that appears at $T<T_{\rm c}$ and was
most distinct in the $x=0.18$ sample, where the effect is still evident at
80 K but absent at 150 K.

For \ybco, the studies of temperature dependence have largely
concentrated on the scattering near $E_r$.  For fully-oxygenated
\YBCO{7}, the intensity at $E_r$ appears fairly abruptly at, or slightly
below, \tc\ and grows with decreasing temperature, with essentially no
shift in $E_r$ \cite{fong96,dai99}.  For underdoped samples, the intensity
at $E_r$ begins to grow below temperatures $T^\ast > T_{\rm c}$, with the
enhancement at \tc\ decreasing with underdoping
\cite{dai99,bour00,stoc04}.  

\section{Effects of perturbations on magnetic correlations}
\label{sc:perturb}

\subsection{Magnetic field}
\label{sc:mag_field}

An important initial study of the impact of an applied magnetic field on
magnetic correlations in a cuprate superconductor was done by Dai {\it et
al.} \cite{dai00} on \YBCO{6.6}\ ($T_{\rm c} = 63$ K).  They showed that
applying a 6.8-T field along the $c$-axis caused a 30\%\ reduction in the
low-temperature intensity of the resonance peak (at 34 meV).  The lost
weight presumably is shifted to other parts of phase space, but it was
not directly detected.  (Applying the field parallel to the CuO$_2$
planes has negligible effect.)  In an earlier study on
\YBCO{7}, Bourges {\it et al.} \cite{bour97b} applied an 11.5 T field and
found that the resonance peak broadened in energy but did not seem to
change its peak intensity.  The difference in response from \YBCO{6.6}\
is likely due to the difference in $H_{\rm c2}$, which is about 5 times
larger in \YBCO{7}\ \cite{ando02b}.

\begin{figure}[t]
\centerline{
\includegraphics[width=2.6in]{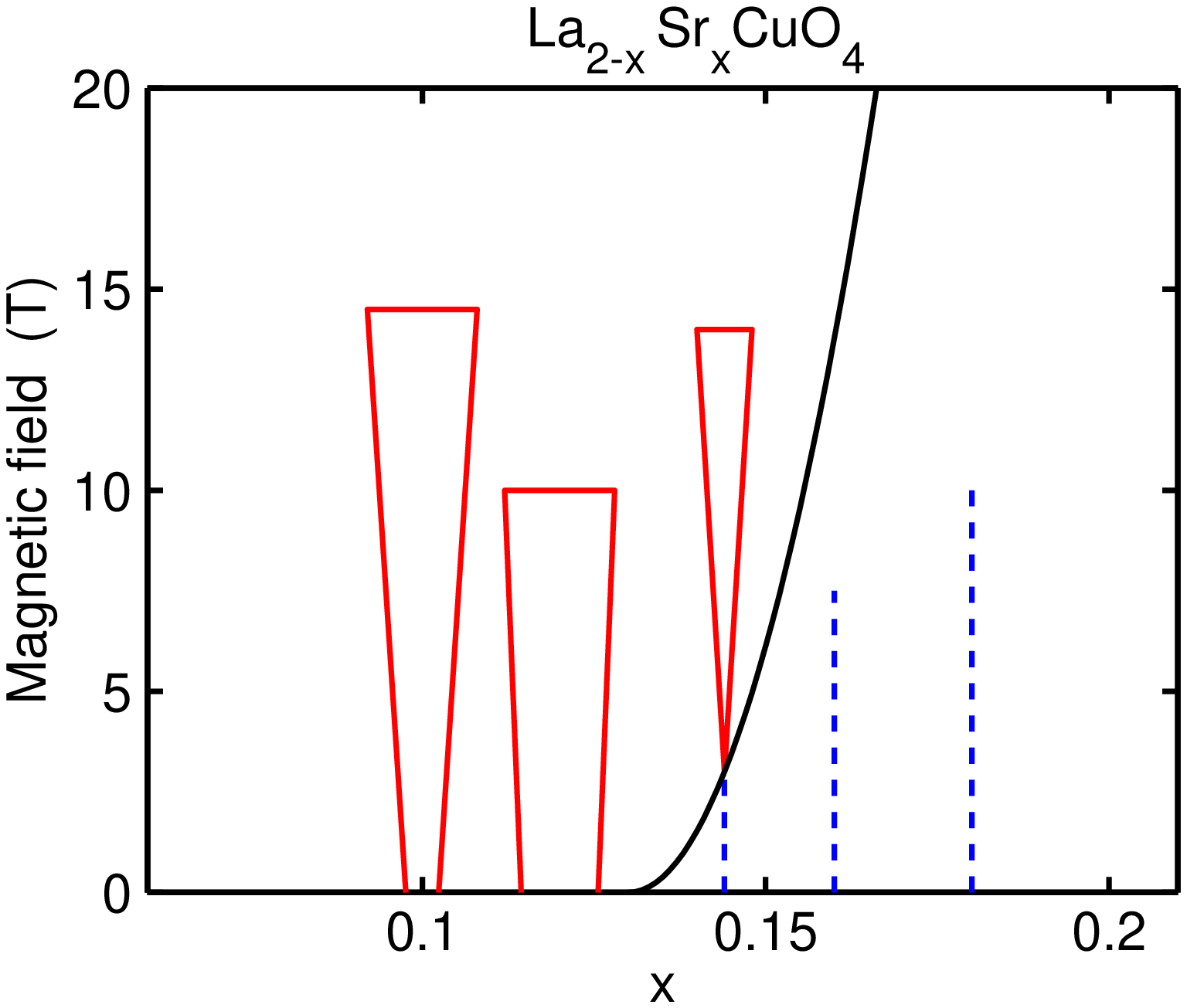}
}
\medskip
\caption{Schematic summary of neutron scattering experiments on \lsco\ in
a magnetic field at $T\ll T_{\rm c}$.  Solid bars indicate observations
of elastic, incommensurate peaks; width indicates variation of peak
intensity with field.  Experiments on $x=0.10$, 0.12, 0.144, 0.163, and
0.18 from \cite{lake02},\cite{kata00}, \cite{khay05}, \cite{lake01}, and
\cite{tran04b}. The solid curve suggests the shape of a boundary between
a state with spin-density-wave order and superconductivity on the left
and superconductivity alone on the right, as first proposed by Demler
{\it et al.} \cite{deml01}.}
\label{fg:lsco_field}
\end{figure}

A series of studies on \lsco\ samples with various dopings have now been
performed \cite{kata00,lake01,lake02,tran04b,kata00}, and a schematic
summary of the results is presented in Fig.~\ref{fg:lsco_field}.  For
samples with lower doping ($x=0.10$ \cite{lake02} and 0.12 \cite{kata00})
there is a small elastic, incommensurate, magnetic peak intensity in zero
field that is substantially enhanced by application of a $c$-axis magnetic
field.  The growth of the intensity with field is consistent with 
\begin{equation}
  I \sim (H/H_{\rm c2}) \ln(H_{\rm c2}/H),
\end{equation}
where $H_{\rm c2}$ is the upper critical field for superconductivity
\cite{lake02}.  This behavior was predicted by Demler {\it et al.}
\cite{deml01} using a model of coexisting but competing phases of
superconductivity and spin-density-wave (SDW) order.  In their model,
local reduction of the superconducting order parameter by magnetic
vortices results in an average increase in the SDW order.  (For an
alternative approach, in which the competing order is restricted to
``halo'' regions centered on vortex cores, see, e.g., \cite{kive02}.)  
Interestingly, the spin-spin correlation length for the induced signal is
$>400$~\AA, which is at least 20 times greater than the radius of a
vortex core
\cite{lake02}.  Very similar results have been obtained on oxygen-doped
\lco\ \cite{khay02,khay03}.  There is an obvious parallel with the
charge-related ``checkerboard'' pattern observed at vortices in
superconducting \bscco\ by scanning tunneling microscopy \cite{hoff02}.

For \lsco\ crystals with $x=0.163$ \cite{lake01} and 0.18 \cite{tran04b}
there is no field-induced static order (at least for the range of fields
studied).  Instead, the field moves spectral weight into the spin gap. 
The study on $x=0.18$ indicated that the increase in weight in the gap is
accompanied by a decrease in the intensity peak above the gap
\cite{tran04b}, the latter result being comparable to the effect seen in
\YBCO{6.6} \cite{dai00}.  For $x=0.163$, an enhancement of the
incommensurate scattering was observed below 10 K for $\hbar\omega=2.5$
meV.

For an intermediate doping concentration of $x=0.144$, Khaykovich {\it et
al.} \cite{khay05} have recently shown that, although no elastic magnetic
peaks are seen at zero field, a static SDW does appear for 
$H>H_c\sim3$~T.  Such behavior was predicted by the model of competing
phases of Demler {\it et al.} \cite{deml01}, and a boundary between
phases with and without SDW order, based on that model, is indicated by
the solid curve in Fig.~\ref{fg:lsco_field}.

Although evidence for field-induced charge-stripe order in \lsco\ has not
yet been reported, it seems likely that the SDW order observed is the
same as the stripe phase found in \lbcoate\ \cite{fuji04} and\linebreak
\lnsco\
\cite{ichi00}.  Consistent with this scenario, it has been shown that an
applied magnetic field has no impact on the Cu magnetic order or the
charge order in the stripe-ordered phase of \lnsco\ with $x=0.15$
\cite{waki03}; however, the field did effect the ordering of the Nd
``spectator'' moments.

Returning to \lsco\ with $x<0.13$, it has been argued in the case of
$x=0.10$ that the zero-field elastic magnetic peak intensity observed at
low temperature is extrinsic \cite{lake02}.  This issue deserves a short
digression.  It is certainly true that crystals of lesser quality can
yield elastic scattering at or near the expected positions of the
incommensurate magnetic peaks; in some cases, this scattering has little
temperature dependence.  Of course, just because spurious signals can
occur does not mean that all signals are spurious.  Let us shift our
attention for a moment to $x=0.12$, where the low-temperature, zero-field
intensity is somewhat larger \cite{suzu98}.   A muon-spin-relaxation
($\mu$SR) study \cite{savi02} on a crystal of good quality has shown that
the magnetic order is not uniform in the sample---at low temperature, only
$\sim20$\%\ of the muons see a static local hyperfine field.  Further
relevant information comes from electron diffraction studies.  The
well-known low-temperature orthorhombic (LTO) phase tends to exhibit twin
domains.  Horibe, Inoue, and Koyama \cite{hori97} have taken dark-field
images using a Bragg peak forbidden in the LTO structure but allowed in
the LTT structure, the phase that pins stripes in \lbcoate\ and \lnsco. 
They find bright lines corresponding to the twin boundaries, indicating
that the structure of the twin boundaries is different from the LTO phase
but similar to the LTT.  (Similar behavior has been studied in \lbco\
\cite{zhu94}.)  The twin boundaries are only a few nanometers wide;
however, given that magnetic vortices can pin spin stripes with a
substantial correlation length, and we will see next that Zn dopants can
also pin spin stripes, it seems likely that LTT-like twin boundaries
should be able to pin stripe order with a significant correlation
length.  Thus, the low-temperature magnetic peaks found in \lsco\ with
$x=0.12$ \cite{suzu98} are likely due to stripes pinned at twinned
boundaries, giving order in only a small volume fraction, consistent with
$\mu$SR \cite{savi02}.  Taking into account the fact that stripe order is
observed in \lnsco\ for a substantial range of $x$ (but with strongest
ordering at $x=0.12$) \cite{ichi00}, it seems reasonable to expect a
small volume fraction of stripe order pinned at twin boundaries in \lsco\
with $x=0.10$.  Is this order extrinsic?  Are twin boundaries extrinsic? 
This may be a matter of semantics.  In any case, I would argue that the
low-temperature zero-field peaks measured in good crystals reflect real
materials physics of the pure compound.

\subsection{Zn substitution}

The effects of Zn substition for Cu are quite similar to those caused by
an applied magnetic field.  For \lsco\ with $x=0.15$, substituting about
1\%\ or less Zn causes the appearance of excitations within the spin gap
of the Zn-free compound \cite{kimu03b,kofu05}.  Substitution of 1.7\%\ Zn
is sufficient to induce weak elastic magnetic peaks.  For $x=0.12$, where
weak elastic magnetic peaks are present without Zn, substitution of Zn
increases the peak intensity, but also increases the $Q$-widths of the
peaks \cite{kimu99,tran99b}.  Wakimoto {\it et al.} \cite{waki05} have
recently found that Zn-substitution into overdoped samples ($x>0.2$)
significantly enhances the low-energy ($<10$ meV) inelastic magnetic
scattering.

In \ybco, Zn substitution causes weight to shift from $E_r$ into the spin
gap \cite{sidi96,kaku93}.  While it causes some increase in the $Q$-width
of the scattering at $E_r$ \cite{sidi00}, it does not make a significant
change in the {\bf Q} dependence of the (unresolved) incommensurate
scattering at lower energies \cite{kaku93}.  Muon-spin rotation studies
indicate that Zn-doping reduces the superfluid density proportional to
the Zn concentration \cite{nach96}, and this provides another parallel
with the properties of the magnetic vortex state.

\subsection{Li-doping}

An alternative way to dope holes into the CuO$_2$ planes is to substitute
Li$^{1+}$ for Cu$^{2+}$.  In this case, the holes are introduced at the
expense of a strong local Coulomb potential that one might expect to
localize the holes.  Surprisingly, the magnetic phase diagram of
La$_{2}$Cu$_{1-x}$Li$_x$O$_4$ is rather similar to that for \lsco\ with
$x<0.06$ \cite{sasa02}.  In particular, the long-range N\'eel order is
destroyed with $\sim0.03$\%\ Li.  The nature of the magnetic correlations
in the paramagnetic phase is different from that in \lsco\ in that the
inelastic magnetic scattering remains commensurate \cite{bao00}.  Studies
of the spin dynamics indicate $\omega/T$ scaling at high temperatures,
but large deviations from such behavior occur at low temperature
\cite{bao03}.

\section{Electron-doped cuprates}
\label{sc:e-doped}

Electron-doped cuprates are very interesting because of their
similarities and differences from the hole-doped materials; however,
considerably less has been done in the way of neutron scattering on the
electron-doped materials, due in part to challenges in growing crystals
of suitable size and quality.  Initial work focused on the systems \ncco\
and \pcco.  A striking difference from hole doping is the fact that the
N\'eel temperature is only gradually reduced by electron-doping.  This
was first demonstrated in a $\mu$SR study of \ncco\ \cite{luke90}, where
it was found that the antiferromagnetic order only disappears at 
$x\approx0.14$ where superconductivity first appears.  The magnetic order
was soon confirmed by neutron diffraction measurements on single crystals
of \pcco\ \cite{thur90} and \ncco\ \cite{zobk91}.

A complication with these materials is
that to obtain the superconducting phase, one must remove a small amount
of oxygen from the as-grown samples.  The challenge of the reduction
process is to obtain a uniform oxygen concentration in the final sample. 
This is more easily done in powders and thin films than in large
crystals.  As grown crystals with $x$ as large as 0.18 are
antiferromagnetic \cite{mats92,mang04a}.  Reducing single crystals can
result in superconductivity; however, it is challenging to completely
eliminate the antiferromagnetic phase \cite{mats92}.  In trying to get a
pure superconducting phase, the reducing conditions can sometimes cause a
crystal to undergo partial decomposition, yielding impurity phases such
as (Nd,Ce)$_2$O$_3$ \cite{mang03,mang04b}.

The effective strength of the spin-spin coupling has been probed through
measurements of the spin correlation length as a function of temperature
in the paramagnetic phase.  The magnitude of the spin stiffness is
clearly observed to decrease with doping \cite{thur90,mats92,mang04a}. 
Mang {\it et al.} \cite{mang04a} have shown that this behavior is
consistent with that found in numerical simulations of a randomly
site-diluted 2D antiferromagnet.  In the model calculations, the
superexchange energy is held constant, and the reduction in spin
stiffness is due purely to the introduction of a finite concentration of
nonmagnetic sites.  To get quantitative agreement, it is necessary to
allow for the concentration of nonmagnetic sties in the model to be about
20\%\ greater than the Ce concentration in the samples.

Yamada and coworkers \cite{yama03} were able to prepare crystals
of\linebreak 
Nd$_{1.85}$Ce$_{0.15}$CuO$_4$ with sufficient quality that it
was possible to study the low-energy magnetic excitations associated with
the superconducting phase.  They found commensurate antiferromagnetic
fluctuations.  In a crystal with $T_{\rm c}=25$~K, they found that a spin
gap of approximately 4 meV developed in the superconducting state. 
Commensurate elastic scattering, with an in-plane correlation length of
150~\AA, was also present for temperatures below $\sim60$~K; however, the
growth of the elastic intensity did not change on crossing the
superconducting \tc.

While the magnetic excitations are commensurate and incompatible with
stripe correlations, there are, nevertheless, other measurements that
suggest electronic inhomogeneity.  Henggeler {\it et al.} \cite{heng95}
used the crystal-field excitations of the Pr ions in \pcco\ as a probe of
the local environment.  They found evidence for several distinct local
environments, and argued that doped regions reached the percolation
limit at $x\approx0.14$, at the phase boundary for superconductivity. 
Recent NMR studies have also found evidence of electronic inhomogeneity
\cite{zamb04, bakh04}.

Motivated by the observation of magnetic-field-induced magnetic
superlattice peaks in hole-doped cuprates (\S\ref{sc:mag_field}), there
has been a series of experiments looking at the effect on electron-doped
cuprates of a field applied along the $c$ axis.  An initial study
\cite{mats02b} on \ncco\ with $x=0.14$ and $T_{\rm c}\sim25$~K found that
applying a field as large as 10~T had no effect on the intensity of an
antiferromagnetic Bragg peak for temperatures down to 15~K.  Shortly after
that came a report \cite{kang03} of large field-induced enhancements of
antiferromagnetic Bragg intensities, as well as new field-induced peaks of
the type $(\frac12,0,0)$, in a crystal of \ncco\ with $x=0.15$ and
$T_{\rm c} = 25$~K.  It was soon pointed out that the new $(\frac12,0,0)$
peaks, as well as most of the effects at antiferromagnetic reflections,
could be explained by the magnetic response of the (Nd,Ce)$_2$O$_3$
impurity phase \cite{mang03,mang04b}.  There now seems to be a consensus
that this is the proper explanation \cite{mats03,mats04}; however, a
modest field-induced intensity enhancement has been seen at
$(\frac12,\frac12,3)$ that is not explained by the impurity-phase model
\cite{mats03}.

In an attempt to clarify the situation, Fujita {\it et al.}
\cite{fuji04b} turned to another electron-doped superconductor,
Pr$_{1-x}$LaCe$_x$CuO$_4$.  This compound also has to be reduced to
obtain superconductivity, and reduced crystals exhibit a (Pr,Ce)$_2$O$_3$
impurity phase; however, the Pr in the impurity phase should not be
magnetic.  They found a weak field-induced enhancement of an
antiferrromagnetic peak intensity for a crystal with $x=0.11$  
($T_{\rm c}=26$~K), but no effect for $x=0.15$ ($T_{\rm c}=16$~K).  The
induced Cu moment for $x=0.11$ at a temperature of 3~K and a field of 5~T
is $\sim10^{-4}$ $\mu_{\rm B}$.  Dai and coworkers \cite{dai05,kang05}
have studied crystals of Pr$_{0.88}$LaCe$_{0.12}$CuO$_{4-\delta}$ in which
they have tuned the superconductivity by adjusting $\delta$.  They have
emphasized the coexistence of the superconductivity with both 3D and
quasi-2D antiferromagnetic order \cite{dai05}.  They report a very slight
enhancement of the quasi-2D antiferromagnetic signal for a $c$-axis
magnetic field \cite{kang05}.  

\section{Discussion}
\label{sc:disc}

\subsection{Summary of experimental trends in hole-doped cuprates}

There are a number of trends in hole-doped cuprates that one can identify
from the results presented in this chapter.  To begin with, the undoped
parent compounds are Mott-Hubbard (or, more properly, charge-transfer)
insulators that exhibit N\'eel order due to antiferromagnetic
superexchange interactions between nearest-neighbor atoms.  The magnitude
of $J$ is material dependent, varying between roughly 100 and 150 meV.  

Doping the CuO$_2$ planes with holes destroys the N\'eel order; in fact,
the presence of holes seems to be incompatible with long-range
antiferromagnetic order.  The observed responses to hole doping indicate
that some sort of phase separation is common.  In some cases, stripe
modulations are found, and in others, finite clusters of
antiferromagnetic order survive.  

In under- and optimally-doped cuprate superconductors, the magnetic
spectrum has an hour-glass-like shape, with an energy scale comparable to
the superexchange energy of the parent insulators.  The strength of the
magnetic scattering, when integrated over momentum and energy, decreases
gradually as one increases the hole concentration from zero to optimal
doping.  A spin gap appears in the superconducting state (at least for
optimal doping), with spectral weight from below the spin gap being
pushed above it.  The magnitude of the spin gap correlates with $T_c$.

Underdoped cuprates with a small or negligible spin gap are very
sensitive to perturbations.  Substituting non-magnetic Zn for Cu or
applying a magnetic field perpendicular to the planes tends to induce
elastic incommensurate magnetic peaks at low temperature.  For samples
with larger spin gaps, the perturbations shift spectral weight from
higher energy into the spin gap.  Breaking the equivalence between
orthogonal Cu-O bonds within a CuO$_2$ plane can result in charge-stripe
order, in addition to the elastic magnetic peaks.

The magnetic correlations within the CuO$_2$ planes are clearly quite
sensitive to hole doping and superconductivity.  While their coexistence
with a metallic normal state is one of the striking characteristics of
the cuprates, their connection to the mechanism of hole-pairing remains a
matter of theoretical speculation.

\subsection{Theoretical interpretations}

The nature and relevance of antiferromagnetic correlations has been a
major theme of much of the theoretical work on cuprate superconductors. 
While some the theoretical concepts are discussed in more detail in other
chapters of this book, it seems appropriate to briefly review some of them
here.

Given that techniques for handling strongly-correlated hole-doped
antiferromagnets continue to be in the development stage, some
researchers choose to rely on a conventional weak-coupling approach to
describing magnetic metals.  This might be appropriate if one imagines
starting out in the very over-doped regime, where Fermi-liquid theory
might be applicable, and then works downward towards optimum doping. 
The magnetic susceptibility can be calculated in terms of electrons
being excited across the Fermi level from filled to empty states. 
Interactions between quasiparticles due to Coulomb or exchange
interactions are assumed to enhance the susceptibility near \qaf, and
this is handled using the random-phase approximation (RPA).  In the
superconducting state, one takes into account the superconducting gap
$\Delta$ with $d$-wave symmetry.  The gapping of states carves holes into
the continuum of electron-hole excitations.  The RPA enhancement can then
pull resonant excitations down into the region below
$2\Delta$ \cite{kao00,norm00,brin99}. With this approach, it has been
possible, with suitable adjustment of the interaction parameter, to
calculate dispersing features in $\chi''$ that resemble those measured in
the superconducting state of optimally-doped
\ybco\ \cite{chub01,erem05a,erem05b}.

The RPA approach runs into difficulties when one considers \lsco, \lbco,
and underdoped \ybco. It predicts that the magnetic excitations
should be highly over damped at energies greater than $2\Delta$; however,
there is no obvious change in the experimental spectra at $E>2\Delta$ in
these materials.  Furthermore, the dispersive features in \lbcoate\ are
observed in the normal state.  Even if one tries to invoke a $d$-wave
pseudogap, the energy scale is likely to be too small, as indicated by
Fig.~\ref{fg:Er}(b).  It is also unclear how one would rationalize, from
a Fermi-liquid perspective, the observation that the energy scale of the
magnetic excitations is of order $J$, as superexchange is an effective
interaction between local moments in a correlated insulator, and has
no direct connection to interactions between quasiparticles \cite{ande97}.

The fact that superexchange seems to remain relevant in the
superconducting phase suggests that it may be profitable to approach the
problem from the perspective of doped antiferromagnets.  The
resonating-valence-bond model was one of the first such attempts
\cite{ande87,ande04,lee06}.  The model is based on the assumption that the
undoped system is a quantum spin liquid.  In such a state, all Cu spins
would be paired into singlets in a manner such that the singlet-triplet
spectrum is gapless.  When a hole is introduced, one singlet is
destroyed, yielding a free spinon; all other Cu spins still couple in
singlet states.  In such a picture, one would expect that the
singlet-triplet excitations would dominate the magnetic excitation
spectrum measured with neutrons; surprisingly, there has been little
effort to make specific theoretical predictions of this spectrum for
comparison with experiment.  Instead, the analysis has been done in terms
of electron-hole excitations \cite{brin99}.  As discussed above, such a
calculation has significant short comings when it comes to understanding
underdoped cuprates.

Another alternative is a spiral spin-density wave, as has been proposed
several times \cite{shra89,hass04,sush04,lind05}.  A spiral state would be
compatible with the incommensurate antiferromagnetic excitations at low
energy \cite{shra89,sush04}, and can also be used to model the full
magnetic spectrum \cite{lind05}.  A look at the experimental record shows
that a spiral phase cannot be the whole story. In the case of
\lbcoate\ and \lnsco, where static magnetic order is observed, charge
order is also found \cite{abba05}. When there is charge order present, it
follows that the spin-density modulation must have a collinear component
in which the magnitude of the local moments is modulated \cite{zach98}. 
There could also be a spiral component, but it is not essential. 
Furthermore, if holes simply cause a
local rotation of the spin direction, then it is unclear why the
ordering temperature of the N\'eel phase is so rapidly reduced by a small
density of holes.

Given that stripe order is observed in certain cuprates
(\S\ref{sc:cu_stripes}) and that the magnetic excitations of the
stripe-ordered phase are consistent with the universal spectrum of good
superconductors (Fig.~\ref{fg:lbco_disp}), the simplest picture that is
compatible with all of the data is to assume that charge stripes (dynamic
ones in the case of the superconducting samples) are a common feature of
the cuprates, at least on the underdoped side of the phase diagram. 
There is certainly plenty of theoretical motivation for stripes
\cite{zaan01,kive03,sach91,mach89}, and the relevance of charge
inhomogeneity to the superconducting mechanism is discussed in the
chapter by Kivelson and Fradkin \cite{kive05}.

One suprising experimental observation is the minimal amount of damping
of the magnetic excitations in underdoped cuprates, especially in the
normal state.  One would expect the continuum of electron-hole excitations
to cause significant damping \cite{lore05}.  Could it be that the
antiphase relationship of spin correlations across a charge stripe acts
to separate the spin and charge excitations in a manner similar to that in
a one-dimensional system \cite{kive96,krui04}?  With over doping, there is
evidence that regions of conventional electronic excitations become more
significant.  This is also the regime where magnetic excitations become
weak.  Could it be that the interaction of conventional electron-hole
excitations with stripe-like patches causes a strong damping of the spin
excitations?  There is clearly plenty of work left to properly understand
the cuprates.

\section*{Acknowledgments}

I am grateful to S. A. Kivelson and M. H\"ucker for critical comments.
My work at Brookhaven is supported by the Office of Science,
U.S. Department of Energy, under Contract No.\ DE-AC02-98CH10886.


\begin{thebibliography}{282}
\expandafter\ifx\csname natexlab\endcsname\relax\def\natexlab#1{#1}\fi
\expandafter\ifx\csname bibnamefont\endcsname\relax
  \def\bibnamefont#1{#1}\fi
\expandafter\ifx\csname bibfnamefont\endcsname\relax
  \def\bibfnamefont#1{#1}\fi
\expandafter\ifx\csname citenamefont\endcsname\relax
  \def\citenamefont#1{#1}\fi
\expandafter\ifx\csname url\endcsname\relax
  \def\url#1{\texttt{#1}}\fi
\expandafter\ifx\csname urlprefix\endcsname\relax\def\urlprefix{URL }\fi
\providecommand{\bibinfo}[2]{#2}
\providecommand{\eprint}[2][]{\url{#2}}

\bibitem[{\citenamefont{Anderson}(1987)}]{ande87}
\bibinfo{author}{\bibfnamefont{P.~W.} \bibnamefont{Anderson}},
  \bibinfo{journal}{Science} \textbf{\bibinfo{volume}{235}},
  \bibinfo{pages}{1169} (\bibinfo{year}{1987}).

\bibitem[{\citenamefont{Vaknin et~al.}(1987)\citenamefont{Vaknin, Sinha,
  Moncton, Johnston, Newsam, Safinya, and H.~E.~King}}]{vakn87}
\bibinfo{author}{\bibfnamefont{D.}~\bibnamefont{Vaknin}},
  \bibinfo{author}{\bibfnamefont{S.~K.} \bibnamefont{Sinha}},
  \bibinfo{author}{\bibfnamefont{D.~E.} \bibnamefont{Moncton}},
  \bibinfo{author}{\bibfnamefont{D.~C.} \bibnamefont{Johnston}},
  \bibinfo{author}{\bibfnamefont{J.~M.} \bibnamefont{Newsam}},
  \bibinfo{author}{\bibfnamefont{C.~R.} \bibnamefont{Safinya}},
  \bibnamefont{and}
  \bibinfo{author}{\bibfnamefont{J.}~\bibnamefont{H.~E.~King}},
  \bibinfo{journal}{Phys. Rev. Lett.} \textbf{\bibinfo{volume}{58}},
  \bibinfo{pages}{2802} (\bibinfo{year}{1987}).

\bibitem[{\citenamefont{Kastner et~al.}(1998)\citenamefont{Kastner, Birgeneau,
  Shirane, and Endoh}}]{kast98}
\bibinfo{author}{\bibfnamefont{M.~A.} \bibnamefont{Kastner}},
  \bibinfo{author}{\bibfnamefont{R.~J.} \bibnamefont{Birgeneau}},
  \bibinfo{author}{\bibfnamefont{G.}~\bibnamefont{Shirane}}, \bibnamefont{and}
  \bibinfo{author}{\bibfnamefont{Y.}~\bibnamefont{Endoh}},
  \bibinfo{journal}{Rev. Mod. Phys.} \textbf{\bibinfo{volume}{70}},
  \bibinfo{pages}{897} (\bibinfo{year}{1998}).

\bibitem[{\citenamefont{Bourges}(1998)}]{bour98}
\bibinfo{author}{\bibfnamefont{P.}~\bibnamefont{Bourges}}, in
  \emph{\bibinfo{booktitle}{The Gap Symmetry and Fluctuations in High
  Temperature Superconductors}}, edited by
  \bibinfo{editor}{\bibfnamefont{J.}~\bibnamefont{Bok}},
  \bibinfo{editor}{\bibfnamefont{G.}~\bibnamefont{Deutscher}},
  \bibinfo{editor}{\bibfnamefont{D.}~\bibnamefont{Pavuna}}, \bibnamefont{and}
  \bibinfo{editor}{\bibfnamefont{S.~A.} \bibnamefont{Wolf}}
  (\bibinfo{publisher}{Plenum}, \bibinfo{address}{New York},
  \bibinfo{year}{1998}), p. \bibinfo{pages}{349}.

\bibitem[{\citenamefont{Regnault et~al.}(1998)\citenamefont{Regnault, Bourges,
  and Burlet}}]{regn98}
\bibinfo{author}{\bibfnamefont{L.~P.} \bibnamefont{Regnault}},
  \bibinfo{author}{\bibfnamefont{P.}~\bibnamefont{Bourges}}, \bibnamefont{and}
  \bibinfo{author}{\bibfnamefont{P.}~\bibnamefont{Burlet}}, in
  \emph{\bibinfo{booktitle}{Neutron Scattering in Layered Copper-Oxide
  Superconductors}}, edited by
  \bibinfo{editor}{\bibfnamefont{A.}~\bibnamefont{Furrer}}
  (\bibinfo{publisher}{Kluwer Academic}, \bibinfo{address}{Dordrecht},
  \bibinfo{year}{1998}), pp. \bibinfo{pages}{85--134}.

\bibitem[{\citenamefont{Hayden}(1998)}]{hayd98}
\bibinfo{author}{\bibfnamefont{S.~M.} \bibnamefont{Hayden}}, in
  \emph{\bibinfo{booktitle}{Neutron Scattering in Layered Copper-Oxide
  Superconductors}}, edited by
  \bibinfo{editor}{\bibfnamefont{A.}~\bibnamefont{Furrer}}
  (\bibinfo{publisher}{Kluwer Academic}, \bibinfo{address}{Dordrecht},
  \bibinfo{year}{1998}), pp. \bibinfo{pages}{135--164}.

\bibitem[{\citenamefont{Mason}(2001)}]{maso01}
\bibinfo{author}{\bibfnamefont{T.~E.} \bibnamefont{Mason}}, in
  \emph{\bibinfo{booktitle}{Handbook on the Physics and Chemistry of Rare
  Earths, Vol.\ 31: High-Temperature Superconductors -- II}}, edited by
  \bibinfo{editor}{\bibfnamefont{J.}~\bibnamefont{K.~A.~Gschneidner}},
  \bibinfo{editor}{\bibfnamefont{L.}~\bibnamefont{Eyring}}, \bibnamefont{and}
  \bibinfo{editor}{\bibfnamefont{M.~B.} \bibnamefont{Maple}}
  (\bibinfo{publisher}{Elsevier}, \bibinfo{address}{Amsterdam},
  \bibinfo{year}{2001}), pp. \bibinfo{pages}{281--314}.

\bibitem[{\citenamefont{Brom and Zaanen}(2003)}]{brom03}
\bibinfo{author}{\bibfnamefont{H.~B.} \bibnamefont{Brom}} \bibnamefont{and}
  \bibinfo{author}{\bibfnamefont{J.}~\bibnamefont{Zaanen}}, in
  \emph{\bibinfo{booktitle}{Handbook of Magnetic Materials, Vol.\ 15}}, edited
  by \bibinfo{editor}{\bibfnamefont{K.~H.~J.} \bibnamefont{Buschow}}
  (\bibinfo{publisher}{Elsevier Science}, \bibinfo{address}{Amsterdam},
  \bibinfo{year}{2003}), pp. \bibinfo{pages}{379--496}.

\bibitem[{\citenamefont{Shirane et~al.}(2002)\citenamefont{Shirane, Shapiro,
  and Tranquada}}]{shir02}
\bibinfo{author}{\bibfnamefont{G.}~\bibnamefont{Shirane}},
  \bibinfo{author}{\bibfnamefont{S.~M.} \bibnamefont{Shapiro}},
  \bibnamefont{and} \bibinfo{author}{\bibfnamefont{J.~M.}
  \bibnamefont{Tranquada}}, \emph{\bibinfo{title}{Neutron Scattering with a
  Triple-Axis Spectrometer: Basic Techniques}} (\bibinfo{publisher}{Cambridge
  University Press}, \bibinfo{address}{Cambridge}, \bibinfo{year}{2002}).

\bibitem[{\citenamefont{Squires}(1996)}]{squi96}
\bibinfo{author}{\bibfnamefont{G.~L.} \bibnamefont{Squires}},
  \emph{\bibinfo{title}{Introduction to the Theory of Thermal Neutron
  Scattering}} (\bibinfo{publisher}{Dover}, \bibinfo{address}{Mineola, NY},
  \bibinfo{year}{1996}).

\bibitem[{\citenamefont{Cheong et~al.}(1991)\citenamefont{Cheong, Aeppli,
  Mason, Mook, Hayden, Canfield, Fisk, Clausen, and Martinez}}]{cheo91}
\bibinfo{author}{\bibfnamefont{S.-W.} \bibnamefont{Cheong}},
  \bibinfo{author}{\bibfnamefont{G.}~\bibnamefont{Aeppli}},
  \bibinfo{author}{\bibfnamefont{T.~E.} \bibnamefont{Mason}},
  \bibinfo{author}{\bibfnamefont{H.}~\bibnamefont{Mook}},
  \bibinfo{author}{\bibfnamefont{S.~M.} \bibnamefont{Hayden}},
  \bibinfo{author}{\bibfnamefont{P.~C.} \bibnamefont{Canfield}},
  \bibinfo{author}{\bibfnamefont{Z.}~\bibnamefont{Fisk}},
  \bibinfo{author}{\bibfnamefont{K.~N.} \bibnamefont{Clausen}},
  \bibnamefont{and} \bibinfo{author}{\bibfnamefont{J.~L.}
  \bibnamefont{Martinez}}, \bibinfo{journal}{Phys. Rev. Lett.}
  \textbf{\bibinfo{volume}{67}}, \bibinfo{pages}{1791} (\bibinfo{year}{1991}).

\bibitem[{\citenamefont{Aeppli et~al.}(1997)\citenamefont{Aeppli, Mason,
  Hayden, Mook, and Kulda}}]{aepp97}
\bibinfo{author}{\bibfnamefont{G.}~\bibnamefont{Aeppli}},
  \bibinfo{author}{\bibfnamefont{T.~E.} \bibnamefont{Mason}},
  \bibinfo{author}{\bibfnamefont{S.~M.} \bibnamefont{Hayden}},
  \bibinfo{author}{\bibfnamefont{H.~A.} \bibnamefont{Mook}}, \bibnamefont{and}
  \bibinfo{author}{\bibfnamefont{J.}~\bibnamefont{Kulda}},
  \bibinfo{journal}{Science} \textbf{\bibinfo{volume}{278}},
  \bibinfo{pages}{1432} (\bibinfo{year}{1997}).

\bibitem[{\citenamefont{Yamada et~al.}(1998)\citenamefont{Yamada, Lee,
  Kurahashi, Wada, Wakimoto, Ueki, Kimura, Endoh, Hosoya, Shirane
  et~al.}}]{yama98a}
\bibinfo{author}{\bibfnamefont{K.}~\bibnamefont{Yamada}},
  \bibinfo{author}{\bibfnamefont{C.~H.} \bibnamefont{Lee}},
  \bibinfo{author}{\bibfnamefont{K.}~\bibnamefont{Kurahashi}},
  \bibinfo{author}{\bibfnamefont{J.}~\bibnamefont{Wada}},
  \bibinfo{author}{\bibfnamefont{S.}~\bibnamefont{Wakimoto}},
  \bibinfo{author}{\bibfnamefont{S.}~\bibnamefont{Ueki}},
  \bibinfo{author}{\bibfnamefont{Y.}~\bibnamefont{Kimura}},
  \bibinfo{author}{\bibfnamefont{Y.}~\bibnamefont{Endoh}},
  \bibinfo{author}{\bibfnamefont{S.}~\bibnamefont{Hosoya}},
  \bibinfo{author}{\bibfnamefont{G.}~\bibnamefont{Shirane}},
  \bibnamefont{et~al.}, \bibinfo{journal}{Phys. Rev. B}
  \textbf{\bibinfo{volume}{57}}, \bibinfo{pages}{6165} (\bibinfo{year}{1998}).

\bibitem[{\citenamefont{Rossat-Mignod et~al.}(1992)\citenamefont{Rossat-Mignod,
  Regnault, Vettier, Bourges, Burlet, Bossy, Henry, and Lapertot}}]{ross92}
\bibinfo{author}{\bibfnamefont{J.}~\bibnamefont{Rossat-Mignod}},
  \bibinfo{author}{\bibfnamefont{L.~P.} \bibnamefont{Regnault}},
  \bibinfo{author}{\bibfnamefont{C.}~\bibnamefont{Vettier}},
  \bibinfo{author}{\bibfnamefont{P.}~\bibnamefont{Bourges}},
  \bibinfo{author}{\bibfnamefont{P.}~\bibnamefont{Burlet}},
  \bibinfo{author}{\bibfnamefont{J.}~\bibnamefont{Bossy}},
  \bibinfo{author}{\bibfnamefont{J.~Y.} \bibnamefont{Henry}}, \bibnamefont{and}
  \bibinfo{author}{\bibfnamefont{G.}~\bibnamefont{Lapertot}},
  \bibinfo{journal}{Physica B} \textbf{\bibinfo{volume}{180\&181}},
  \bibinfo{pages}{383} (\bibinfo{year}{1992}).

\bibitem[{\citenamefont{Mook et~al.}(1993)\citenamefont{Mook, Yethiraj, Aeppli,
  Mason, and Armstrong}}]{mook93}
\bibinfo{author}{\bibfnamefont{H.~A.} \bibnamefont{Mook}},
  \bibinfo{author}{\bibfnamefont{M.}~\bibnamefont{Yethiraj}},
  \bibinfo{author}{\bibfnamefont{G.}~\bibnamefont{Aeppli}},
  \bibinfo{author}{\bibfnamefont{T.~E.} \bibnamefont{Mason}}, \bibnamefont{and}
  \bibinfo{author}{\bibfnamefont{T.}~\bibnamefont{Armstrong}},
  \bibinfo{journal}{Phys. Rev. Lett.} \textbf{\bibinfo{volume}{70}},
  \bibinfo{pages}{3490} (\bibinfo{year}{1993}).

\bibitem[{\citenamefont{Fong et~al.}(1996)\citenamefont{Fong, Keimer, Reznik,
  Milius, and Aksay}}]{fong96}
\bibinfo{author}{\bibfnamefont{H.~F.} \bibnamefont{Fong}},
  \bibinfo{author}{\bibfnamefont{B.}~\bibnamefont{Keimer}},
  \bibinfo{author}{\bibfnamefont{D.}~\bibnamefont{Reznik}},
  \bibinfo{author}{\bibfnamefont{D.~L.} \bibnamefont{Milius}},
  \bibnamefont{and} \bibinfo{author}{\bibfnamefont{I.~A.} \bibnamefont{Aksay}},
  \bibinfo{journal}{Phys. Rev. B} \textbf{\bibinfo{volume}{54}},
  \bibinfo{pages}{6708} (\bibinfo{year}{1996}).

\bibitem[{\citenamefont{Dai et~al.}(2001)\citenamefont{Dai, Mook, Hunt, and
  Do\u{g}an}}]{dai01}
\bibinfo{author}{\bibfnamefont{P.}~\bibnamefont{Dai}},
  \bibinfo{author}{\bibfnamefont{H.~A.} \bibnamefont{Mook}},
  \bibinfo{author}{\bibfnamefont{R.~D.} \bibnamefont{Hunt}}, \bibnamefont{and}
  \bibinfo{author}{\bibfnamefont{F.}~\bibnamefont{Do\u{g}an}},
  \bibinfo{journal}{Phys. Rev. B} \textbf{\bibinfo{volume}{63}},
  \bibinfo{pages}{054525} (\bibinfo{year}{2001}).

\bibitem[{\citenamefont{Sidis et~al.}(2004)\citenamefont{Sidis, Pailh{\`e}s,
  Keimer, Ulrich, and Regnault}}]{sidi04}
\bibinfo{author}{\bibfnamefont{Y.}~\bibnamefont{Sidis}},
  \bibinfo{author}{\bibfnamefont{S.}~\bibnamefont{Pailh{\`e}s}},
  \bibinfo{author}{\bibfnamefont{B.}~\bibnamefont{Keimer}},
  \bibinfo{author}{\bibfnamefont{C.}~\bibnamefont{Ulrich}}, \bibnamefont{and}
  \bibinfo{author}{\bibfnamefont{L.~P.} \bibnamefont{Regnault}},
  \bibinfo{journal}{Phys. Stat. Sol. (b)} \textbf{\bibinfo{volume}{241}},
  \bibinfo{pages}{1204} (\bibinfo{year}{2004}).

\bibitem[{\citenamefont{Bourges et~al.}(1999)\citenamefont{Bourges, Sidis,
  Fong, Keimer, Regnault, Bossy, Ivanov, Milius, and Aksay}}]{bour99}
\bibinfo{author}{\bibfnamefont{P.}~\bibnamefont{Bourges}},
  \bibinfo{author}{\bibfnamefont{Y.}~\bibnamefont{Sidis}},
  \bibinfo{author}{\bibfnamefont{H.~F.} \bibnamefont{Fong}},
  \bibinfo{author}{\bibfnamefont{B.}~\bibnamefont{Keimer}},
  \bibinfo{author}{\bibfnamefont{L.~P.} \bibnamefont{Regnault}},
  \bibinfo{author}{\bibfnamefont{J.}~\bibnamefont{Bossy}},
  \bibinfo{author}{\bibfnamefont{A.~S.} \bibnamefont{Ivanov}},
  \bibinfo{author}{\bibfnamefont{D.~L.} \bibnamefont{Milius}},
  \bibnamefont{and} \bibinfo{author}{\bibfnamefont{I.~A.} \bibnamefont{Aksay}},
  in \emph{\bibinfo{booktitle}{High Temperature Superconductivity}}, edited by
  \bibinfo{editor}{\bibfnamefont{S.~E.} \bibnamefont{Barnes}},
  \bibinfo{editor}{\bibfnamefont{J.}~\bibnamefont{Ashkenazi}},
  \bibinfo{editor}{\bibfnamefont{J.~L.} \bibnamefont{Cohn}}, \bibnamefont{and}
  \bibinfo{editor}{\bibfnamefont{F.}~\bibnamefont{Zuo}}
  (\bibinfo{publisher}{American Institute of Physics},
  \bibinfo{address}{Woodbury, NY}, \bibinfo{year}{1999}), pp.
  \bibinfo{pages}{207--212}.

\bibitem[{\citenamefont{Fong et~al.}(1999)\citenamefont{Fong, Bourges, Sidis,
  Regnault, Ivanov, Gu, Koshizuka, and Keimer}}]{fong99}
\bibinfo{author}{\bibfnamefont{H.~F.} \bibnamefont{Fong}},
  \bibinfo{author}{\bibfnamefont{P.}~\bibnamefont{Bourges}},
  \bibinfo{author}{\bibfnamefont{Y.}~\bibnamefont{Sidis}},
  \bibinfo{author}{\bibfnamefont{L.~P.} \bibnamefont{Regnault}},
  \bibinfo{author}{\bibfnamefont{A.}~\bibnamefont{Ivanov}},
  \bibinfo{author}{\bibfnamefont{G.~D.} \bibnamefont{Gu}},
  \bibinfo{author}{\bibfnamefont{N.}~\bibnamefont{Koshizuka}},
  \bibnamefont{and} \bibinfo{author}{\bibfnamefont{B.}~\bibnamefont{Keimer}},
  \bibinfo{journal}{Nature} \textbf{\bibinfo{volume}{398}},
  \bibinfo{pages}{588} (\bibinfo{year}{1999}).

\bibitem[{\citenamefont{Mesot et~al.}(2000)\citenamefont{Mesot, Metoki, Bohm,
  Hiess, and Kadowaki}}]{meso00}
\bibinfo{author}{\bibfnamefont{J.}~\bibnamefont{Mesot}},
  \bibinfo{author}{\bibfnamefont{N.}~\bibnamefont{Metoki}},
  \bibinfo{author}{\bibfnamefont{M.}~\bibnamefont{Bohm}},
  \bibinfo{author}{\bibfnamefont{A.}~\bibnamefont{Hiess}}, \bibnamefont{and}
  \bibinfo{author}{\bibfnamefont{K.}~\bibnamefont{Kadowaki}},
  \bibinfo{journal}{Physica C} \textbf{\bibinfo{volume}{341}},
  \bibinfo{pages}{2105} (\bibinfo{year}{2000}).

\bibitem[{\citenamefont{He et~al.}(2001)\citenamefont{He, Sidis, Bourges, Gu,
  Ivanov, Koshizuka, Liang, Lin, Regnault, Schoenherr et~al.}}]{he01}
\bibinfo{author}{\bibfnamefont{H.}~\bibnamefont{He}},
  \bibinfo{author}{\bibfnamefont{Y.}~\bibnamefont{Sidis}},
  \bibinfo{author}{\bibfnamefont{P.}~\bibnamefont{Bourges}},
  \bibinfo{author}{\bibfnamefont{G.~D.} \bibnamefont{Gu}},
  \bibinfo{author}{\bibfnamefont{A.}~\bibnamefont{Ivanov}},
  \bibinfo{author}{\bibfnamefont{N.}~\bibnamefont{Koshizuka}},
  \bibinfo{author}{\bibfnamefont{B.}~\bibnamefont{Liang}},
  \bibinfo{author}{\bibfnamefont{C.~T.} \bibnamefont{Lin}},
  \bibinfo{author}{\bibfnamefont{L.~P.} \bibnamefont{Regnault}},
  \bibinfo{author}{\bibfnamefont{E.}~\bibnamefont{Schoenherr}},
  \bibnamefont{et~al.}, \bibinfo{journal}{Phys. Rev. Lett.}
  \textbf{\bibinfo{volume}{86}}, \bibinfo{pages}{1610} (\bibinfo{year}{2001}).

\bibitem[{\citenamefont{He et~al.}(2002)\citenamefont{He, Bourges, Sidis,
  Ulrich, Regnault, Pailh\`es, Berzigiarova, Kolesnikov, and Keimer}}]{he02}
\bibinfo{author}{\bibfnamefont{H.}~\bibnamefont{He}},
  \bibinfo{author}{\bibfnamefont{P.}~\bibnamefont{Bourges}},
  \bibinfo{author}{\bibfnamefont{Y.}~\bibnamefont{Sidis}},
  \bibinfo{author}{\bibfnamefont{C.}~\bibnamefont{Ulrich}},
  \bibinfo{author}{\bibfnamefont{L.~P.} \bibnamefont{Regnault}},
  \bibinfo{author}{\bibfnamefont{S.}~\bibnamefont{Pailh\`es}},
  \bibinfo{author}{\bibfnamefont{N.~S.} \bibnamefont{Berzigiarova}},
  \bibinfo{author}{\bibfnamefont{N.~N.} \bibnamefont{Kolesnikov}},
  \bibnamefont{and} \bibinfo{author}{\bibfnamefont{B.}~\bibnamefont{Keimer}},
  \bibinfo{journal}{Science} \textbf{\bibinfo{volume}{295}},
  \bibinfo{pages}{1045} (\bibinfo{year}{2002}).

\bibitem[{\citenamefont{Eschrig and Norman}(2000)}]{esch00}
\bibinfo{author}{\bibfnamefont{M.}~\bibnamefont{Eschrig}} \bibnamefont{and}
  \bibinfo{author}{\bibfnamefont{M.~R.} \bibnamefont{Norman}},
  \bibinfo{journal}{Phys. Rev. Lett.} \textbf{\bibinfo{volume}{85}},
  \bibinfo{pages}{3261} (\bibinfo{year}{2000}).

\bibitem[{\citenamefont{Abanov et~al.}(2001)\citenamefont{Abanov, Chubukov, and
  Schmalian}}]{aban01}
\bibinfo{author}{\bibfnamefont{A.}~\bibnamefont{Abanov}},
  \bibinfo{author}{\bibfnamefont{A.~V.} \bibnamefont{Chubukov}},
  \bibnamefont{and}
  \bibinfo{author}{\bibfnamefont{J.}~\bibnamefont{Schmalian}},
  \bibinfo{journal}{J. Electron. Spectrosc. Relat. Phenom.}
  \textbf{\bibinfo{volume}{117--118}}, \bibinfo{pages}{129}
  (\bibinfo{year}{2001}).

\bibitem[{\citenamefont{Kee et~al.}(2002)\citenamefont{Kee, Kivelson, and
  Aeppli}}]{kee02}
\bibinfo{author}{\bibfnamefont{H.-Y.} \bibnamefont{Kee}},
  \bibinfo{author}{\bibfnamefont{S.~A.} \bibnamefont{Kivelson}},
  \bibnamefont{and} \bibinfo{author}{\bibfnamefont{G.}~\bibnamefont{Aeppli}},
  \bibinfo{journal}{Phys. Rev. Lett.} \textbf{\bibinfo{volume}{88}},
  \bibinfo{pages}{257002} (\bibinfo{year}{2002}).

\bibitem[{\citenamefont{Hayden et~al.}(2004)\citenamefont{Hayden, Mook, Dai,
  Perring, and Do\u{g}an}}]{hayd04}
\bibinfo{author}{\bibfnamefont{S.~M.} \bibnamefont{Hayden}},
  \bibinfo{author}{\bibfnamefont{H.~A.} \bibnamefont{Mook}},
  \bibinfo{author}{\bibfnamefont{P.}~\bibnamefont{Dai}},
  \bibinfo{author}{\bibfnamefont{T.~G.} \bibnamefont{Perring}},
  \bibnamefont{and}
  \bibinfo{author}{\bibfnamefont{F.}~\bibnamefont{Do\u{g}an}},
  \bibinfo{journal}{Nature} \textbf{\bibinfo{volume}{429}},
  \bibinfo{pages}{531} (\bibinfo{year}{2004}).

\bibitem[{\citenamefont{Stock et~al.}(2005{\natexlab{a}})\citenamefont{Stock,
  Buyers, Cowley, Clegg, Coldea, Frost, Liang, Peets, Bonn, Hardy
  et~al.}}]{stoc05}
\bibinfo{author}{\bibfnamefont{C.}~\bibnamefont{Stock}},
  \bibinfo{author}{\bibfnamefont{W.~J.~L.} \bibnamefont{Buyers}},
  \bibinfo{author}{\bibfnamefont{R.~A.} \bibnamefont{Cowley}},
  \bibinfo{author}{\bibfnamefont{P.~S.} \bibnamefont{Clegg}},
  \bibinfo{author}{\bibfnamefont{R.}~\bibnamefont{Coldea}},
  \bibinfo{author}{\bibfnamefont{C.~D.} \bibnamefont{Frost}},
  \bibinfo{author}{\bibfnamefont{R.}~\bibnamefont{Liang}},
  \bibinfo{author}{\bibfnamefont{D.}~\bibnamefont{Peets}},
  \bibinfo{author}{\bibfnamefont{D.}~\bibnamefont{Bonn}},
  \bibinfo{author}{\bibfnamefont{W.~N.} \bibnamefont{Hardy}},
  \bibnamefont{et~al.}, \bibinfo{journal}{Phys. Rev. B}
  \textbf{\bibinfo{volume}{71}}, \bibinfo{pages}{024522}
  (\bibinfo{year}{2005}{\natexlab{a}}).

\bibitem[{\citenamefont{Hinkov et~al.}(2004)\citenamefont{Hinkov, Pailh\`{e}s,
  Bourges, Sidis, Ivanov, Kulakov, Lin, Chen, Bernhard, and Keimer}}]{hink04}
\bibinfo{author}{\bibfnamefont{V.}~\bibnamefont{Hinkov}},
  \bibinfo{author}{\bibfnamefont{S.}~\bibnamefont{Pailh\`{e}s}},
  \bibinfo{author}{\bibfnamefont{P.}~\bibnamefont{Bourges}},
  \bibinfo{author}{\bibfnamefont{Y.}~\bibnamefont{Sidis}},
  \bibinfo{author}{\bibfnamefont{A.}~\bibnamefont{Ivanov}},
  \bibinfo{author}{\bibfnamefont{A.}~\bibnamefont{Kulakov}},
  \bibinfo{author}{\bibfnamefont{C.~T.} \bibnamefont{Lin}},
  \bibinfo{author}{\bibfnamefont{D.~P.} \bibnamefont{Chen}},
  \bibinfo{author}{\bibfnamefont{C.}~\bibnamefont{Bernhard}}, \bibnamefont{and}
  \bibinfo{author}{\bibfnamefont{B.}~\bibnamefont{Keimer}},
  \bibinfo{journal}{Nature} \textbf{\bibinfo{volume}{430}},
  \bibinfo{pages}{650} (\bibinfo{year}{2004}).

\bibitem[{\citenamefont{Bourges
  et~al.}(1997{\natexlab{a}})\citenamefont{Bourges, Fong, Regnault, Bossy,
  Vettier, Milius, Aksay, and Keimer}}]{bour97}
\bibinfo{author}{\bibfnamefont{P.}~\bibnamefont{Bourges}},
  \bibinfo{author}{\bibfnamefont{H.~F.} \bibnamefont{Fong}},
  \bibinfo{author}{\bibfnamefont{L.~P.} \bibnamefont{Regnault}},
  \bibinfo{author}{\bibfnamefont{J.}~\bibnamefont{Bossy}},
  \bibinfo{author}{\bibfnamefont{C.}~\bibnamefont{Vettier}},
  \bibinfo{author}{\bibfnamefont{D.~L.} \bibnamefont{Milius}},
  \bibinfo{author}{\bibfnamefont{I.~A.} \bibnamefont{Aksay}}, \bibnamefont{and}
  \bibinfo{author}{\bibfnamefont{B.}~\bibnamefont{Keimer}},
  \bibinfo{journal}{Phys. Rev. B} \textbf{\bibinfo{volume}{56}},
  \bibinfo{pages}{R11 439} (\bibinfo{year}{1997}{\natexlab{a}}).

\bibitem[{\citenamefont{Dai et~al.}(1999)\citenamefont{Dai, Mook, Hayden,
  Aeppli, Perring, Hunt, and {Do\u gan}}}]{dai99}
\bibinfo{author}{\bibfnamefont{P.}~\bibnamefont{Dai}},
  \bibinfo{author}{\bibfnamefont{H.~A.} \bibnamefont{Mook}},
  \bibinfo{author}{\bibfnamefont{S.~M.} \bibnamefont{Hayden}},
  \bibinfo{author}{\bibfnamefont{G.}~\bibnamefont{Aeppli}},
  \bibinfo{author}{\bibfnamefont{T.~G.} \bibnamefont{Perring}},
  \bibinfo{author}{\bibfnamefont{R.~D.} \bibnamefont{Hunt}}, \bibnamefont{and}
  \bibinfo{author}{\bibfnamefont{F.}~\bibnamefont{{Do\u gan}}},
  \bibinfo{journal}{Science} \textbf{\bibinfo{volume}{284}},
  \bibinfo{pages}{1344} (\bibinfo{year}{1999}).

\bibitem[{\citenamefont{Hayden et~al.}(1996)\citenamefont{Hayden, Aeppli,
  Perring, Mook, and {Do\u gan}}}]{hayd96b}
\bibinfo{author}{\bibfnamefont{S.~M.} \bibnamefont{Hayden}},
  \bibinfo{author}{\bibfnamefont{G.}~\bibnamefont{Aeppli}},
  \bibinfo{author}{\bibfnamefont{T.~G.} \bibnamefont{Perring}},
  \bibinfo{author}{\bibfnamefont{H.~A.} \bibnamefont{Mook}}, \bibnamefont{and}
  \bibinfo{author}{\bibfnamefont{F.}~\bibnamefont{{Do\u gan}}},
  \bibinfo{journal}{Phys. Rev. B} \textbf{\bibinfo{volume}{54}},
  \bibinfo{pages}{R6905} (\bibinfo{year}{1996}).

\bibitem[{\citenamefont{Reznik et~al.}(1996)\citenamefont{Reznik, Bourges,
  Fong, Regnault, Bossy, Vettier, Milius, Aksay, and Keimer}}]{rezn96}
\bibinfo{author}{\bibfnamefont{D.}~\bibnamefont{Reznik}},
  \bibinfo{author}{\bibfnamefont{P.}~\bibnamefont{Bourges}},
  \bibinfo{author}{\bibfnamefont{H.~F.} \bibnamefont{Fong}},
  \bibinfo{author}{\bibfnamefont{L.~P.} \bibnamefont{Regnault}},
  \bibinfo{author}{\bibfnamefont{J.}~\bibnamefont{Bossy}},
  \bibinfo{author}{\bibfnamefont{C.}~\bibnamefont{Vettier}},
  \bibinfo{author}{\bibfnamefont{D.~L.} \bibnamefont{Milius}},
  \bibinfo{author}{\bibfnamefont{I.~A.} \bibnamefont{Aksay}}, \bibnamefont{and}
  \bibinfo{author}{\bibfnamefont{B.}~\bibnamefont{Keimer}},
  \bibinfo{journal}{Phys. Rev. B} \textbf{\bibinfo{volume}{53}},
  \bibinfo{pages}{R14741} (\bibinfo{year}{1996}).

\bibitem[{\citenamefont{Mook et~al.}(1998)\citenamefont{Mook, Dai, Hayden,
  Aeppli, Perring, and {Do\u gan}}}]{mook98}
\bibinfo{author}{\bibfnamefont{H.~A.} \bibnamefont{Mook}},
  \bibinfo{author}{\bibfnamefont{P.}~\bibnamefont{Dai}},
  \bibinfo{author}{\bibfnamefont{S.~M.} \bibnamefont{Hayden}},
  \bibinfo{author}{\bibfnamefont{G.}~\bibnamefont{Aeppli}},
  \bibinfo{author}{\bibfnamefont{T.~G.} \bibnamefont{Perring}},
  \bibnamefont{and} \bibinfo{author}{\bibfnamefont{F.}~\bibnamefont{{Do\u
  gan}}}, \bibinfo{journal}{Nature} \textbf{\bibinfo{volume}{395}},
  \bibinfo{pages}{580} (\bibinfo{year}{1998}).

\bibitem[{\citenamefont{Arai et~al.}(1999)\citenamefont{Arai, Nishijima, Endoh,
  Egami, Tajima, Tomimoto, Shiohara, Takahashi, Garrett, and
  Bennington}}]{arai99}
\bibinfo{author}{\bibfnamefont{M.}~\bibnamefont{Arai}},
  \bibinfo{author}{\bibfnamefont{T.}~\bibnamefont{Nishijima}},
  \bibinfo{author}{\bibfnamefont{Y.}~\bibnamefont{Endoh}},
  \bibinfo{author}{\bibfnamefont{T.}~\bibnamefont{Egami}},
  \bibinfo{author}{\bibfnamefont{S.}~\bibnamefont{Tajima}},
  \bibinfo{author}{\bibfnamefont{K.}~\bibnamefont{Tomimoto}},
  \bibinfo{author}{\bibfnamefont{Y.}~\bibnamefont{Shiohara}},
  \bibinfo{author}{\bibfnamefont{M.}~\bibnamefont{Takahashi}},
  \bibinfo{author}{\bibfnamefont{A.}~\bibnamefont{Garrett}}, \bibnamefont{and}
  \bibinfo{author}{\bibfnamefont{S.~M.} \bibnamefont{Bennington}},
  \bibinfo{journal}{Phys. Rev. Lett.} \textbf{\bibinfo{volume}{83}},
  \bibinfo{pages}{608} (\bibinfo{year}{1999}).

\bibitem[{\citenamefont{Bourges et~al.}(2000)\citenamefont{Bourges, Sidis,
  Fong, Regnault, Bossy, Ivanov, and Keimer}}]{bour00}
\bibinfo{author}{\bibfnamefont{P.}~\bibnamefont{Bourges}},
  \bibinfo{author}{\bibfnamefont{Y.}~\bibnamefont{Sidis}},
  \bibinfo{author}{\bibfnamefont{H.~F.} \bibnamefont{Fong}},
  \bibinfo{author}{\bibfnamefont{L.~P.} \bibnamefont{Regnault}},
  \bibinfo{author}{\bibfnamefont{J.}~\bibnamefont{Bossy}},
  \bibinfo{author}{\bibfnamefont{A.}~\bibnamefont{Ivanov}}, \bibnamefont{and}
  \bibinfo{author}{\bibfnamefont{B.}~\bibnamefont{Keimer}},
  \bibinfo{journal}{Science} \textbf{\bibinfo{volume}{288}},
  \bibinfo{pages}{1234} (\bibinfo{year}{2000}).

\bibitem[{\citenamefont{Pailh\`es et~al.}(2004)\citenamefont{Pailh\`es, Sidis,
  Bourges, Hinkov, Ivanov, Ulrich, Regnault, and Keimer}}]{pail04}
\bibinfo{author}{\bibfnamefont{S.}~\bibnamefont{Pailh\`es}},
  \bibinfo{author}{\bibfnamefont{Y.}~\bibnamefont{Sidis}},
  \bibinfo{author}{\bibfnamefont{P.}~\bibnamefont{Bourges}},
  \bibinfo{author}{\bibfnamefont{V.}~\bibnamefont{Hinkov}},
  \bibinfo{author}{\bibfnamefont{A.}~\bibnamefont{Ivanov}},
  \bibinfo{author}{\bibfnamefont{C.}~\bibnamefont{Ulrich}},
  \bibinfo{author}{\bibfnamefont{L.~P.} \bibnamefont{Regnault}},
  \bibnamefont{and} \bibinfo{author}{\bibfnamefont{B.}~\bibnamefont{Keimer}},
  \bibinfo{journal}{Phys. Rev. Lett.} \textbf{\bibinfo{volume}{93}},
  \bibinfo{pages}{167001} (\bibinfo{year}{2004}).

\bibitem[{\citenamefont{Reznik et~al.}(2004)\citenamefont{Reznik, Bourges,
  Pintschovius, Endoh, Sidis, Matsui, and Tajima}}]{rezn04}
\bibinfo{author}{\bibfnamefont{D.}~\bibnamefont{Reznik}},
  \bibinfo{author}{\bibfnamefont{P.}~\bibnamefont{Bourges}},
  \bibinfo{author}{\bibfnamefont{L.}~\bibnamefont{Pintschovius}},
  \bibinfo{author}{\bibfnamefont{Y.}~\bibnamefont{Endoh}},
  \bibinfo{author}{\bibfnamefont{Y.}~\bibnamefont{Sidis}},
  \bibinfo{author}{\bibfnamefont{T.}~\bibnamefont{Matsui}}, \bibnamefont{and}
  \bibinfo{author}{\bibfnamefont{S.}~\bibnamefont{Tajima}},
  \bibinfo{journal}{Phys. Rev. Lett.} \textbf{\bibinfo{volume}{93}},
  \bibinfo{pages}{207003} (\bibinfo{year}{2004}).

\bibitem[{\citenamefont{Ito et~al.}(2002)\citenamefont{Ito, Harashina, Yasui,
  Kanada, Iikubo, Sato, Kobayashi, and Kakurai}}]{ito02}
\bibinfo{author}{\bibfnamefont{M.}~\bibnamefont{Ito}},
  \bibinfo{author}{\bibfnamefont{H.}~\bibnamefont{Harashina}},
  \bibinfo{author}{\bibfnamefont{Y.}~\bibnamefont{Yasui}},
  \bibinfo{author}{\bibfnamefont{M.}~\bibnamefont{Kanada}},
  \bibinfo{author}{\bibfnamefont{S.}~\bibnamefont{Iikubo}},
  \bibinfo{author}{\bibfnamefont{M.}~\bibnamefont{Sato}},
  \bibinfo{author}{\bibfnamefont{A.}~\bibnamefont{Kobayashi}},
  \bibnamefont{and} \bibinfo{author}{\bibfnamefont{K.}~\bibnamefont{Kakurai}},
  \bibinfo{journal}{J. Phys. Soc. Japan} \textbf{\bibinfo{volume}{71}},
  \bibinfo{pages}{265} (\bibinfo{year}{2002}).

\bibitem[{\citenamefont{Christensen et~al.}(2004)\citenamefont{Christensen,
  McMorrow, R{\o}nnow, Lake, Hayden, Aeppli, Perring, Mangkorntong, Nohara, and
  Tagaki}}]{chri04}
\bibinfo{author}{\bibfnamefont{N.~B.} \bibnamefont{Christensen}},
  \bibinfo{author}{\bibfnamefont{D.~F.} \bibnamefont{McMorrow}},
  \bibinfo{author}{\bibfnamefont{H.~M.} \bibnamefont{R{\o}nnow}},
  \bibinfo{author}{\bibfnamefont{B.}~\bibnamefont{Lake}},
  \bibinfo{author}{\bibfnamefont{S.~M.} \bibnamefont{Hayden}},
  \bibinfo{author}{\bibfnamefont{G.}~\bibnamefont{Aeppli}},
  \bibinfo{author}{\bibfnamefont{T.~G.} \bibnamefont{Perring}},
  \bibinfo{author}{\bibfnamefont{M.}~\bibnamefont{Mangkorntong}},
  \bibinfo{author}{\bibfnamefont{M.}~\bibnamefont{Nohara}}, \bibnamefont{and}
  \bibinfo{author}{\bibfnamefont{H.}~\bibnamefont{Tagaki}},
  \bibinfo{journal}{Phys. Rev. Lett.} \textbf{\bibinfo{volume}{93}},
  \bibinfo{pages}{147002} (\bibinfo{year}{2004}).

\bibitem[{\citenamefont{Fujita et~al.}(2004)\citenamefont{Fujita, Goka, Yamada,
  Tranquada, and Regnault}}]{fuji04}
\bibinfo{author}{\bibfnamefont{M.}~\bibnamefont{Fujita}},
  \bibinfo{author}{\bibfnamefont{H.}~\bibnamefont{Goka}},
  \bibinfo{author}{\bibfnamefont{K.}~\bibnamefont{Yamada}},
  \bibinfo{author}{\bibfnamefont{J.~M.} \bibnamefont{Tranquada}},
  \bibnamefont{and} \bibinfo{author}{\bibfnamefont{L.~P.}
  \bibnamefont{Regnault}}, \bibinfo{journal}{Phys. Rev. B}
  \textbf{\bibinfo{volume}{70}}, \bibinfo{pages}{104517}
  (\bibinfo{year}{2004}).

\bibitem[{\citenamefont{Tranquada
  et~al.}(2004{\natexlab{a}})\citenamefont{Tranquada, Woo, Perring, Goka, Gu,
  Xu, Fujita, and Yamada}}]{tran04}
\bibinfo{author}{\bibfnamefont{J.~M.} \bibnamefont{Tranquada}},
  \bibinfo{author}{\bibfnamefont{H.}~\bibnamefont{Woo}},
  \bibinfo{author}{\bibfnamefont{T.~G.} \bibnamefont{Perring}},
  \bibinfo{author}{\bibfnamefont{H.}~\bibnamefont{Goka}},
  \bibinfo{author}{\bibfnamefont{G.~D.} \bibnamefont{Gu}},
  \bibinfo{author}{\bibfnamefont{G.}~\bibnamefont{Xu}},
  \bibinfo{author}{\bibfnamefont{M.}~\bibnamefont{Fujita}}, \bibnamefont{and}
  \bibinfo{author}{\bibfnamefont{K.}~\bibnamefont{Yamada}},
  \bibinfo{journal}{Nature} \textbf{\bibinfo{volume}{429}},
  \bibinfo{pages}{534} (\bibinfo{year}{2004}{\natexlab{a}}).

\bibitem[{\citenamefont{Tranquada
  et~al.}(1995{\natexlab{a}})\citenamefont{Tranquada, Sternlieb, Axe, Nakamura,
  and Uchida}}]{tran95a}
\bibinfo{author}{\bibfnamefont{J.~M.} \bibnamefont{Tranquada}},
  \bibinfo{author}{\bibfnamefont{B.~J.} \bibnamefont{Sternlieb}},
  \bibinfo{author}{\bibfnamefont{J.~D.} \bibnamefont{Axe}},
  \bibinfo{author}{\bibfnamefont{Y.}~\bibnamefont{Nakamura}}, \bibnamefont{and}
  \bibinfo{author}{\bibfnamefont{S.}~\bibnamefont{Uchida}},
  \bibinfo{journal}{Nature} \textbf{\bibinfo{volume}{375}},
  \bibinfo{pages}{561} (\bibinfo{year}{1995}{\natexlab{a}}).

\bibitem[{\citenamefont{Tranquada et~al.}()\citenamefont{Tranquada, Woo,
  Perring, Goka, Gu, Xu, Fujita, and Yamada}}]{tran05a}
\bibinfo{author}{\bibfnamefont{J.~M.} \bibnamefont{Tranquada}},
  \bibinfo{author}{\bibfnamefont{H.}~\bibnamefont{Woo}},
  \bibinfo{author}{\bibfnamefont{T.~G.} \bibnamefont{Perring}},
  \bibinfo{author}{\bibfnamefont{H.}~\bibnamefont{Goka}},
  \bibinfo{author}{\bibfnamefont{G.~D.} \bibnamefont{Gu}},
  \bibinfo{author}{\bibfnamefont{G.}~\bibnamefont{Xu}},
  \bibinfo{author}{\bibfnamefont{M.}~\bibnamefont{Fujita}}, \bibnamefont{and}
  \bibinfo{author}{\bibfnamefont{K.}~\bibnamefont{Yamada}},
  \bibinfo{note}{cond-mat/0411082}.

\bibitem[{\citenamefont{Barnes and Riera}(1994)}]{barn94}
\bibinfo{author}{\bibfnamefont{T.}~\bibnamefont{Barnes}} \bibnamefont{and}
  \bibinfo{author}{\bibfnamefont{J.}~\bibnamefont{Riera}},
  \bibinfo{journal}{Phys. Rev. B} \textbf{\bibinfo{volume}{50}},
  \bibinfo{pages}{6817} (\bibinfo{year}{1994}).

\bibitem[{\citenamefont{Mook et~al.}(2000)\citenamefont{Mook, Dai, {Do\u gan},
  and Hunt}}]{mook00}
\bibinfo{author}{\bibfnamefont{H.~A.} \bibnamefont{Mook}},
  \bibinfo{author}{\bibfnamefont{P.}~\bibnamefont{Dai}},
  \bibinfo{author}{\bibfnamefont{F.}~\bibnamefont{{Do\u gan}}},
  \bibnamefont{and} \bibinfo{author}{\bibfnamefont{R.~D.} \bibnamefont{Hunt}},
  \bibinfo{journal}{Nature} \textbf{\bibinfo{volume}{404}},
  \bibinfo{pages}{729} (\bibinfo{year}{2000}).

\bibitem[{\citenamefont{Stock et~al.}(2004)\citenamefont{Stock, Buyers, Liang,
  Peets, Tun, Bonn, Hardy, and Birgeneau}}]{stoc04}
\bibinfo{author}{\bibfnamefont{C.}~\bibnamefont{Stock}},
  \bibinfo{author}{\bibfnamefont{W.~J.~L.} \bibnamefont{Buyers}},
  \bibinfo{author}{\bibfnamefont{R.}~\bibnamefont{Liang}},
  \bibinfo{author}{\bibfnamefont{D.}~\bibnamefont{Peets}},
  \bibinfo{author}{\bibfnamefont{Z.}~\bibnamefont{Tun}},
  \bibinfo{author}{\bibfnamefont{D.}~\bibnamefont{Bonn}},
  \bibinfo{author}{\bibfnamefont{W.~N.} \bibnamefont{Hardy}}, \bibnamefont{and}
  \bibinfo{author}{\bibfnamefont{R.~J.} \bibnamefont{Birgeneau}},
  \bibinfo{journal}{Phys. Rev. B} \textbf{\bibinfo{volume}{69}},
  \bibinfo{pages}{014502} (\bibinfo{year}{2004}).

\bibitem[{\citenamefont{Lake et~al.}(1999)\citenamefont{Lake, Aeppli, Mason,
  Schr\"oder, McMorrow, Lefmann, Isshiki, Nohara, Takagi, and Hayden}}]{lake99}
\bibinfo{author}{\bibfnamefont{B.}~\bibnamefont{Lake}},
  \bibinfo{author}{\bibfnamefont{G.}~\bibnamefont{Aeppli}},
  \bibinfo{author}{\bibfnamefont{T.~E.} \bibnamefont{Mason}},
  \bibinfo{author}{\bibfnamefont{A.}~\bibnamefont{Schr\"oder}},
  \bibinfo{author}{\bibfnamefont{D.~F.} \bibnamefont{McMorrow}},
  \bibinfo{author}{\bibfnamefont{K.}~\bibnamefont{Lefmann}},
  \bibinfo{author}{\bibfnamefont{M.}~\bibnamefont{Isshiki}},
  \bibinfo{author}{\bibfnamefont{M.}~\bibnamefont{Nohara}},
  \bibinfo{author}{\bibfnamefont{H.}~\bibnamefont{Takagi}}, \bibnamefont{and}
  \bibinfo{author}{\bibfnamefont{S.~M.} \bibnamefont{Hayden}},
  \bibinfo{journal}{Nature} \textbf{\bibinfo{volume}{400}}, \bibinfo{pages}{43}
  (\bibinfo{year}{1999}).

\bibitem[{\citenamefont{Lu}(1992)}]{lu92}
\bibinfo{author}{\bibfnamefont{J.~P.} \bibnamefont{Lu}},
  \bibinfo{journal}{Phys. Rev. Lett.} \textbf{\bibinfo{volume}{68}},
  \bibinfo{pages}{125} (\bibinfo{year}{1992}).

\bibitem[{\citenamefont{Dai et~al.}(2000)\citenamefont{Dai, Mook, Aeppli,
  Hayden, and Do\u{g}an}}]{dai00}
\bibinfo{author}{\bibfnamefont{P.}~\bibnamefont{Dai}},
  \bibinfo{author}{\bibfnamefont{H.~A.} \bibnamefont{Mook}},
  \bibinfo{author}{\bibfnamefont{G.}~\bibnamefont{Aeppli}},
  \bibinfo{author}{\bibfnamefont{S.~M.} \bibnamefont{Hayden}},
  \bibnamefont{and}
  \bibinfo{author}{\bibfnamefont{F.}~\bibnamefont{Do\u{g}an}},
  \bibinfo{journal}{Nature} \textbf{\bibinfo{volume}{406}},
  \bibinfo{pages}{965} (\bibinfo{year}{2000}).

\bibitem[{\citenamefont{Tranquada
  et~al.}(2004{\natexlab{b}})\citenamefont{Tranquada, Lee, Yamada, Lee,
  Regnault, and R{\o}nnow}}]{tran04b}
\bibinfo{author}{\bibfnamefont{J.~M.} \bibnamefont{Tranquada}},
  \bibinfo{author}{\bibfnamefont{C.~H.} \bibnamefont{Lee}},
  \bibinfo{author}{\bibfnamefont{K.}~\bibnamefont{Yamada}},
  \bibinfo{author}{\bibfnamefont{Y.~S.} \bibnamefont{Lee}},
  \bibinfo{author}{\bibfnamefont{L.~P.} \bibnamefont{Regnault}},
  \bibnamefont{and} \bibinfo{author}{\bibfnamefont{H.~M.}
  \bibnamefont{R{\o}nnow}}, \bibinfo{journal}{Phys. Rev. B}
  \textbf{\bibinfo{volume}{69}}, \bibinfo{pages}{174507}
  (\bibinfo{year}{2004}{\natexlab{b}}).

\bibitem[{\citenamefont{Lake et~al.}(2001)\citenamefont{Lake, Aeppli, Clausen,
  McMorrow, Lefmann, Hussey, Mangkorntong, Nohara, Takagi, Mason
  et~al.}}]{lake01}
\bibinfo{author}{\bibfnamefont{B.}~\bibnamefont{Lake}},
  \bibinfo{author}{\bibfnamefont{G.}~\bibnamefont{Aeppli}},
  \bibinfo{author}{\bibfnamefont{K.~N.} \bibnamefont{Clausen}},
  \bibinfo{author}{\bibfnamefont{D.~F.} \bibnamefont{McMorrow}},
  \bibinfo{author}{\bibfnamefont{K.}~\bibnamefont{Lefmann}},
  \bibinfo{author}{\bibfnamefont{N.~E.} \bibnamefont{Hussey}},
  \bibinfo{author}{\bibfnamefont{N.}~\bibnamefont{Mangkorntong}},
  \bibinfo{author}{\bibfnamefont{M.}~\bibnamefont{Nohara}},
  \bibinfo{author}{\bibfnamefont{H.}~\bibnamefont{Takagi}},
  \bibinfo{author}{\bibfnamefont{T.~E.} \bibnamefont{Mason}},
  \bibnamefont{et~al.}, \bibinfo{journal}{Science}
  \textbf{\bibinfo{volume}{291}}, \bibinfo{pages}{1759} (\bibinfo{year}{2001}).

\bibitem[{\citenamefont{Anderson}(1997)}]{ande97}
\bibinfo{author}{\bibfnamefont{P.~W.} \bibnamefont{Anderson}},
  \bibinfo{journal}{Adv. Phys.} \textbf{\bibinfo{volume}{46}},
  \bibinfo{pages}{3} (\bibinfo{year}{1997}).

\bibitem[{\citenamefont{Tranquada
  et~al.}(1988{\natexlab{a}})\citenamefont{Tranquada, Moudden, Goldman,
  Zolliker, Cox, Shirane, Sinha, Vaknin, Johnston, Alvarez et~al.}}]{tran88}
\bibinfo{author}{\bibfnamefont{J.~M.} \bibnamefont{Tranquada}},
  \bibinfo{author}{\bibfnamefont{A.~H.} \bibnamefont{Moudden}},
  \bibinfo{author}{\bibfnamefont{A.~I.} \bibnamefont{Goldman}},
  \bibinfo{author}{\bibfnamefont{P.}~\bibnamefont{Zolliker}},
  \bibinfo{author}{\bibfnamefont{D.~E.} \bibnamefont{Cox}},
  \bibinfo{author}{\bibfnamefont{G.}~\bibnamefont{Shirane}},
  \bibinfo{author}{\bibfnamefont{S.~K.} \bibnamefont{Sinha}},
  \bibinfo{author}{\bibfnamefont{D.}~\bibnamefont{Vaknin}},
  \bibinfo{author}{\bibfnamefont{D.~C.} \bibnamefont{Johnston}},
  \bibinfo{author}{\bibfnamefont{M.~S.} \bibnamefont{Alvarez}},
  \bibnamefont{et~al.}, \bibinfo{journal}{Phys. Rev. B}
  \textbf{\bibinfo{volume}{38}}, \bibinfo{pages}{2477}
  (\bibinfo{year}{1988}{\natexlab{a}}).

\bibitem[{\citenamefont{Lee et~al.}(1999{\natexlab{a}})\citenamefont{Lee,
  Birgeneau, Kastner, Endoh, Wakimoto, Yamada, Erwin, Lee, and
  Shirane}}]{lee99}
\bibinfo{author}{\bibfnamefont{Y.~S.} \bibnamefont{Lee}},
  \bibinfo{author}{\bibfnamefont{R.~J.} \bibnamefont{Birgeneau}},
  \bibinfo{author}{\bibfnamefont{M.~A.} \bibnamefont{Kastner}},
  \bibinfo{author}{\bibfnamefont{Y.}~\bibnamefont{Endoh}},
  \bibinfo{author}{\bibfnamefont{S.}~\bibnamefont{Wakimoto}},
  \bibinfo{author}{\bibfnamefont{K.}~\bibnamefont{Yamada}},
  \bibinfo{author}{\bibfnamefont{R.~W.} \bibnamefont{Erwin}},
  \bibinfo{author}{\bibfnamefont{S.-H.} \bibnamefont{Lee}}, \bibnamefont{and}
  \bibinfo{author}{\bibfnamefont{G.}~\bibnamefont{Shirane}},
  \bibinfo{journal}{Phys. Rev. B} \textbf{\bibinfo{volume}{60}},
  \bibinfo{pages}{3643} (\bibinfo{year}{1999}{\natexlab{a}}).

\bibitem[{\citenamefont{Tranquada
  et~al.}(1988{\natexlab{b}})\citenamefont{Tranquada, Cox, Kunnmann, Moudden,
  Shirane, Suenaga, Zolliker, Vaknin, Sinha, Alvarez et~al.}}]{tran88a}
\bibinfo{author}{\bibfnamefont{J.~M.} \bibnamefont{Tranquada}},
  \bibinfo{author}{\bibfnamefont{D.~E.} \bibnamefont{Cox}},
  \bibinfo{author}{\bibfnamefont{W.}~\bibnamefont{Kunnmann}},
  \bibinfo{author}{\bibfnamefont{H.}~\bibnamefont{Moudden}},
  \bibinfo{author}{\bibfnamefont{G.}~\bibnamefont{Shirane}},
  \bibinfo{author}{\bibfnamefont{M.}~\bibnamefont{Suenaga}},
  \bibinfo{author}{\bibfnamefont{P.}~\bibnamefont{Zolliker}},
  \bibinfo{author}{\bibfnamefont{D.}~\bibnamefont{Vaknin}},
  \bibinfo{author}{\bibfnamefont{S.~K.} \bibnamefont{Sinha}},
  \bibinfo{author}{\bibfnamefont{M.~S.} \bibnamefont{Alvarez}},
  \bibnamefont{et~al.}, \bibinfo{journal}{Phys. Rev. Lett.}
  \textbf{\bibinfo{volume}{60}}, \bibinfo{pages}{156}
  (\bibinfo{year}{1988}{\natexlab{b}}).

\bibitem[{\citenamefont{Freltoft et~al.}(1987)\citenamefont{Freltoft, Fischer,
  Shirane, Moncton, Sinha, Vaknin, Remeikas, Cooper, and Harshman}}]{frel87}
\bibinfo{author}{\bibfnamefont{T.}~\bibnamefont{Freltoft}},
  \bibinfo{author}{\bibfnamefont{J.~E.} \bibnamefont{Fischer}},
  \bibinfo{author}{\bibfnamefont{G.}~\bibnamefont{Shirane}},
  \bibinfo{author}{\bibfnamefont{D.~E.} \bibnamefont{Moncton}},
  \bibinfo{author}{\bibfnamefont{S.~K.} \bibnamefont{Sinha}},
  \bibinfo{author}{\bibfnamefont{D.}~\bibnamefont{Vaknin}},
  \bibinfo{author}{\bibfnamefont{J.~P.} \bibnamefont{Remeikas}},
  \bibinfo{author}{\bibfnamefont{A.~S.} \bibnamefont{Cooper}},
  \bibnamefont{and} \bibinfo{author}{\bibfnamefont{D.}~\bibnamefont{Harshman}},
  \bibinfo{journal}{Phys. Rev. B} \textbf{\bibinfo{volume}{36}},
  \bibinfo{pages}{826} (\bibinfo{year}{1987}).

\bibitem[{\citenamefont{Kastner et~al.}(1988)\citenamefont{Kastner, Birgeneau,
  Thurston, Picone, Jensen, Gabbe, Sato, Fukuda, Shamoto, Endoh
  et~al.}}]{kast88}
\bibinfo{author}{\bibfnamefont{M.~A.} \bibnamefont{Kastner}},
  \bibinfo{author}{\bibfnamefont{R.~J.} \bibnamefont{Birgeneau}},
  \bibinfo{author}{\bibfnamefont{T.~R.} \bibnamefont{Thurston}},
  \bibinfo{author}{\bibfnamefont{P.~J.} \bibnamefont{Picone}},
  \bibinfo{author}{\bibfnamefont{H.~P.} \bibnamefont{Jensen}},
  \bibinfo{author}{\bibfnamefont{D.~R.} \bibnamefont{Gabbe}},
  \bibinfo{author}{\bibfnamefont{M.}~\bibnamefont{Sato}},
  \bibinfo{author}{\bibfnamefont{K.}~\bibnamefont{Fukuda}},
  \bibinfo{author}{\bibfnamefont{S.}~\bibnamefont{Shamoto}},
  \bibinfo{author}{\bibfnamefont{Y.}~\bibnamefont{Endoh}},
  \bibnamefont{et~al.}, \bibinfo{journal}{Phys. Rev. B}
  \textbf{\bibinfo{volume}{38}}, \bibinfo{pages}{6636} (\bibinfo{year}{1988}).

\bibitem[{\citenamefont{Thio and Aharony}(1994)}]{thio94}
\bibinfo{author}{\bibfnamefont{T.}~\bibnamefont{Thio}} \bibnamefont{and}
  \bibinfo{author}{\bibfnamefont{A.}~\bibnamefont{Aharony}},
  \bibinfo{journal}{Phys. Rev. Lett.} \textbf{\bibinfo{volume}{73}},
  \bibinfo{pages}{894} (\bibinfo{year}{1994}).

\bibitem[{\citenamefont{Zheludev et~al.}(1998)\citenamefont{Zheludev, Maslov,
  Zaliznyak, Regnault, Masuda, Uchinokura, Erwin, and Shirane}}]{zhel98}
\bibinfo{author}{\bibfnamefont{A.}~\bibnamefont{Zheludev}},
  \bibinfo{author}{\bibfnamefont{S.}~\bibnamefont{Maslov}},
  \bibinfo{author}{\bibfnamefont{I.}~\bibnamefont{Zaliznyak}},
  \bibinfo{author}{\bibfnamefont{L.~P.} \bibnamefont{Regnault}},
  \bibinfo{author}{\bibfnamefont{T.}~\bibnamefont{Masuda}},
  \bibinfo{author}{\bibfnamefont{K.}~\bibnamefont{Uchinokura}},
  \bibinfo{author}{\bibfnamefont{R.}~\bibnamefont{Erwin}}, \bibnamefont{and}
  \bibinfo{author}{\bibfnamefont{G.}~\bibnamefont{Shirane}},
  \bibinfo{journal}{Phys. Rev. Lett.} \textbf{\bibinfo{volume}{81}},
  \bibinfo{pages}{5410} (\bibinfo{year}{1998}).

\bibitem[{\citenamefont{Coffey et~al.}(1991)\citenamefont{Coffey, Rice, and
  Zhang}}]{coff91}
\bibinfo{author}{\bibfnamefont{D.}~\bibnamefont{Coffey}},
  \bibinfo{author}{\bibfnamefont{T.~M.} \bibnamefont{Rice}}, \bibnamefont{and}
  \bibinfo{author}{\bibfnamefont{F.~C.} \bibnamefont{Zhang}},
  \bibinfo{journal}{Phys. Rev. B} \textbf{\bibinfo{volume}{44}},
  \bibinfo{pages}{10112} (\bibinfo{year}{1991}).

\bibitem[{\citenamefont{Stein et~al.}(1996)\citenamefont{Stein, Entin-Wohlman,
  and Aharony}}]{stei96}
\bibinfo{author}{\bibfnamefont{J.}~\bibnamefont{Stein}},
  \bibinfo{author}{\bibfnamefont{O.}~\bibnamefont{Entin-Wohlman}},
  \bibnamefont{and} \bibinfo{author}{\bibfnamefont{A.}~\bibnamefont{Aharony}},
  \bibinfo{journal}{Phys. Rev. B} \textbf{\bibinfo{volume}{53}},
  \bibinfo{pages}{775} (\bibinfo{year}{1996}).

\bibitem[{\citenamefont{Wells et~al.}(1997)\citenamefont{Wells, Lee, Kastner,
  Christianson, Birgeneau, Yamada, Endoh, and Shirane}}]{well97}
\bibinfo{author}{\bibfnamefont{B.~O.} \bibnamefont{Wells}},
  \bibinfo{author}{\bibfnamefont{Y.~S.} \bibnamefont{Lee}},
  \bibinfo{author}{\bibfnamefont{M.~A.} \bibnamefont{Kastner}},
  \bibinfo{author}{\bibfnamefont{R.~J.} \bibnamefont{Christianson}},
  \bibinfo{author}{\bibfnamefont{R.~J.} \bibnamefont{Birgeneau}},
  \bibinfo{author}{\bibfnamefont{K.}~\bibnamefont{Yamada}},
  \bibinfo{author}{\bibfnamefont{Y.}~\bibnamefont{Endoh}}, \bibnamefont{and}
  \bibinfo{author}{\bibfnamefont{G.}~\bibnamefont{Shirane}},
  \bibinfo{journal}{Science} \textbf{\bibinfo{volume}{277}},
  \bibinfo{pages}{1067} (\bibinfo{year}{1997}).

\bibitem[{\citenamefont{Keimer et~al.}(1992)\citenamefont{Keimer, Aharony,
  Auerbach, Birgeneau, Cassanho, Endoh, Erwin, Kastner, and Shirane}}]{keim92a}
\bibinfo{author}{\bibfnamefont{B.}~\bibnamefont{Keimer}},
  \bibinfo{author}{\bibfnamefont{A.}~\bibnamefont{Aharony}},
  \bibinfo{author}{\bibfnamefont{A.}~\bibnamefont{Auerbach}},
  \bibinfo{author}{\bibfnamefont{R.~J.} \bibnamefont{Birgeneau}},
  \bibinfo{author}{\bibfnamefont{A.}~\bibnamefont{Cassanho}},
  \bibinfo{author}{\bibfnamefont{Y.}~\bibnamefont{Endoh}},
  \bibinfo{author}{\bibfnamefont{R.~W.} \bibnamefont{Erwin}},
  \bibinfo{author}{\bibfnamefont{M.~A.} \bibnamefont{Kastner}},
  \bibnamefont{and} \bibinfo{author}{\bibfnamefont{G.}~\bibnamefont{Shirane}},
  \bibinfo{journal}{Phys. Rev. B} \textbf{\bibinfo{volume}{45}},
  \bibinfo{pages}{7430} (\bibinfo{year}{1992}).

\bibitem[{\citenamefont{Yamada et~al.}(1987)\citenamefont{Yamada, Kudo, Endoh,
  Hidaka, Oda, Suzuki, and Murakami}}]{yama87}
\bibinfo{author}{\bibfnamefont{K.}~\bibnamefont{Yamada}},
  \bibinfo{author}{\bibfnamefont{E.}~\bibnamefont{Kudo}},
  \bibinfo{author}{\bibfnamefont{Y.}~\bibnamefont{Endoh}},
  \bibinfo{author}{\bibfnamefont{Y.}~\bibnamefont{Hidaka}},
  \bibinfo{author}{\bibfnamefont{M.}~\bibnamefont{Oda}},
  \bibinfo{author}{\bibfnamefont{M.}~\bibnamefont{Suzuki}}, \bibnamefont{and}
  \bibinfo{author}{\bibfnamefont{T.}~\bibnamefont{Murakami}},
  \bibinfo{journal}{Solid State Commun.} \textbf{\bibinfo{volume}{64}},
  \bibinfo{pages}{753} (\bibinfo{year}{1987}).

\bibitem[{\citenamefont{Yildirim
  et~al.}(1994{\natexlab{a}})\citenamefont{Yildirim, Harris, Entin-Wohlman, and
  Aharony}}]{yild94a}
\bibinfo{author}{\bibfnamefont{T.}~\bibnamefont{Yildirim}},
  \bibinfo{author}{\bibfnamefont{A.~B.} \bibnamefont{Harris}},
  \bibinfo{author}{\bibfnamefont{O.}~\bibnamefont{Entin-Wohlman}},
  \bibnamefont{and} \bibinfo{author}{\bibfnamefont{A.}~\bibnamefont{Aharony}},
  \bibinfo{journal}{Phys. Rev. Lett.} \textbf{\bibinfo{volume}{72}},
  \bibinfo{pages}{3710} (\bibinfo{year}{1994}{\natexlab{a}}).

\bibitem[{\citenamefont{Yildirim
  et~al.}(1994{\natexlab{b}})\citenamefont{Yildirim, Harris, Entin-Wohlman, and
  Aharony}}]{yild94b}
\bibinfo{author}{\bibfnamefont{T.}~\bibnamefont{Yildirim}},
  \bibinfo{author}{\bibfnamefont{A.~B.} \bibnamefont{Harris}},
  \bibinfo{author}{\bibfnamefont{O.}~\bibnamefont{Entin-Wohlman}},
  \bibnamefont{and} \bibinfo{author}{\bibfnamefont{A.}~\bibnamefont{Aharony}},
  \bibinfo{journal}{Phys. Rev. Lett.} \textbf{\bibinfo{volume}{73}},
  \bibinfo{pages}{2919} (\bibinfo{year}{1994}{\natexlab{b}}).

\bibitem[{\citenamefont{Coldea et~al.}(2001)\citenamefont{Coldea, Hayden,
  Aeppli, Perring, Frost, Mason, Cheong, and Fisk}}]{cold01}
\bibinfo{author}{\bibfnamefont{R.}~\bibnamefont{Coldea}},
  \bibinfo{author}{\bibfnamefont{S.~M.} \bibnamefont{Hayden}},
  \bibinfo{author}{\bibfnamefont{G.}~\bibnamefont{Aeppli}},
  \bibinfo{author}{\bibfnamefont{T.~G.} \bibnamefont{Perring}},
  \bibinfo{author}{\bibfnamefont{C.~D.} \bibnamefont{Frost}},
  \bibinfo{author}{\bibfnamefont{T.~E.} \bibnamefont{Mason}},
  \bibinfo{author}{\bibfnamefont{S.-W.} \bibnamefont{Cheong}},
  \bibnamefont{and} \bibinfo{author}{\bibfnamefont{Z.}~\bibnamefont{Fisk}},
  \bibinfo{journal}{Phys. Rev. Lett.} \textbf{\bibinfo{volume}{86}},
  \bibinfo{pages}{5377} (\bibinfo{year}{2001}).

\bibitem[{\citenamefont{Vaknin et~al.}(1990)\citenamefont{Vaknin, Sinha,
  Stassis, Miller, and Johnston}}]{vakn90}
\bibinfo{author}{\bibfnamefont{D.}~\bibnamefont{Vaknin}},
  \bibinfo{author}{\bibfnamefont{S.~K.} \bibnamefont{Sinha}},
  \bibinfo{author}{\bibfnamefont{C.}~\bibnamefont{Stassis}},
  \bibinfo{author}{\bibfnamefont{L.~L.} \bibnamefont{Miller}},
  \bibnamefont{and} \bibinfo{author}{\bibfnamefont{D.~C.}
  \bibnamefont{Johnston}}, \bibinfo{journal}{Phys. Rev. B}
  \textbf{\bibinfo{volume}{41}}, \bibinfo{pages}{1926} (\bibinfo{year}{1990}).

\bibitem[{\citenamefont{Greven et~al.}(1995)\citenamefont{Greven, Birgeneau,
  Endoh, Kastner, Matsuda, and Shirane}}]{grev95}
\bibinfo{author}{\bibfnamefont{M.}~\bibnamefont{Greven}},
  \bibinfo{author}{\bibfnamefont{R.~J.} \bibnamefont{Birgeneau}},
  \bibinfo{author}{\bibfnamefont{Y.}~\bibnamefont{Endoh}},
  \bibinfo{author}{\bibfnamefont{M.~A.} \bibnamefont{Kastner}},
  \bibinfo{author}{\bibfnamefont{M.}~\bibnamefont{Matsuda}}, \bibnamefont{and}
  \bibinfo{author}{\bibfnamefont{G.}~\bibnamefont{Shirane}},
  \bibinfo{journal}{Z. Phys. B} \textbf{\bibinfo{volume}{96}},
  \bibinfo{pages}{465} (\bibinfo{year}{1995}).

\bibitem[{\citenamefont{Tokura et~al.}(1990)\citenamefont{Tokura, Koshihara,
  Arima, Takagi, Ishibashi, Ido, and Uchida}}]{toku90}
\bibinfo{author}{\bibfnamefont{Y.}~\bibnamefont{Tokura}},
  \bibinfo{author}{\bibfnamefont{S.}~\bibnamefont{Koshihara}},
  \bibinfo{author}{\bibfnamefont{T.}~\bibnamefont{Arima}},
  \bibinfo{author}{\bibfnamefont{H.}~\bibnamefont{Takagi}},
  \bibinfo{author}{\bibfnamefont{S.}~\bibnamefont{Ishibashi}},
  \bibinfo{author}{\bibfnamefont{T.}~\bibnamefont{Ido}}, \bibnamefont{and}
  \bibinfo{author}{\bibfnamefont{S.}~\bibnamefont{Uchida}},
  \bibinfo{journal}{Phys. Rev. B} \textbf{\bibinfo{volume}{41}},
  \bibinfo{pages}{11657} (\bibinfo{year}{1990}).

\bibitem[{\citenamefont{Vaknin et~al.}(1997)\citenamefont{Vaknin, Miller, and
  Zarestky}}]{vakn97}
\bibinfo{author}{\bibfnamefont{D.}~\bibnamefont{Vaknin}},
  \bibinfo{author}{\bibfnamefont{L.~L.} \bibnamefont{Miller}},
  \bibnamefont{and} \bibinfo{author}{\bibfnamefont{J.~L.}
  \bibnamefont{Zarestky}}, \bibinfo{journal}{Phys. Rev. B}
  \textbf{\bibinfo{volume}{56}}, \bibinfo{pages}{8351} (\bibinfo{year}{1997}).

\bibitem[{\citenamefont{Matsuda et~al.}(1990)\citenamefont{Matsuda, Yamada,
  Kakurai, Kadowaki, Thurston, Endoh, Hidaka, Birgeneau, Kastner, Gehring
  et~al.}}]{mats90}
\bibinfo{author}{\bibfnamefont{M.}~\bibnamefont{Matsuda}},
  \bibinfo{author}{\bibfnamefont{K.}~\bibnamefont{Yamada}},
  \bibinfo{author}{\bibfnamefont{K.}~\bibnamefont{Kakurai}},
  \bibinfo{author}{\bibfnamefont{H.}~\bibnamefont{Kadowaki}},
  \bibinfo{author}{\bibfnamefont{T.~R.} \bibnamefont{Thurston}},
  \bibinfo{author}{\bibfnamefont{Y.}~\bibnamefont{Endoh}},
  \bibinfo{author}{\bibfnamefont{Y.}~\bibnamefont{Hidaka}},
  \bibinfo{author}{\bibfnamefont{R.~J.} \bibnamefont{Birgeneau}},
  \bibinfo{author}{\bibfnamefont{M.~A.} \bibnamefont{Kastner}},
  \bibinfo{author}{\bibfnamefont{P.~M.} \bibnamefont{Gehring}},
  \bibnamefont{et~al.}, \bibinfo{journal}{Phys. Rev. B}
  \textbf{\bibinfo{volume}{42}}, \bibinfo{pages}{10098} (\bibinfo{year}{1990}).

\bibitem[{\citenamefont{Skanthakumar et~al.}(1993)\citenamefont{Skanthakumar,
  Lynn, Peng, and Li}}]{skan93}
\bibinfo{author}{\bibfnamefont{S.}~\bibnamefont{Skanthakumar}},
  \bibinfo{author}{\bibfnamefont{J.~W.} \bibnamefont{Lynn}},
  \bibinfo{author}{\bibfnamefont{J.~L.} \bibnamefont{Peng}}, \bibnamefont{and}
  \bibinfo{author}{\bibfnamefont{Z.~Y.} \bibnamefont{Li}},
  \bibinfo{journal}{Phys. Rev. B} \textbf{\bibinfo{volume}{47}},
  \bibinfo{pages}{6173} (\bibinfo{year}{1993}).

\bibitem[{\citenamefont{Lynn and Skanthakumar}(2001)}]{lynn01}
\bibinfo{author}{\bibfnamefont{J.~W.} \bibnamefont{Lynn}} \bibnamefont{and}
  \bibinfo{author}{\bibfnamefont{S.}~\bibnamefont{Skanthakumar}}, in
  \emph{\bibinfo{booktitle}{Handbood on the Physics and Chemistry of Rare
  Earths, Vol. 31}}, edited by
  \bibinfo{editor}{\bibfnamefont{J.}~\bibnamefont{K.~A.~Gschneidner}},
  \bibinfo{editor}{\bibfnamefont{L.}~\bibnamefont{Eyring}}, \bibnamefont{and}
  \bibinfo{editor}{\bibfnamefont{M.~B.} \bibnamefont{Maple}}
  (\bibinfo{publisher}{Elsevier Science}, \bibinfo{address}{Amsterdam},
  \bibinfo{year}{2001}), pp. \bibinfo{pages}{315--350}.

\bibitem[{\citenamefont{Bourges
  et~al.}(1997{\natexlab{b}})\citenamefont{Bourges, Casalta, Ivanov, and
  Petitgrand}}]{bour97c}
\bibinfo{author}{\bibfnamefont{P.}~\bibnamefont{Bourges}},
  \bibinfo{author}{\bibfnamefont{H.}~\bibnamefont{Casalta}},
  \bibinfo{author}{\bibfnamefont{A.~S.} \bibnamefont{Ivanov}},
  \bibnamefont{and}
  \bibinfo{author}{\bibfnamefont{D.}~\bibnamefont{Petitgrand}},
  \bibinfo{journal}{Phys. Rev. Lett.} \textbf{\bibinfo{volume}{79}},
  \bibinfo{pages}{4906} (\bibinfo{year}{1997}{\natexlab{b}}).

\bibitem[{\citenamefont{Sumarlin et~al.}(1995)\citenamefont{Sumarlin, Lynn,
  Chattopadhyay, Barilo, Zhigunov, and Peng}}]{suma95}
\bibinfo{author}{\bibfnamefont{I.~W.} \bibnamefont{Sumarlin}},
  \bibinfo{author}{\bibfnamefont{J.~W.} \bibnamefont{Lynn}},
  \bibinfo{author}{\bibfnamefont{T.}~\bibnamefont{Chattopadhyay}},
  \bibinfo{author}{\bibfnamefont{S.~N.} \bibnamefont{Barilo}},
  \bibinfo{author}{\bibfnamefont{D.~I.} \bibnamefont{Zhigunov}},
  \bibnamefont{and} \bibinfo{author}{\bibfnamefont{J.~L.} \bibnamefont{Peng}},
  \bibinfo{journal}{Phys. Rev. B} \textbf{\bibinfo{volume}{51}},
  \bibinfo{pages}{5824} (\bibinfo{year}{1995}).

\bibitem[{\citenamefont{Casalta et~al.}(1994)\citenamefont{Casalta, Schleger,
  Brecht, Montfrooij, Andersen, Lebech, Schmahl, Fuess, Liang, Hardy
  et~al.}}]{casa94}
\bibinfo{author}{\bibfnamefont{H.}~\bibnamefont{Casalta}},
  \bibinfo{author}{\bibfnamefont{P.}~\bibnamefont{Schleger}},
  \bibinfo{author}{\bibfnamefont{E.}~\bibnamefont{Brecht}},
  \bibinfo{author}{\bibfnamefont{W.}~\bibnamefont{Montfrooij}},
  \bibinfo{author}{\bibfnamefont{N.~H.} \bibnamefont{Andersen}},
  \bibinfo{author}{\bibfnamefont{B.}~\bibnamefont{Lebech}},
  \bibinfo{author}{\bibfnamefont{W.~W.} \bibnamefont{Schmahl}},
  \bibinfo{author}{\bibfnamefont{H.}~\bibnamefont{Fuess}},
  \bibinfo{author}{\bibfnamefont{R.}~\bibnamefont{Liang}},
  \bibinfo{author}{\bibfnamefont{W.~N.} \bibnamefont{Hardy}},
  \bibnamefont{et~al.}, \bibinfo{journal}{Phys. Rev. B}
  \textbf{\bibinfo{volume}{50}}, \bibinfo{pages}{9688} (\bibinfo{year}{1994}).

\bibitem[{\citenamefont{Mizuki et~al.}(1988)\citenamefont{Mizuki, Kubo, Manako,
  Shimakawa, Igarashi, Tranquada, Fujii, Rebelsky, and Shirane}}]{mizu88}
\bibinfo{author}{\bibfnamefont{J.}~\bibnamefont{Mizuki}},
  \bibinfo{author}{\bibfnamefont{Y.}~\bibnamefont{Kubo}},
  \bibinfo{author}{\bibfnamefont{T.}~\bibnamefont{Manako}},
  \bibinfo{author}{\bibfnamefont{Y.}~\bibnamefont{Shimakawa}},
  \bibinfo{author}{\bibfnamefont{H.}~\bibnamefont{Igarashi}},
  \bibinfo{author}{\bibfnamefont{J.~M.} \bibnamefont{Tranquada}},
  \bibinfo{author}{\bibfnamefont{Y.}~\bibnamefont{Fujii}},
  \bibinfo{author}{\bibfnamefont{L.}~\bibnamefont{Rebelsky}}, \bibnamefont{and}
  \bibinfo{author}{\bibfnamefont{G.}~\bibnamefont{Shirane}},
  \bibinfo{journal}{Physica C} \textbf{\bibinfo{volume}{156}},
  \bibinfo{pages}{781} (\bibinfo{year}{1988}).

\bibitem[{\citenamefont{Vaknin et~al.}(1989)\citenamefont{Vaknin, Caignol,
  Davies, Fischer, Johnston, and Goshorn}}]{vakn89}
\bibinfo{author}{\bibfnamefont{D.}~\bibnamefont{Vaknin}},
  \bibinfo{author}{\bibfnamefont{E.}~\bibnamefont{Caignol}},
  \bibinfo{author}{\bibfnamefont{P.~K.} \bibnamefont{Davies}},
  \bibinfo{author}{\bibfnamefont{J.~E.} \bibnamefont{Fischer}},
  \bibinfo{author}{\bibfnamefont{D.~C.} \bibnamefont{Johnston}},
  \bibnamefont{and} \bibinfo{author}{\bibfnamefont{D.~P.}
  \bibnamefont{Goshorn}}, \bibinfo{journal}{Phys. Rev. B}
  \textbf{\bibinfo{volume}{39}}, \bibinfo{pages}{9122} (\bibinfo{year}{1989}).

\bibitem[{\citenamefont{Sachidanandam et~al.}(1997)\citenamefont{Sachidanandam,
  Yildirim, Harris, Aharony, and Entin-Wohlman}}]{sach97}
\bibinfo{author}{\bibfnamefont{R.}~\bibnamefont{Sachidanandam}},
  \bibinfo{author}{\bibfnamefont{T.}~\bibnamefont{Yildirim}},
  \bibinfo{author}{\bibfnamefont{A.~B.} \bibnamefont{Harris}},
  \bibinfo{author}{\bibfnamefont{A.}~\bibnamefont{Aharony}}, \bibnamefont{and}
  \bibinfo{author}{\bibfnamefont{O.}~\bibnamefont{Entin-Wohlman}},
  \bibinfo{journal}{Phys. Rev. B} \textbf{\bibinfo{volume}{56}},
  \bibinfo{pages}{260} (\bibinfo{year}{1997}).

\bibitem[{\citenamefont{Petitgrand et~al.}(1999)\citenamefont{Petitgrand,
  Maleyev, Bourges, and Ivanov}}]{peti99}
\bibinfo{author}{\bibfnamefont{D.}~\bibnamefont{Petitgrand}},
  \bibinfo{author}{\bibfnamefont{S.~V.} \bibnamefont{Maleyev}},
  \bibinfo{author}{\bibfnamefont{P.}~\bibnamefont{Bourges}}, \bibnamefont{and}
  \bibinfo{author}{\bibfnamefont{A.~S.} \bibnamefont{Ivanov}},
  \bibinfo{journal}{Phys. Rev. B} \textbf{\bibinfo{volume}{59}},
  \bibinfo{pages}{1079} (\bibinfo{year}{1999}).

\bibitem[{\citenamefont{Burlet et~al.}(1998)\citenamefont{Burlet, Henry, and
  Regnault}}]{burl98}
\bibinfo{author}{\bibfnamefont{P.}~\bibnamefont{Burlet}},
  \bibinfo{author}{\bibfnamefont{J.~Y.} \bibnamefont{Henry}}, \bibnamefont{and}
  \bibinfo{author}{\bibfnamefont{L.~P.} \bibnamefont{Regnault}},
  \bibinfo{journal}{Physica C} \textbf{\bibinfo{volume}{296}},
  \bibinfo{pages}{205} (\bibinfo{year}{1998}).

\bibitem[{\citenamefont{{J\'anossy} et~al.}(1999)\citenamefont{{J\'anossy},
  Simon, {Feh\'er}, Rockenbauer, Korecz, Chen, Chowdhury, and Hodby}}]{jano99}
\bibinfo{author}{\bibfnamefont{A.}~\bibnamefont{{J\'anossy}}},
  \bibinfo{author}{\bibfnamefont{F.}~\bibnamefont{Simon}},
  \bibinfo{author}{\bibfnamefont{T.}~\bibnamefont{{Feh\'er}}},
  \bibinfo{author}{\bibfnamefont{A.}~\bibnamefont{Rockenbauer}},
  \bibinfo{author}{\bibfnamefont{L.}~\bibnamefont{Korecz}},
  \bibinfo{author}{\bibfnamefont{C.}~\bibnamefont{Chen}},
  \bibinfo{author}{\bibfnamefont{A.~J.~S.} \bibnamefont{Chowdhury}},
  \bibnamefont{and} \bibinfo{author}{\bibfnamefont{J.~W.} \bibnamefont{Hodby}},
  \bibinfo{journal}{Phys. Rev. B} \textbf{\bibinfo{volume}{59}},
  \bibinfo{pages}{1176} (\bibinfo{year}{1999}).

\bibitem[{\citenamefont{Kadowaki et~al.}(1988)\citenamefont{Kadowaki, Nishi,
  Yamada, Takeya, Takei, Shapiro, and Shirane}}]{kado88}
\bibinfo{author}{\bibfnamefont{H.}~\bibnamefont{Kadowaki}},
  \bibinfo{author}{\bibfnamefont{M.}~\bibnamefont{Nishi}},
  \bibinfo{author}{\bibfnamefont{Y.}~\bibnamefont{Yamada}},
  \bibinfo{author}{\bibfnamefont{H.}~\bibnamefont{Takeya}},
  \bibinfo{author}{\bibfnamefont{H.}~\bibnamefont{Takei}},
  \bibinfo{author}{\bibfnamefont{S.~M.} \bibnamefont{Shapiro}},
  \bibnamefont{and} \bibinfo{author}{\bibfnamefont{G.}~\bibnamefont{Shirane}},
  \bibinfo{journal}{Phys. Rev. B} \textbf{\bibinfo{volume}{37}},
  \bibinfo{pages}{7932} (\bibinfo{year}{1988}).

\bibitem[{\citenamefont{Brecht et~al.}(1995)\citenamefont{Brecht, Schmahl,
  Fuess, Casalta, Schleger, Lebech, Andersen, and Wolf}}]{brec95}
\bibinfo{author}{\bibfnamefont{E.}~\bibnamefont{Brecht}},
  \bibinfo{author}{\bibfnamefont{W.~W.} \bibnamefont{Schmahl}},
  \bibinfo{author}{\bibfnamefont{H.}~\bibnamefont{Fuess}},
  \bibinfo{author}{\bibfnamefont{H.}~\bibnamefont{Casalta}},
  \bibinfo{author}{\bibfnamefont{P.}~\bibnamefont{Schleger}},
  \bibinfo{author}{\bibfnamefont{B.}~\bibnamefont{Lebech}},
  \bibinfo{author}{\bibfnamefont{N.~H.} \bibnamefont{Andersen}},
  \bibnamefont{and} \bibinfo{author}{\bibfnamefont{T.}~\bibnamefont{Wolf}},
  \bibinfo{journal}{Phys. Rev. B} \textbf{\bibinfo{volume}{52}},
  \bibinfo{pages}{9601} (\bibinfo{year}{1995}).

\bibitem[{\citenamefont{Andersen and Uimin}(1997)}]{ande97b}
\bibinfo{author}{\bibfnamefont{N.~H.} \bibnamefont{Andersen}} \bibnamefont{and}
  \bibinfo{author}{\bibfnamefont{G.}~\bibnamefont{Uimin}},
  \bibinfo{journal}{Phys. Rev. B} \textbf{\bibinfo{volume}{56}},
  \bibinfo{pages}{10840} (\bibinfo{year}{1997}).

\bibitem[{\citenamefont{Shamoto et~al.}(1993)\citenamefont{Shamoto, Sato,
  Tranquada, Sternlieb, and Shirane}}]{sham93}
\bibinfo{author}{\bibfnamefont{S.}~\bibnamefont{Shamoto}},
  \bibinfo{author}{\bibfnamefont{M.}~\bibnamefont{Sato}},
  \bibinfo{author}{\bibfnamefont{J.~M.} \bibnamefont{Tranquada}},
  \bibinfo{author}{\bibfnamefont{B.~J.} \bibnamefont{Sternlieb}},
  \bibnamefont{and} \bibinfo{author}{\bibfnamefont{G.}~\bibnamefont{Shirane}},
  \bibinfo{journal}{Phys. Rev. B} \textbf{\bibinfo{volume}{48}},
  \bibinfo{pages}{13817} (\bibinfo{year}{1993}).

\bibitem[{\citenamefont{Lines}(1970)}]{line70}
\bibinfo{author}{\bibfnamefont{M.~E.} \bibnamefont{Lines}},
  \bibinfo{journal}{J. Phys. Chem. Solids} \textbf{\bibinfo{volume}{31}},
  \bibinfo{pages}{101} (\bibinfo{year}{1970}).

\bibitem[{\citenamefont{Lorenzana et~al.}(2005)\citenamefont{Lorenzana,
  Seibold, and Coldea}}]{lore05}
\bibinfo{author}{\bibfnamefont{J.}~\bibnamefont{Lorenzana}},
  \bibinfo{author}{\bibfnamefont{G.}~\bibnamefont{Seibold}}, \bibnamefont{and}
  \bibinfo{author}{\bibfnamefont{R.}~\bibnamefont{Coldea}}
  (\bibinfo{year}{2005}), \eprint{cond-mat/0507131}.

\bibitem[{\citenamefont{Capriotta et~al.}(2005)\citenamefont{Capriotta,
  L\"auchli, and Paramekanti}}]{capr05}
\bibinfo{author}{\bibfnamefont{L.}~\bibnamefont{Capriotta}},
  \bibinfo{author}{\bibfnamefont{A.}~\bibnamefont{L\"auchli}},
  \bibnamefont{and}
  \bibinfo{author}{\bibfnamefont{A.}~\bibnamefont{Paramekanti}}
  (\bibinfo{year}{2005}), \eprint{cond-mat/0406188}.

\bibitem[{\citenamefont{Yasuda et~al.}(2005)\citenamefont{Yasuda, Todo,
  Hukushima, Alet, Keller, Troyer, and Takayama}}]{yasu05}
\bibinfo{author}{\bibfnamefont{C.}~\bibnamefont{Yasuda}},
  \bibinfo{author}{\bibfnamefont{S.}~\bibnamefont{Todo}},
  \bibinfo{author}{\bibfnamefont{K.}~\bibnamefont{Hukushima}},
  \bibinfo{author}{\bibfnamefont{F.}~\bibnamefont{Alet}},
  \bibinfo{author}{\bibfnamefont{M.}~\bibnamefont{Keller}},
  \bibinfo{author}{\bibfnamefont{M.}~\bibnamefont{Troyer}}, \bibnamefont{and}
  \bibinfo{author}{\bibfnamefont{H.}~\bibnamefont{Takayama}},
  \bibinfo{journal}{Phys. Rev. Lett.} \textbf{\bibinfo{volume}{94}},
  \bibinfo{pages}{217201} (\bibinfo{year}{2005}).

\bibitem[{\citenamefont{Kojima et~al.}(1997)\citenamefont{Kojima, Fudamoto,
  Larkin, Luke, Merrin, Nachumi, Uemura, Motoyama, Eisaki, Uchida
  et~al.}}]{koji97}
\bibinfo{author}{\bibfnamefont{K.~M.} \bibnamefont{Kojima}},
  \bibinfo{author}{\bibfnamefont{Y.}~\bibnamefont{Fudamoto}},
  \bibinfo{author}{\bibfnamefont{M.}~\bibnamefont{Larkin}},
  \bibinfo{author}{\bibfnamefont{G.~M.} \bibnamefont{Luke}},
  \bibinfo{author}{\bibfnamefont{J.}~\bibnamefont{Merrin}},
  \bibinfo{author}{\bibfnamefont{B.}~\bibnamefont{Nachumi}},
  \bibinfo{author}{\bibfnamefont{Y.~J.} \bibnamefont{Uemura}},
  \bibinfo{author}{\bibfnamefont{N.}~\bibnamefont{Motoyama}},
  \bibinfo{author}{\bibfnamefont{H.}~\bibnamefont{Eisaki}},
  \bibinfo{author}{\bibfnamefont{S.}~\bibnamefont{Uchida}},
  \bibnamefont{et~al.}, \bibinfo{journal}{Phys. Rev. Lett.}
  \textbf{\bibinfo{volume}{78}}, \bibinfo{pages}{1787} (\bibinfo{year}{1997}).

\bibitem[{\citenamefont{Anderson}(1959)}]{ande59}
\bibinfo{author}{\bibfnamefont{P.~W.} \bibnamefont{Anderson}},
  \bibinfo{journal}{Phys. Rev.} \textbf{\bibinfo{volume}{115}},
  \bibinfo{pages}{2} (\bibinfo{year}{1959}).

\bibitem[{\citenamefont{Oguchi}(1960)}]{oguc60}
\bibinfo{author}{\bibfnamefont{T.}~\bibnamefont{Oguchi}},
  \bibinfo{journal}{Phys. Rev.} \textbf{\bibinfo{volume}{117}},
  \bibinfo{pages}{117} (\bibinfo{year}{1960}).

\bibitem[{\citenamefont{Singh}(1989)}]{sing89}
\bibinfo{author}{\bibfnamefont{R.~R.~P.} \bibnamefont{Singh}},
  \bibinfo{journal}{Phys. Rev. B} \textbf{\bibinfo{volume}{39}},
  \bibinfo{pages}{9760} (\bibinfo{year}{1989}).

\bibitem[{\citenamefont{Lyons et~al.}(1988)\citenamefont{Lyons, Fleury,
  Remeika, Cooper, and Negran}}]{lyon88}
\bibinfo{author}{\bibfnamefont{K.~B.} \bibnamefont{Lyons}},
  \bibinfo{author}{\bibfnamefont{P.~A.} \bibnamefont{Fleury}},
  \bibinfo{author}{\bibfnamefont{J.~P.} \bibnamefont{Remeika}},
  \bibinfo{author}{\bibfnamefont{A.~S.} \bibnamefont{Cooper}},
  \bibnamefont{and} \bibinfo{author}{\bibfnamefont{T.~J.}
  \bibnamefont{Negran}}, \bibinfo{journal}{Phys. Rev. B}
  \textbf{\bibinfo{volume}{37}}, \bibinfo{pages}{2353} (\bibinfo{year}{1988}).

\bibitem[{\citenamefont{Emery and Reiter}(1988)}]{emer88}
\bibinfo{author}{\bibfnamefont{V.~J.} \bibnamefont{Emery}} \bibnamefont{and}
  \bibinfo{author}{\bibfnamefont{G.}~\bibnamefont{Reiter}},
  \bibinfo{journal}{Phys. Rev. B} \textbf{\bibinfo{volume}{38}},
  \bibinfo{pages}{4547} (\bibinfo{year}{1988}).

\bibitem[{\citenamefont{{Van Oosten} et~al.}(1996)\citenamefont{{Van Oosten},
  Broer, and Nieupoort}}]{oost96}
\bibinfo{author}{\bibfnamefont{A.~B.} \bibnamefont{{Van Oosten}}},
  \bibinfo{author}{\bibfnamefont{R.}~\bibnamefont{Broer}}, \bibnamefont{and}
  \bibinfo{author}{\bibfnamefont{W.~C.} \bibnamefont{Nieupoort}},
  \bibinfo{journal}{Chem. Phys. Lett.} \textbf{\bibinfo{volume}{257}},
  \bibinfo{pages}{207} (\bibinfo{year}{1996}).

\bibitem[{\citenamefont{{Mu\~noz} et~al.}(2000)\citenamefont{{Mu\~noz}, Illas,
  and Moreira}}]{muno00}
\bibinfo{author}{\bibfnamefont{D.}~\bibnamefont{{Mu\~noz}}},
  \bibinfo{author}{\bibfnamefont{F.}~\bibnamefont{Illas}}, \bibnamefont{and}
  \bibinfo{author}{\bibfnamefont{I.~P.~R.} \bibnamefont{Moreira}},
  \bibinfo{journal}{Phys. Rev. Lett.} \textbf{\bibinfo{volume}{84}},
  \bibinfo{pages}{1579} (\bibinfo{year}{2000}).

\bibitem[{\citenamefont{Zaliznyak et~al.}(2004)\citenamefont{Zaliznyak, Woo,
  Perring, Broholm, Frost, and Takagi}}]{zali04}
\bibinfo{author}{\bibfnamefont{I.~A.} \bibnamefont{Zaliznyak}},
  \bibinfo{author}{\bibfnamefont{H.}~\bibnamefont{Woo}},
  \bibinfo{author}{\bibfnamefont{T.~G.} \bibnamefont{Perring}},
  \bibinfo{author}{\bibfnamefont{C.~L.} \bibnamefont{Broholm}},
  \bibinfo{author}{\bibfnamefont{C.~D.} \bibnamefont{Frost}}, \bibnamefont{and}
  \bibinfo{author}{\bibfnamefont{H.}~\bibnamefont{Takagi}},
  \bibinfo{journal}{Phys. Rev. Lett.} \textbf{\bibinfo{volume}{93}},
  \bibinfo{pages}{087202} (\bibinfo{year}{2004}).

\bibitem[{\citenamefont{Takahashi}(1977)}]{taka77}
\bibinfo{author}{\bibfnamefont{M.}~\bibnamefont{Takahashi}},
  \bibinfo{journal}{J. Phys. C} \textbf{\bibinfo{volume}{10}},
  \bibinfo{pages}{1289} (\bibinfo{year}{1977}).

\bibitem[{\citenamefont{Roger and Delrieu}(1989)}]{roge89}
\bibinfo{author}{\bibfnamefont{M.}~\bibnamefont{Roger}} \bibnamefont{and}
  \bibinfo{author}{\bibfnamefont{J.~M.} \bibnamefont{Delrieu}},
  \bibinfo{journal}{Phys. Rev. B} \textbf{\bibinfo{volume}{39}},
  \bibinfo{pages}{2299} (\bibinfo{year}{1989}).

\bibitem[{\citenamefont{MacDonald et~al.}(1988)\citenamefont{MacDonald, Girvin,
  and Yoshioka}}]{macd88}
\bibinfo{author}{\bibfnamefont{A.~H.} \bibnamefont{MacDonald}},
  \bibinfo{author}{\bibfnamefont{S.~M.} \bibnamefont{Girvin}},
  \bibnamefont{and} \bibinfo{author}{\bibfnamefont{D.}~\bibnamefont{Yoshioka}},
  \bibinfo{journal}{Phys. Rev. B} \textbf{\bibinfo{volume}{37}},
  \bibinfo{pages}{9753} (\bibinfo{year}{1988}).

\bibitem[{\citenamefont{Morr}(1998)}]{morr98b}
\bibinfo{author}{\bibfnamefont{D.~K.} \bibnamefont{Morr}},
  \bibinfo{journal}{Phys. Rev. B} \textbf{\bibinfo{volume}{58}},
  \bibinfo{pages}{R587} (\bibinfo{year}{1998}).

\bibitem[{\citenamefont{Annett et~al.}(1989)\citenamefont{Annett, Martin,
  McMahan, and Satpathy}}]{anne89}
\bibinfo{author}{\bibfnamefont{J.~F.} \bibnamefont{Annett}},
  \bibinfo{author}{\bibfnamefont{R.~M.} \bibnamefont{Martin}},
  \bibinfo{author}{\bibfnamefont{A.~K.} \bibnamefont{McMahan}},
  \bibnamefont{and} \bibinfo{author}{\bibfnamefont{S.}~\bibnamefont{Satpathy}},
  \bibinfo{journal}{Phys. Rev. B} \textbf{\bibinfo{volume}{40}},
  \bibinfo{pages}{2620} (\bibinfo{year}{1989}).

\bibitem[{\citenamefont{Tranquada et~al.}(1989)\citenamefont{Tranquada,
  Shirane, Keimer, Shamoto, and Sato}}]{tran89}
\bibinfo{author}{\bibfnamefont{J.~M.} \bibnamefont{Tranquada}},
  \bibinfo{author}{\bibfnamefont{G.}~\bibnamefont{Shirane}},
  \bibinfo{author}{\bibfnamefont{B.}~\bibnamefont{Keimer}},
  \bibinfo{author}{\bibfnamefont{S.}~\bibnamefont{Shamoto}}, \bibnamefont{and}
  \bibinfo{author}{\bibfnamefont{M.}~\bibnamefont{Sato}},
  \bibinfo{journal}{Phys. Rev. B} \textbf{\bibinfo{volume}{40}},
  \bibinfo{pages}{4503} (\bibinfo{year}{1989}).

\bibitem[{\citenamefont{Jorgensen et~al.}(1990)\citenamefont{Jorgensen, Veal,
  Paulikas, Nowicki, Crabtree, Claus, and Kwok}}]{jorg90}
\bibinfo{author}{\bibfnamefont{J.~D.} \bibnamefont{Jorgensen}},
  \bibinfo{author}{\bibfnamefont{B.~W.} \bibnamefont{Veal}},
  \bibinfo{author}{\bibfnamefont{A.~P.} \bibnamefont{Paulikas}},
  \bibinfo{author}{\bibfnamefont{L.~J.} \bibnamefont{Nowicki}},
  \bibinfo{author}{\bibfnamefont{G.~W.} \bibnamefont{Crabtree}},
  \bibinfo{author}{\bibfnamefont{H.}~\bibnamefont{Claus}}, \bibnamefont{and}
  \bibinfo{author}{\bibfnamefont{W.~K.} \bibnamefont{Kwok}},
  \bibinfo{journal}{Phys. Rev. B} \textbf{\bibinfo{volume}{41}},
  \bibinfo{pages}{1863} (\bibinfo{year}{1990}).

\bibitem[{\citenamefont{Shender}(1982)}]{shen82}
\bibinfo{author}{\bibfnamefont{E.~F.} \bibnamefont{Shender}},
  \bibinfo{journal}{Sov. Phys. JETP} \textbf{\bibinfo{volume}{56}}
  (\bibinfo{year}{1982}).

\bibitem[{\citenamefont{Keimer et~al.}(1993)\citenamefont{Keimer, Birgeneau,
  Cassanho, Endoh, Greven, Kastner, and Shirane}}]{keim93}
\bibinfo{author}{\bibfnamefont{B.}~\bibnamefont{Keimer}},
  \bibinfo{author}{\bibfnamefont{R.~J.} \bibnamefont{Birgeneau}},
  \bibinfo{author}{\bibfnamefont{A.}~\bibnamefont{Cassanho}},
  \bibinfo{author}{\bibfnamefont{Y.}~\bibnamefont{Endoh}},
  \bibinfo{author}{\bibfnamefont{M.}~\bibnamefont{Greven}},
  \bibinfo{author}{\bibfnamefont{M.~A.} \bibnamefont{Kastner}},
  \bibnamefont{and} \bibinfo{author}{\bibfnamefont{G.}~\bibnamefont{Shirane}},
  \bibinfo{journal}{Z. Phys. B} \textbf{\bibinfo{volume}{91}},
  \bibinfo{pages}{373} (\bibinfo{year}{1993}).

\bibitem[{\citenamefont{Peters et~al.}(1988)\citenamefont{Peters, Birgeneau,
  Kastner, Yoshizawa, Endoh, Tranquada, Shirane, Hidaka, Oda, Suzuki
  et~al.}}]{pete88}
\bibinfo{author}{\bibfnamefont{C.~J.} \bibnamefont{Peters}},
  \bibinfo{author}{\bibfnamefont{R.~J.} \bibnamefont{Birgeneau}},
  \bibinfo{author}{\bibfnamefont{M.~A.} \bibnamefont{Kastner}},
  \bibinfo{author}{\bibfnamefont{H.}~\bibnamefont{Yoshizawa}},
  \bibinfo{author}{\bibfnamefont{Y.}~\bibnamefont{Endoh}},
  \bibinfo{author}{\bibfnamefont{J.}~\bibnamefont{Tranquada}},
  \bibinfo{author}{\bibfnamefont{G.}~\bibnamefont{Shirane}},
  \bibinfo{author}{\bibfnamefont{Y.}~\bibnamefont{Hidaka}},
  \bibinfo{author}{\bibfnamefont{M.}~\bibnamefont{Oda}},
  \bibinfo{author}{\bibfnamefont{M.}~\bibnamefont{Suzuki}},
  \bibnamefont{et~al.}, \bibinfo{journal}{Phys. Rev. B}
  \textbf{\bibinfo{volume}{37}}, \bibinfo{pages}{9761} (\bibinfo{year}{1988}).

\bibitem[{\citenamefont{Shekhtman et~al.}(1992)\citenamefont{Shekhtman,
  Entin-Wohlman, and Aharony}}]{shek92}
\bibinfo{author}{\bibfnamefont{L.}~\bibnamefont{Shekhtman}},
  \bibinfo{author}{\bibfnamefont{O.}~\bibnamefont{Entin-Wohlman}},
  \bibnamefont{and} \bibinfo{author}{\bibfnamefont{A.}~\bibnamefont{Aharony}},
  \bibinfo{journal}{Phys. Rev. Lett.} \textbf{\bibinfo{volume}{69}},
  \bibinfo{pages}{836} (\bibinfo{year}{1992}).

\bibitem[{\citenamefont{Vettier et~al.}(1989)\citenamefont{Vettier, Burlet,
  Henry, Jurgens, Lapertot, Regnault, and Rossat-Mignod}}]{vett89}
\bibinfo{author}{\bibfnamefont{C.}~\bibnamefont{Vettier}},
  \bibinfo{author}{\bibfnamefont{P.}~\bibnamefont{Burlet}},
  \bibinfo{author}{\bibfnamefont{J.~Y.} \bibnamefont{Henry}},
  \bibinfo{author}{\bibfnamefont{M.~J.} \bibnamefont{Jurgens}},
  \bibinfo{author}{\bibfnamefont{G.}~\bibnamefont{Lapertot}},
  \bibinfo{author}{\bibfnamefont{L.~P.} \bibnamefont{Regnault}},
  \bibnamefont{and}
  \bibinfo{author}{\bibfnamefont{J.}~\bibnamefont{Rossat-Mignod}},
  \bibinfo{journal}{Physica Scripta} \textbf{\bibinfo{volume}{T29}},
  \bibinfo{pages}{110} (\bibinfo{year}{1989}).

\bibitem[{\citenamefont{Rossat-Mignod et~al.}(1990)\citenamefont{Rossat-Mignod,
  Regnault, Jurgens, Vettier, Burlet, Henry, and Lapertot}}]{ross90}
\bibinfo{author}{\bibfnamefont{J.}~\bibnamefont{Rossat-Mignod}},
  \bibinfo{author}{\bibfnamefont{L.~P.} \bibnamefont{Regnault}},
  \bibinfo{author}{\bibfnamefont{M.~J.} \bibnamefont{Jurgens}},
  \bibinfo{author}{\bibfnamefont{C.}~\bibnamefont{Vettier}},
  \bibinfo{author}{\bibfnamefont{P.}~\bibnamefont{Burlet}},
  \bibinfo{author}{\bibfnamefont{J.~Y.} \bibnamefont{Henry}}, \bibnamefont{and}
  \bibinfo{author}{\bibfnamefont{G.}~\bibnamefont{Lapertot}},
  \bibinfo{journal}{Physica B} \textbf{\bibinfo{volume}{163}},
  \bibinfo{pages}{4} (\bibinfo{year}{1990}).

\bibitem[{\citenamefont{Shirane et~al.}(1987)\citenamefont{Shirane, Endoh,
  Birgeneau, Kastner, Hidaka, Oda, Suzuki, and Murakami}}]{shir87}
\bibinfo{author}{\bibfnamefont{G.}~\bibnamefont{Shirane}},
  \bibinfo{author}{\bibfnamefont{Y.}~\bibnamefont{Endoh}},
  \bibinfo{author}{\bibfnamefont{R.~J.} \bibnamefont{Birgeneau}},
  \bibinfo{author}{\bibfnamefont{M.~A.} \bibnamefont{Kastner}},
  \bibinfo{author}{\bibfnamefont{Y.}~\bibnamefont{Hidaka}},
  \bibinfo{author}{\bibfnamefont{M.}~\bibnamefont{Oda}},
  \bibinfo{author}{\bibfnamefont{M.}~\bibnamefont{Suzuki}}, \bibnamefont{and}
  \bibinfo{author}{\bibfnamefont{T.}~\bibnamefont{Murakami}},
  \bibinfo{journal}{Phys. Rev. Lett.} \textbf{\bibinfo{volume}{59}},
  \bibinfo{pages}{1613} (\bibinfo{year}{1987}).

\bibitem[{\citenamefont{Birgeneau et~al.}(1999)\citenamefont{Birgeneau, Greven,
  Kastner, Lee, Wells, Endoh, Yamada, and Shirane}}]{birg99}
\bibinfo{author}{\bibfnamefont{R.~J.} \bibnamefont{Birgeneau}},
  \bibinfo{author}{\bibfnamefont{M.}~\bibnamefont{Greven}},
  \bibinfo{author}{\bibfnamefont{M.~A.} \bibnamefont{Kastner}},
  \bibinfo{author}{\bibfnamefont{Y.~S.} \bibnamefont{Lee}},
  \bibinfo{author}{\bibfnamefont{B.~O.} \bibnamefont{Wells}},
  \bibinfo{author}{\bibfnamefont{Y.}~\bibnamefont{Endoh}},
  \bibinfo{author}{\bibfnamefont{K.}~\bibnamefont{Yamada}}, \bibnamefont{and}
  \bibinfo{author}{\bibfnamefont{G.}~\bibnamefont{Shirane}},
  \bibinfo{journal}{Phys. Rev. B} \textbf{\bibinfo{volume}{59}},
  \bibinfo{pages}{13788} (\bibinfo{year}{1999}).

\bibitem[{\citenamefont{Greven et~al.}(1994)\citenamefont{Greven, Birgeneau,
  Endoh, Kastner, Keimer, Matsuda, Shirane, and Thurston}}]{grev94}
\bibinfo{author}{\bibfnamefont{M.}~\bibnamefont{Greven}},
  \bibinfo{author}{\bibfnamefont{R.~J.} \bibnamefont{Birgeneau}},
  \bibinfo{author}{\bibfnamefont{Y.}~\bibnamefont{Endoh}},
  \bibinfo{author}{\bibfnamefont{M.~A.} \bibnamefont{Kastner}},
  \bibinfo{author}{\bibfnamefont{B.}~\bibnamefont{Keimer}},
  \bibinfo{author}{\bibfnamefont{M.}~\bibnamefont{Matsuda}},
  \bibinfo{author}{\bibfnamefont{G.}~\bibnamefont{Shirane}}, \bibnamefont{and}
  \bibinfo{author}{\bibfnamefont{T.~R.} \bibnamefont{Thurston}},
  \bibinfo{journal}{Phys. Rev. Lett.} \textbf{\bibinfo{volume}{72}},
  \bibinfo{pages}{1096} (\bibinfo{year}{1994}).

\bibitem[{\citenamefont{Chakravarty et~al.}(1989)\citenamefont{Chakravarty,
  Halperin, and Nelson}}]{chak89}
\bibinfo{author}{\bibfnamefont{S.}~\bibnamefont{Chakravarty}},
  \bibinfo{author}{\bibfnamefont{B.~I.} \bibnamefont{Halperin}},
  \bibnamefont{and} \bibinfo{author}{\bibfnamefont{D.~R.}
  \bibnamefont{Nelson}}, \bibinfo{journal}{Phys. Rev. B}
  \textbf{\bibinfo{volume}{39}}, \bibinfo{pages}{2344} (\bibinfo{year}{1989}).

\bibitem[{\citenamefont{Hasenfratz and Niedermayer}(1991)}]{hase91}
\bibinfo{author}{\bibfnamefont{P.}~\bibnamefont{Hasenfratz}} \bibnamefont{and}
  \bibinfo{author}{\bibfnamefont{F.}~\bibnamefont{Niedermayer}},
  \bibinfo{journal}{Phys. Lett. B} \textbf{\bibinfo{volume}{268}},
  \bibinfo{pages}{231} (\bibinfo{year}{1991}).

\bibitem[{\citenamefont{Chakravarty et~al.}(1988)\citenamefont{Chakravarty,
  Halperin, and Nelson}}]{chak88}
\bibinfo{author}{\bibfnamefont{S.}~\bibnamefont{Chakravarty}},
  \bibinfo{author}{\bibfnamefont{B.~I.} \bibnamefont{Halperin}},
  \bibnamefont{and} \bibinfo{author}{\bibfnamefont{D.~R.}
  \bibnamefont{Nelson}}, \bibinfo{journal}{Phys. Rev. Lett.}
  \textbf{\bibinfo{volume}{60}}, \bibinfo{pages}{1057} (\bibinfo{year}{1988}).

\bibitem[{\citenamefont{Suh et~al.}(1995)\citenamefont{Suh, Borsa, Miller,
  Corti, Johnston, and Torgeson}}]{suh95}
\bibinfo{author}{\bibfnamefont{B.~J.} \bibnamefont{Suh}},
  \bibinfo{author}{\bibfnamefont{F.}~\bibnamefont{Borsa}},
  \bibinfo{author}{\bibfnamefont{L.~L.} \bibnamefont{Miller}},
  \bibinfo{author}{\bibfnamefont{M.}~\bibnamefont{Corti}},
  \bibinfo{author}{\bibfnamefont{D.~C.} \bibnamefont{Johnston}},
  \bibnamefont{and} \bibinfo{author}{\bibfnamefont{D.~R.}
  \bibnamefont{Torgeson}}, \bibinfo{journal}{Phys. Rev. Lett.}
  \textbf{\bibinfo{volume}{75}}, \bibinfo{pages}{2212} (\bibinfo{year}{1995}).

\bibitem[{\citenamefont{Cuccoli et~al.}(2003)\citenamefont{Cuccoli, Roscilde,
  Vaia, and Verrucchi}}]{cucc03}
\bibinfo{author}{\bibfnamefont{A.}~\bibnamefont{Cuccoli}},
  \bibinfo{author}{\bibfnamefont{T.}~\bibnamefont{Roscilde}},
  \bibinfo{author}{\bibfnamefont{R.}~\bibnamefont{Vaia}}, \bibnamefont{and}
  \bibinfo{author}{\bibfnamefont{P.}~\bibnamefont{Verrucchi}},
  \bibinfo{journal}{Phys. Rev. Lett.} \textbf{\bibinfo{volume}{90}},
  \bibinfo{pages}{167205} (\bibinfo{year}{2003}).

\bibitem[{\citenamefont{Kim et~al.}(2001)\citenamefont{Kim, Birgeneau, Chou,
  Erwin, and Kastner}}]{kim01}
\bibinfo{author}{\bibfnamefont{Y.~J.} \bibnamefont{Kim}},
  \bibinfo{author}{\bibfnamefont{R.~J.} \bibnamefont{Birgeneau}},
  \bibinfo{author}{\bibfnamefont{F.~C.} \bibnamefont{Chou}},
  \bibinfo{author}{\bibfnamefont{R.~W.} \bibnamefont{Erwin}}, \bibnamefont{and}
  \bibinfo{author}{\bibfnamefont{M.~A.} \bibnamefont{Kastner}},
  \bibinfo{journal}{Phys. Rev. Lett.} \textbf{\bibinfo{volume}{86}},
  \bibinfo{pages}{3144} (\bibinfo{year}{2001}).

\bibitem[{\citenamefont{Hohenberg and Halperin}(1977)}]{hohe77}
\bibinfo{author}{\bibfnamefont{P.~C.} \bibnamefont{Hohenberg}}
  \bibnamefont{and} \bibinfo{author}{\bibfnamefont{B.~I.}
  \bibnamefont{Halperin}}, \bibinfo{journal}{Rev. Mod. Phys.}
  \textbf{\bibinfo{volume}{49}}, \bibinfo{pages}{435} (\bibinfo{year}{1977}).

\bibitem[{\citenamefont{Niedermayer et~al.}(1998)\citenamefont{Niedermayer,
  Bernhard, Blasius, Golnik, Moodenbaugh, and Budnick}}]{nied98}
\bibinfo{author}{\bibfnamefont{C.}~\bibnamefont{Niedermayer}},
  \bibinfo{author}{\bibfnamefont{C.}~\bibnamefont{Bernhard}},
  \bibinfo{author}{\bibfnamefont{T.}~\bibnamefont{Blasius}},
  \bibinfo{author}{\bibfnamefont{A.}~\bibnamefont{Golnik}},
  \bibinfo{author}{\bibfnamefont{A.}~\bibnamefont{Moodenbaugh}},
  \bibnamefont{and} \bibinfo{author}{\bibfnamefont{J.~I.}
  \bibnamefont{Budnick}}, \bibinfo{journal}{Phys. Rev. Lett.}
  \textbf{\bibinfo{volume}{80}}, \bibinfo{pages}{3843} (\bibinfo{year}{1998}).

\bibitem[{\citenamefont{Vajk et~al.}(2002)\citenamefont{Vajk, Mang, Greven,
  Gehring, and Lynn}}]{vajk02}
\bibinfo{author}{\bibfnamefont{O.~P.} \bibnamefont{Vajk}},
  \bibinfo{author}{\bibfnamefont{P.~K.} \bibnamefont{Mang}},
  \bibinfo{author}{\bibfnamefont{M.}~\bibnamefont{Greven}},
  \bibinfo{author}{\bibfnamefont{P.~M.} \bibnamefont{Gehring}},
  \bibnamefont{and} \bibinfo{author}{\bibfnamefont{J.~W.} \bibnamefont{Lynn}},
  \bibinfo{journal}{Science} \textbf{\bibinfo{volume}{295}},
  \bibinfo{pages}{1691} (\bibinfo{year}{2002}).

\bibitem[{\citenamefont{Matsuda
  et~al.}(2002{\natexlab{a}})\citenamefont{Matsuda, Fujita, Yamada, Birgeneau,
  Endoh, and Shirane}}]{mats02}
\bibinfo{author}{\bibfnamefont{M.}~\bibnamefont{Matsuda}},
  \bibinfo{author}{\bibfnamefont{M.}~\bibnamefont{Fujita}},
  \bibinfo{author}{\bibfnamefont{K.}~\bibnamefont{Yamada}},
  \bibinfo{author}{\bibfnamefont{R.~J.} \bibnamefont{Birgeneau}},
  \bibinfo{author}{\bibfnamefont{Y.}~\bibnamefont{Endoh}}, \bibnamefont{and}
  \bibinfo{author}{\bibfnamefont{G.}~\bibnamefont{Shirane}},
  \bibinfo{journal}{Phys. Rev. B} \textbf{\bibinfo{volume}{65}},
  \bibinfo{pages}{134515} (\bibinfo{year}{2002}{\natexlab{a}}).

\bibitem[{\citenamefont{Wakimoto
  et~al.}(2000{\natexlab{a}})\citenamefont{Wakimoto, Birgeneau, Kastner, Lee,
  Erwin, Gehring, Lee, Fujita, Yamada, Endoh et~al.}}]{waki00}
\bibinfo{author}{\bibfnamefont{S.}~\bibnamefont{Wakimoto}},
  \bibinfo{author}{\bibfnamefont{R.~J.} \bibnamefont{Birgeneau}},
  \bibinfo{author}{\bibfnamefont{M.~A.} \bibnamefont{Kastner}},
  \bibinfo{author}{\bibfnamefont{Y.~S.} \bibnamefont{Lee}},
  \bibinfo{author}{\bibfnamefont{R.}~\bibnamefont{Erwin}},
  \bibinfo{author}{\bibfnamefont{P.~M.} \bibnamefont{Gehring}},
  \bibinfo{author}{\bibfnamefont{S.~H.} \bibnamefont{Lee}},
  \bibinfo{author}{\bibfnamefont{M.}~\bibnamefont{Fujita}},
  \bibinfo{author}{\bibfnamefont{K.}~\bibnamefont{Yamada}},
  \bibinfo{author}{\bibfnamefont{Y.}~\bibnamefont{Endoh}},
  \bibnamefont{et~al.}, \bibinfo{journal}{Phys. Rev. B}
  \textbf{\bibinfo{volume}{61}}, \bibinfo{pages}{3699}
  (\bibinfo{year}{2000}{\natexlab{a}}).

\bibitem[{\citenamefont{Matsuda et~al.}(2000)\citenamefont{Matsuda, Fujita,
  Yamada, Birgeneau, Kastner, Hiraka, Endoh, Wakimoto, and Shirane}}]{mats00b}
\bibinfo{author}{\bibfnamefont{M.}~\bibnamefont{Matsuda}},
  \bibinfo{author}{\bibfnamefont{M.}~\bibnamefont{Fujita}},
  \bibinfo{author}{\bibfnamefont{K.}~\bibnamefont{Yamada}},
  \bibinfo{author}{\bibfnamefont{R.~J.} \bibnamefont{Birgeneau}},
  \bibinfo{author}{\bibfnamefont{M.~A.} \bibnamefont{Kastner}},
  \bibinfo{author}{\bibfnamefont{H.}~\bibnamefont{Hiraka}},
  \bibinfo{author}{\bibfnamefont{Y.}~\bibnamefont{Endoh}},
  \bibinfo{author}{\bibfnamefont{S.}~\bibnamefont{Wakimoto}}, \bibnamefont{and}
  \bibinfo{author}{\bibfnamefont{G.}~\bibnamefont{Shirane}},
  \bibinfo{journal}{Phys. Rev. B} \textbf{\bibinfo{volume}{62}},
  \bibinfo{pages}{9148} (\bibinfo{year}{2000}).

\bibitem[{\citenamefont{Wakimoto et~al.}(2001)\citenamefont{Wakimoto,
  Birgeneau, Lee, and Shirane}}]{waki01b}
\bibinfo{author}{\bibfnamefont{S.}~\bibnamefont{Wakimoto}},
  \bibinfo{author}{\bibfnamefont{R.~J.} \bibnamefont{Birgeneau}},
  \bibinfo{author}{\bibfnamefont{Y.~S.} \bibnamefont{Lee}}, \bibnamefont{and}
  \bibinfo{author}{\bibfnamefont{G.}~\bibnamefont{Shirane}},
  \bibinfo{journal}{Phys. Rev. B} \textbf{\bibinfo{volume}{63}},
  \bibinfo{pages}{172501} (\bibinfo{year}{2001}).

\bibitem[{\citenamefont{Fujita et~al.}(2002{\natexlab{a}})\citenamefont{Fujita,
  Yamada, Hiraka, Gehring, Lee, Wakimoto, and Shirane}}]{fuji02c}
\bibinfo{author}{\bibfnamefont{M.}~\bibnamefont{Fujita}},
  \bibinfo{author}{\bibfnamefont{K.}~\bibnamefont{Yamada}},
  \bibinfo{author}{\bibfnamefont{H.}~\bibnamefont{Hiraka}},
  \bibinfo{author}{\bibfnamefont{P.~M.} \bibnamefont{Gehring}},
  \bibinfo{author}{\bibfnamefont{S.~H.} \bibnamefont{Lee}},
  \bibinfo{author}{\bibfnamefont{S.}~\bibnamefont{Wakimoto}}, \bibnamefont{and}
  \bibinfo{author}{\bibfnamefont{G.}~\bibnamefont{Shirane}},
  \bibinfo{journal}{Phys. Rev. B} \textbf{\bibinfo{volume}{65}},
  \bibinfo{pages}{064505} (\bibinfo{year}{2002}{\natexlab{a}}).

\bibitem[{\citenamefont{Chou et~al.}(1995)\citenamefont{Chou, Belk, Kastner,
  Birgeneau, and Aharony}}]{chou95}
\bibinfo{author}{\bibfnamefont{F.~C.} \bibnamefont{Chou}},
  \bibinfo{author}{\bibfnamefont{N.~R.} \bibnamefont{Belk}},
  \bibinfo{author}{\bibfnamefont{M.~A.} \bibnamefont{Kastner}},
  \bibinfo{author}{\bibfnamefont{R.~J.} \bibnamefont{Birgeneau}},
  \bibnamefont{and} \bibinfo{author}{\bibfnamefont{A.}~\bibnamefont{Aharony}},
  \bibinfo{journal}{Phys. Rev. Lett.} \textbf{\bibinfo{volume}{75}},
  \bibinfo{pages}{2204} (\bibinfo{year}{1995}).

\bibitem[{\citenamefont{Wakimoto
  et~al.}(2000{\natexlab{b}})\citenamefont{Wakimoto, Ueki, Endoh, and
  Yamada}}]{waki00b}
\bibinfo{author}{\bibfnamefont{S.}~\bibnamefont{Wakimoto}},
  \bibinfo{author}{\bibfnamefont{S.}~\bibnamefont{Ueki}},
  \bibinfo{author}{\bibfnamefont{Y.}~\bibnamefont{Endoh}}, \bibnamefont{and}
  \bibinfo{author}{\bibfnamefont{K.}~\bibnamefont{Yamada}},
  \bibinfo{journal}{Phys. Rev. B} \textbf{\bibinfo{volume}{62}},
  \bibinfo{pages}{3547} (\bibinfo{year}{2000}{\natexlab{b}}).

\bibitem[{\citenamefont{Tranquada}(1998{\natexlab{a}})}]{tran98b}
\bibinfo{author}{\bibfnamefont{J.~M.} \bibnamefont{Tranquada}}, in
  \emph{\bibinfo{booktitle}{Neutron Scattering in Layered Copper-Oxide
  Superconductors}}, edited by
  \bibinfo{editor}{\bibfnamefont{A.}~\bibnamefont{Furrer}}
  (\bibinfo{publisher}{Kluwer}, \bibinfo{address}{Dordrecht, The Netherlands},
  \bibinfo{year}{1998}{\natexlab{a}}), p. \bibinfo{pages}{225}.

\bibitem[{\citenamefont{Jorgensen et~al.}(1988)\citenamefont{Jorgensen,
  Dabrowski, Pei, Hinks, Soderholm, Morosin, Schirber, Venturini, and
  Ginley}}]{jorg88}
\bibinfo{author}{\bibfnamefont{J.~D.} \bibnamefont{Jorgensen}},
  \bibinfo{author}{\bibfnamefont{B.}~\bibnamefont{Dabrowski}},
  \bibinfo{author}{\bibfnamefont{S.}~\bibnamefont{Pei}},
  \bibinfo{author}{\bibfnamefont{D.~G.} \bibnamefont{Hinks}},
  \bibinfo{author}{\bibfnamefont{L.}~\bibnamefont{Soderholm}},
  \bibinfo{author}{\bibfnamefont{B.}~\bibnamefont{Morosin}},
  \bibinfo{author}{\bibfnamefont{J.~E.} \bibnamefont{Schirber}},
  \bibinfo{author}{\bibfnamefont{E.~L.} \bibnamefont{Venturini}},
  \bibnamefont{and} \bibinfo{author}{\bibfnamefont{D.~S.}
  \bibnamefont{Ginley}}, \bibinfo{journal}{Phys. Rev. B}
  \textbf{\bibinfo{volume}{38}}, \bibinfo{pages}{11337} (\bibinfo{year}{1988}).

\bibitem[{\citenamefont{Hammel et~al.}(1990)\citenamefont{Hammel, Reyes, Fisk,
  Takigawa, Thompson, Heffner, Cheong, and Schirber}}]{hamm90}
\bibinfo{author}{\bibfnamefont{P.~C.} \bibnamefont{Hammel}},
  \bibinfo{author}{\bibfnamefont{A.~P.} \bibnamefont{Reyes}},
  \bibinfo{author}{\bibfnamefont{Z.}~\bibnamefont{Fisk}},
  \bibinfo{author}{\bibfnamefont{M.}~\bibnamefont{Takigawa}},
  \bibinfo{author}{\bibfnamefont{J.~D.} \bibnamefont{Thompson}},
  \bibinfo{author}{\bibfnamefont{R.~H.} \bibnamefont{Heffner}},
  \bibinfo{author}{\bibfnamefont{S.-W.} \bibnamefont{Cheong}},
  \bibnamefont{and} \bibinfo{author}{\bibfnamefont{J.~E.}
  \bibnamefont{Schirber}}, \bibinfo{journal}{Phys. Rev. B}
  \textbf{\bibinfo{volume}{42}}, \bibinfo{pages}{6781} (\bibinfo{year}{1990}).

\bibitem[{\citenamefont{Gnezdilov et~al.}(2004)\citenamefont{Gnezdilov,
  Pashkevich, Tranquada, Lemmens, GŸntherodt, Yeremenko, Barilo, Shiryaev,
  Kurnevich, and Gehring}}]{gnez04}
\bibinfo{author}{\bibfnamefont{V.~P.} \bibnamefont{Gnezdilov}},
  \bibinfo{author}{\bibfnamefont{Y.~G.} \bibnamefont{Pashkevich}},
  \bibinfo{author}{\bibfnamefont{J.~M.} \bibnamefont{Tranquada}},
  \bibinfo{author}{\bibfnamefont{P.}~\bibnamefont{Lemmens}},
  \bibinfo{author}{\bibfnamefont{G.}~\bibnamefont{GŸntherodt}},
  \bibinfo{author}{\bibfnamefont{A.~V.} \bibnamefont{Yeremenko}},
  \bibinfo{author}{\bibfnamefont{S.~N.} \bibnamefont{Barilo}},
  \bibinfo{author}{\bibfnamefont{S.~V.} \bibnamefont{Shiryaev}},
  \bibinfo{author}{\bibfnamefont{L.~A.} \bibnamefont{Kurnevich}},
  \bibnamefont{and} \bibinfo{author}{\bibfnamefont{P.~M.}
  \bibnamefont{Gehring}}, \bibinfo{journal}{Phys. Rev. B}
  \textbf{\bibinfo{volume}{69}}, \bibinfo{pages}{174508}
  (\bibinfo{year}{2004}).

\bibitem[{\citenamefont{Renault et~al.}(1987)\citenamefont{Renault, Burdett,
  and Pouget}}]{rena87}
\bibinfo{author}{\bibfnamefont{A.}~\bibnamefont{Renault}},
  \bibinfo{author}{\bibfnamefont{J.~K.} \bibnamefont{Burdett}},
  \bibnamefont{and} \bibinfo{author}{\bibfnamefont{J.-P.}
  \bibnamefont{Pouget}}, \bibinfo{journal}{J. Solid State Chem.}
  \textbf{\bibinfo{volume}{71}}, \bibinfo{pages}{587} (\bibinfo{year}{1987}).

\bibitem[{\citenamefont{Maletta et~al.}(1989)\citenamefont{Maletta, P\"orschke,
  Rupp, and Meuffels}}]{male89}
\bibinfo{author}{\bibfnamefont{H.}~\bibnamefont{Maletta}},
  \bibinfo{author}{\bibfnamefont{E.}~\bibnamefont{P\"orschke}},
  \bibinfo{author}{\bibfnamefont{B.}~\bibnamefont{Rupp}}, \bibnamefont{and}
  \bibinfo{author}{\bibfnamefont{P.}~\bibnamefont{Meuffels}},
  \bibinfo{journal}{Z. Phys. B} \textbf{\bibinfo{volume}{77}},
  \bibinfo{pages}{181} (\bibinfo{year}{1989}).

\bibitem[{\citenamefont{Cava et~al.}(1990)\citenamefont{Cava, Hewat, Hewat,
  Batlogg, Marezio, Rabe, Krajewski, W.~F.~Peck, and L.~W.~Rupp}}]{cava90}
\bibinfo{author}{\bibfnamefont{R.~J.} \bibnamefont{Cava}},
  \bibinfo{author}{\bibfnamefont{A.~W.} \bibnamefont{Hewat}},
  \bibinfo{author}{\bibfnamefont{E.~A.} \bibnamefont{Hewat}},
  \bibinfo{author}{\bibfnamefont{B.}~\bibnamefont{Batlogg}},
  \bibinfo{author}{\bibfnamefont{M.}~\bibnamefont{Marezio}},
  \bibinfo{author}{\bibfnamefont{K.~M.} \bibnamefont{Rabe}},
  \bibinfo{author}{\bibfnamefont{J.~J.} \bibnamefont{Krajewski}},
  \bibinfo{author}{\bibfnamefont{J.}~\bibnamefont{W.~F.~Peck}},
  \bibnamefont{and}
  \bibinfo{author}{\bibfnamefont{J.}~\bibnamefont{L.~W.~Rupp}},
  \bibinfo{journal}{Physica C} \textbf{\bibinfo{volume}{165}},
  \bibinfo{pages}{419} (\bibinfo{year}{1990}).

\bibitem[{\citenamefont{Uimin and Rossat-Mignod}(1992)}]{uimi92}
\bibinfo{author}{\bibfnamefont{G.}~\bibnamefont{Uimin}} \bibnamefont{and}
  \bibinfo{author}{\bibfnamefont{J.}~\bibnamefont{Rossat-Mignod}},
  \bibinfo{journal}{Physica C} \textbf{\bibinfo{volume}{199}},
  \bibinfo{pages}{251} (\bibinfo{year}{1992}).

\bibitem[{\citenamefont{Rossat-Mignod et~al.}(1988)\citenamefont{Rossat-Mignod,
  Burlet, Jurgens, Vettier, Regnault, Henry, Ayache, Forro, Noel, Potel
  et~al.}}]{ross88}
\bibinfo{author}{\bibfnamefont{J.}~\bibnamefont{Rossat-Mignod}},
  \bibinfo{author}{\bibfnamefont{P.}~\bibnamefont{Burlet}},
  \bibinfo{author}{\bibfnamefont{M.~J.} \bibnamefont{Jurgens}},
  \bibinfo{author}{\bibfnamefont{C.}~\bibnamefont{Vettier}},
  \bibinfo{author}{\bibfnamefont{L.~P.} \bibnamefont{Regnault}},
  \bibinfo{author}{\bibfnamefont{J.~Y.} \bibnamefont{Henry}},
  \bibinfo{author}{\bibfnamefont{C.}~\bibnamefont{Ayache}},
  \bibinfo{author}{\bibfnamefont{L.}~\bibnamefont{Forro}},
  \bibinfo{author}{\bibfnamefont{H.}~\bibnamefont{Noel}},
  \bibinfo{author}{\bibfnamefont{M.}~\bibnamefont{Potel}},
  \bibnamefont{et~al.}, \bibinfo{journal}{J. Phys. (Paris)}
  \textbf{\bibinfo{volume}{49}}, \bibinfo{pages}{C8} (\bibinfo{year}{1988}).

\bibitem[{\citenamefont{Stock et~al.}(2005{\natexlab{b}})\citenamefont{Stock,
  Buyers, Yamani, Broholm, Chung, Tun, Liang, Bonn, Hardy, and
  Birgeneau}}]{stoc05b}
\bibinfo{author}{\bibfnamefont{C.}~\bibnamefont{Stock}},
  \bibinfo{author}{\bibfnamefont{W.~J.~L.} \bibnamefont{Buyers}},
  \bibinfo{author}{\bibfnamefont{Z.}~\bibnamefont{Yamani}},
  \bibinfo{author}{\bibfnamefont{C.~L.} \bibnamefont{Broholm}},
  \bibinfo{author}{\bibfnamefont{J.-H.} \bibnamefont{Chung}},
  \bibinfo{author}{\bibfnamefont{Z.}~\bibnamefont{Tun}},
  \bibinfo{author}{\bibfnamefont{R.}~\bibnamefont{Liang}},
  \bibinfo{author}{\bibfnamefont{D.}~\bibnamefont{Bonn}},
  \bibinfo{author}{\bibfnamefont{W.~N.} \bibnamefont{Hardy}}, \bibnamefont{and}
  \bibinfo{author}{\bibfnamefont{R.~J.} \bibnamefont{Birgeneau}}
  (\bibinfo{year}{2005}{\natexlab{b}}), \eprint{cond-mat/0505083}.

\bibitem[{\citenamefont{Ulrich et~al.}(2002)\citenamefont{Ulrich, Kondo,
  Reehuis, He, Bernhard, Niedarmayer, Bour\'ee, Bourges, Ohl, R{\o}nnow
  et~al.}}]{ulri02}
\bibinfo{author}{\bibfnamefont{C.}~\bibnamefont{Ulrich}},
  \bibinfo{author}{\bibfnamefont{S.}~\bibnamefont{Kondo}},
  \bibinfo{author}{\bibfnamefont{M.}~\bibnamefont{Reehuis}},
  \bibinfo{author}{\bibfnamefont{H.}~\bibnamefont{He}},
  \bibinfo{author}{\bibfnamefont{C.}~\bibnamefont{Bernhard}},
  \bibinfo{author}{\bibfnamefont{C.}~\bibnamefont{Niedarmayer}},
  \bibinfo{author}{\bibfnamefont{F.}~\bibnamefont{Bour\'ee}},
  \bibinfo{author}{\bibfnamefont{P.}~\bibnamefont{Bourges}},
  \bibinfo{author}{\bibfnamefont{M.}~\bibnamefont{Ohl}},
  \bibinfo{author}{\bibfnamefont{H.~M.} \bibnamefont{R{\o}nnow}},
  \bibnamefont{et~al.}, \bibinfo{journal}{Phys. Rev. B}
  \textbf{\bibinfo{volume}{65}}, \bibinfo{pages}{220507}
  (\bibinfo{year}{2002}).

\bibitem[{\citenamefont{H\"ucker et~al.}(2005)\citenamefont{H\"ucker, Kim, Gu,
  Tranquada, Gaulin, and Lynn}}]{huck05}
\bibinfo{author}{\bibfnamefont{M.}~\bibnamefont{H\"ucker}},
  \bibinfo{author}{\bibfnamefont{Y.-J.} \bibnamefont{Kim}},
  \bibinfo{author}{\bibfnamefont{G.~D.} \bibnamefont{Gu}},
  \bibinfo{author}{\bibfnamefont{J.~M.} \bibnamefont{Tranquada}},
  \bibinfo{author}{\bibfnamefont{B.~D.} \bibnamefont{Gaulin}},
  \bibnamefont{and} \bibinfo{author}{\bibfnamefont{J.~W.} \bibnamefont{Lynn}},
  \bibinfo{journal}{Phys. Rev. B} \textbf{\bibinfo{volume}{71}},
  \bibinfo{pages}{094510} (\bibinfo{year}{2005}).

\bibitem[{\citenamefont{Wang et~al.}(2003)\citenamefont{Wang, Zheng, Feng, Gu,
  Homes, Tranquada, Gaulin, and Timusk}}]{wang03}
\bibinfo{author}{\bibfnamefont{N.~L.} \bibnamefont{Wang}},
  \bibinfo{author}{\bibfnamefont{P.}~\bibnamefont{Zheng}},
  \bibinfo{author}{\bibfnamefont{T.}~\bibnamefont{Feng}},
  \bibinfo{author}{\bibfnamefont{G.~D.} \bibnamefont{Gu}},
  \bibinfo{author}{\bibfnamefont{C.~C.} \bibnamefont{Homes}},
  \bibinfo{author}{\bibfnamefont{J.~M.} \bibnamefont{Tranquada}},
  \bibinfo{author}{\bibfnamefont{B.~D.} \bibnamefont{Gaulin}},
  \bibnamefont{and} \bibinfo{author}{\bibfnamefont{T.}~\bibnamefont{Timusk}},
  \bibinfo{journal}{Phys. Rev. B} \textbf{\bibinfo{volume}{67}},
  \bibinfo{pages}{134526} (\bibinfo{year}{2003}).

\bibitem[{\citenamefont{Cava et~al.}(1991)\citenamefont{Cava, Batlogg, Palstra,
  Krajewski, Jr., Ramirez, and Jr.}}]{cava91}
\bibinfo{author}{\bibfnamefont{R.~J.} \bibnamefont{Cava}},
  \bibinfo{author}{\bibfnamefont{B.}~\bibnamefont{Batlogg}},
  \bibinfo{author}{\bibfnamefont{T.~T.} \bibnamefont{Palstra}},
  \bibinfo{author}{\bibfnamefont{J.~J.} \bibnamefont{Krajewski}},
  \bibinfo{author}{\bibfnamefont{W.~F.~P.} \bibnamefont{Jr.}},
  \bibinfo{author}{\bibfnamefont{A.~P.} \bibnamefont{Ramirez}},
  \bibnamefont{and} \bibinfo{author}{\bibfnamefont{L.~W.~R.}
  \bibnamefont{Jr.}}, \bibinfo{journal}{Phys. Rev. B}
  \textbf{\bibinfo{volume}{43}}, \bibinfo{pages}{1229} (\bibinfo{year}{1991}).

\bibitem[{\citenamefont{Shinomori et~al.}(2002)\citenamefont{Shinomori,
  Okimoto, Kawasaki, and Tokura}}]{shin02}
\bibinfo{author}{\bibfnamefont{S.}~\bibnamefont{Shinomori}},
  \bibinfo{author}{\bibfnamefont{Y.}~\bibnamefont{Okimoto}},
  \bibinfo{author}{\bibfnamefont{M.}~\bibnamefont{Kawasaki}}, \bibnamefont{and}
  \bibinfo{author}{\bibfnamefont{Y.}~\bibnamefont{Tokura}},
  \bibinfo{journal}{J. Phys. Soc. Japan} \textbf{\bibinfo{volume}{71}},
  \bibinfo{pages}{705} (\bibinfo{year}{2002}).

\bibitem[{\citenamefont{Cheong et~al.}(1994)\citenamefont{Cheong, Hwang, Chen,
  Batlogg, L.~W.~Rupp, and Carter}}]{cheo94}
\bibinfo{author}{\bibfnamefont{S.-W.} \bibnamefont{Cheong}},
  \bibinfo{author}{\bibfnamefont{H.~Y.} \bibnamefont{Hwang}},
  \bibinfo{author}{\bibfnamefont{C.~H.} \bibnamefont{Chen}},
  \bibinfo{author}{\bibfnamefont{B.}~\bibnamefont{Batlogg}},
  \bibinfo{author}{\bibfnamefont{J.}~\bibnamefont{L.~W.~Rupp}},
  \bibnamefont{and} \bibinfo{author}{\bibfnamefont{S.~A.}
  \bibnamefont{Carter}}, \bibinfo{journal}{Phys. Rev. B}
  \textbf{\bibinfo{volume}{49}}, \bibinfo{pages}{7088} (\bibinfo{year}{1994}).

\bibitem[{\citenamefont{Yoshizawa et~al.}(2000)\citenamefont{Yoshizawa,
  Kakeshita, Kajimoto, Tanabe, Katsufuji, and Tokura}}]{yosh00}
\bibinfo{author}{\bibfnamefont{H.}~\bibnamefont{Yoshizawa}},
  \bibinfo{author}{\bibfnamefont{T.}~\bibnamefont{Kakeshita}},
  \bibinfo{author}{\bibfnamefont{R.}~\bibnamefont{Kajimoto}},
  \bibinfo{author}{\bibfnamefont{T.}~\bibnamefont{Tanabe}},
  \bibinfo{author}{\bibfnamefont{T.}~\bibnamefont{Katsufuji}},
  \bibnamefont{and} \bibinfo{author}{\bibfnamefont{Y.}~\bibnamefont{Tokura}},
  \bibinfo{journal}{Phys. Rev. B} \textbf{\bibinfo{volume}{61}},
  \bibinfo{pages}{R854} (\bibinfo{year}{2000}).

\bibitem[{\citenamefont{Tranquada
  et~al.}(1994{\natexlab{a}})\citenamefont{Tranquada, Buttrey, Sachan, and
  Lorenzo}}]{tran94a}
\bibinfo{author}{\bibfnamefont{J.~M.} \bibnamefont{Tranquada}},
  \bibinfo{author}{\bibfnamefont{D.~J.} \bibnamefont{Buttrey}},
  \bibinfo{author}{\bibfnamefont{V.}~\bibnamefont{Sachan}}, \bibnamefont{and}
  \bibinfo{author}{\bibfnamefont{J.~E.} \bibnamefont{Lorenzo}},
  \bibinfo{journal}{Phys. Rev. Lett.} \textbf{\bibinfo{volume}{73}},
  \bibinfo{pages}{1003} (\bibinfo{year}{1994}{\natexlab{a}}).

\bibitem[{\citenamefont{Sachan et~al.}(1995)\citenamefont{Sachan, Buttrey,
  Tranquada, Lorenzo, and Shirane}}]{sach95}
\bibinfo{author}{\bibfnamefont{V.}~\bibnamefont{Sachan}},
  \bibinfo{author}{\bibfnamefont{D.~J.} \bibnamefont{Buttrey}},
  \bibinfo{author}{\bibfnamefont{J.~M.} \bibnamefont{Tranquada}},
  \bibinfo{author}{\bibfnamefont{J.~E.} \bibnamefont{Lorenzo}},
  \bibnamefont{and} \bibinfo{author}{\bibfnamefont{G.}~\bibnamefont{Shirane}},
  \bibinfo{journal}{Phys. Rev. B} \textbf{\bibinfo{volume}{51}},
  \bibinfo{pages}{12742} (\bibinfo{year}{1995}).

\bibitem[{\citenamefont{Tranquada
  et~al.}(1996{\natexlab{a}})\citenamefont{Tranquada, Buttrey, and
  Sachan}}]{tran96a}
\bibinfo{author}{\bibfnamefont{J.~M.} \bibnamefont{Tranquada}},
  \bibinfo{author}{\bibfnamefont{D.~J.} \bibnamefont{Buttrey}},
  \bibnamefont{and} \bibinfo{author}{\bibfnamefont{V.}~\bibnamefont{Sachan}},
  \bibinfo{journal}{Phys. Rev. B} \textbf{\bibinfo{volume}{54}},
  \bibinfo{pages}{12318} (\bibinfo{year}{1996}{\natexlab{a}}).

\bibitem[{\citenamefont{Wochner et~al.}(1998)\citenamefont{Wochner, Tranquada,
  Buttrey, and Sachan}}]{woch98}
\bibinfo{author}{\bibfnamefont{P.}~\bibnamefont{Wochner}},
  \bibinfo{author}{\bibfnamefont{J.~M.} \bibnamefont{Tranquada}},
  \bibinfo{author}{\bibfnamefont{D.~J.} \bibnamefont{Buttrey}},
  \bibnamefont{and} \bibinfo{author}{\bibfnamefont{V.}~\bibnamefont{Sachan}},
  \bibinfo{journal}{Phys. Rev. B} \textbf{\bibinfo{volume}{57}},
  \bibinfo{pages}{1066} (\bibinfo{year}{1998}).

\bibitem[{\citenamefont{Yamada et~al.}(1992)\citenamefont{Yamada, Omata,
  Nakajima, Hosoya, Sumida, and Endoh}}]{yama92}
\bibinfo{author}{\bibfnamefont{K.}~\bibnamefont{Yamada}},
  \bibinfo{author}{\bibfnamefont{T.}~\bibnamefont{Omata}},
  \bibinfo{author}{\bibfnamefont{K.}~\bibnamefont{Nakajima}},
  \bibinfo{author}{\bibfnamefont{S.}~\bibnamefont{Hosoya}},
  \bibinfo{author}{\bibfnamefont{T.}~\bibnamefont{Sumida}}, \bibnamefont{and}
  \bibinfo{author}{\bibfnamefont{Y.}~\bibnamefont{Endoh}},
  \bibinfo{journal}{Physica C} \textbf{\bibinfo{volume}{191}},
  \bibinfo{pages}{15} (\bibinfo{year}{1992}).

\bibitem[{\citenamefont{Tranquada
  et~al.}(1994{\natexlab{b}})\citenamefont{Tranquada, Kong, Lorenzo, Buttrey,
  Rice, and Sachan}}]{tran94b}
\bibinfo{author}{\bibfnamefont{J.~M.} \bibnamefont{Tranquada}},
  \bibinfo{author}{\bibfnamefont{Y.}~\bibnamefont{Kong}},
  \bibinfo{author}{\bibfnamefont{J.~E.} \bibnamefont{Lorenzo}},
  \bibinfo{author}{\bibfnamefont{D.~J.} \bibnamefont{Buttrey}},
  \bibinfo{author}{\bibfnamefont{D.~E.} \bibnamefont{Rice}}, \bibnamefont{and}
  \bibinfo{author}{\bibfnamefont{V.}~\bibnamefont{Sachan}},
  \bibinfo{journal}{Phys. Rev. B} \textbf{\bibinfo{volume}{50}},
  \bibinfo{pages}{6340} (\bibinfo{year}{1994}{\natexlab{b}}).

\bibitem[{\citenamefont{Tranquada
  et~al.}(1995{\natexlab{b}})\citenamefont{Tranquada, Lorenzo, Buttrey, and
  Sachan}}]{tran95b}
\bibinfo{author}{\bibfnamefont{J.~M.} \bibnamefont{Tranquada}},
  \bibinfo{author}{\bibfnamefont{J.~E.} \bibnamefont{Lorenzo}},
  \bibinfo{author}{\bibfnamefont{D.~J.} \bibnamefont{Buttrey}},
  \bibnamefont{and} \bibinfo{author}{\bibfnamefont{V.}~\bibnamefont{Sachan}},
  \bibinfo{journal}{Phys. Rev. B} \textbf{\bibinfo{volume}{52}},
  \bibinfo{pages}{3581} (\bibinfo{year}{1995}{\natexlab{b}}).

\bibitem[{\citenamefont{Li et~al.}(2003)\citenamefont{Li, Zhu, Tranquada,
  Yamada, and Buttrey}}]{li03}
\bibinfo{author}{\bibfnamefont{J.}~\bibnamefont{Li}},
  \bibinfo{author}{\bibfnamefont{Y.}~\bibnamefont{Zhu}},
  \bibinfo{author}{\bibfnamefont{J.~M.} \bibnamefont{Tranquada}},
  \bibinfo{author}{\bibfnamefont{K.}~\bibnamefont{Yamada}}, \bibnamefont{and}
  \bibinfo{author}{\bibfnamefont{D.~J.} \bibnamefont{Buttrey}},
  \bibinfo{journal}{Phys. Rev. B} \textbf{\bibinfo{volume}{67}},
  \bibinfo{pages}{012404} (\bibinfo{year}{2003}).

\bibitem[{\citenamefont{Yoshinari et~al.}(1999)\citenamefont{Yoshinari, Hammel,
  and Cheong}}]{yosh99}
\bibinfo{author}{\bibfnamefont{Y.}~\bibnamefont{Yoshinari}},
  \bibinfo{author}{\bibfnamefont{P.~C.} \bibnamefont{Hammel}},
  \bibnamefont{and} \bibinfo{author}{\bibfnamefont{S.-W.}
  \bibnamefont{Cheong}}, \bibinfo{journal}{Phys. Rev. Lett.}
  \textbf{\bibinfo{volume}{82}}, \bibinfo{pages}{3536} (\bibinfo{year}{1999}).

\bibitem[{\citenamefont{Abu-Shiekah et~al.}(1999)\citenamefont{Abu-Shiekah,
  Bernal, Menovsky, Brom, and Zaanen}}]{abu99}
\bibinfo{author}{\bibfnamefont{I.~M.} \bibnamefont{Abu-Shiekah}},
  \bibinfo{author}{\bibfnamefont{O.~O.} \bibnamefont{Bernal}},
  \bibinfo{author}{\bibfnamefont{A.~A.} \bibnamefont{Menovsky}},
  \bibinfo{author}{\bibfnamefont{H.~B.} \bibnamefont{Brom}}, \bibnamefont{and}
  \bibinfo{author}{\bibfnamefont{J.}~\bibnamefont{Zaanen}},
  \bibinfo{journal}{Phys. Rev. Lett.} \textbf{\bibinfo{volume}{83}},
  \bibinfo{pages}{3309} (\bibinfo{year}{1999}).

\bibitem[{\citenamefont{Lee and Cheong}(1997)}]{lee97}
\bibinfo{author}{\bibfnamefont{S.-H.} \bibnamefont{Lee}} \bibnamefont{and}
  \bibinfo{author}{\bibfnamefont{S.-W.} \bibnamefont{Cheong}},
  \bibinfo{journal}{Phys. Rev. Lett.} \textbf{\bibinfo{volume}{79}},
  \bibinfo{pages}{2514} (\bibinfo{year}{1997}).

\bibitem[{\citenamefont{Ishizaka et~al.}(2004)\citenamefont{Ishizaka, Arima,
  Murakami, Kajimoto, Yoshizawa, Nagaosa, and Tokura}}]{ishi04}
\bibinfo{author}{\bibfnamefont{K.}~\bibnamefont{Ishizaka}},
  \bibinfo{author}{\bibfnamefont{T.}~\bibnamefont{Arima}},
  \bibinfo{author}{\bibfnamefont{Y.}~\bibnamefont{Murakami}},
  \bibinfo{author}{\bibfnamefont{R.}~\bibnamefont{Kajimoto}},
  \bibinfo{author}{\bibfnamefont{H.}~\bibnamefont{Yoshizawa}},
  \bibinfo{author}{\bibfnamefont{N.}~\bibnamefont{Nagaosa}}, \bibnamefont{and}
  \bibinfo{author}{\bibfnamefont{Y.}~\bibnamefont{Tokura}},
  \bibinfo{journal}{Phys. Rev. Lett.} \textbf{\bibinfo{volume}{92}},
  \bibinfo{pages}{196404} (\bibinfo{year}{2004}).

\bibitem[{\citenamefont{Bourges et~al.}(2003)\citenamefont{Bourges, Sidis,
  Braden, Nakajima, and Tranquada}}]{bour03}
\bibinfo{author}{\bibfnamefont{P.}~\bibnamefont{Bourges}},
  \bibinfo{author}{\bibfnamefont{Y.}~\bibnamefont{Sidis}},
  \bibinfo{author}{\bibfnamefont{M.}~\bibnamefont{Braden}},
  \bibinfo{author}{\bibfnamefont{K.}~\bibnamefont{Nakajima}}, \bibnamefont{and}
  \bibinfo{author}{\bibfnamefont{J.~M.} \bibnamefont{Tranquada}},
  \bibinfo{journal}{Phys. Rev. Lett.} \textbf{\bibinfo{volume}{90}},
  \bibinfo{pages}{147202} (\bibinfo{year}{2003}).

\bibitem[{\citenamefont{Boothroyd
  et~al.}(2003{\natexlab{a}})\citenamefont{Boothroyd, Prabhakaran, Freeman,
  Lister, Enderle, Hiess, and Kulda}}]{boot03}
\bibinfo{author}{\bibfnamefont{A.~T.} \bibnamefont{Boothroyd}},
  \bibinfo{author}{\bibfnamefont{D.}~\bibnamefont{Prabhakaran}},
  \bibinfo{author}{\bibfnamefont{P.~G.} \bibnamefont{Freeman}},
  \bibinfo{author}{\bibfnamefont{S.~J.~S.} \bibnamefont{Lister}},
  \bibinfo{author}{\bibfnamefont{M.}~\bibnamefont{Enderle}},
  \bibinfo{author}{\bibfnamefont{A.}~\bibnamefont{Hiess}}, \bibnamefont{and}
  \bibinfo{author}{\bibfnamefont{J.}~\bibnamefont{Kulda}},
  \bibinfo{journal}{Phys. Rev. B} \textbf{\bibinfo{volume}{67}},
  \bibinfo{pages}{100407(R)} (\bibinfo{year}{2003}{\natexlab{a}}).

\bibitem[{\citenamefont{Woo et~al.}(2005)\citenamefont{Woo, Boothroyd,
  Nakajima, Perring, Frost, Freeman, Prabhakaran, Yamada, and
  Tranquada}}]{woo05}
\bibinfo{author}{\bibfnamefont{H.}~\bibnamefont{Woo}},
  \bibinfo{author}{\bibfnamefont{A.~T.} \bibnamefont{Boothroyd}},
  \bibinfo{author}{\bibfnamefont{K.}~\bibnamefont{Nakajima}},
  \bibinfo{author}{\bibfnamefont{T.~G.} \bibnamefont{Perring}},
  \bibinfo{author}{\bibfnamefont{C.~D.} \bibnamefont{Frost}},
  \bibinfo{author}{\bibfnamefont{P.~G.} \bibnamefont{Freeman}},
  \bibinfo{author}{\bibfnamefont{D.}~\bibnamefont{Prabhakaran}},
  \bibinfo{author}{\bibfnamefont{K.}~\bibnamefont{Yamada}}, \bibnamefont{and}
  \bibinfo{author}{\bibfnamefont{J.~M.} \bibnamefont{Tranquada}},
  \bibinfo{journal}{Phys. Rev. B} \textbf{\bibinfo{volume}{72}},
  \bibinfo{pages}{064437} (\bibinfo{year}{2005}).

\bibitem[{\citenamefont{Yamada et~al.}(1991)\citenamefont{Yamada, Arai, Endoh,
  Hosoya, Nakajima, Perring, and Taylor}}]{yama91}
\bibinfo{author}{\bibfnamefont{K.}~\bibnamefont{Yamada}},
  \bibinfo{author}{\bibfnamefont{M.}~\bibnamefont{Arai}},
  \bibinfo{author}{\bibfnamefont{Y.}~\bibnamefont{Endoh}},
  \bibinfo{author}{\bibfnamefont{S.}~\bibnamefont{Hosoya}},
  \bibinfo{author}{\bibfnamefont{K.}~\bibnamefont{Nakajima}},
  \bibinfo{author}{\bibfnamefont{T.}~\bibnamefont{Perring}}, \bibnamefont{and}
  \bibinfo{author}{\bibfnamefont{A.}~\bibnamefont{Taylor}},
  \bibinfo{journal}{J. Phys. Soc. Jpn.} \textbf{\bibinfo{volume}{60}},
  \bibinfo{pages}{1197} (\bibinfo{year}{1991}).

\bibitem[{\citenamefont{Boothroyd
  et~al.}(2003{\natexlab{b}})\citenamefont{Boothroyd, Freeman, Prabhakaran,
  Hiess, Enderle, Kulda, and Altorfer}}]{boot03b}
\bibinfo{author}{\bibfnamefont{A.~T.} \bibnamefont{Boothroyd}},
  \bibinfo{author}{\bibfnamefont{P.~G.} \bibnamefont{Freeman}},
  \bibinfo{author}{\bibfnamefont{D.}~\bibnamefont{Prabhakaran}},
  \bibinfo{author}{\bibfnamefont{A.}~\bibnamefont{Hiess}},
  \bibinfo{author}{\bibfnamefont{M.}~\bibnamefont{Enderle}},
  \bibinfo{author}{\bibfnamefont{J.}~\bibnamefont{Kulda}}, \bibnamefont{and}
  \bibinfo{author}{\bibfnamefont{F.}~\bibnamefont{Altorfer}},
  \bibinfo{journal}{Phys. Rev. Lett.} \textbf{\bibinfo{volume}{91}},
  \bibinfo{pages}{257201} (\bibinfo{year}{2003}{\natexlab{b}}).

\bibitem[{\citenamefont{Lee et~al.}(2002)\citenamefont{Lee, Tranquada, Yamada,
  Buttrey, Li, and Cheong}}]{lee02}
\bibinfo{author}{\bibfnamefont{S.-H.} \bibnamefont{Lee}},
  \bibinfo{author}{\bibfnamefont{J.~M.} \bibnamefont{Tranquada}},
  \bibinfo{author}{\bibfnamefont{K.}~\bibnamefont{Yamada}},
  \bibinfo{author}{\bibfnamefont{D.~J.} \bibnamefont{Buttrey}},
  \bibinfo{author}{\bibfnamefont{Q.}~\bibnamefont{Li}}, \bibnamefont{and}
  \bibinfo{author}{\bibfnamefont{S.-W.} \bibnamefont{Cheong}},
  \bibinfo{journal}{Phys. Rev. Lett.} \textbf{\bibinfo{volume}{88}},
  \bibinfo{pages}{126401} (\bibinfo{year}{2002}).

\bibitem[{\citenamefont{Katsufuji et~al.}(1996)\citenamefont{Katsufuji, Tanabe,
  Ishikawa, Fukuda, Arima, and Tokura}}]{kats96}
\bibinfo{author}{\bibfnamefont{T.}~\bibnamefont{Katsufuji}},
  \bibinfo{author}{\bibfnamefont{T.}~\bibnamefont{Tanabe}},
  \bibinfo{author}{\bibfnamefont{T.}~\bibnamefont{Ishikawa}},
  \bibinfo{author}{\bibfnamefont{Y.}~\bibnamefont{Fukuda}},
  \bibinfo{author}{\bibfnamefont{T.}~\bibnamefont{Arima}}, \bibnamefont{and}
  \bibinfo{author}{\bibfnamefont{Y.}~\bibnamefont{Tokura}},
  \bibinfo{journal}{Phys. Rev. B} \textbf{\bibinfo{volume}{54}},
  \bibinfo{pages}{R14230} (\bibinfo{year}{1996}).

\bibitem[{\citenamefont{Homes et~al.}(2003)\citenamefont{Homes, Tranquada, Li,
  Moodenbaugh, and Buttrey}}]{home03}
\bibinfo{author}{\bibfnamefont{C.~C.} \bibnamefont{Homes}},
  \bibinfo{author}{\bibfnamefont{J.~M.} \bibnamefont{Tranquada}},
  \bibinfo{author}{\bibfnamefont{Q.}~\bibnamefont{Li}},
  \bibinfo{author}{\bibfnamefont{A.~R.} \bibnamefont{Moodenbaugh}},
  \bibnamefont{and} \bibinfo{author}{\bibfnamefont{D.~J.}
  \bibnamefont{Buttrey}}, \bibinfo{journal}{Phys. Rev. B}
  \textbf{\bibinfo{volume}{67}}, \bibinfo{pages}{184516}
  (\bibinfo{year}{2003}).

\bibitem[{\citenamefont{Tranquada}(1998{\natexlab{b}})}]{tran98c}
\bibinfo{author}{\bibfnamefont{J.~M.} \bibnamefont{Tranquada}},
  \bibinfo{journal}{J. Phys. Chem. Solids} \textbf{\bibinfo{volume}{59}},
  \bibinfo{pages}{2150} (\bibinfo{year}{1998}{\natexlab{b}}).

\bibitem[{\citenamefont{Ichikawa et~al.}(2000)\citenamefont{Ichikawa, Uchida,
  Tranquada, Niem\"oller, Gehring, Lee, and Schneider}}]{ichi00}
\bibinfo{author}{\bibfnamefont{N.}~\bibnamefont{Ichikawa}},
  \bibinfo{author}{\bibfnamefont{S.}~\bibnamefont{Uchida}},
  \bibinfo{author}{\bibfnamefont{J.~M.} \bibnamefont{Tranquada}},
  \bibinfo{author}{\bibfnamefont{T.}~\bibnamefont{Niem\"oller}},
  \bibinfo{author}{\bibfnamefont{P.~M.} \bibnamefont{Gehring}},
  \bibinfo{author}{\bibfnamefont{S.-H.} \bibnamefont{Lee}}, \bibnamefont{and}
  \bibinfo{author}{\bibfnamefont{J.~R.} \bibnamefont{Schneider}},
  \bibinfo{journal}{Phys. Rev. Lett.} \textbf{\bibinfo{volume}{85}},
  \bibinfo{pages}{1738} (\bibinfo{year}{2000}).

\bibitem[{\citenamefont{Axe and Crawford}(1994)}]{axe94}
\bibinfo{author}{\bibfnamefont{J.~D.} \bibnamefont{Axe}} \bibnamefont{and}
  \bibinfo{author}{\bibfnamefont{M.~K.} \bibnamefont{Crawford}},
  \bibinfo{journal}{J. Low Temp. Phys.} \textbf{\bibinfo{volume}{95}},
  \bibinfo{pages}{271} (\bibinfo{year}{1994}).

\bibitem[{\citenamefont{Fujita et~al.}(2002{\natexlab{b}})\citenamefont{Fujita,
  Goka, Yamada, and Matsuda}}]{fuji02}
\bibinfo{author}{\bibfnamefont{M.}~\bibnamefont{Fujita}},
  \bibinfo{author}{\bibfnamefont{H.}~\bibnamefont{Goka}},
  \bibinfo{author}{\bibfnamefont{K.}~\bibnamefont{Yamada}}, \bibnamefont{and}
  \bibinfo{author}{\bibfnamefont{M.}~\bibnamefont{Matsuda}},
  \bibinfo{journal}{Phys. Rev. Lett.} \textbf{\bibinfo{volume}{88}},
  \bibinfo{pages}{167008} (\bibinfo{year}{2002}{\natexlab{b}}).

\bibitem[{\citenamefont{Fujita et~al.}(2002{\natexlab{c}})\citenamefont{Fujita,
  Goka, Yamada, and Matsuda}}]{fuji02b}
\bibinfo{author}{\bibfnamefont{M.}~\bibnamefont{Fujita}},
  \bibinfo{author}{\bibfnamefont{H.}~\bibnamefont{Goka}},
  \bibinfo{author}{\bibfnamefont{K.}~\bibnamefont{Yamada}}, \bibnamefont{and}
  \bibinfo{author}{\bibfnamefont{M.}~\bibnamefont{Matsuda}},
  \bibinfo{journal}{Phys. Rev. B} \textbf{\bibinfo{volume}{66}},
  \bibinfo{pages}{184503} (\bibinfo{year}{2002}{\natexlab{c}}).

\bibitem[{\citenamefont{Kimura et~al.}(2003{\natexlab{a}})\citenamefont{Kimura,
  Goka, Fujita, Noda, Yamada, and Ikeda}}]{kimu03}
\bibinfo{author}{\bibfnamefont{H.}~\bibnamefont{Kimura}},
  \bibinfo{author}{\bibfnamefont{H.}~\bibnamefont{Goka}},
  \bibinfo{author}{\bibfnamefont{M.}~\bibnamefont{Fujita}},
  \bibinfo{author}{\bibfnamefont{Y.}~\bibnamefont{Noda}},
  \bibinfo{author}{\bibfnamefont{K.}~\bibnamefont{Yamada}}, \bibnamefont{and}
  \bibinfo{author}{\bibfnamefont{N.}~\bibnamefont{Ikeda}},
  \bibinfo{journal}{Phys. Rev. B} \textbf{\bibinfo{volume}{67}},
  \bibinfo{pages}{140503(R)} (\bibinfo{year}{2003}{\natexlab{a}}).

\bibitem[{\citenamefont{Fine}(2004)}]{fine04}
\bibinfo{author}{\bibfnamefont{B.~V.} \bibnamefont{Fine}},
  \bibinfo{journal}{Phys. Rev. B} \textbf{\bibinfo{volume}{70}},
  \bibinfo{pages}{224508} (\bibinfo{year}{2004}).

\bibitem[{\citenamefont{Tranquada et~al.}(1999)\citenamefont{Tranquada,
  Ichikawa, Kakurai, and Uchida}}]{tran99b}
\bibinfo{author}{\bibfnamefont{J.~M.} \bibnamefont{Tranquada}},
  \bibinfo{author}{\bibfnamefont{N.}~\bibnamefont{Ichikawa}},
  \bibinfo{author}{\bibfnamefont{K.}~\bibnamefont{Kakurai}}, \bibnamefont{and}
  \bibinfo{author}{\bibfnamefont{S.}~\bibnamefont{Uchida}},
  \bibinfo{journal}{J. Phys. Chem. Solids} \textbf{\bibinfo{volume}{60}},
  \bibinfo{pages}{1019} (\bibinfo{year}{1999}).

\bibitem[{\citenamefont{v.~Zimmermann et~al.}(1998)\citenamefont{v.~Zimmermann,
  Vigliante, Niem\"oller, Ichikawa, Frello, Uchida, Andersen, Madsen, Wochner,
  Tranquada et~al.}}]{vonz98}
\bibinfo{author}{\bibfnamefont{M.}~\bibnamefont{v.~Zimmermann}},
  \bibinfo{author}{\bibfnamefont{A.}~\bibnamefont{Vigliante}},
  \bibinfo{author}{\bibfnamefont{T.}~\bibnamefont{Niem\"oller}},
  \bibinfo{author}{\bibfnamefont{N.}~\bibnamefont{Ichikawa}},
  \bibinfo{author}{\bibfnamefont{T.}~\bibnamefont{Frello}},
  \bibinfo{author}{\bibfnamefont{S.}~\bibnamefont{Uchida}},
  \bibinfo{author}{\bibfnamefont{N.~H.} \bibnamefont{Andersen}},
  \bibinfo{author}{\bibfnamefont{J.}~\bibnamefont{Madsen}},
  \bibinfo{author}{\bibfnamefont{P.}~\bibnamefont{Wochner}},
  \bibinfo{author}{\bibfnamefont{J.~M.} \bibnamefont{Tranquada}},
  \bibnamefont{et~al.}, \bibinfo{journal}{Europhys. Lett.}
  \textbf{\bibinfo{volume}{41}}, \bibinfo{pages}{629} (\bibinfo{year}{1998}).

\bibitem[{\citenamefont{Mook et~al.}(2002{\natexlab{a}})\citenamefont{Mook,
  Dai, and Do\u{g}an}}]{mook02}
\bibinfo{author}{\bibfnamefont{H.~A.} \bibnamefont{Mook}},
  \bibinfo{author}{\bibfnamefont{P.}~\bibnamefont{Dai}}, \bibnamefont{and}
  \bibinfo{author}{\bibfnamefont{F.}~\bibnamefont{Do\u{g}an}},
  \bibinfo{journal}{Phys. Rev. Lett.} \textbf{\bibinfo{volume}{88}},
  \bibinfo{pages}{097004} (\bibinfo{year}{2002}{\natexlab{a}}).

\bibitem[{\citenamefont{Sanna et~al.}(2004)\citenamefont{Sanna, Allodi, Concas,
  Hillier, and Renzi}}]{sann04}
\bibinfo{author}{\bibfnamefont{S.}~\bibnamefont{Sanna}},
  \bibinfo{author}{\bibfnamefont{G.}~\bibnamefont{Allodi}},
  \bibinfo{author}{\bibfnamefont{G.}~\bibnamefont{Concas}},
  \bibinfo{author}{\bibfnamefont{A.~D.} \bibnamefont{Hillier}},
  \bibnamefont{and} \bibinfo{author}{\bibfnamefont{R.~D.} \bibnamefont{Renzi}},
  \bibinfo{journal}{Phys. Rev. Lett.} \textbf{\bibinfo{volume}{93}},
  \bibinfo{pages}{207001} (\bibinfo{year}{2004}).

\bibitem[{\citenamefont{Savici et~al.}(2002)\citenamefont{Savici, Fudamoto,
  Gat, Ito, Larkin, Uemura, Luke, Kojima, Lee, Kastner et~al.}}]{savi02}
\bibinfo{author}{\bibfnamefont{A.~T.} \bibnamefont{Savici}},
  \bibinfo{author}{\bibfnamefont{Y.}~\bibnamefont{Fudamoto}},
  \bibinfo{author}{\bibfnamefont{I.~M.} \bibnamefont{Gat}},
  \bibinfo{author}{\bibfnamefont{T.}~\bibnamefont{Ito}},
  \bibinfo{author}{\bibfnamefont{M.~I.} \bibnamefont{Larkin}},
  \bibinfo{author}{\bibfnamefont{Y.~J.} \bibnamefont{Uemura}},
  \bibinfo{author}{\bibfnamefont{G.~M.} \bibnamefont{Luke}},
  \bibinfo{author}{\bibfnamefont{K.~M.} \bibnamefont{Kojima}},
  \bibinfo{author}{\bibfnamefont{Y.~S.} \bibnamefont{Lee}},
  \bibinfo{author}{\bibfnamefont{M.~A.} \bibnamefont{Kastner}},
  \bibnamefont{et~al.}, \bibinfo{journal}{Phys. Rev. B}
  \textbf{\bibinfo{volume}{66}}, \bibinfo{pages}{014524}
  (\bibinfo{year}{2002}).

\bibitem[{\citenamefont{Batista et~al.}(2001)\citenamefont{Batista, Ortiz, and
  Balatsky}}]{bati01}
\bibinfo{author}{\bibfnamefont{C.~D.} \bibnamefont{Batista}},
  \bibinfo{author}{\bibfnamefont{G.}~\bibnamefont{Ortiz}}, \bibnamefont{and}
  \bibinfo{author}{\bibfnamefont{A.~V.} \bibnamefont{Balatsky}},
  \bibinfo{journal}{Phys. Rev. B} \textbf{\bibinfo{volume}{64}},
  \bibinfo{pages}{172508} (\bibinfo{year}{2001}).

\bibitem[{\citenamefont{Kr\"uger and Scheidl}(2003)}]{krug03}
\bibinfo{author}{\bibfnamefont{F.}~\bibnamefont{Kr\"uger}} \bibnamefont{and}
  \bibinfo{author}{\bibfnamefont{S.}~\bibnamefont{Scheidl}},
  \bibinfo{journal}{Phys. Rev. B} \textbf{\bibinfo{volume}{67}},
  \bibinfo{pages}{134512} (\bibinfo{year}{2003}).

\bibitem[{\citenamefont{Carlson et~al.}(2004)\citenamefont{Carlson, Yao, and
  Campbell}}]{carl04}
\bibinfo{author}{\bibfnamefont{E.~W.} \bibnamefont{Carlson}},
  \bibinfo{author}{\bibfnamefont{D.~X.} \bibnamefont{Yao}}, \bibnamefont{and}
  \bibinfo{author}{\bibfnamefont{D.~K.} \bibnamefont{Campbell}},
  \bibinfo{journal}{Phys. Rev. B} \textbf{\bibinfo{volume}{70}},
  \bibinfo{pages}{064505} (\bibinfo{year}{2004}).

\bibitem[{\citenamefont{Vojta and Ulbricht}(2004)}]{vojt04}
\bibinfo{author}{\bibfnamefont{M.}~\bibnamefont{Vojta}} \bibnamefont{and}
  \bibinfo{author}{\bibfnamefont{T.}~\bibnamefont{Ulbricht}},
  \bibinfo{journal}{Phys. Rev. Lett.} \textbf{\bibinfo{volume}{93}},
  \bibinfo{pages}{127002} (\bibinfo{year}{2004}).

\bibitem[{\citenamefont{Uhrig et~al.}(2004)\citenamefont{Uhrig, Schmidt, and
  Gr\"uninger}}]{uhri04}
\bibinfo{author}{\bibfnamefont{G.~S.} \bibnamefont{Uhrig}},
  \bibinfo{author}{\bibfnamefont{K.~P.} \bibnamefont{Schmidt}},
  \bibnamefont{and}
  \bibinfo{author}{\bibfnamefont{M.}~\bibnamefont{Gr\"uninger}},
  \bibinfo{journal}{Phys. Rev. Lett.} \textbf{\bibinfo{volume}{93}},
  \bibinfo{pages}{267003} (\bibinfo{year}{2004}).

\bibitem[{\citenamefont{Seibold and Lorenzana}(2005)}]{seib05}
\bibinfo{author}{\bibfnamefont{G.}~\bibnamefont{Seibold}} \bibnamefont{and}
  \bibinfo{author}{\bibfnamefont{J.}~\bibnamefont{Lorenzana}},
  \bibinfo{journal}{Phys. Rev. Lett.} \textbf{\bibinfo{volume}{94}},
  \bibinfo{pages}{107006} (\bibinfo{year}{2005}).

\bibitem[{\citenamefont{Ito et~al.}(2003)\citenamefont{Ito, Yasui, Iikubo,
  Sato, Kobayashi, and Kakurai}}]{ito03}
\bibinfo{author}{\bibfnamefont{M.}~\bibnamefont{Ito}},
  \bibinfo{author}{\bibfnamefont{Y.}~\bibnamefont{Yasui}},
  \bibinfo{author}{\bibfnamefont{S.}~\bibnamefont{Iikubo}},
  \bibinfo{author}{\bibfnamefont{M.}~\bibnamefont{Sato}},
  \bibinfo{author}{\bibfnamefont{A.}~\bibnamefont{Kobayashi}},
  \bibnamefont{and} \bibinfo{author}{\bibfnamefont{K.}~\bibnamefont{Kakurai}},
  \bibinfo{journal}{J. Phys. Soc. Japan} \textbf{\bibinfo{volume}{72}},
  \bibinfo{pages}{1627} (\bibinfo{year}{2003}).

\bibitem[{\citenamefont{Fawcett et~al.}(1994)\citenamefont{Fawcett, Alberts,
  Galkin, Noakes, and Yakhmi}}]{fawc94}
\bibinfo{author}{\bibfnamefont{E.}~\bibnamefont{Fawcett}},
  \bibinfo{author}{\bibfnamefont{H.~L.} \bibnamefont{Alberts}},
  \bibinfo{author}{\bibfnamefont{V.~Y.} \bibnamefont{Galkin}},
  \bibinfo{author}{\bibfnamefont{D.~R.} \bibnamefont{Noakes}},
  \bibnamefont{and} \bibinfo{author}{\bibfnamefont{J.~V.}
  \bibnamefont{Yakhmi}}, \bibinfo{journal}{Rev. Mod. Phys.}
  \textbf{\bibinfo{volume}{66}}, \bibinfo{pages}{25} (\bibinfo{year}{1994}).

\bibitem[{\citenamefont{Fawcett}(1988)}]{fawc88}
\bibinfo{author}{\bibfnamefont{E.}~\bibnamefont{Fawcett}},
  \bibinfo{journal}{Rev. Mod. Phys.} \textbf{\bibinfo{volume}{60}},
  \bibinfo{pages}{209} (\bibinfo{year}{1988}).

\bibitem[{\citenamefont{Overhauser and Arrott}(1960)}]{over60}
\bibinfo{author}{\bibfnamefont{A.~W.} \bibnamefont{Overhauser}}
  \bibnamefont{and} \bibinfo{author}{\bibfnamefont{A.}~\bibnamefont{Arrott}},
  \bibinfo{journal}{Phys. Rev. Lett.} \textbf{\bibinfo{volume}{4}},
  \bibinfo{pages}{226} (\bibinfo{year}{1960}).

\bibitem[{\citenamefont{Lomer}(1962)}]{lome62}
\bibinfo{author}{\bibfnamefont{W.~M.} \bibnamefont{Lomer}},
  \bibinfo{journal}{Proc. Phys. Soc. London} \textbf{\bibinfo{volume}{80}},
  \bibinfo{pages}{489} (\bibinfo{year}{1962}).

\bibitem[{\citenamefont{Sinha et~al.}(1977)\citenamefont{Sinha, Kline, Stassis,
  Chesser, and Wakabayashi}}]{sinh77}
\bibinfo{author}{\bibfnamefont{S.~K.} \bibnamefont{Sinha}},
  \bibinfo{author}{\bibfnamefont{G.~R.} \bibnamefont{Kline}},
  \bibinfo{author}{\bibfnamefont{C.}~\bibnamefont{Stassis}},
  \bibinfo{author}{\bibfnamefont{N.}~\bibnamefont{Chesser}}, \bibnamefont{and}
  \bibinfo{author}{\bibfnamefont{N.}~\bibnamefont{Wakabayashi}},
  \bibinfo{journal}{Phys. Rev. B} \textbf{\bibinfo{volume}{15}},
  \bibinfo{pages}{1415} (\bibinfo{year}{1977}).

\bibitem[{\citenamefont{Fukuda et~al.}(1996)\citenamefont{Fukuda, Endoh,
  Yamada, Takeda, Itoh, Arai, and Otomo}}]{fuku96b}
\bibinfo{author}{\bibfnamefont{T.}~\bibnamefont{Fukuda}},
  \bibinfo{author}{\bibfnamefont{Y.}~\bibnamefont{Endoh}},
  \bibinfo{author}{\bibfnamefont{K.}~\bibnamefont{Yamada}},
  \bibinfo{author}{\bibfnamefont{M.}~\bibnamefont{Takeda}},
  \bibinfo{author}{\bibfnamefont{S.}~\bibnamefont{Itoh}},
  \bibinfo{author}{\bibfnamefont{M.}~\bibnamefont{Arai}}, \bibnamefont{and}
  \bibinfo{author}{\bibfnamefont{T.}~\bibnamefont{Otomo}}, \bibinfo{journal}{J.
  Phys. Soc. Japan} \textbf{\bibinfo{volume}{65}}, \bibinfo{pages}{1418}
  (\bibinfo{year}{1996}).

\bibitem[{\citenamefont{Kaneshita et~al.}(2001)\citenamefont{Kaneshita,
  Ichioka, and Machida}}]{kane01}
\bibinfo{author}{\bibfnamefont{E.}~\bibnamefont{Kaneshita}},
  \bibinfo{author}{\bibfnamefont{M.}~\bibnamefont{Ichioka}}, \bibnamefont{and}
  \bibinfo{author}{\bibfnamefont{K.}~\bibnamefont{Machida}},
  \bibinfo{journal}{J. Phys. Soc. Jpn.} \textbf{\bibinfo{volume}{70}},
  \bibinfo{pages}{866} (\bibinfo{year}{2001}).

\bibitem[{\citenamefont{Fishman and Liu}(1996)}]{fish96}
\bibinfo{author}{\bibfnamefont{R.~S.} \bibnamefont{Fishman}} \bibnamefont{and}
  \bibinfo{author}{\bibfnamefont{S.~H.} \bibnamefont{Liu}},
  \bibinfo{journal}{Phys. Rev. Lett.} \textbf{\bibinfo{volume}{76}},
  \bibinfo{pages}{2398} (\bibinfo{year}{1996}).

\bibitem[{\citenamefont{Hayden et~al.}(2000)\citenamefont{Hayden, Doubble,
  Aeppli, Perring, and Fawcett}}]{hayd00}
\bibinfo{author}{\bibfnamefont{S.~M.} \bibnamefont{Hayden}},
  \bibinfo{author}{\bibfnamefont{R.}~\bibnamefont{Doubble}},
  \bibinfo{author}{\bibfnamefont{G.}~\bibnamefont{Aeppli}},
  \bibinfo{author}{\bibfnamefont{T.~G.} \bibnamefont{Perring}},
  \bibnamefont{and} \bibinfo{author}{\bibfnamefont{E.}~\bibnamefont{Fawcett}},
  \bibinfo{journal}{Phys. Rev. Lett.} \textbf{\bibinfo{volume}{84}},
  \bibinfo{pages}{999} (\bibinfo{year}{2000}).

\bibitem[{\citenamefont{Staunton et~al.}(1999)\citenamefont{Staunton, Poulter,
  Ginatempo, Bruno, and Johnson}}]{stau99}
\bibinfo{author}{\bibfnamefont{J.~B.} \bibnamefont{Staunton}},
  \bibinfo{author}{\bibfnamefont{J.}~\bibnamefont{Poulter}},
  \bibinfo{author}{\bibfnamefont{B.}~\bibnamefont{Ginatempo}},
  \bibinfo{author}{\bibfnamefont{E.}~\bibnamefont{Bruno}}, \bibnamefont{and}
  \bibinfo{author}{\bibfnamefont{D.~D.} \bibnamefont{Johnson}},
  \bibinfo{journal}{Phys. Rev. Lett.} \textbf{\bibinfo{volume}{82}},
  \bibinfo{pages}{3340} (\bibinfo{year}{1999}).

\bibitem[{\citenamefont{Pynn et~al.}(1976)\citenamefont{Pynn, Press, Shapiro,
  and Werner}}]{pynn76}
\bibinfo{author}{\bibfnamefont{R.}~\bibnamefont{Pynn}},
  \bibinfo{author}{\bibfnamefont{W.}~\bibnamefont{Press}},
  \bibinfo{author}{\bibfnamefont{S.~M.} \bibnamefont{Shapiro}},
  \bibnamefont{and} \bibinfo{author}{\bibfnamefont{S.~A.}
  \bibnamefont{Werner}}, \bibinfo{journal}{Phys. Rev. B}
  \textbf{\bibinfo{volume}{13}}, \bibinfo{pages}{295} (\bibinfo{year}{1976}).

\bibitem[{\citenamefont{Tranquada
  et~al.}(1996{\natexlab{b}})\citenamefont{Tranquada, Axe, Ichikawa, Nakamura,
  Uchida, and Nachumi}}]{tran96b}
\bibinfo{author}{\bibfnamefont{J.~M.} \bibnamefont{Tranquada}},
  \bibinfo{author}{\bibfnamefont{J.~D.} \bibnamefont{Axe}},
  \bibinfo{author}{\bibfnamefont{N.}~\bibnamefont{Ichikawa}},
  \bibinfo{author}{\bibfnamefont{Y.}~\bibnamefont{Nakamura}},
  \bibinfo{author}{\bibfnamefont{S.}~\bibnamefont{Uchida}}, \bibnamefont{and}
  \bibinfo{author}{\bibfnamefont{B.}~\bibnamefont{Nachumi}},
  \bibinfo{journal}{Phys. Rev. B} \textbf{\bibinfo{volume}{54}},
  \bibinfo{pages}{7489} (\bibinfo{year}{1996}{\natexlab{b}}).

\bibitem[{\citenamefont{Zachar et~al.}(1998)\citenamefont{Zachar, Kivelson, and
  Emery}}]{zach98}
\bibinfo{author}{\bibfnamefont{O.}~\bibnamefont{Zachar}},
  \bibinfo{author}{\bibfnamefont{S.~A.} \bibnamefont{Kivelson}},
  \bibnamefont{and} \bibinfo{author}{\bibfnamefont{V.~J.} \bibnamefont{Emery}},
  \bibinfo{journal}{Phys. Rev. B} \textbf{\bibinfo{volume}{57}},
  \bibinfo{pages}{1422} (\bibinfo{year}{1998}).

\bibitem[{\citenamefont{Kivelson et~al.}(2003)\citenamefont{Kivelson, Bindloss,
  Fradkin, Oganesyan, Tranquada, Kapitulnik, and Howald}}]{kive03}
\bibinfo{author}{\bibfnamefont{S.~A.} \bibnamefont{Kivelson}},
  \bibinfo{author}{\bibfnamefont{I.~P.} \bibnamefont{Bindloss}},
  \bibinfo{author}{\bibfnamefont{E.}~\bibnamefont{Fradkin}},
  \bibinfo{author}{\bibfnamefont{V.}~\bibnamefont{Oganesyan}},
  \bibinfo{author}{\bibfnamefont{J.~M.} \bibnamefont{Tranquada}},
  \bibinfo{author}{\bibfnamefont{A.}~\bibnamefont{Kapitulnik}},
  \bibnamefont{and} \bibinfo{author}{\bibfnamefont{C.}~\bibnamefont{Howald}},
  \bibinfo{journal}{Rev. Mod. Phys.} \textbf{\bibinfo{volume}{75}},
  \bibinfo{pages}{1201} (\bibinfo{year}{2003}).

\bibitem[{\citenamefont{Chakravarty et~al.}(2001)\citenamefont{Chakravarty,
  Laughlin, Morr, and Nayak}}]{chak01}
\bibinfo{author}{\bibfnamefont{S.}~\bibnamefont{Chakravarty}},
  \bibinfo{author}{\bibfnamefont{R.~B.} \bibnamefont{Laughlin}},
  \bibinfo{author}{\bibfnamefont{D.~K.} \bibnamefont{Morr}}, \bibnamefont{and}
  \bibinfo{author}{\bibfnamefont{C.}~\bibnamefont{Nayak}},
  \bibinfo{journal}{Phys. Rev. B} \textbf{\bibinfo{volume}{63}},
  \bibinfo{pages}{094503} (\bibinfo{year}{2001}).

\bibitem[{\citenamefont{Wen and Lee}(1996)}]{wen96}
\bibinfo{author}{\bibfnamefont{X.-G.} \bibnamefont{Wen}} \bibnamefont{and}
  \bibinfo{author}{\bibfnamefont{P.~A.} \bibnamefont{Lee}},
  \bibinfo{journal}{Phys. Rev. Lett.} \textbf{\bibinfo{volume}{76}},
  \bibinfo{pages}{503} (\bibinfo{year}{1996}).

\bibitem[{\citenamefont{Mook et~al.}(2002{\natexlab{b}})\citenamefont{Mook,
  Dai, Hayden, Hiess, Lynn, Lee, and Do\u{g}an}}]{mook02b}
\bibinfo{author}{\bibfnamefont{H.~A.} \bibnamefont{Mook}},
  \bibinfo{author}{\bibfnamefont{P.~C.} \bibnamefont{Dai}},
  \bibinfo{author}{\bibfnamefont{S.~M.} \bibnamefont{Hayden}},
  \bibinfo{author}{\bibfnamefont{A.}~\bibnamefont{Hiess}},
  \bibinfo{author}{\bibfnamefont{J.~W.} \bibnamefont{Lynn}},
  \bibinfo{author}{\bibfnamefont{S.-H.} \bibnamefont{Lee}}, \bibnamefont{and}
  \bibinfo{author}{\bibfnamefont{F.}~\bibnamefont{Do\u{g}an}},
  \bibinfo{journal}{Phys. Rev. B} \textbf{\bibinfo{volume}{66}},
  \bibinfo{pages}{144513} (\bibinfo{year}{2002}{\natexlab{b}}).

\bibitem[{\citenamefont{Stock et~al.}(2002)\citenamefont{Stock, Buyers, Tun,
  Liang, Peets, Bonn, Hardy, and Taillefer}}]{stoc02}
\bibinfo{author}{\bibfnamefont{C.}~\bibnamefont{Stock}},
  \bibinfo{author}{\bibfnamefont{W.~J.~L.} \bibnamefont{Buyers}},
  \bibinfo{author}{\bibfnamefont{Z.}~\bibnamefont{Tun}},
  \bibinfo{author}{\bibfnamefont{R.}~\bibnamefont{Liang}},
  \bibinfo{author}{\bibfnamefont{D.}~\bibnamefont{Peets}},
  \bibinfo{author}{\bibfnamefont{D.}~\bibnamefont{Bonn}},
  \bibinfo{author}{\bibfnamefont{W.~N.} \bibnamefont{Hardy}}, \bibnamefont{and}
  \bibinfo{author}{\bibfnamefont{L.}~\bibnamefont{Taillefer}},
  \bibinfo{journal}{Phys. Rev. B} \textbf{\bibinfo{volume}{66}},
  \bibinfo{pages}{024505} (\bibinfo{year}{2002}).

\bibitem[{\citenamefont{Mook et~al.}(2004)\citenamefont{Mook, Dai, Hayden,
  Hiess, Lee, and Do\u{g}an}}]{mook04}
\bibinfo{author}{\bibfnamefont{H.~A.} \bibnamefont{Mook}},
  \bibinfo{author}{\bibfnamefont{P.~C.} \bibnamefont{Dai}},
  \bibinfo{author}{\bibfnamefont{S.~M.} \bibnamefont{Hayden}},
  \bibinfo{author}{\bibfnamefont{A.}~\bibnamefont{Hiess}},
  \bibinfo{author}{\bibfnamefont{S.-H.} \bibnamefont{Lee}}, \bibnamefont{and}
  \bibinfo{author}{\bibfnamefont{F.}~\bibnamefont{Do\u{g}an}},
  \bibinfo{journal}{Phys. Rev. B} \textbf{\bibinfo{volume}{69}},
  \bibinfo{pages}{134509} (\bibinfo{year}{2004}).

\bibitem[{\citenamefont{Varma}(1997)}]{varm97}
\bibinfo{author}{\bibfnamefont{C.~M.} \bibnamefont{Varma}},
  \bibinfo{journal}{Phys. Rev. B} \textbf{\bibinfo{volume}{55}},
  \bibinfo{pages}{14554} (\bibinfo{year}{1997}).

\bibitem[{\citenamefont{Lee et~al.}(1999{\natexlab{b}})\citenamefont{Lee,
  Majkrzak, Sinha, Stassis, Kawano, Lander, Brown, Fong, Cheong, Matsushita
  et~al.}}]{lee99b}
\bibinfo{author}{\bibfnamefont{S.-H.} \bibnamefont{Lee}},
  \bibinfo{author}{\bibfnamefont{C.~F.} \bibnamefont{Majkrzak}},
  \bibinfo{author}{\bibfnamefont{S.~K.} \bibnamefont{Sinha}},
  \bibinfo{author}{\bibfnamefont{C.}~\bibnamefont{Stassis}},
  \bibinfo{author}{\bibfnamefont{H.}~\bibnamefont{Kawano}},
  \bibinfo{author}{\bibfnamefont{G.~H.} \bibnamefont{Lander}},
  \bibinfo{author}{\bibfnamefont{P.~J.} \bibnamefont{Brown}},
  \bibinfo{author}{\bibfnamefont{H.~F.} \bibnamefont{Fong}},
  \bibinfo{author}{\bibfnamefont{S.-W.} \bibnamefont{Cheong}},
  \bibinfo{author}{\bibfnamefont{H.}~\bibnamefont{Matsushita}},
  \bibnamefont{et~al.}, \bibinfo{journal}{Phys. Rev. B}
  \textbf{\bibinfo{volume}{60}}, \bibinfo{pages}{10405}
  (\bibinfo{year}{1999}{\natexlab{b}}).

\bibitem[{\citenamefont{Simon and Varma}(2002)}]{simo02}
\bibinfo{author}{\bibfnamefont{M.~E.} \bibnamefont{Simon}} \bibnamefont{and}
  \bibinfo{author}{\bibfnamefont{C.~M.} \bibnamefont{Varma}},
  \bibinfo{journal}{Phys. Rev. Lett.} \textbf{\bibinfo{volume}{89}},
  \bibinfo{pages}{247003} (\bibinfo{year}{2002}).

\bibitem[{\citenamefont{Fauqu{\'e} et~al.}(2005)\citenamefont{Fauqu{\'e},
  Sidis, Hinkov, Pailh{\`e}s, Lin, Chaud, and Bourges}}]{fauq05}
\bibinfo{author}{\bibfnamefont{B.}~\bibnamefont{Fauqu{\'e}}},
  \bibinfo{author}{\bibfnamefont{Y.}~\bibnamefont{Sidis}},
  \bibinfo{author}{\bibfnamefont{V.}~\bibnamefont{Hinkov}},
  \bibinfo{author}{\bibfnamefont{S.}~\bibnamefont{Pailh{\`e}s}},
  \bibinfo{author}{\bibfnamefont{C.~T.} \bibnamefont{Lin}},
  \bibinfo{author}{\bibfnamefont{X.}~\bibnamefont{Chaud}}, \bibnamefont{and}
  \bibinfo{author}{\bibfnamefont{P.}~\bibnamefont{Bourges}}
  (\bibinfo{year}{2005}), \eprint{cond-mat/0509210}.

\bibitem[{\citenamefont{Wakimoto et~al.}(2004)\citenamefont{Wakimoto, Zhang,
  Yamada, Swainson, Kim, and Birgeneau}}]{waki04}
\bibinfo{author}{\bibfnamefont{S.}~\bibnamefont{Wakimoto}},
  \bibinfo{author}{\bibfnamefont{H.}~\bibnamefont{Zhang}},
  \bibinfo{author}{\bibfnamefont{K.}~\bibnamefont{Yamada}},
  \bibinfo{author}{\bibfnamefont{I.}~\bibnamefont{Swainson}},
  \bibinfo{author}{\bibfnamefont{H.}~\bibnamefont{Kim}}, \bibnamefont{and}
  \bibinfo{author}{\bibfnamefont{R.~J.} \bibnamefont{Birgeneau}},
  \bibinfo{journal}{Phys. Rev. Lett.} \textbf{\bibinfo{volume}{92}},
  \bibinfo{pages}{217004} (\bibinfo{year}{2004}).

\bibitem[{\citenamefont{Balatsky and Bourges}(1999)}]{bala99}
\bibinfo{author}{\bibfnamefont{A.~V.} \bibnamefont{Balatsky}} \bibnamefont{and}
  \bibinfo{author}{\bibfnamefont{P.}~\bibnamefont{Bourges}},
  \bibinfo{journal}{Phys. Rev. Lett.} \textbf{\bibinfo{volume}{82}},
  \bibinfo{pages}{5337} (\bibinfo{year}{1999}).

\bibitem[{\citenamefont{Hiraka et~al.}(2001)\citenamefont{Hiraka, Endoh,
  Fujita, Lee, Kulda, Ivanov, and Birgeneau}}]{hira01}
\bibinfo{author}{\bibfnamefont{H.}~\bibnamefont{Hiraka}},
  \bibinfo{author}{\bibfnamefont{Y.}~\bibnamefont{Endoh}},
  \bibinfo{author}{\bibfnamefont{M.}~\bibnamefont{Fujita}},
  \bibinfo{author}{\bibfnamefont{Y.~S.} \bibnamefont{Lee}},
  \bibinfo{author}{\bibfnamefont{J.}~\bibnamefont{Kulda}},
  \bibinfo{author}{\bibfnamefont{A.}~\bibnamefont{Ivanov}}, \bibnamefont{and}
  \bibinfo{author}{\bibfnamefont{R.~J.} \bibnamefont{Birgeneau}},
  \bibinfo{journal}{J. Phys. Soc. Japan} \textbf{\bibinfo{volume}{70}},
  \bibinfo{pages}{853} (\bibinfo{year}{2001}).

\bibitem[{\citenamefont{Nakano et~al.}(1998)\citenamefont{Nakano, Momono, Oda,
  and Ido}}]{naka98b}
\bibinfo{author}{\bibfnamefont{T.}~\bibnamefont{Nakano}},
  \bibinfo{author}{\bibfnamefont{N.}~\bibnamefont{Momono}},
  \bibinfo{author}{\bibfnamefont{M.}~\bibnamefont{Oda}}, \bibnamefont{and}
  \bibinfo{author}{\bibfnamefont{M.}~\bibnamefont{Ido}}, \bibinfo{journal}{J.
  Phys. Soc. Japan} \textbf{\bibinfo{volume}{67}}, \bibinfo{pages}{2622}
  (\bibinfo{year}{1998}).

\bibitem[{\citenamefont{Kato et~al.}(2005)\citenamefont{Kato, Okitsu, and
  Sakata}}]{kato05}
\bibinfo{author}{\bibfnamefont{T.}~\bibnamefont{Kato}},
  \bibinfo{author}{\bibfnamefont{S.}~\bibnamefont{Okitsu}}, \bibnamefont{and}
  \bibinfo{author}{\bibfnamefont{H.}~\bibnamefont{Sakata}},
  \bibinfo{journal}{Phys. Rev. B} \textbf{\bibinfo{volume}{72}},
  \bibinfo{pages}{144518} (\bibinfo{year}{2005}).

\bibitem[{\citenamefont{Ino et~al.}(2002)\citenamefont{Ino, Kim, Nakamura,
  Yoshida, Mizokawa, Shen, Fujimori, Kakeshita, Eisaki, and Uchida}}]{ino02}
\bibinfo{author}{\bibfnamefont{A.}~\bibnamefont{Ino}},
  \bibinfo{author}{\bibfnamefont{C.}~\bibnamefont{Kim}},
  \bibinfo{author}{\bibfnamefont{M.}~\bibnamefont{Nakamura}},
  \bibinfo{author}{\bibfnamefont{T.}~\bibnamefont{Yoshida}},
  \bibinfo{author}{\bibfnamefont{T.}~\bibnamefont{Mizokawa}},
  \bibinfo{author}{\bibfnamefont{Z.-X.} \bibnamefont{Shen}},
  \bibinfo{author}{\bibfnamefont{A.}~\bibnamefont{Fujimori}},
  \bibinfo{author}{\bibfnamefont{T.}~\bibnamefont{Kakeshita}},
  \bibinfo{author}{\bibfnamefont{H.}~\bibnamefont{Eisaki}}, \bibnamefont{and}
  \bibinfo{author}{\bibfnamefont{S.}~\bibnamefont{Uchida}},
  \bibinfo{journal}{Phys. Rev. B} \textbf{\bibinfo{volume}{65}},
  \bibinfo{pages}{094504} (\bibinfo{year}{2002}).

\bibitem[{\citenamefont{Lee et~al.}(2000)\citenamefont{Lee, Yamada, Endoh,
  Shirane, Birgeneau, Kastner, Greven, and Kim}}]{lee00}
\bibinfo{author}{\bibfnamefont{C.-H.} \bibnamefont{Lee}},
  \bibinfo{author}{\bibfnamefont{K.}~\bibnamefont{Yamada}},
  \bibinfo{author}{\bibfnamefont{Y.}~\bibnamefont{Endoh}},
  \bibinfo{author}{\bibfnamefont{G.}~\bibnamefont{Shirane}},
  \bibinfo{author}{\bibfnamefont{R.~J.} \bibnamefont{Birgeneau}},
  \bibinfo{author}{\bibfnamefont{M.~A.} \bibnamefont{Kastner}},
  \bibinfo{author}{\bibfnamefont{M.}~\bibnamefont{Greven}}, \bibnamefont{and}
  \bibinfo{author}{\bibfnamefont{Y.-J.} \bibnamefont{Kim}},
  \bibinfo{journal}{J. Phys. Soc. Japan} \textbf{\bibinfo{volume}{69}},
  \bibinfo{pages}{1170} (\bibinfo{year}{2000}).

\bibitem[{\citenamefont{Lee et~al.}(2003)\citenamefont{Lee, Yamada, Hiraka,
  {Venkateswara Rao}, and Endoh}}]{lee03}
\bibinfo{author}{\bibfnamefont{C.~H.} \bibnamefont{Lee}},
  \bibinfo{author}{\bibfnamefont{K.}~\bibnamefont{Yamada}},
  \bibinfo{author}{\bibfnamefont{H.}~\bibnamefont{Hiraka}},
  \bibinfo{author}{\bibfnamefont{C.~R.} \bibnamefont{{Venkateswara Rao}}},
  \bibnamefont{and} \bibinfo{author}{\bibfnamefont{Y.}~\bibnamefont{Endoh}},
  \bibinfo{journal}{Phys. Rev. B} \textbf{\bibinfo{volume}{67}},
  \bibinfo{pages}{134521} (\bibinfo{year}{2003}).

\bibitem[{\citenamefont{Bourges
  et~al.}(1997{\natexlab{c}})\citenamefont{Bourges, Casalta, Regnault, Bossy,
  Burlet, Vettier, Beaugnon, Gautier-Picard, and Tournier}}]{bour97b}
\bibinfo{author}{\bibfnamefont{P.}~\bibnamefont{Bourges}},
  \bibinfo{author}{\bibfnamefont{H.}~\bibnamefont{Casalta}},
  \bibinfo{author}{\bibfnamefont{L.~P.} \bibnamefont{Regnault}},
  \bibinfo{author}{\bibfnamefont{J.}~\bibnamefont{Bossy}},
  \bibinfo{author}{\bibfnamefont{P.}~\bibnamefont{Burlet}},
  \bibinfo{author}{\bibfnamefont{C.}~\bibnamefont{Vettier}},
  \bibinfo{author}{\bibfnamefont{E.}~\bibnamefont{Beaugnon}},
  \bibinfo{author}{\bibfnamefont{P.}~\bibnamefont{Gautier-Picard}},
  \bibnamefont{and} \bibinfo{author}{\bibfnamefont{R.}~\bibnamefont{Tournier}},
  \bibinfo{journal}{Physica B} \textbf{\bibinfo{volume}{234-236}},
  \bibinfo{pages}{830} (\bibinfo{year}{1997}{\natexlab{c}}).

\bibitem[{\citenamefont{Ando and Segawa}(2002)}]{ando02b}
\bibinfo{author}{\bibfnamefont{Y.}~\bibnamefont{Ando}} \bibnamefont{and}
  \bibinfo{author}{\bibfnamefont{K.}~\bibnamefont{Segawa}},
  \bibinfo{journal}{Phys. Rev. Lett.} \textbf{\bibinfo{volume}{88}},
  \bibinfo{pages}{167005} (\bibinfo{year}{2002}).

\bibitem[{\citenamefont{Katano et~al.}(2000)\citenamefont{Katano, Sato, Yamada,
  Suzuki, and Fukase}}]{kata00}
\bibinfo{author}{\bibfnamefont{S.}~\bibnamefont{Katano}},
  \bibinfo{author}{\bibfnamefont{M.}~\bibnamefont{Sato}},
  \bibinfo{author}{\bibfnamefont{K.}~\bibnamefont{Yamada}},
  \bibinfo{author}{\bibfnamefont{T.}~\bibnamefont{Suzuki}}, \bibnamefont{and}
  \bibinfo{author}{\bibfnamefont{T.}~\bibnamefont{Fukase}},
  \bibinfo{journal}{Phys. Rev. B} \textbf{\bibinfo{volume}{62}},
  \bibinfo{pages}{R14677} (\bibinfo{year}{2000}).

\bibitem[{\citenamefont{Lake et~al.}(2002)\citenamefont{Lake, {R\o nnow},
  Christensen, Aeppli, Lefmann, McMorrow, Vorderwisch, Smeibidl, Mangkorntong,
  Sasagawa et~al.}}]{lake02}
\bibinfo{author}{\bibfnamefont{B.}~\bibnamefont{Lake}},
  \bibinfo{author}{\bibfnamefont{H.~M.} \bibnamefont{{R\o nnow}}},
  \bibinfo{author}{\bibfnamefont{N.~B.} \bibnamefont{Christensen}},
  \bibinfo{author}{\bibfnamefont{G.}~\bibnamefont{Aeppli}},
  \bibinfo{author}{\bibfnamefont{K.}~\bibnamefont{Lefmann}},
  \bibinfo{author}{\bibfnamefont{D.~F.} \bibnamefont{McMorrow}},
  \bibinfo{author}{\bibfnamefont{P.}~\bibnamefont{Vorderwisch}},
  \bibinfo{author}{\bibfnamefont{P.}~\bibnamefont{Smeibidl}},
  \bibinfo{author}{\bibfnamefont{N.}~\bibnamefont{Mangkorntong}},
  \bibinfo{author}{\bibfnamefont{T.}~\bibnamefont{Sasagawa}},
  \bibnamefont{et~al.}, \bibinfo{journal}{Nature}
  \textbf{\bibinfo{volume}{415}}, \bibinfo{pages}{299} (\bibinfo{year}{2002}).

\bibitem[{\citenamefont{Demler et~al.}(2001)\citenamefont{Demler, Sachdev, and
  Zhang}}]{deml01}
\bibinfo{author}{\bibfnamefont{E.}~\bibnamefont{Demler}},
  \bibinfo{author}{\bibfnamefont{S.}~\bibnamefont{Sachdev}}, \bibnamefont{and}
  \bibinfo{author}{\bibfnamefont{Y.}~\bibnamefont{Zhang}},
  \bibinfo{journal}{Phys. Rev. Lett.} \textbf{\bibinfo{volume}{87}},
  \bibinfo{pages}{067202} (\bibinfo{year}{2001}).

\bibitem[{\citenamefont{Kivelson et~al.}(2002)\citenamefont{Kivelson, Lee,
  Fradkin, and Oganesyan}}]{kive02}
\bibinfo{author}{\bibfnamefont{S.~A.} \bibnamefont{Kivelson}},
  \bibinfo{author}{\bibfnamefont{D.-H.} \bibnamefont{Lee}},
  \bibinfo{author}{\bibfnamefont{E.}~\bibnamefont{Fradkin}}, \bibnamefont{and}
  \bibinfo{author}{\bibfnamefont{V.}~\bibnamefont{Oganesyan}},
  \bibinfo{journal}{Phys. Rev. B} \textbf{\bibinfo{volume}{66}},
  \bibinfo{pages}{144516} (\bibinfo{year}{2002}).

\bibitem[{\citenamefont{Khaykovich et~al.}(2002)\citenamefont{Khaykovich, Lee,
  Erwin, Lee, Wakimoto, Thomas, Kastner, and Birgeneau}}]{khay02}
\bibinfo{author}{\bibfnamefont{B.}~\bibnamefont{Khaykovich}},
  \bibinfo{author}{\bibfnamefont{Y.~S.} \bibnamefont{Lee}},
  \bibinfo{author}{\bibfnamefont{R.~W.} \bibnamefont{Erwin}},
  \bibinfo{author}{\bibfnamefont{S.-H.} \bibnamefont{Lee}},
  \bibinfo{author}{\bibfnamefont{S.}~\bibnamefont{Wakimoto}},
  \bibinfo{author}{\bibfnamefont{K.~J.} \bibnamefont{Thomas}},
  \bibinfo{author}{\bibfnamefont{M.~A.} \bibnamefont{Kastner}},
  \bibnamefont{and} \bibinfo{author}{\bibfnamefont{R.~J.}
  \bibnamefont{Birgeneau}}, \bibinfo{journal}{Phys. Rev. B}
  \textbf{\bibinfo{volume}{66}}, \bibinfo{pages}{014528}
  (\bibinfo{year}{2002}).

\bibitem[{\citenamefont{Khaykovich et~al.}(2003)\citenamefont{Khaykovich,
  Birgeneau, Chou, Erwin, Kastner, Lee, Lee, Smeibidl, Vorderwisch, and
  Wakimoto}}]{khay03}
\bibinfo{author}{\bibfnamefont{B.}~\bibnamefont{Khaykovich}},
  \bibinfo{author}{\bibfnamefont{R.~J.} \bibnamefont{Birgeneau}},
  \bibinfo{author}{\bibfnamefont{F.~C.} \bibnamefont{Chou}},
  \bibinfo{author}{\bibfnamefont{R.~W.} \bibnamefont{Erwin}},
  \bibinfo{author}{\bibfnamefont{M.~A.} \bibnamefont{Kastner}},
  \bibinfo{author}{\bibfnamefont{S.-H.} \bibnamefont{Lee}},
  \bibinfo{author}{\bibfnamefont{Y.~S.} \bibnamefont{Lee}},
  \bibinfo{author}{\bibfnamefont{P.}~\bibnamefont{Smeibidl}},
  \bibinfo{author}{\bibfnamefont{P.}~\bibnamefont{Vorderwisch}},
  \bibnamefont{and} \bibinfo{author}{\bibfnamefont{S.}~\bibnamefont{Wakimoto}},
  \bibinfo{journal}{Phys. Rev. B} \textbf{\bibinfo{volume}{67}},
  \bibinfo{pages}{054501} (\bibinfo{year}{2003}).

\bibitem[{\citenamefont{Hoffman et~al.}(2002)\citenamefont{Hoffman, Hudson,
  Lang, Madhavan, Eisaki, Uchida, and Davis}}]{hoff02}
\bibinfo{author}{\bibfnamefont{J.~E.} \bibnamefont{Hoffman}},
  \bibinfo{author}{\bibfnamefont{E.~W.} \bibnamefont{Hudson}},
  \bibinfo{author}{\bibfnamefont{K.~M.} \bibnamefont{Lang}},
  \bibinfo{author}{\bibfnamefont{V.}~\bibnamefont{Madhavan}},
  \bibinfo{author}{\bibfnamefont{H.}~\bibnamefont{Eisaki}},
  \bibinfo{author}{\bibfnamefont{S.}~\bibnamefont{Uchida}}, \bibnamefont{and}
  \bibinfo{author}{\bibfnamefont{J.~C.} \bibnamefont{Davis}},
  \bibinfo{journal}{Science} \textbf{\bibinfo{volume}{295}},
  \bibinfo{pages}{466} (\bibinfo{year}{2002}).

\bibitem[{\citenamefont{Khaykovich et~al.}(2005)\citenamefont{Khaykovich,
  Wakimoto, Birgeneau, Kastner, Lee, Smeibidl, Vorderwisch, and
  Yamada}}]{khay05}
\bibinfo{author}{\bibfnamefont{B.}~\bibnamefont{Khaykovich}},
  \bibinfo{author}{\bibfnamefont{S.}~\bibnamefont{Wakimoto}},
  \bibinfo{author}{\bibfnamefont{R.~J.} \bibnamefont{Birgeneau}},
  \bibinfo{author}{\bibfnamefont{M.~A.} \bibnamefont{Kastner}},
  \bibinfo{author}{\bibfnamefont{Y.~S.} \bibnamefont{Lee}},
  \bibinfo{author}{\bibfnamefont{P.}~\bibnamefont{Smeibidl}},
  \bibinfo{author}{\bibfnamefont{P.}~\bibnamefont{Vorderwisch}},
  \bibnamefont{and} \bibinfo{author}{\bibfnamefont{K.}~\bibnamefont{Yamada}},
  \bibinfo{journal}{Phys. Rev. B} \textbf{\bibinfo{volume}{71}},
  \bibinfo{pages}{220508(R)} (\bibinfo{year}{2005}).

\bibitem[{\citenamefont{Wakimoto et~al.}(2003)\citenamefont{Wakimoto,
  Birgeneau, Fujimaki, Ichikawa, Kasuga, Kim, Kojima, Lee, Niko, Tranquada
  et~al.}}]{waki03}
\bibinfo{author}{\bibfnamefont{S.}~\bibnamefont{Wakimoto}},
  \bibinfo{author}{\bibfnamefont{R.~J.} \bibnamefont{Birgeneau}},
  \bibinfo{author}{\bibfnamefont{Y.}~\bibnamefont{Fujimaki}},
  \bibinfo{author}{\bibfnamefont{N.}~\bibnamefont{Ichikawa}},
  \bibinfo{author}{\bibfnamefont{T.}~\bibnamefont{Kasuga}},
  \bibinfo{author}{\bibfnamefont{Y.~J.} \bibnamefont{Kim}},
  \bibinfo{author}{\bibfnamefont{K.~M.} \bibnamefont{Kojima}},
  \bibinfo{author}{\bibfnamefont{S.-H.} \bibnamefont{Lee}},
  \bibinfo{author}{\bibfnamefont{H.}~\bibnamefont{Niko}},
  \bibinfo{author}{\bibfnamefont{J.~M.} \bibnamefont{Tranquada}},
  \bibnamefont{et~al.}, \bibinfo{journal}{Phys. Rev. B}
  \textbf{\bibinfo{volume}{67}}, \bibinfo{pages}{184419}
  (\bibinfo{year}{2003}).

\bibitem[{\citenamefont{Suzuki et~al.}(1998)\citenamefont{Suzuki, Goto, Chiba,
  Shinoda, Fukase, Kimura, Yamada, Ohashi, and Yamaguchi}}]{suzu98}
\bibinfo{author}{\bibfnamefont{T.}~\bibnamefont{Suzuki}},
  \bibinfo{author}{\bibfnamefont{T.}~\bibnamefont{Goto}},
  \bibinfo{author}{\bibfnamefont{K.}~\bibnamefont{Chiba}},
  \bibinfo{author}{\bibfnamefont{T.}~\bibnamefont{Shinoda}},
  \bibinfo{author}{\bibfnamefont{T.}~\bibnamefont{Fukase}},
  \bibinfo{author}{\bibfnamefont{H.}~\bibnamefont{Kimura}},
  \bibinfo{author}{\bibfnamefont{K.}~\bibnamefont{Yamada}},
  \bibinfo{author}{\bibfnamefont{M.}~\bibnamefont{Ohashi}}, \bibnamefont{and}
  \bibinfo{author}{\bibfnamefont{Y.}~\bibnamefont{Yamaguchi}},
  \bibinfo{journal}{Phys. Rev. B} \textbf{\bibinfo{volume}{57}},
  \bibinfo{pages}{R3229} (\bibinfo{year}{1998}).

\bibitem[{\citenamefont{Horibe et~al.}(1997)\citenamefont{Horibe, Inoue, and
  Koyama}}]{hori97}
\bibinfo{author}{\bibfnamefont{Y.}~\bibnamefont{Horibe}},
  \bibinfo{author}{\bibfnamefont{Y.}~\bibnamefont{Inoue}}, \bibnamefont{and}
  \bibinfo{author}{\bibfnamefont{Y.}~\bibnamefont{Koyama}},
  \bibinfo{journal}{Physica C} \textbf{\bibinfo{volume}{282--287}},
  \bibinfo{pages}{1071} (\bibinfo{year}{1997}).

\bibitem[{\citenamefont{Zhu et~al.}(1994)\citenamefont{Zhu, Moodenbaugh, Cai,
  Tafto, Suenaga, and Welch}}]{zhu94}
\bibinfo{author}{\bibfnamefont{Y.}~\bibnamefont{Zhu}},
  \bibinfo{author}{\bibfnamefont{A.~R.} \bibnamefont{Moodenbaugh}},
  \bibinfo{author}{\bibfnamefont{Z.~X.} \bibnamefont{Cai}},
  \bibinfo{author}{\bibfnamefont{J.}~\bibnamefont{Tafto}},
  \bibinfo{author}{\bibfnamefont{M.}~\bibnamefont{Suenaga}}, \bibnamefont{and}
  \bibinfo{author}{\bibfnamefont{D.~O.} \bibnamefont{Welch}},
  \bibinfo{journal}{Phys. Rev. Lett.} \textbf{\bibinfo{volume}{73}},
  \bibinfo{pages}{3026} (\bibinfo{year}{1994}).

\bibitem[{\citenamefont{Kimura et~al.}(2003{\natexlab{b}})\citenamefont{Kimura,
  Kofu, Matsumoto, and Hirota}}]{kimu03b}
\bibinfo{author}{\bibfnamefont{H.}~\bibnamefont{Kimura}},
  \bibinfo{author}{\bibfnamefont{M.}~\bibnamefont{Kofu}},
  \bibinfo{author}{\bibfnamefont{Y.}~\bibnamefont{Matsumoto}},
  \bibnamefont{and} \bibinfo{author}{\bibfnamefont{K.}~\bibnamefont{Hirota}},
  \bibinfo{journal}{Phys. Rev. Lett.} \textbf{\bibinfo{volume}{91}},
  \bibinfo{pages}{067002} (\bibinfo{year}{2003}{\natexlab{b}}).

\bibitem[{\citenamefont{Kofu et~al.}(2005)\citenamefont{Kofu, Kimura, and
  Hirota}}]{kofu05}
\bibinfo{author}{\bibfnamefont{M.}~\bibnamefont{Kofu}},
  \bibinfo{author}{\bibfnamefont{H.}~\bibnamefont{Kimura}}, \bibnamefont{and}
  \bibinfo{author}{\bibfnamefont{K.}~\bibnamefont{Hirota}},
  \bibinfo{journal}{Phys. Rev. B} \textbf{\bibinfo{volume}{72}},
  \bibinfo{pages}{064502} (\bibinfo{year}{2005}).

\bibitem[{\citenamefont{Kimura et~al.}(1999)\citenamefont{Kimura, Hirota,
  Matsushita, Yamada, Endoh, Lee, Majkrzak, Erwin, Shirane, Greven
  et~al.}}]{kimu99}
\bibinfo{author}{\bibfnamefont{H.}~\bibnamefont{Kimura}},
  \bibinfo{author}{\bibfnamefont{K.}~\bibnamefont{Hirota}},
  \bibinfo{author}{\bibfnamefont{H.}~\bibnamefont{Matsushita}},
  \bibinfo{author}{\bibfnamefont{K.}~\bibnamefont{Yamada}},
  \bibinfo{author}{\bibfnamefont{Y.}~\bibnamefont{Endoh}},
  \bibinfo{author}{\bibfnamefont{S.-H.} \bibnamefont{Lee}},
  \bibinfo{author}{\bibfnamefont{C.~F.} \bibnamefont{Majkrzak}},
  \bibinfo{author}{\bibfnamefont{R.}~\bibnamefont{Erwin}},
  \bibinfo{author}{\bibfnamefont{G.}~\bibnamefont{Shirane}},
  \bibinfo{author}{\bibfnamefont{M.}~\bibnamefont{Greven}},
  \bibnamefont{et~al.}, \bibinfo{journal}{Phys. Rev. B}
  \textbf{\bibinfo{volume}{59}}, \bibinfo{pages}{6517} (\bibinfo{year}{1999}).

\bibitem[{\citenamefont{Wakimoto et~al.}(2005)\citenamefont{Wakimoto,
  Birgeneau, Kagedan, Kim, Swainson, Yamada, and Zhang}}]{waki05}
\bibinfo{author}{\bibfnamefont{S.}~\bibnamefont{Wakimoto}},
  \bibinfo{author}{\bibfnamefont{R.~J.} \bibnamefont{Birgeneau}},
  \bibinfo{author}{\bibfnamefont{A.}~\bibnamefont{Kagedan}},
  \bibinfo{author}{\bibfnamefont{H.}~\bibnamefont{Kim}},
  \bibinfo{author}{\bibfnamefont{I.}~\bibnamefont{Swainson}},
  \bibinfo{author}{\bibfnamefont{K.}~\bibnamefont{Yamada}}, \bibnamefont{and}
  \bibinfo{author}{\bibfnamefont{H.}~\bibnamefont{Zhang}},
  \bibinfo{journal}{Phys. Rev. B} \textbf{\bibinfo{volume}{72}},
  \bibinfo{pages}{064521} (\bibinfo{year}{2005}).

\bibitem[{\citenamefont{Sidis et~al.}(1996)\citenamefont{Sidis, Bourges,
  Hennion, Regnault, Villeneuve, Collin, and Marucco}}]{sidi96}
\bibinfo{author}{\bibfnamefont{Y.}~\bibnamefont{Sidis}},
  \bibinfo{author}{\bibfnamefont{P.}~\bibnamefont{Bourges}},
  \bibinfo{author}{\bibfnamefont{B.}~\bibnamefont{Hennion}},
  \bibinfo{author}{\bibfnamefont{L.~P.} \bibnamefont{Regnault}},
  \bibinfo{author}{\bibfnamefont{R.}~\bibnamefont{Villeneuve}},
  \bibinfo{author}{\bibfnamefont{G.}~\bibnamefont{Collin}}, \bibnamefont{and}
  \bibinfo{author}{\bibfnamefont{J.~F.} \bibnamefont{Marucco}},
  \bibinfo{journal}{Phys. Rev. B} \textbf{\bibinfo{volume}{53}},
  \bibinfo{pages}{6811} (\bibinfo{year}{1996}).

\bibitem[{\citenamefont{Kakurai et~al.}(1993)\citenamefont{Kakurai, Shamoto,
  Kiyokura, Sato, Tranquada, and Shirane}}]{kaku93}
\bibinfo{author}{\bibfnamefont{K.}~\bibnamefont{Kakurai}},
  \bibinfo{author}{\bibfnamefont{S.}~\bibnamefont{Shamoto}},
  \bibinfo{author}{\bibfnamefont{T.}~\bibnamefont{Kiyokura}},
  \bibinfo{author}{\bibfnamefont{M.}~\bibnamefont{Sato}},
  \bibinfo{author}{\bibfnamefont{J.~M.} \bibnamefont{Tranquada}},
  \bibnamefont{and} \bibinfo{author}{\bibfnamefont{G.}~\bibnamefont{Shirane}},
  \bibinfo{journal}{Phys. Rev. B} \textbf{\bibinfo{volume}{48}},
  \bibinfo{pages}{3485} (\bibinfo{year}{1993}).

\bibitem[{\citenamefont{Sidis et~al.}(2000)\citenamefont{Sidis, Bourges, Fong,
  Keimer, Regnault, Bossy, Ivanov, Hennion, Gautier-Picard, Collin
  et~al.}}]{sidi00}
\bibinfo{author}{\bibfnamefont{Y.}~\bibnamefont{Sidis}},
  \bibinfo{author}{\bibfnamefont{P.}~\bibnamefont{Bourges}},
  \bibinfo{author}{\bibfnamefont{H.~F.} \bibnamefont{Fong}},
  \bibinfo{author}{\bibfnamefont{B.}~\bibnamefont{Keimer}},
  \bibinfo{author}{\bibfnamefont{L.~P.} \bibnamefont{Regnault}},
  \bibinfo{author}{\bibfnamefont{J.}~\bibnamefont{Bossy}},
  \bibinfo{author}{\bibfnamefont{A.}~\bibnamefont{Ivanov}},
  \bibinfo{author}{\bibfnamefont{B.}~\bibnamefont{Hennion}},
  \bibinfo{author}{\bibfnamefont{P.}~\bibnamefont{Gautier-Picard}},
  \bibinfo{author}{\bibfnamefont{G.}~\bibnamefont{Collin}},
  \bibnamefont{et~al.}, \bibinfo{journal}{Phys. Rev. Lett.}
  \textbf{\bibinfo{volume}{84}}, \bibinfo{pages}{5900} (\bibinfo{year}{2000}).

\bibitem[{\citenamefont{Nachumi et~al.}(1996)\citenamefont{Nachumi, Keren,
  Kojima, Larkin, Luke, Merrin, Tchernysh\"ov, Uemura, Ichikawa, Goto
  et~al.}}]{nach96}
\bibinfo{author}{\bibfnamefont{B.}~\bibnamefont{Nachumi}},
  \bibinfo{author}{\bibfnamefont{A.}~\bibnamefont{Keren}},
  \bibinfo{author}{\bibfnamefont{K.}~\bibnamefont{Kojima}},
  \bibinfo{author}{\bibfnamefont{M.}~\bibnamefont{Larkin}},
  \bibinfo{author}{\bibfnamefont{G.~M.} \bibnamefont{Luke}},
  \bibinfo{author}{\bibfnamefont{J.}~\bibnamefont{Merrin}},
  \bibinfo{author}{\bibfnamefont{O.}~\bibnamefont{Tchernysh\"ov}},
  \bibinfo{author}{\bibfnamefont{Y.~J.} \bibnamefont{Uemura}},
  \bibinfo{author}{\bibfnamefont{N.}~\bibnamefont{Ichikawa}},
  \bibinfo{author}{\bibfnamefont{M.}~\bibnamefont{Goto}}, \bibnamefont{et~al.},
  \bibinfo{journal}{Phys. Rev. Lett.} \textbf{\bibinfo{volume}{77}},
  \bibinfo{pages}{5421} (\bibinfo{year}{1996}).

\bibitem[{\citenamefont{Sasagawa et~al.}(2002)\citenamefont{Sasagawa, Mang,
  Vajk, Kapitulnik, and Greven}}]{sasa02}
\bibinfo{author}{\bibfnamefont{T.}~\bibnamefont{Sasagawa}},
  \bibinfo{author}{\bibfnamefont{P.~K.} \bibnamefont{Mang}},
  \bibinfo{author}{\bibfnamefont{O.~P.} \bibnamefont{Vajk}},
  \bibinfo{author}{\bibfnamefont{A.}~\bibnamefont{Kapitulnik}},
  \bibnamefont{and} \bibinfo{author}{\bibfnamefont{M.}~\bibnamefont{Greven}},
  \bibinfo{journal}{Phys. Rev. B} \textbf{\bibinfo{volume}{66}},
  \bibinfo{pages}{184512} (\bibinfo{year}{2002}).

\bibitem[{\citenamefont{Bao et~al.}(2000)\citenamefont{Bao, McQueeney, Heffner,
  Sarrao, Dai, and Zarestky}}]{bao00}
\bibinfo{author}{\bibfnamefont{W.}~\bibnamefont{Bao}},
  \bibinfo{author}{\bibfnamefont{R.~J.} \bibnamefont{McQueeney}},
  \bibinfo{author}{\bibfnamefont{R.}~\bibnamefont{Heffner}},
  \bibinfo{author}{\bibfnamefont{J.~L.} \bibnamefont{Sarrao}},
  \bibinfo{author}{\bibfnamefont{P.}~\bibnamefont{Dai}}, \bibnamefont{and}
  \bibinfo{author}{\bibfnamefont{J.~L.} \bibnamefont{Zarestky}},
  \bibinfo{journal}{Phys. Rev. Lett.} \textbf{\bibinfo{volume}{84}},
  \bibinfo{pages}{3978} (\bibinfo{year}{2000}).

\bibitem[{\citenamefont{Bao et~al.}(2003)\citenamefont{Bao, Chen, Qiu, and
  Sarrao}}]{bao03}
\bibinfo{author}{\bibfnamefont{W.}~\bibnamefont{Bao}},
  \bibinfo{author}{\bibfnamefont{Y.}~\bibnamefont{Chen}},
  \bibinfo{author}{\bibfnamefont{Y.}~\bibnamefont{Qiu}}, \bibnamefont{and}
  \bibinfo{author}{\bibfnamefont{J.~L.} \bibnamefont{Sarrao}},
  \bibinfo{journal}{Phys. Rev. Lett.} \textbf{\bibinfo{volume}{91}},
  \bibinfo{pages}{127005} (\bibinfo{year}{2003}).

\bibitem[{\citenamefont{{\it et al.}}(1990)}]{luke90}
\bibinfo{author}{\bibfnamefont{G.~M.~L.} \bibnamefont{{\it et al.}}},
  \bibinfo{journal}{Phys. Rev. B} \textbf{\bibinfo{volume}{42}},
  \bibinfo{pages}{7981} (\bibinfo{year}{1990}).

\bibitem[{\citenamefont{Thurston et~al.}(1990)\citenamefont{Thurston, Matsuda,
  Kakurai, Yamada, Endoh, Birgeneau, Gehring, Hidaka, Kastner, Murakami
  et~al.}}]{thur90}
\bibinfo{author}{\bibfnamefont{T.~R.} \bibnamefont{Thurston}},
  \bibinfo{author}{\bibfnamefont{M.}~\bibnamefont{Matsuda}},
  \bibinfo{author}{\bibfnamefont{K.}~\bibnamefont{Kakurai}},
  \bibinfo{author}{\bibfnamefont{K.}~\bibnamefont{Yamada}},
  \bibinfo{author}{\bibfnamefont{Y.}~\bibnamefont{Endoh}},
  \bibinfo{author}{\bibfnamefont{R.~J.} \bibnamefont{Birgeneau}},
  \bibinfo{author}{\bibfnamefont{P.~M.} \bibnamefont{Gehring}},
  \bibinfo{author}{\bibfnamefont{Y.}~\bibnamefont{Hidaka}},
  \bibinfo{author}{\bibfnamefont{M.~A.} \bibnamefont{Kastner}},
  \bibinfo{author}{\bibfnamefont{T.}~\bibnamefont{Murakami}},
  \bibnamefont{et~al.}, \bibinfo{journal}{Phys. Rev. Lett.}
  \textbf{\bibinfo{volume}{65}}, \bibinfo{pages}{263} (\bibinfo{year}{1990}).

\bibitem[{\citenamefont{Zobkalo et~al.}(1991)\citenamefont{Zobkalo, Gukasov,
  Kokovin, Barilo, and Zhigunov}}]{zobk91}
\bibinfo{author}{\bibfnamefont{I.~A.} \bibnamefont{Zobkalo}},
  \bibinfo{author}{\bibfnamefont{A.~G.} \bibnamefont{Gukasov}},
  \bibinfo{author}{\bibfnamefont{S.~Y.} \bibnamefont{Kokovin}},
  \bibinfo{author}{\bibfnamefont{S.~N.} \bibnamefont{Barilo}},
  \bibnamefont{and} \bibinfo{author}{\bibfnamefont{D.~I.}
  \bibnamefont{Zhigunov}}, \bibinfo{journal}{Solid State Commun.}
  \textbf{\bibinfo{volume}{80}}, \bibinfo{pages}{921} (\bibinfo{year}{1991}).

\bibitem[{\citenamefont{Matsuda et~al.}(1992)\citenamefont{Matsuda, Endoh,
  Yamada, Kojima, Tanaka, Birgeneau, Kastner, and Shirane}}]{mats92}
\bibinfo{author}{\bibfnamefont{M.}~\bibnamefont{Matsuda}},
  \bibinfo{author}{\bibfnamefont{Y.}~\bibnamefont{Endoh}},
  \bibinfo{author}{\bibfnamefont{K.}~\bibnamefont{Yamada}},
  \bibinfo{author}{\bibfnamefont{H.}~\bibnamefont{Kojima}},
  \bibinfo{author}{\bibfnamefont{I.}~\bibnamefont{Tanaka}},
  \bibinfo{author}{\bibfnamefont{R.~J.} \bibnamefont{Birgeneau}},
  \bibinfo{author}{\bibfnamefont{M.~A.} \bibnamefont{Kastner}},
  \bibnamefont{and} \bibinfo{author}{\bibfnamefont{G.}~\bibnamefont{Shirane}},
  \bibinfo{journal}{Phys. Rev. B} \textbf{\bibinfo{volume}{45}},
  \bibinfo{pages}{12548} (\bibinfo{year}{1992}).

\bibitem[{\citenamefont{Mang et~al.}(2004{\natexlab{a}})\citenamefont{Mang,
  Vajk, Arvanitaki, Lynn, and Greven}}]{mang04a}
\bibinfo{author}{\bibfnamefont{P.~K.} \bibnamefont{Mang}},
  \bibinfo{author}{\bibfnamefont{O.~P.} \bibnamefont{Vajk}},
  \bibinfo{author}{\bibfnamefont{A.}~\bibnamefont{Arvanitaki}},
  \bibinfo{author}{\bibfnamefont{J.~W.} \bibnamefont{Lynn}}, \bibnamefont{and}
  \bibinfo{author}{\bibfnamefont{M.}~\bibnamefont{Greven}},
  \bibinfo{journal}{Phys. Rev. Lett.} \textbf{\bibinfo{volume}{93}},
  \bibinfo{pages}{027002} (\bibinfo{year}{2004}{\natexlab{a}}).

\bibitem[{\citenamefont{Mang et~al.}(2003)\citenamefont{Mang, Larochelle, and
  Greven}}]{mang03}
\bibinfo{author}{\bibfnamefont{P.~K.} \bibnamefont{Mang}},
  \bibinfo{author}{\bibfnamefont{S.}~\bibnamefont{Larochelle}},
  \bibnamefont{and} \bibinfo{author}{\bibfnamefont{M.}~\bibnamefont{Greven}},
  \bibinfo{journal}{Nature} \textbf{\bibinfo{volume}{426}},
  \bibinfo{pages}{139} (\bibinfo{year}{2003}).

\bibitem[{\citenamefont{Mang et~al.}(2004{\natexlab{b}})\citenamefont{Mang,
  Larochelle, Mehta, Vajk, Erickson, Lu, Buyers, Marshall, Prokes, and
  Greven}}]{mang04b}
\bibinfo{author}{\bibfnamefont{P.~K.} \bibnamefont{Mang}},
  \bibinfo{author}{\bibfnamefont{S.}~\bibnamefont{Larochelle}},
  \bibinfo{author}{\bibfnamefont{A.}~\bibnamefont{Mehta}},
  \bibinfo{author}{\bibfnamefont{O.~P.} \bibnamefont{Vajk}},
  \bibinfo{author}{\bibfnamefont{A.~S.} \bibnamefont{Erickson}},
  \bibinfo{author}{\bibfnamefont{L.}~\bibnamefont{Lu}},
  \bibinfo{author}{\bibfnamefont{W.~L.} \bibnamefont{Buyers}},
  \bibinfo{author}{\bibfnamefont{A.~F.} \bibnamefont{Marshall}},
  \bibinfo{author}{\bibfnamefont{K.}~\bibnamefont{Prokes}}, \bibnamefont{and}
  \bibinfo{author}{\bibfnamefont{M.}~\bibnamefont{Greven}},
  \bibinfo{journal}{Phys. Rev. B} \textbf{\bibinfo{volume}{70}},
  \bibinfo{pages}{094507} (\bibinfo{year}{2004}{\natexlab{b}}).

\bibitem[{\citenamefont{Yamada et~al.}(2003)\citenamefont{Yamada, Kurahashi,
  Uefuji, Fujita, Park, Lee, and Endoh}}]{yama03}
\bibinfo{author}{\bibfnamefont{K.}~\bibnamefont{Yamada}},
  \bibinfo{author}{\bibfnamefont{K.}~\bibnamefont{Kurahashi}},
  \bibinfo{author}{\bibfnamefont{T.}~\bibnamefont{Uefuji}},
  \bibinfo{author}{\bibfnamefont{M.}~\bibnamefont{Fujita}},
  \bibinfo{author}{\bibfnamefont{S.}~\bibnamefont{Park}},
  \bibinfo{author}{\bibfnamefont{S.-H.} \bibnamefont{Lee}}, \bibnamefont{and}
  \bibinfo{author}{\bibfnamefont{Y.}~\bibnamefont{Endoh}},
  \bibinfo{journal}{Phys. Rev. Lett.} \textbf{\bibinfo{volume}{90}},
  \bibinfo{pages}{137004} (\bibinfo{year}{2003}).

\bibitem[{\citenamefont{Henggeler et~al.}(1995)\citenamefont{Henggeler, Cuntze,
  Mesot, Klauda, Saemann-Ischenko, and Furrer}}]{heng95}
\bibinfo{author}{\bibfnamefont{W.}~\bibnamefont{Henggeler}},
  \bibinfo{author}{\bibfnamefont{G.}~\bibnamefont{Cuntze}},
  \bibinfo{author}{\bibfnamefont{J.}~\bibnamefont{Mesot}},
  \bibinfo{author}{\bibfnamefont{M.}~\bibnamefont{Klauda}},
  \bibinfo{author}{\bibfnamefont{G.}~\bibnamefont{Saemann-Ischenko}},
  \bibnamefont{and} \bibinfo{author}{\bibfnamefont{A.}~\bibnamefont{Furrer}},
  \bibinfo{journal}{Europhys. Lett.} \textbf{\bibinfo{volume}{29}},
  \bibinfo{pages}{233} (\bibinfo{year}{1995}).

\bibitem[{\citenamefont{Zamborszky et~al.}(2004)\citenamefont{Zamborszky, Wu,
  Shinagawa, Yu, Balci, Greene, Clark, and Brown}}]{zamb04}
\bibinfo{author}{\bibfnamefont{F.}~\bibnamefont{Zamborszky}},
  \bibinfo{author}{\bibfnamefont{G.}~\bibnamefont{Wu}},
  \bibinfo{author}{\bibfnamefont{J.}~\bibnamefont{Shinagawa}},
  \bibinfo{author}{\bibfnamefont{W.}~\bibnamefont{Yu}},
  \bibinfo{author}{\bibfnamefont{H.}~\bibnamefont{Balci}},
  \bibinfo{author}{\bibfnamefont{R.~L.} \bibnamefont{Greene}},
  \bibinfo{author}{\bibfnamefont{W.~G.} \bibnamefont{Clark}}, \bibnamefont{and}
  \bibinfo{author}{\bibfnamefont{S.~E.} \bibnamefont{Brown}},
  \bibinfo{journal}{Phys. Rev. Lett.} \textbf{\bibinfo{volume}{92}},
  \bibinfo{pages}{047003} (\bibinfo{year}{2004}).

\bibitem[{\citenamefont{Bakharev et~al.}(2004)\citenamefont{Bakharev,
  Abu-Shiekah, Brom, Nugroho, McCulloch, and Zaanen}}]{bakh04}
\bibinfo{author}{\bibfnamefont{O.~N.} \bibnamefont{Bakharev}},
  \bibinfo{author}{\bibfnamefont{I.~M.} \bibnamefont{Abu-Shiekah}},
  \bibinfo{author}{\bibfnamefont{H.~B.} \bibnamefont{Brom}},
  \bibinfo{author}{\bibfnamefont{A.~A.} \bibnamefont{Nugroho}},
  \bibinfo{author}{\bibfnamefont{I.~P.} \bibnamefont{McCulloch}},
  \bibnamefont{and} \bibinfo{author}{\bibfnamefont{J.}~\bibnamefont{Zaanen}},
  \bibinfo{journal}{Phys. Rev. Lett.} \textbf{\bibinfo{volume}{93}},
  \bibinfo{pages}{037002} (\bibinfo{year}{2004}).

\bibitem[{\citenamefont{Matsuda
  et~al.}(2002{\natexlab{b}})\citenamefont{Matsuda, Katano, Uefuji, Fujita, and
  Yamada}}]{mats02b}
\bibinfo{author}{\bibfnamefont{M.}~\bibnamefont{Matsuda}},
  \bibinfo{author}{\bibfnamefont{S.}~\bibnamefont{Katano}},
  \bibinfo{author}{\bibfnamefont{T.}~\bibnamefont{Uefuji}},
  \bibinfo{author}{\bibfnamefont{M.}~\bibnamefont{Fujita}}, \bibnamefont{and}
  \bibinfo{author}{\bibfnamefont{K.}~\bibnamefont{Yamada}},
  \bibinfo{journal}{Phys. Rev. B} \textbf{\bibinfo{volume}{66}},
  \bibinfo{pages}{172509} (\bibinfo{year}{2002}{\natexlab{b}}).

\bibitem[{\citenamefont{Kang et~al.}(2003)\citenamefont{Kang, Dai, Lynn,
  Matsuura, Thompson, Zhang, Argyriou, Onose, and Tokura}}]{kang03}
\bibinfo{author}{\bibfnamefont{H.~J.} \bibnamefont{Kang}},
  \bibinfo{author}{\bibfnamefont{P.}~\bibnamefont{Dai}},
  \bibinfo{author}{\bibfnamefont{J.~W.} \bibnamefont{Lynn}},
  \bibinfo{author}{\bibfnamefont{M.}~\bibnamefont{Matsuura}},
  \bibinfo{author}{\bibfnamefont{J.~R.} \bibnamefont{Thompson}},
  \bibinfo{author}{\bibfnamefont{S.-C.} \bibnamefont{Zhang}},
  \bibinfo{author}{\bibfnamefont{D.~N.} \bibnamefont{Argyriou}},
  \bibinfo{author}{\bibfnamefont{Y.}~\bibnamefont{Onose}}, \bibnamefont{and}
  \bibinfo{author}{\bibfnamefont{Y.}~\bibnamefont{Tokura}},
  \bibinfo{journal}{Nature} \textbf{\bibinfo{volume}{423}},
  \bibinfo{pages}{522} (\bibinfo{year}{2003}).

\bibitem[{\citenamefont{Matsuura et~al.}(2003)\citenamefont{Matsuura, Dai,
  Kang, Lynn, Argyriou, Prokes, Onose, and Tokura}}]{mats03}
\bibinfo{author}{\bibfnamefont{M.}~\bibnamefont{Matsuura}},
  \bibinfo{author}{\bibfnamefont{P.}~\bibnamefont{Dai}},
  \bibinfo{author}{\bibfnamefont{H.~J.} \bibnamefont{Kang}},
  \bibinfo{author}{\bibfnamefont{J.~W.} \bibnamefont{Lynn}},
  \bibinfo{author}{\bibfnamefont{D.~N.} \bibnamefont{Argyriou}},
  \bibinfo{author}{\bibfnamefont{K.}~\bibnamefont{Prokes}},
  \bibinfo{author}{\bibfnamefont{Y.}~\bibnamefont{Onose}}, \bibnamefont{and}
  \bibinfo{author}{\bibfnamefont{Y.}~\bibnamefont{Tokura}},
  \bibinfo{journal}{Phys. Rev. B} \textbf{\bibinfo{volume}{68}},
  \bibinfo{pages}{144503} (\bibinfo{year}{2003}).

\bibitem[{\citenamefont{Matsuura et~al.}(2004)\citenamefont{Matsuura, Dai,
  Kang, Lynn, Argyriou, Onose, and Tokura}}]{mats04}
\bibinfo{author}{\bibfnamefont{M.}~\bibnamefont{Matsuura}},
  \bibinfo{author}{\bibfnamefont{P.}~\bibnamefont{Dai}},
  \bibinfo{author}{\bibfnamefont{H.~J.} \bibnamefont{Kang}},
  \bibinfo{author}{\bibfnamefont{J.~W.} \bibnamefont{Lynn}},
  \bibinfo{author}{\bibfnamefont{D.~N.} \bibnamefont{Argyriou}},
  \bibinfo{author}{\bibfnamefont{Y.}~\bibnamefont{Onose}}, \bibnamefont{and}
  \bibinfo{author}{\bibfnamefont{Y.}~\bibnamefont{Tokura}},
  \bibinfo{journal}{Phys. Rev. B} \textbf{\bibinfo{volume}{69}},
  \bibinfo{pages}{104510} (\bibinfo{year}{2004}).

\bibitem[{\citenamefont{Fujita et~al.}(2005)\citenamefont{Fujita, Matsuda,
  Katano, and Yamada}}]{fuji04b}
\bibinfo{author}{\bibfnamefont{M.}~\bibnamefont{Fujita}},
  \bibinfo{author}{\bibfnamefont{M.}~\bibnamefont{Matsuda}},
  \bibinfo{author}{\bibfnamefont{S.}~\bibnamefont{Katano}}, \bibnamefont{and}
  \bibinfo{author}{\bibfnamefont{K.}~\bibnamefont{Yamada}},
  \bibinfo{journal}{Phys. Rev. Lett.} \textbf{\bibinfo{volume}{93}},
  \bibinfo{pages}{147003} (\bibinfo{year}{2005}).

\bibitem[{\citenamefont{Dai et~al.}(2005)\citenamefont{Dai, Kang, Mook,
  Matsuura, Lynn, Kurita, Komiya, and Ando}}]{dai05}
\bibinfo{author}{\bibfnamefont{P.}~\bibnamefont{Dai}},
  \bibinfo{author}{\bibfnamefont{H.~J.} \bibnamefont{Kang}},
  \bibinfo{author}{\bibfnamefont{H.~A.} \bibnamefont{Mook}},
  \bibinfo{author}{\bibfnamefont{M.}~\bibnamefont{Matsuura}},
  \bibinfo{author}{\bibfnamefont{J.~W.} \bibnamefont{Lynn}},
  \bibinfo{author}{\bibfnamefont{Y.}~\bibnamefont{Kurita}},
  \bibinfo{author}{\bibfnamefont{S.}~\bibnamefont{Komiya}}, \bibnamefont{and}
  \bibinfo{author}{\bibfnamefont{Y.}~\bibnamefont{Ando}},
  \bibinfo{journal}{Phys. Rev. B} \textbf{\bibinfo{volume}{71}},
  \bibinfo{pages}{100502(R)} (\bibinfo{year}{2005}).

\bibitem[{\citenamefont{Kang et~al.}(2005)\citenamefont{Kang, Dai, Mook,
  Argyriou, Sikolenko, Lynn, Kurita, Komiya, and Ando}}]{kang05}
\bibinfo{author}{\bibfnamefont{H.~J.} \bibnamefont{Kang}},
  \bibinfo{author}{\bibfnamefont{P.}~\bibnamefont{Dai}},
  \bibinfo{author}{\bibfnamefont{H.~A.} \bibnamefont{Mook}},
  \bibinfo{author}{\bibfnamefont{D.~N.} \bibnamefont{Argyriou}},
  \bibinfo{author}{\bibfnamefont{V.}~\bibnamefont{Sikolenko}},
  \bibinfo{author}{\bibfnamefont{J.~W.} \bibnamefont{Lynn}},
  \bibinfo{author}{\bibfnamefont{Y.}~\bibnamefont{Kurita}},
  \bibinfo{author}{\bibfnamefont{S.}~\bibnamefont{Komiya}}, \bibnamefont{and}
  \bibinfo{author}{\bibfnamefont{Y.}~\bibnamefont{Ando}},
  \bibinfo{journal}{Phys. Rev. B} \textbf{\bibinfo{volume}{71}},
  \bibinfo{pages}{214512} (\bibinfo{year}{2005}).

\bibitem[{\citenamefont{Kao et~al.}(2000)\citenamefont{Kao, Si, and
  Levin}}]{kao00}
\bibinfo{author}{\bibfnamefont{Y.-J.} \bibnamefont{Kao}},
  \bibinfo{author}{\bibfnamefont{Q.}~\bibnamefont{Si}}, \bibnamefont{and}
  \bibinfo{author}{\bibfnamefont{K.}~\bibnamefont{Levin}},
  \bibinfo{journal}{Phys. Rev. B} \textbf{\bibinfo{volume}{61}},
  \bibinfo{pages}{R11898} (\bibinfo{year}{2000}).

\bibitem[{\citenamefont{Norman}(2000)}]{norm00}
\bibinfo{author}{\bibfnamefont{M.~R.} \bibnamefont{Norman}},
  \bibinfo{journal}{Phys. Rev. B} \textbf{\bibinfo{volume}{61}},
  \bibinfo{pages}{14751} (\bibinfo{year}{2000}).

\bibitem[{\citenamefont{Brinckmann and Lee}(1999)}]{brin99}
\bibinfo{author}{\bibfnamefont{J.}~\bibnamefont{Brinckmann}} \bibnamefont{and}
  \bibinfo{author}{\bibfnamefont{P.~A.} \bibnamefont{Lee}},
  \bibinfo{journal}{Phys. Rev. Lett.} \textbf{\bibinfo{volume}{82}},
  \bibinfo{pages}{2915} (\bibinfo{year}{1999}).

\bibitem[{\citenamefont{Chubukov et~al.}(2001)\citenamefont{Chubukov, Jank\'o,
  and Tchernyshyov}}]{chub01}
\bibinfo{author}{\bibfnamefont{A.~V.} \bibnamefont{Chubukov}},
  \bibinfo{author}{\bibfnamefont{B.}~\bibnamefont{Jank\'o}}, \bibnamefont{and}
  \bibinfo{author}{\bibfnamefont{O.}~\bibnamefont{Tchernyshyov}},
  \bibinfo{journal}{Phys. Rev. B} \textbf{\bibinfo{volume}{63}},
  \bibinfo{pages}{180507R} (\bibinfo{year}{2001}).

\bibitem[{\citenamefont{Eremin and Manske}(2005)}]{erem05a}
\bibinfo{author}{\bibfnamefont{I.}~\bibnamefont{Eremin}} \bibnamefont{and}
  \bibinfo{author}{\bibfnamefont{D.}~\bibnamefont{Manske}},
  \bibinfo{journal}{Phys. Rev. Lett.} \textbf{\bibinfo{volume}{94}},
  \bibinfo{pages}{067006} (\bibinfo{year}{2005}).

\bibitem[{\citenamefont{Eremin et~al.}(2005)\citenamefont{Eremin, Morr,
  Chubukov, Bennemann, and Norman}}]{erem05b}
\bibinfo{author}{\bibfnamefont{I.}~\bibnamefont{Eremin}},
  \bibinfo{author}{\bibfnamefont{D.~K.} \bibnamefont{Morr}},
  \bibinfo{author}{\bibfnamefont{A.~V.} \bibnamefont{Chubukov}},
  \bibinfo{author}{\bibfnamefont{K.~H.} \bibnamefont{Bennemann}},
  \bibnamefont{and} \bibinfo{author}{\bibfnamefont{M.~R.}
  \bibnamefont{Norman}}, \bibinfo{journal}{Phys. Rev. Lett.}
  \textbf{\bibinfo{volume}{94}}, \bibinfo{pages}{147001}
  (\bibinfo{year}{2005}).

\bibitem[{\citenamefont{Anderson et~al.}(2004)\citenamefont{Anderson, Lee,
  Randeria, Rice, Trivedi, and Zhang}}]{ande04}
\bibinfo{author}{\bibfnamefont{P.~W.} \bibnamefont{Anderson}},
  \bibinfo{author}{\bibfnamefont{P.~A.} \bibnamefont{Lee}},
  \bibinfo{author}{\bibfnamefont{M.}~\bibnamefont{Randeria}},
  \bibinfo{author}{\bibfnamefont{T.~M.} \bibnamefont{Rice}},
  \bibinfo{author}{\bibfnamefont{N.}~\bibnamefont{Trivedi}}, \bibnamefont{and}
  \bibinfo{author}{\bibfnamefont{F.~C.} \bibnamefont{Zhang}},
  \bibinfo{journal}{J. Phys. Condens. Matter} \textbf{\bibinfo{volume}{16}},
  \bibinfo{pages}{R755} (\bibinfo{year}{2004}).

\bibitem[{\citenamefont{Lee et~al.}(2006)\citenamefont{Lee, Nagaosa, and
  Wen}}]{lee06}
\bibinfo{author}{\bibfnamefont{P.~A.} \bibnamefont{Lee}},
  \bibinfo{author}{\bibfnamefont{N.}~\bibnamefont{Nagaosa}}, \bibnamefont{and}
  \bibinfo{author}{\bibfnamefont{X.-G.} \bibnamefont{Wen}},
  \bibinfo{journal}{Rev. Mod. Phys.} \textbf{\bibinfo{volume}{78}}
  (\bibinfo{year}{2006}), \eprint{cond-mat/0410445}.

\bibitem[{\citenamefont{Shraiman and Siggia}(1989)}]{shra89}
\bibinfo{author}{\bibfnamefont{B.~I.} \bibnamefont{Shraiman}} \bibnamefont{and}
  \bibinfo{author}{\bibfnamefont{E.~D.} \bibnamefont{Siggia}},
  \bibinfo{journal}{Phys. Rev. Lett.} \textbf{\bibinfo{volume}{62}},
  \bibinfo{pages}{1564} (\bibinfo{year}{1989}).

\bibitem[{\citenamefont{Hasselmann et~al.}(2004)\citenamefont{Hasselmann,
  {Castro Neto}, and {Morais Smith}}}]{hass04}
\bibinfo{author}{\bibfnamefont{N.}~\bibnamefont{Hasselmann}},
  \bibinfo{author}{\bibfnamefont{A.~H.} \bibnamefont{{Castro Neto}}},
  \bibnamefont{and} \bibinfo{author}{\bibfnamefont{C.}~\bibnamefont{{Morais
  Smith}}}, \bibinfo{journal}{Phys. Rev. B} \textbf{\bibinfo{volume}{69}},
  \bibinfo{pages}{014424} (\bibinfo{year}{2004}).

\bibitem[{\citenamefont{Sushkov and Kotov}(2004)}]{sush04}
\bibinfo{author}{\bibfnamefont{O.~P.} \bibnamefont{Sushkov}} \bibnamefont{and}
  \bibinfo{author}{\bibfnamefont{V.~N.} \bibnamefont{Kotov}},
  \bibinfo{journal}{Phys. Rev. B} \textbf{\bibinfo{volume}{70}},
  \bibinfo{pages}{024503} (\bibinfo{year}{2004}).

\bibitem[{\citenamefont{Lindg{\aa}rd}(2005)}]{lind05}
\bibinfo{author}{\bibfnamefont{P.-A.} \bibnamefont{Lindg{\aa}rd}},
  \bibinfo{journal}{Phys. Rev. Lett.} \textbf{\bibinfo{volume}{95}},
  \bibinfo{pages}{217001} (\bibinfo{year}{2005}).

\bibitem[{\citenamefont{Abbamonte et~al.}(2005)\citenamefont{Abbamonte, Rusydi,
  Smadici, Gu, Sawatzky, and Feng}}]{abba05}
\bibinfo{author}{\bibfnamefont{P.}~\bibnamefont{Abbamonte}},
  \bibinfo{author}{\bibfnamefont{A.}~\bibnamefont{Rusydi}},
  \bibinfo{author}{\bibfnamefont{S.}~\bibnamefont{Smadici}},
  \bibinfo{author}{\bibfnamefont{G.~D.} \bibnamefont{Gu}},
  \bibinfo{author}{\bibfnamefont{G.~A.} \bibnamefont{Sawatzky}},
  \bibnamefont{and} \bibinfo{author}{\bibfnamefont{D.~L.} \bibnamefont{Feng}},
  \bibinfo{journal}{Nature Physics} \textbf{\bibinfo{volume}{1}},
  \bibinfo{pages}{155} (\bibinfo{year}{2005}).

\bibitem[{\citenamefont{Zaanen et~al.}(2001)\citenamefont{Zaanen, Osman, Kruis,
  Nussinov, and {Tworzyd\l o}}}]{zaan01}
\bibinfo{author}{\bibfnamefont{J.}~\bibnamefont{Zaanen}},
  \bibinfo{author}{\bibfnamefont{O.~Y.} \bibnamefont{Osman}},
  \bibinfo{author}{\bibfnamefont{H.~V.} \bibnamefont{Kruis}},
  \bibinfo{author}{\bibfnamefont{Z.}~\bibnamefont{Nussinov}}, \bibnamefont{and}
  \bibinfo{author}{\bibfnamefont{J.}~\bibnamefont{{Tworzyd\l o}}},
  \bibinfo{journal}{Phil. Mag. B} \textbf{\bibinfo{volume}{81}},
  \bibinfo{pages}{1485} (\bibinfo{year}{2001}).

\bibitem[{\citenamefont{Sachdev and Read}(1991)}]{sach91}
\bibinfo{author}{\bibfnamefont{S.}~\bibnamefont{Sachdev}} \bibnamefont{and}
  \bibinfo{author}{\bibfnamefont{N.}~\bibnamefont{Read}},
  \bibinfo{journal}{Int. J. Mod. Phys. B} \textbf{\bibinfo{volume}{5}},
  \bibinfo{pages}{219} (\bibinfo{year}{1991}).

\bibitem[{\citenamefont{Machida}(1989)}]{mach89}
\bibinfo{author}{\bibfnamefont{K.}~\bibnamefont{Machida}},
  \bibinfo{journal}{Physica C} \textbf{\bibinfo{volume}{158}},
  \bibinfo{pages}{192} (\bibinfo{year}{1989}).

\bibitem[{\citenamefont{Kivelson and Fradkin}(2005)}]{kive05}
\bibinfo{author}{\bibfnamefont{S.~A.} \bibnamefont{Kivelson}} \bibnamefont{and}
  \bibinfo{author}{\bibfnamefont{E.}~\bibnamefont{Fradkin}}
  (\bibinfo{year}{2005}), \eprint{cond-mat/0507459}.

\bibitem[{\citenamefont{Kivelson and Emery}(1996)}]{kive96}
\bibinfo{author}{\bibfnamefont{S.~A.} \bibnamefont{Kivelson}} \bibnamefont{and}
  \bibinfo{author}{\bibfnamefont{V.~J.} \bibnamefont{Emery}},
  \bibinfo{journal}{Synth. Met.} \textbf{\bibinfo{volume}{80}},
  \bibinfo{pages}{151} (\bibinfo{year}{1996}).

\bibitem[{\citenamefont{Kruis et~al.}(2004)\citenamefont{Kruis, McCulloch,
  Nussinov, and Zaanen}}]{krui04}
\bibinfo{author}{\bibfnamefont{H.~V.} \bibnamefont{Kruis}},
  \bibinfo{author}{\bibfnamefont{I.~P.} \bibnamefont{McCulloch}},
  \bibinfo{author}{\bibfnamefont{Z.}~\bibnamefont{Nussinov}}, \bibnamefont{and}
  \bibinfo{author}{\bibfnamefont{J.}~\bibnamefont{Zaanen}},
  \bibinfo{journal}{Phys. Rev. B} \textbf{\bibinfo{volume}{70}},
  \bibinfo{pages}{075109} (\bibinfo{year}{2004}).

\end{thebibliography}

\end{document}